\documentclass[prb,aps,reprint,superscriptaddress,longbibliography, 10pt]{revtex4-1}
\usepackage[pretty,uselistings]{revquantum}
\usepackage{graphicx}
\usepackage{textcomp}
\usepackage{amsmath}
\usepackage{dcolumn}
\usepackage{graphicx}
\usepackage{bm}
\usepackage{graphicx}
\usepackage{dcolumn}
\usepackage{threeparttable}
\usepackage{gensymb}
\usepackage[english]{babel}
\usepackage{inputenc}  
\usepackage{bm} 
\usepackage{amsmath}  
\usepackage{amsfonts} 
\usepackage{siunitx}
\usepackage{mathtools}
\DeclarePairedDelimiter\abs{\lvert}{\rvert}
\begin{document}
\title{\LARGE Nanoantenna Enhanced Terahertz Interaction of Biomolecules}
\author{Subham Adak}
\affiliation{Department of Physics, Birla Institute of Technology, Mesra, Ranchi-835215, Jharkhand, India.}
\author{Laxmi Narayan Tripathi}
\email{nara.laxmi@gmail.com}
\affiliation{Department of Physics, Birla Institute of Technology, Mesra, Ranchi-835215, Jharkhand, India.}

\date{\today}

\begin{abstract}
Terahertz time-domain spectroscopy (THz-TDS) is a non-invasive, non-contact and label-free technique for biological and chemical sensing as THz-spectra is less energetic and lies in the characteristic vibration frequency regime of proteins and DNA molecules. However, THz-TDS is less sensitive for detection of micro-organisms of size equal to or less than $ \lambda/100 $ (where, $ \lambda $  is wavelength of incident THz wave) and, molecules in extremely low concentrated solutions (like, a few femtomolar). After successful high-throughput fabrication of nanostructures, nanoantennas and metamaterials were found to be indispensable in enhancing the sensitivity of conventional THz-TDS. These nanostructures lead to strong THz field enhancement which when in resonance with absorption spectrum of absorptive molecules, causing significant changes in the magnitude of the transmission spectrum, therefore, enhancing the sensitivity and allowing detection of molecules and biomaterials in extremely low concentrated solutions. Hereby, we review the recent developments in ultra-sensitive and selective nanogap biosensors. We have also provided an in-depth review of various high-throughput nanofabrication techniques. We also discussed the physics behind the field enhancements in sub-skin depth as well as sub-nanometer sized nanogaps. We introduce finite-difference time-domain (FDTD) and molecular dynamics (MD) simulations tools to study THz biomolecular interactions. Finally, we provide a comprehensive account of nanoantenna enhanced sensing of viruses (like, H1N1) and biomolecules such as artificial sweeteners which are addictive and carcinogenic.
\end{abstract}
\maketitle
\tableofcontents
\section{Introduction}
\begin{figure*}
	\centering
	\includegraphics[scale=0.48]{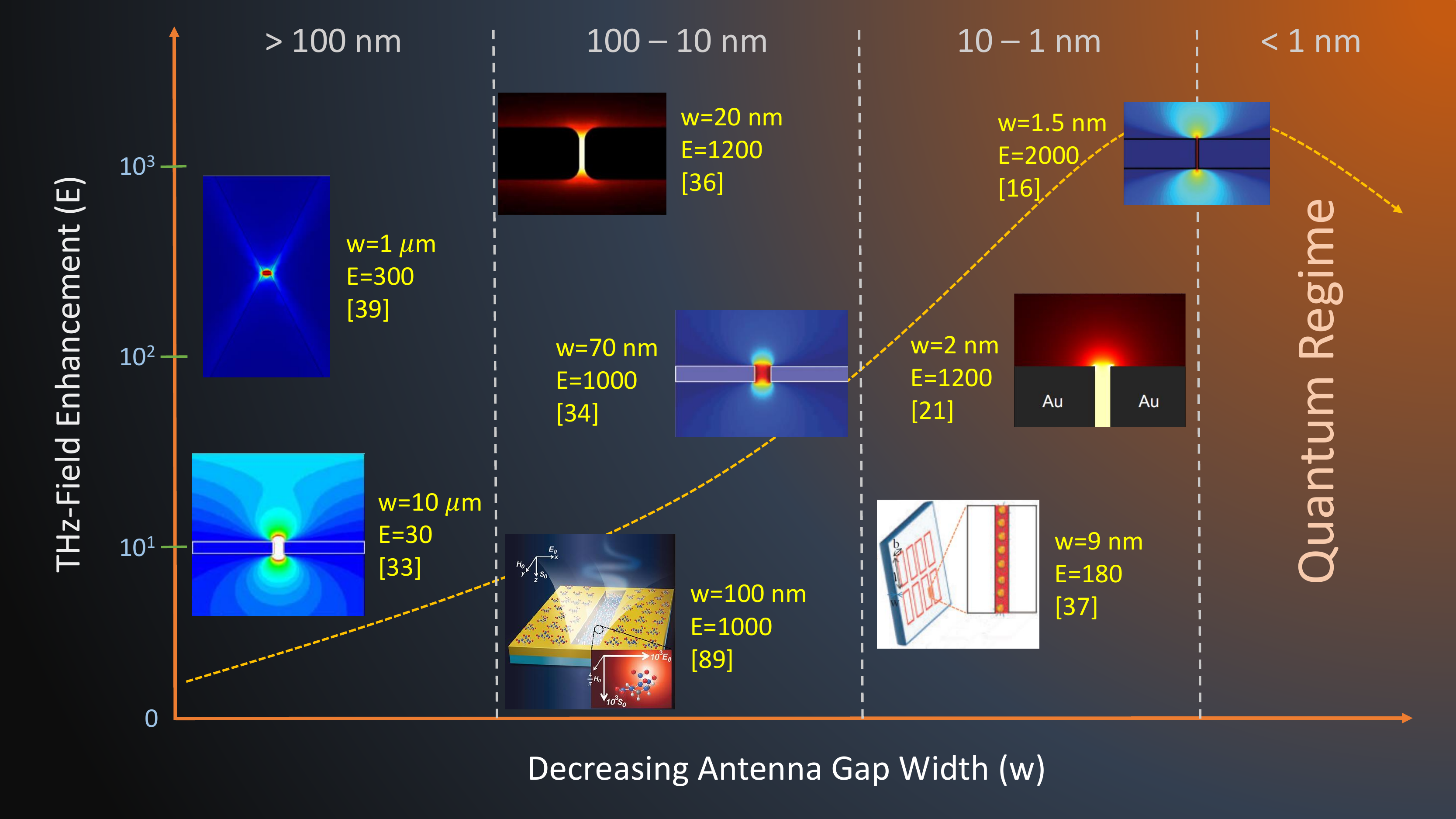}
	\caption{THz-field enhancement profile with decreasing gap width.{\label{fig1}}}
\end{figure*}
Optical antennas,\cite{Bharadwaj2009,Muhlschlegel2005} devices converting free optical radiations into localized ones, and vice versa are important components in modern nanophotonic devices. In the recent decade, metallic nanoantenna has attracted much consideration with the significant advancement of nanofabrication technologies, which provide the feasibility of fabrication at nanometre resolution, enabling fascinating nano-optoelectronic devices. Since metals exhibit finite conductivities and plasmonic effect at optical frequencies, therefore the optical field response is dictated by the resonance oscillation of electron clouds (surface plasmons, or SP) in the metal, coupling with the incident photon and sustaining surface plasmon polaritons (SSP’s).\cite{Alda2005,Ross2009} Henceforth, at optical frequencies a nanoantenna, depending upon the material properties, resonates for considerably shorter effective wavelength, rather than half-wavelength of free-space.\cite{Alu2008,Novotny2007}

Electromagnetic interaction of nanoantennas has lead to huge local field enhancement and confinement in nano-size volumes, enabling broadband application in nano-optics.\cite{Kneipp1997,Nie1997,Xu1999,Juan2011,Marago2013,Ward2010,Kauranen2012,Bahk2014,Lassiter2014} Terahertz (THz) wave funnelling through nanogaps/nanoantennas has resulted huge field enhancement.\cite{Bahk2017,Bahk2018,Bonod2008,Cao2004,Chen2006,Chen2014,Jeong2014,Kang2018,Kim2015,Kim2017,Lee2007,Moser2005,Novitsky2012,Novitsky2011,Park2010,Qu2004a,Rivas2003,Seo2008,Seo2009,Shalaby2011,Toma2014,Tripathi2016,Werley2012,Park2018,Park2011,Park2010} Figure \ref{fig1} summarizes the various investigations performed to study the THz field enhancement with different gap sizes. Such investigations have lead to the development of the terahertz technology, which are now exploited in electronics,\cite{Lindquist2012,Hoffmann2010,Tanoto2012,Ju2011,Shur2010,Upadhyaya2008} photonics,\cite{Bahramipanah2014,Mittleman2013,Williams2008} medical sciences,\cite{Fitzgerald2002,Parrott2011,Handley2002} military,\cite{Tu2016,Li2015,Semashkin2015} security\cite{Zimdars2006,Han2016,Han2016} and conservation of cultural heritages.\cite{Fukunaga2010,Manceau2008} These studies also lead to the improvisation of terahertz devices.\cite{Lee2006a,Libon2000,Mendis2010,Kaliteevski2008,Lan2014,Heshmat2012,Rahm2012,Choi2011,Unlu2014,Li2007,Gao2014,Liu2013,Astley2012,Alves2012,Hassani2008,Miyamaru2006,Xu2016}

Terahertz time-domain spectroscopy (THz-TDS)\cite{Lee2008,Baxter2011,Xi-ChengZhang2009,Dexheimer2007,Nagel2006,Kar2018} is a non-invasive and one of the most widely used technique in biosensing\cite{Fischer2002,Dexheimer2007,Markelz2002,Arora2012,Yang2016,Yang2016a} because the THz-spectra are less energetic (about a few milli-electronvolts) lies in the characteristic vibration frequency regime of proteins and DNA molecules. Terahertz nanoantennas provide us label-free and ultra-sensitive sensing of molecules\cite{Lee2015,Park2013} and biomolecules\cite{Nagel2006,Limaj2016,Rodrigo2015,Tripathi2016,Lee2017,Park2014,Park2017} as the nanoantennas focus the THz waves into an extremely confined volume where we can place our analytes at very low concentrations. Significant changes in the magnitude of the transmission spectrum were observed in the absorptive molecules when the resonance mode of the waveguide matches its absorption spectrum,\cite{Toma2014,Fischer2002} enhancing the sensitivity of THz-resonators based sensing devices, therefore making it feasible for detection of biomolecules at extremely low concentration.

In this review work, we describe the high-throughput fabrication techniques of THz nanoantenna via. photolithography, atomic layer lithography, pattern and peel method, self-assembly lithography etc. Then we discuss the THz field enhancement using metal nanogap antenna where we discuss the field enhancement beyond the skin depth of the metal up to the quantum regime. Then we briefly discuss finite-difference time-domain (FDTD) methods for calculation of field enhancement across the nanogap. At the last, we describe the THz-TDS for sensing the biomolecules due to the field enhancement across the nanogap. We also showed how a molecular dynamics (MD) simulation can be a versatile tool to estimate the terahertz absorption and vibrational density of states (VDOS). Then we have taken real examples from various literature that how THz nanoantenna and metamaterials can be used to enhance the terahertz biomolecular sensing.
\begin{figure}
	\centering
	\includegraphics[scale=0.32]{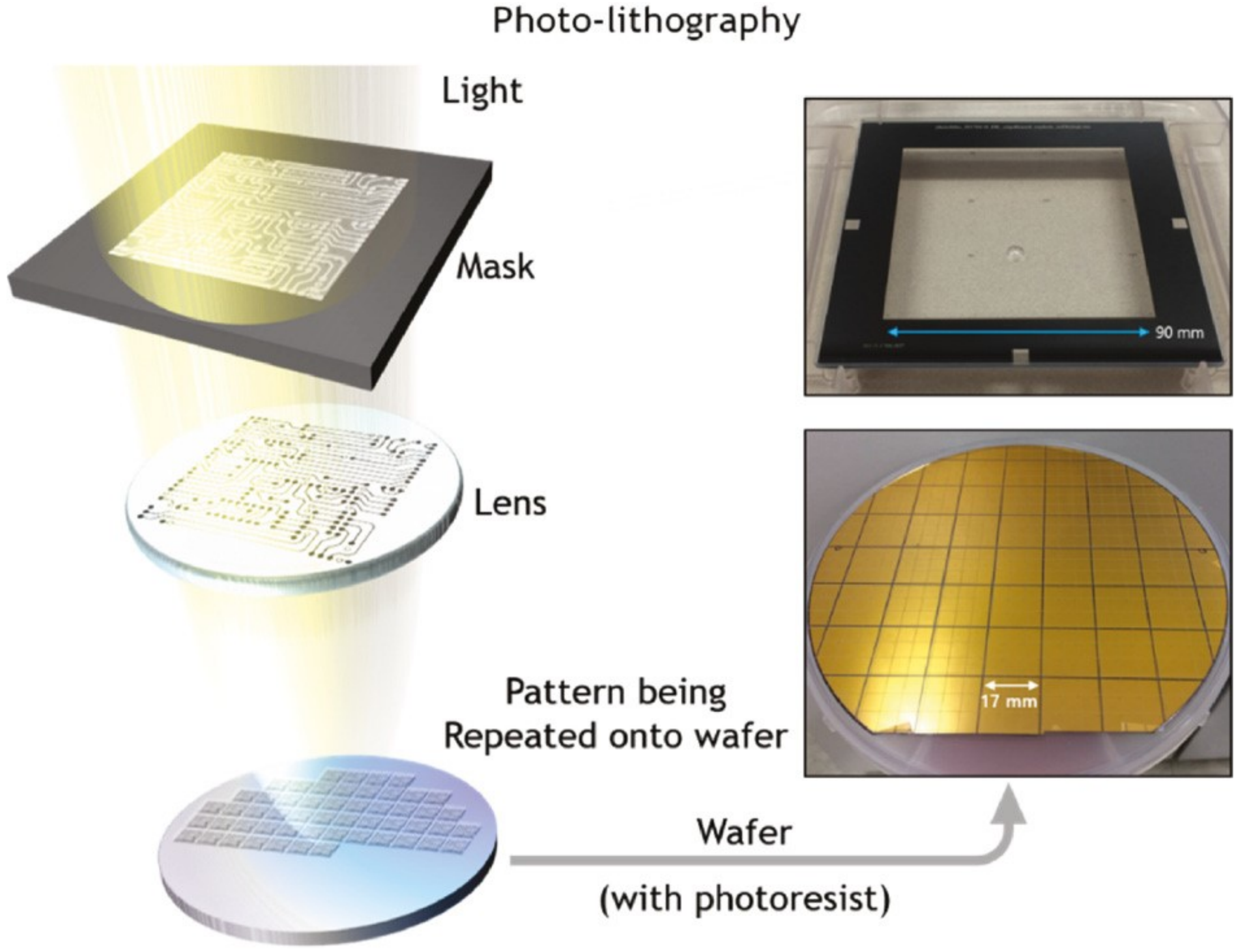}
	\caption{Photolithography : A high resolution and high throughput technique for nanogap structures. Reprinted with permission from Ref.~\cite{Kang2018}. \textcopyright~2018, De Gruyter.{\label{fig2}}}
\end{figure}
\begin{figure*}
	\includegraphics[scale=0.7]{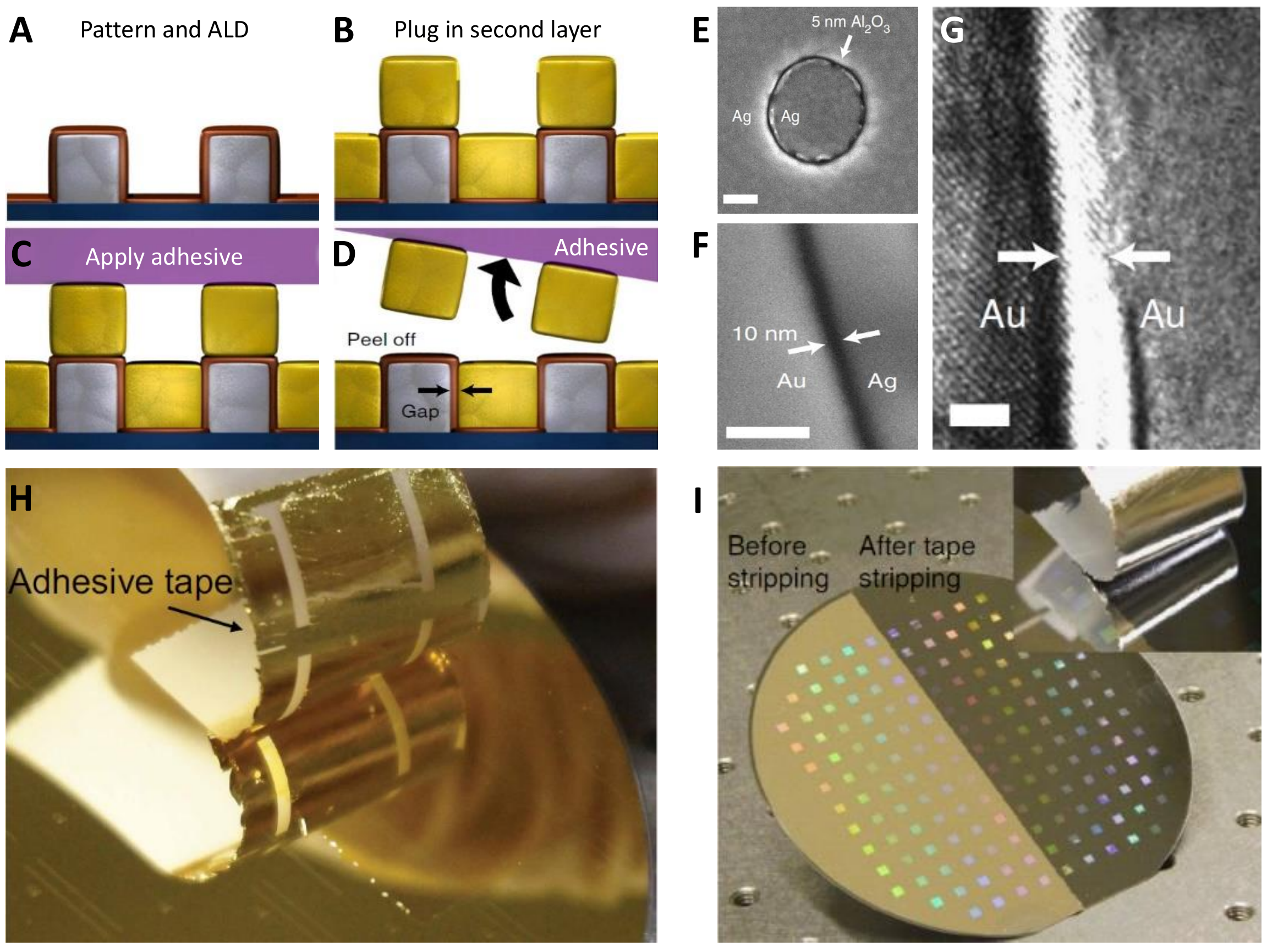}
	\caption{{\label{fig3}}Atomic layer lithography. \textbf{(A)} Thin Al$_{2}$O$_{3}$ layer coated on a patterened substrate using ALD. \textbf{(B)} Metal evaporation plugging second metal layer. \textbf{(C,D)} Removing excess metal by using an adhesive. Scanning electron micrograph (SEM) of \textbf{(E)} 5 nm gap in Ag-film (F) 10 nm gap between Au and Ag, and \textbf{(G)} 9.9 \AA~Au - Al$_{2}$O$_{3}$ -Au vertical nanogap. Reprinted with permission from Ref.~\cite{Chen2013}. \textcopyright~2013, Nature Communications. \textbf{(H)} Adhesive tape planarization of nanogap. Adapted with permission from Ref.~\cite{Chen2014}. \textcopyright~2014, Springer Nature. \textbf{(I)} Planarization of half Si-wafer by using an adhesive. Reprinted with permission from Ref.~\cite{Chen2013}. \textcopyright~2013, Nature Communications.}
\end{figure*}

\section{High-Throughput Wafer-scale Fabrication Techniques}
A nanogap must have a reliable THz-field interaction and plasmonic characteristics (like, field enhancement and non-linearity). But fabricating high throughput wafer-scale nanogaps is one of the most challenging part of the job. To achieve sufficiently intense far-field signals, nanogap arrays should be fabricated over large sample area. Several methods like: fs-laser beam machining,\cite{Bahk2018,Lee2006,Lee2006b,Ward2006} focussed ion beam,\cite{Bahk2018,Enkrich2005,Ocelic2004,Ghaemi1998,Ebbesen1998,Park2011a,Park2011,Bell2009} photolithography,\cite{Bahk2018,Chen2008,Singh2009,Choi2011a,Willson1997} electron-beam (e-beam) lithography,\cite{Bahk2018,Lin2004,Park2015,Fursina2008,Vieu2000,Broers1996,Chen2015,Tseng2003} nanosphere lithography,\cite{Bahk2018,Hulteen1995} sidewall lithography,\cite{Vettiger1989} nanoimprint,\cite{Bahk2018,Chou1996,Chou1996a,Austin2004,Guo2007,Guo1997} pattern and peel method\cite{Ambhire2018} and atomic layer lithography\cite{Bahk2018,Feuillet-Palma2013,Suwal2017,Lindquist2012,Park2016,Chen2013} have been developed for high resolution patterning over large area of sample.

Photolithography\cite{Bahk2018,Kang2018} (Fig.~\ref{fig2}) is one of the most high-throughput methods for fabricating nanogap structures. In photolithography, a desired polymer pattern is patterned using ultraviolet (UV) light and a photomask. This polymer pattering is formed by the physical change of a photo-sensitive material, known as photoresist (PR). Polymers added to PR decides the solubility under UV exposure. A PR which consist of an insoluble polymer which on UV exposure, changes into a soluble polymer, is called a positive PR and a PR which consist of a soluble polymer which on UV exposure, polymerizes and becomes insoluble, is called a negative PR. Therefore by using the desired PR, wafer-scale metal structures are fabricated after performing a metal deposition process on patterned resists and lift-off of the unwanted metal film.
\begin{figure}
	\includegraphics[scale=0.35]{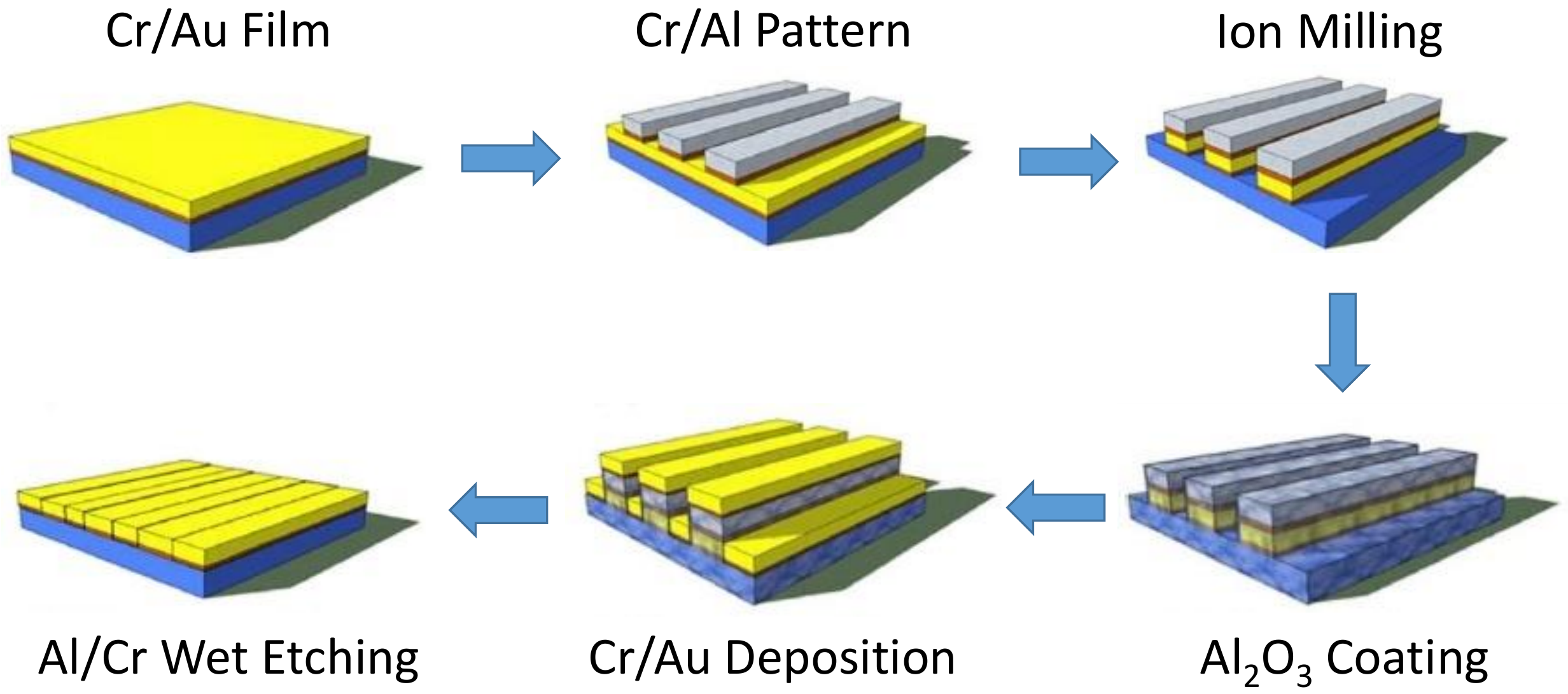}
	\caption{{\label{fig4}}High-throughput fabrication technique of nanometer-scale gaps introduced by Jeyong et al. Adapted with permission from Ref.~\cite{Jeong2014}. \textcopyright~2014, Springer Nature.}
\end{figure}

The lithography method discussed earlier are limited to fabrication of nanometer-gaps up to a few nanometers. To study the nano- and sub-nanoscale optics, we need better lithography techniques to fabricate sub-10 nm wide gap. Many researchers have reported the significance of deposition of an atomically thin insulating layer or sacrificial layer in fabricating small gaps between two metal structures.\cite{Im2010,Miyazaki2006,Zhu2011,Fan2005,Ji2017,Ding2015} Chen et al.\cite{Chen2013} developed a pioneering method for fabricating vertical nanogaps known as atomic layer lithography (Fig.~\ref{fig3} (A - D)). Figures \ref{fig3} (E - G) are the scanning electron micrographs (SEM) of 5 nm gap in Ag film, 10 nm gap between Au and Ag, and 9.9 \AA~Au - Al$_{2}$O$_{3}$ - Au vertical nanogap, respectively, which are fabricated using atomic layer lithography technique. This method also includes a new planarization scheme which eliminates the background light transmission, enabling background-free transmission measurements. This lithography technique is the combination of atomic layer deposition (ALD) and \textquoteleft plug-and-peel\textquoteright~adhesive tape planarization and is a three-step process. The first step includes the pre-patterning of metal using conventional lithography techniques, the second step includes the sidewall formation of materials like Al$_{2}$O$_{3}$ or SiO$_{2}$ by ALD, and the third step includes the \textquoteleft plug-and-peel\textquoteright~taping process for planarization of metal plugs. Therefore, atomic layer lithography is a high-resolution patterning method with large scale uniformity.
\begin{figure*}
	\centering
	\includegraphics[scale=0.6]{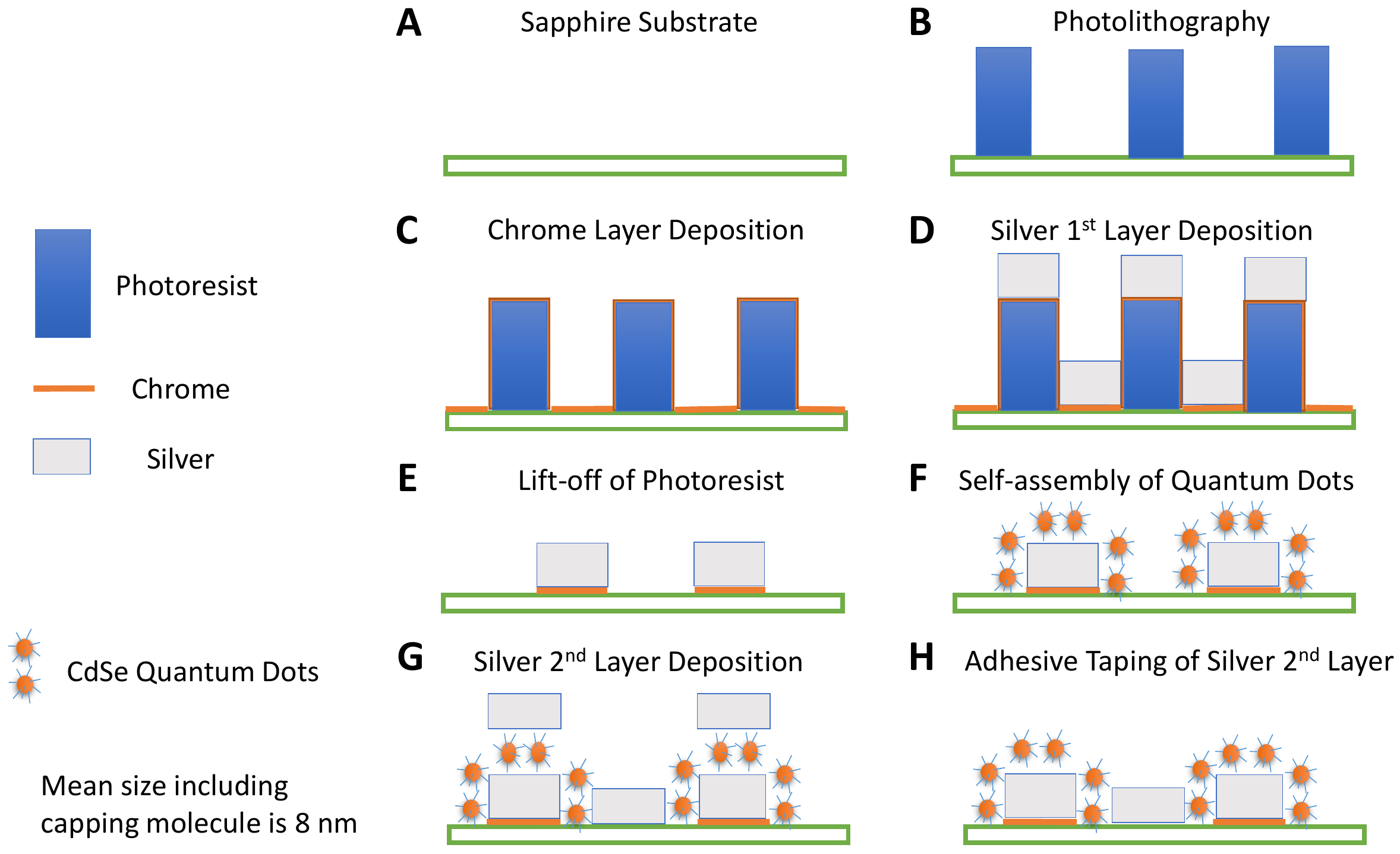}
	\caption{Self-assembly lithography technique. \textbf{(A)} Clean and dried sapphire substrate. \textbf{(B)} PR patterned using photolithography. \textbf{(C)} 10 m chrome layer deposition over PR pattern. \textbf{(D)} Evaporation of silver film of thickness 200 nm over the chrome layer. \textbf{(E)} Sonication and lift-off of PR after first silver layer deposition. \textbf{(F)} Self-assembly of QD monolayer over the first layer of metal. \textbf{(G)} Second silver layer deposition after self-assembly of QD monolayer. \textbf{(H)} QDs filled nanogap metamaterial after taping-off second silver layer. Adapted with permission from Ref.~\cite{Tripathi2015}. \textcopyright~2015, Optical Society of America.{\label{fig5}}}
\end{figure*}

Jeong et al.\cite{Jeong2014} improvised the atomic layer lithography technique which is demonstrated in Fig.~\ref{fig4}. In this lithography technique, conventional photolithography and liftoff process were used to pattern 30 nm Cr/150 nm Al layer on the 3 nm Cr/100 nm gold-sapphire substrate. The Cr-Al double layer acts as a sacrificial layer which will be etched out later, removing excess metal and making the structure planar. As compared to Al, Au is less-resistant to ion beam therefore, excluding the Au film beneath Al-Cr layer, the exposed Au layer is removed using ion milling process. Uniform ALD of Al$_{2}$O$_{3}$ is performed over the whole structure with a thickness of nanometer accuracy. After that, the second layer of Au with adhesive Cr is deposited, filling the trenches. Finally, Al/Cr wet etching is performed which removes the overhanging Au and Al layers, exposing the dielectric gaps. This lithography technique can be used to fabricate large-scale nanogaps with ultra-high aspect ratio.

Tripathi et al.\cite{Tripathi2015} developed self-assembled lithography process by which they fabricated quantum dots (QDs) nanogap metamaterials (Fig.~\ref{fig5}). In this lithography process, photolithography was used to pattern PR on a clean and dried sapphire substrate. With the help of a thermal evaporator (evaporation rate = 1 \AA/sec), a chrome layer of thickness 10 nm was deposited over the PR pattern. Following the process, a silver film of thickness 200 nm was evaporated over the chrome layer. After the evaporation of first silver layer, the whole sample was then dipped in acetone sonicated for 2-minutes at 150 W and 36 kHz, lifting off the PR pattern. After sonication, the sample was washed with isopropyl alcohol and dried using N$_{2}$ gas. The dried sample was then dipped into a toluene solution of OT (1-8 octanedithiol) functionalized QDs, resulting in the formation of a self-assembled monolayer of QDs. The resulting sample was again dried using N$_{2}$ gas and a second silver layer of same/higher thickness was deposited over the self-assembled monolayer. Then by using a scotch tape\cite{Chen2013}, the second layer of silver was taped-off, leaving a vertical metal nanogap filled with a monolayer of QDs.

\section{Terahertz-Field Enhancement using Nanoantennas}
In THz frequency, a metal film is not a perfect conductor as well as infinitely flat. Equivalence of THz-electric field enhancement in metal nanogap\cite{Seo2009,Bonod2008,Kang2009,Novitsky2012,Choe2012,Garcia-Vidal2006,Shalaby2011} and magnetic field enhancement around metallic nanoantennas\cite{Park2012,Yang2011,Maksymov2012,Taminiau2008} can be shown by using Babinet\textquotesingle s principle.\cite{Pelosi2017,Booker1946,Mushiake2004,Born1981,Koo2009} This principle is still used implicitly in different areas of THz research.\cite{Hoffmann2010,Razzari2012,Toma2014,Jeong2014,Park2014,Jeong2013,Shu2011,Park2011a,Merbold2011,Dai2009,Werley2012,Oeguet2011,Yang2010,Hu2013,Iwaszczuk2012,Bulgarevich2012,Gadalla2014,Fan2016,Low2014,Fan2013,Panaretos2016,Yu2010,Ishikawa2009,Chern2008,Martin-Cano2011,Lee2013,Choi2011a,Choi2011a,Moser2005,Wang2015,Shi2014,Takano2010} Sommerfeld\textquotesingle s half-plane problem\cite{Sommerfeld2004,Sheppard2013,Marathay2004} is also used to study the light-aperture interaction theoretically. Bethe\cite{Bethe1944} and Bouwkamp\cite{Bouwkamp1954,Bouwkamp1954a} reconsidered the Sommerfeld\textquotesingle s half-plane problem for small apertures in infinitely thin perfect conducting plates. Further studies were performed on the light interaction with system of periodic apertures.\cite{Ulrich1967} Later, extensive studies were performed on enhanced light transmission through the apertures in various spectral regime.\cite{Cao2004,Ebbesen1998,Kim1999,Kim2003,Minhas2002,Rivas2003,Naweed2003,Qu2004,Qu2004a} Theoretical approaches like: coupling-mode theory,\cite{Choe2012,Kang2009a,Gordon2005,Bravo-Abad2007,Takakura2001,Delgado2011,Galindo1966,Sheng1982,Garcia-Vidal2010} microscopic models,\cite{Novitsky2012,Liu2008,Novitsky2011} transfer-matrix method\cite{He2009,Bell1995} and capacitor model,\cite{Kang2009} were combined with numerical methods\cite{Xu2016,Li1997,Salomon2001,Salomon2001} to solve the enhanced light transmission and huge field enhancement around metallic apertures.
\begin{figure}
	\centering
	\includegraphics[scale=0.51]{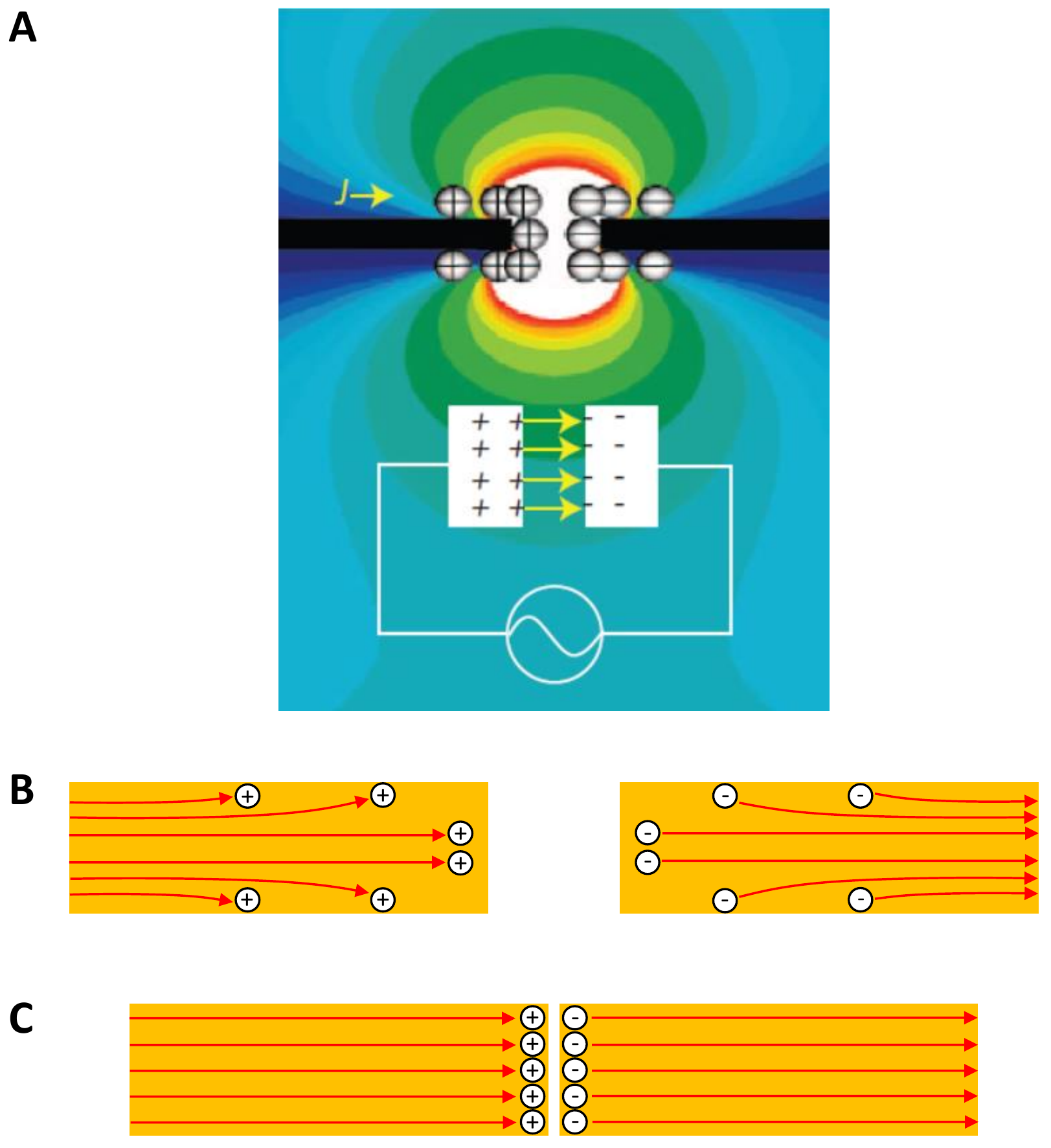}
	\caption{Capacitor-model representation of a nanogap. \textbf{(A)} Nanogap gets charged just like a line-capacitor due to light-induced alternating current \textit{J}. Hence, the electric field enhances which is shown by gradual colour contour. Reprinted with permission from Ref.~\cite{Seo2009}. \textcopyright~2009, Springer Nature. Schematics of charge accumulation near metal edges and corresponding current density induced inside metal film of two gap-sizes \textbf{(B)} $ w\sim h (\ll\lambda) $ and \textbf{(C)} $ w\ll h (\ll\lambda) $. Adapted with permission from Ref.~\cite{Bahk2017}. \textcopyright~2017, American Physical Society.{\label{fig6}}}
\end{figure}

When an electromagnetic wave is incident normally on a perfectly conducting metallic plane, an induced current is developed on the surface, which reflects light back, with no charge accumulation anywhere.\cite{Seo2009} When this plane is divided into two perfectly conducting Sommerfeld half planes, macroscopic accumulation of charges take place at the edges with a length scale of one wavelength, such that the surface charge density ($ \sigma $) depends as a time function for small values of x $ \ll $ $ \lambda $,\cite{Seo2009} given by
\begin{equation}
	\sigma(x,t)=\frac{\epsilon_{0}}{\sqrt{2}\pi}\sqrt{\frac{\lambda}{x}}E_{0}e^{-i(\omega t+\pi/4)}
\end{equation}
Here, $ \epsilon_{0} $, $ E_{0} $, $ \omega $ and x describes the permittivity of vacuum, the incident electric field, the angular frequency and distance from the edge, respectively.\cite{Born1981} The charge singularity, at x = 0, for this half plane is very feeble and disappears with integration. When the two metallic half planes are brought close together, an electrostatic force will be experienced by the charges due to their counter members across the gap, pulling the charges towards the edge, and developing a strong electric field across the gap. As the gap shrinks, more charges gets accumulated at the edges as the light-induced currents becomes more and more stronger. This increases the surface charge density at the edges as shown in Fig.\ref{fig6} B - C. This system can be portrayed as a line-capacitor driven by the light-induced alternating current as shown in Fig.~\ref{fig6} (A).

Numerous studies have reported the monotonic enhancement of the electric field with decreasing gap size.\cite{Park2014,Seo2009,Koo2009,Chen2013} Further studies have also reported the decrease in field enhancement when the gap size decreases and enters a quantum regime.\cite{Zuloaga2009,Esteban2012,Marinica2012,Ciraci2012,Tame2013,Tan2014,Tan2014,Kim2015} Increase in field enhancement beyond sub-skin depth is also reported.\cite{Seo2009,Bahk2017,Kang2018} In this section of the review, we discuss the THz field enhancement for gaps beyond skin-depth and also in the quantum regime. We also discuss the classical limit of field enhancement of single nanoslit before entering quantum regime, and about finite-difference time-domain algorithm, which is a computational approach developed for the quantitative estimation of far- and near-field enhancement measurements.
\begin{figure}
	\centering
	\includegraphics[scale=0.5]{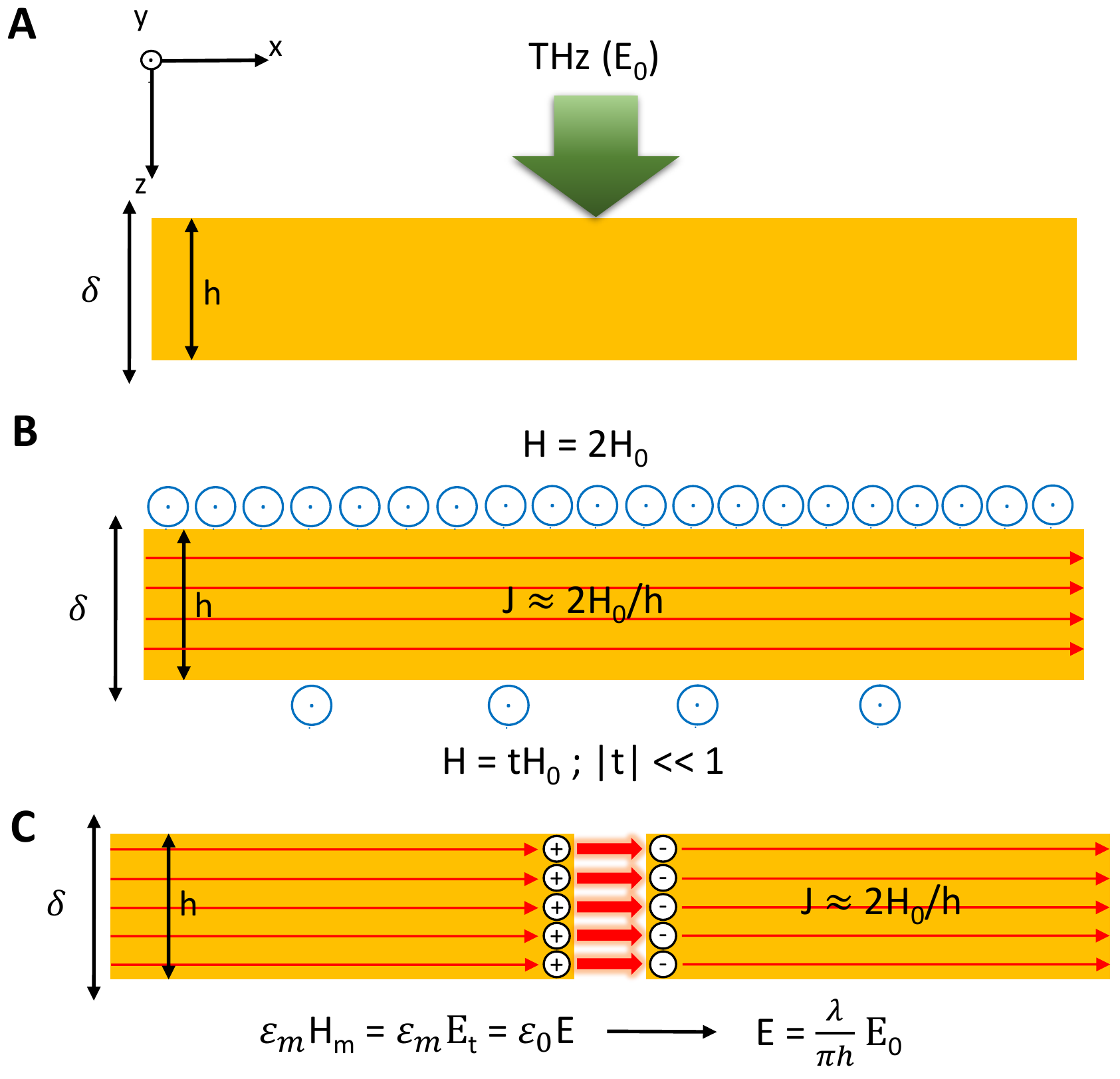}
	\caption{Field enhancement in metal film of thickness below skin-depth. \textbf{(A)} THz field incident normally on a metal film of thickness less than THz skin-depth. \textbf{(B)} Magnetic field of light (represented by blue encircled dots) and current density (represented by red lines) near the thin metal film. At incident surface (up), the magnetic field is twice that of the incident field and a transmission side (down), the magnetic field is much smaller as compared to the incident side. \textbf{(C)} Displacement current and \textit{E} around metal gap. Adapted with permission from Ref.~\cite{Kang2018}. \textcopyright~2018, De Gruyter.{\label{fig7}}}
\end{figure}

\subsection{THz Field Enhancement in Metal Nanogaps at Sub Skin-Depth Regime}
In this part of the section, we discuss the skin-depth physics\cite{Kang2018} and the results reported in various literatures\cite{Seo2009,Bahk2017}. Consider two perfectly conducting metallic planes of thickness \textit{h}\cite{Bahk2017,Kang2009,Kang2009} kept at a distance \textit{w} from each other, forming a metal-air-metal nanogap. An electromagnetic wave is incident normally on the metal nanogap whose electric field has a magnitude of E$_{0}$. Then the ultimate field enhancement in a high-aspect ratio (\textit{w/h}~$\gg$~1) is given by\cite{Kang2018}
\begin{equation}
	E_{enhancement}=\abs*{\frac{E}{E_{0}}}=\frac{\lambda}{\pi h}
	\label{2}
\end{equation} 
where, \textit{E} denotes the electric field at the gap and $\lambda$ denotes the wavelength of the incident electromagnetic wave in vacuum. If a dielectric material of permittivity $\epsilon$ is filled inside the gap, Eq.~\ref{2} modifies to
\begin{equation}
	E_{enhancement}=\abs*{\frac{E}{E_{0}}}=\frac{\lambda}{\pi h}\frac{\epsilon_{0}}{\epsilon}
	\label{3}
\end{equation}
where, $ \epsilon_{0} $ denotes the permittivity of vacuum. From the above equations, it can be said that the field enhancement ($ E_{enhancement} $) is independent of metal characteristics.
\begin{figure}
	\centering
	\includegraphics[scale=0.52]{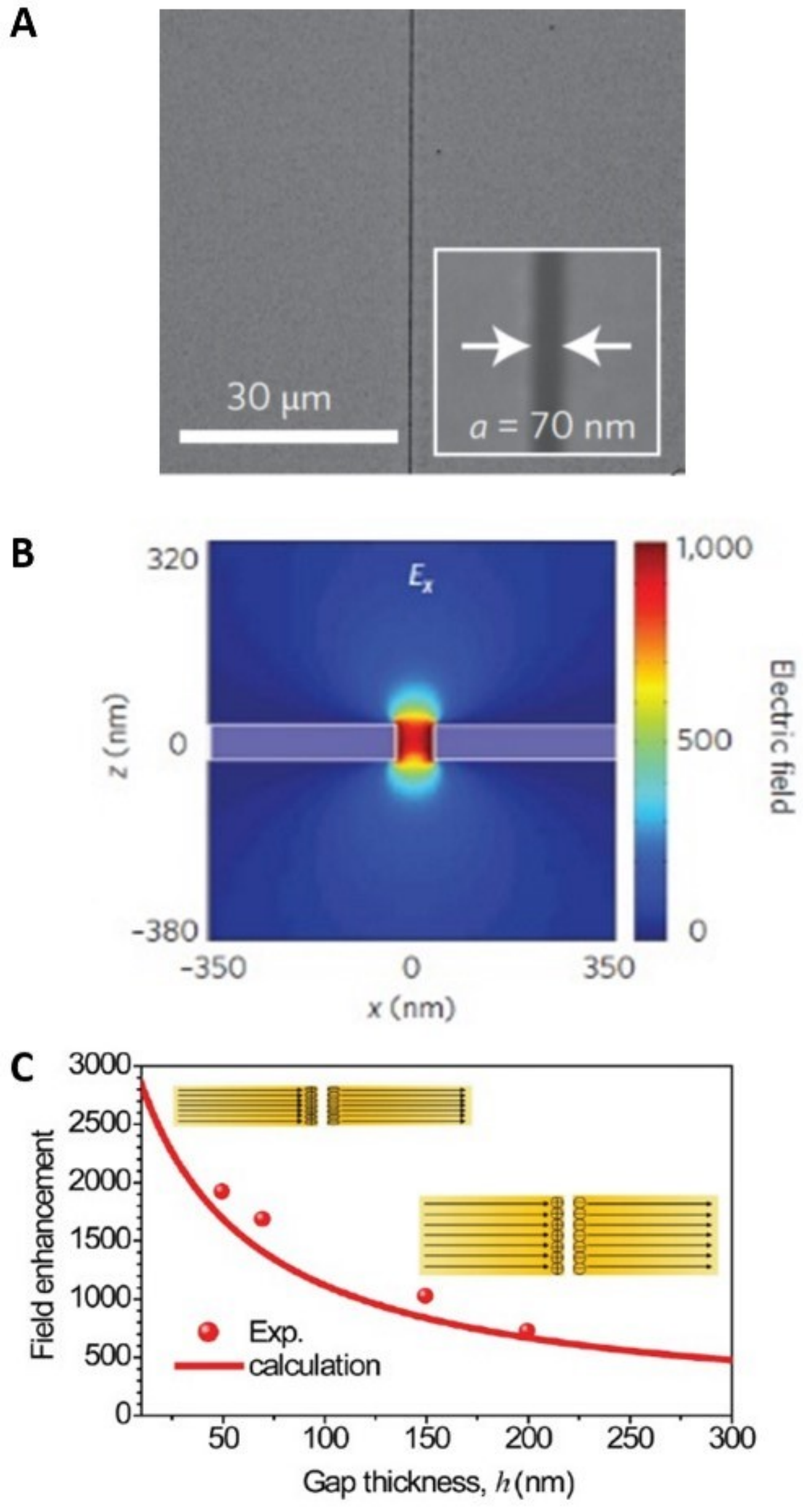}
	\caption{THz field enhancement in metal nanogaps at sub-skin depth regime. \textbf{(A)} SEM of a 70 nm wide gold nanogap fabricated using focused ion beam. \textbf{(B)} Horizontal electric field enhancement in a 70 nm gap, obtained from FDTD simulation. Reprinted with permission from Ref.~\cite{Seo2009}. \textcopyright~2009, Springer Nature. \textbf{(C)} Variation of electric field enhancement with respect to gap thickness, indicating the $ 1/h $ dependence of the field enhancement. Adapted with permission from Ref.~\cite{Bahk2017}. \textcopyright~2017, American Physical Society.{\label{fig8}}}
\end{figure} 

Considering an electromagnetic wave incident normally on a thin metal film of thickness \textit{h} which is smaller than skin-depth $ \delta=\sqrt{\frac{2}{\mu_{0}\sigma_{m}\omega}} $,~but larger the characteristic thickness $ h_{0}=\frac{2\epsilon_{0}c}{\sigma_{m}} $; ($ h_{0}\ll h \leq \delta $), shown in Fig.~\ref{fig7} (A). Here, $ \sigma_{m} $ denotes the conductivity of the metal, $ \mu_{0} $ denotes the magnetic permeability in vacuum and $\omega$ denotes the angular frequency of electromagnetic wave. The thickness at which the absorption loss by metal is 50\%, is known as the characteristic thickness ($ h_{0} $). For a metal with conductivity 10$ ^{7} $~\si{\ohm}$ ^{-1} $m$^{-1}$, the characteristic thickness is 0.53 nm. At 1 THz, the skin-depth is about 100 nm or more for good metals. Therefore, the thickness range for transitional metal film is 5-100 nm. The electric field amplitude transmission (\textit{t}) and reflection coefficient (\textit{r}) for such thin films are given by
\begin{equation}
	t=\abs*{\frac{E_{t}}{E_{0}}}\approx\frac{2\epsilon_{0}c}{\sigma_{m}h}=\frac{h_{0}}{h}\ll 1;
	~r\approx 1-\frac{h_{0}}{h}\approx 1
	\label{4}
\end{equation}
where, $ E_{t} $ denotes the transmitted electric field.\cite{Kang2018}
\begin{figure}
	\centering
	\includegraphics[scale=0.52]{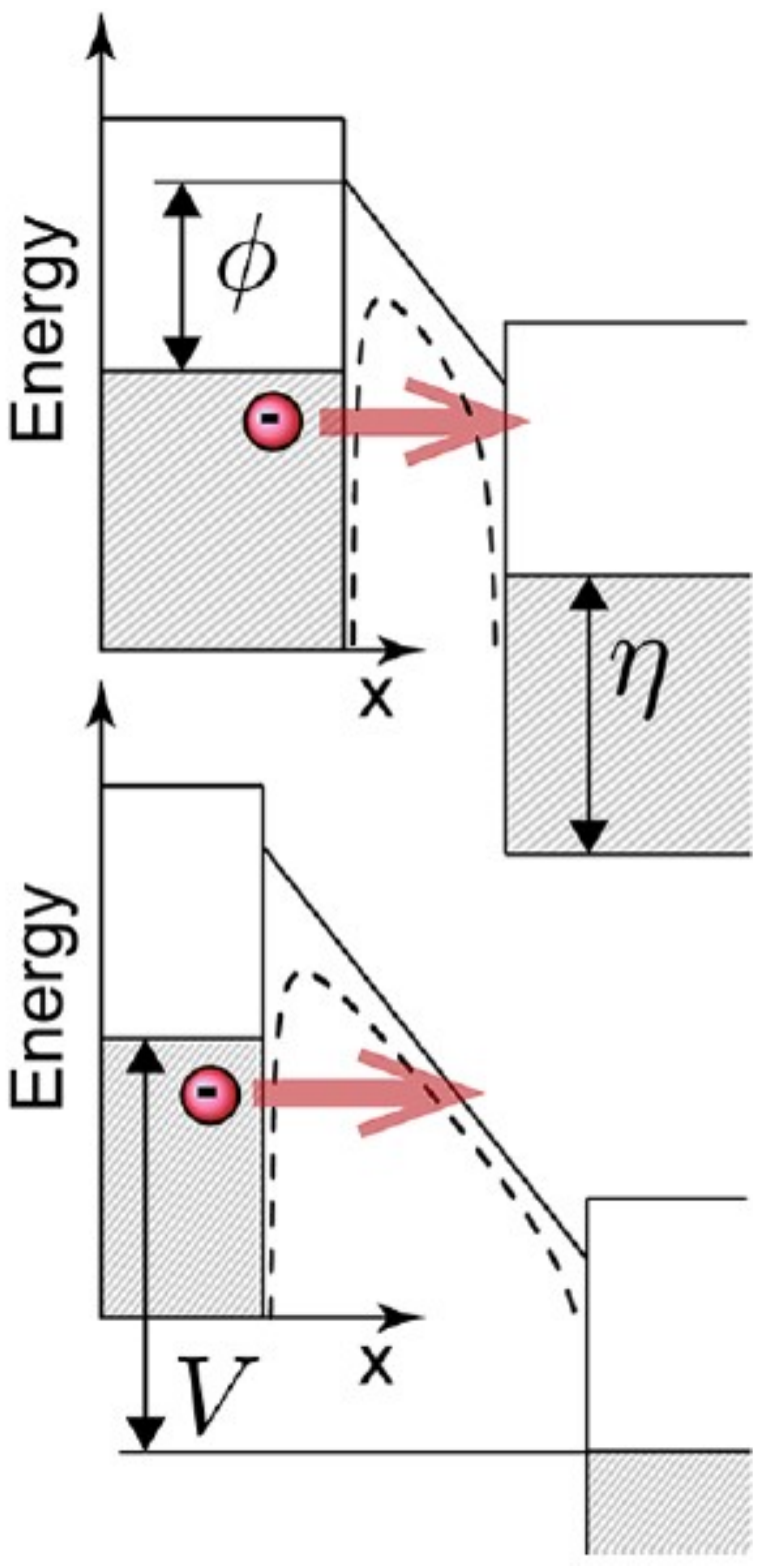}
	\caption{Electron tunnelling through thick and thin barriers. Due to the high gap electric field \textit{E}, separation of Fermi level $\eta$ in two metals by potential energy $ V=eEw $ takes place. $\phi$ represents the rectangular barrier height and the dotted line describes the barrier including image forces. Reprinted with permission from Ref.~\cite{Kim2015}. \textcopyright~2015, American Physical Society.{\label{fig9}}}
\end{figure}

Considering a transverse magnetic polarized light (E$ _{x} $, H$ _{y} $) is incident normally on a thin metal film. Due to reflection from the incident surface, the magnetic field of the light near the incident surface becomes twice that of the incident magnetic field. But the magnetic field on the transmission side is much smaller than the incident field (Fig.~\ref{fig7} (B)). Assuming that a constant electric field/current density is developed inside the thin film, using Ampere\textquotesingle s law, the expression of the current density inside the thin film is $ J=\frac{2H_{0}}{h} $ (neglecting vacuum displacement current).\cite{Kang2018} Here, \textit{J} denotes the current density developed inside the thin metal film and $ H_{0} $ denotes the incident magnetic field. The tangential component of electric fields at the air-metal interface are continuous and is given by
\begin{equation}
	E_{t}=E_{m}=\frac{J}{\sigma_{m}}=\frac{2H_{0}}{\sigma_{m}h}=\frac{2\epsilon_{0}c}{\sigma_{m}h}E_{0}
	\label{5}
\end{equation}
therefore, reproducing Eq.~\ref{4}. At transmitting side, electric field just inside the metal surface is denoted as $ E_{m} $. Across the air nanogap, the normal component of the displacement current will be same (Fig.~\ref{fig7} (C)), therefore electric field inside the air nanogap will be
\begin{equation}
	E=\abs*{\frac{\epsilon_{m}}{\epsilon_{0}}}E_{m}\approx\frac{\sigma_{m}}{\epsilon_{0}\omega}\frac{2\epsilon_{0}c}{\sigma_{m}h}E_{0}=\frac{\lambda}{\pi h}E_{0}
	\label{6}
\end{equation}
where, $ \epsilon_{m} $ denotes the terahertz metal dielectric constant given by
\begin{equation}
	\epsilon_{m}=\epsilon_{\infty}+i\dfrac{\sigma_{m}}{\omega\bigg(1-i\dfrac{\omega}{\gamma}\bigg)}\approx i\frac{\sigma_{m}}{\omega}
\end{equation}
where, $ \epsilon_{\infty} $ denotes the dielectric constant of metal at high-frequency and $\gamma$ denotes the Drude damping constant.

Through this analogy, it can be said that with an increase in the conductivity of the metal, weaker electric fields are induced inside the metal. This is compensated by the metal\textquotesingle s high dielectric constant when the displacement boundary condition is considered. Therefore, the field enhancement is independent of the conductivity of the metal, which is mathematically proven above. Fig.~\ref{fig8} (A) is the SEM of a gold nanogap sample of gap width 70 nm, fabricated by Seo et al.\cite{Seo2009} using a focused ion beam technique. Fig.~\ref{fig8} (B) shows the x-z plane horizontal electric field enhancement in 70 nm gap, analyzed from FDTD simulation at 0.1 THz. In the figure, it can also be seen that the electric field does not penetrate into the metal, rather it is seen to be completely concentrated inside the nanogap, even if the gap size is much smaller than the skin depth (250 nm). Fig.~\ref{fig8} (C) shows the varying electric field enhancement of 5 nm gap with respect to the gap thickness (\textit{h}) at a frequency of 0.3 THz. The red dots signify the experimental data and the red solid line shows the calculated data from the modal expansion. the figure also signifies the $ 1/h $ dependence of the field enhancement.
\begin{figure}
	\centering
	\includegraphics[scale=0.52]{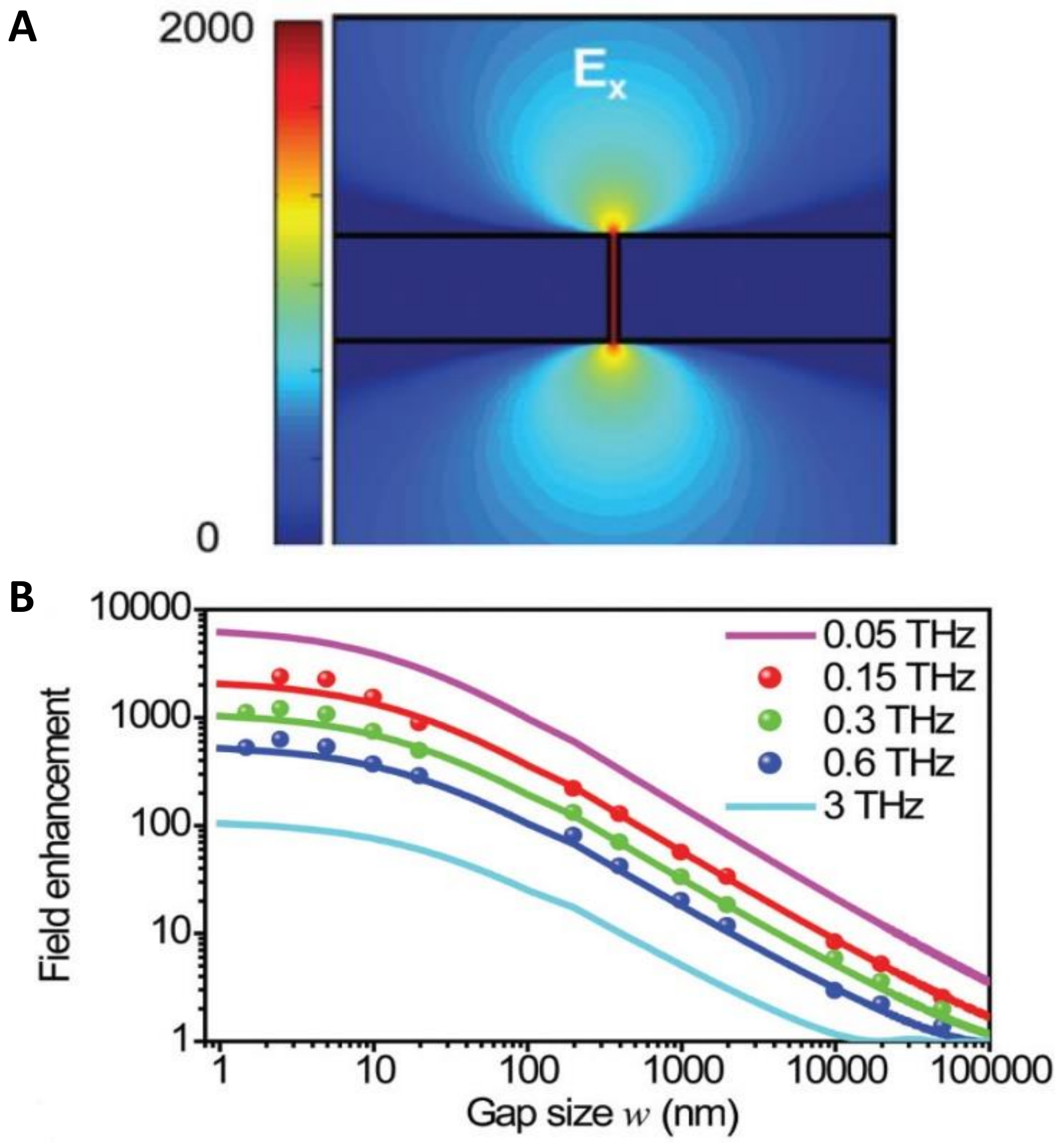}
	\caption{Classical limit of THz field enhancement before entering quantum regime inside gold nanogap. \textbf{(A)} FDTD simulated horizontal electric field around gold nanogap of width 1.5 nm. \textbf{(B)} Variation of field enhancement in different gap widths for various THz frequencies. Solid lines represents the theoretically calculated values from modal expansion formalism. Adapted with permission from Ref.~\cite{Bahk2017}. \textcopyright~2017, American Physical Society.{\label{fig10}}}
\end{figure}

\subsection{THz Field Enhancement in Sub-Nanometer and \AA ngstrom Metallic gaps: Quantum Regime}
Recently, Bahk et al.\cite{Bahk2017} discussed about the field enhancement in thin ($w\sim h<\delta$) and thick ($w<\delta<h$) narrow gaps, and in wide ($\delta$, $ h<w $) gaps. In wide gaps, the charges are mostly spread over the surface outside the gap rather than at metal edges (Fig.~\ref{fig6} (B)), which results in a decrease of field enhancement. In narrow gaps, charges mostly accumulate at the metal edges of the gap (Fig.~\ref{fig6} (C)), resulting in stronger field enhancement. As the width of the gap (\textit{w}) decreases further entering the sub-nanometer and \AA ngstrom regime, the charge distribution becomes insensitive to gap size and electron tunnels through the potential barrier of the nanogap, showing quantum effects.\cite{Savage2012,Scholl2013,Tan2014,Zhu2014,Zhu2016,Nijs2017,Cristofolini2012,Mar2011,Cazier2016,Uskov2016,Stolz2014,Ciraci2012,Kravtsov2014} When THz electromagnetic waves are incident on the nanogap, a transient voltage is developed in the dielectric gap which bends the conduction band of the dielectric toward the Fermi energy of metals (Fig.~\ref{fig9}).\cite{Kim2015} This increases the chances for electron tunneling through the potential barrier, causing non-linear transmissions.

Bahk et al. also investigated the classical limit of field enhancement before the gap-width entered the quantum regime.\cite{Bahk2017} They reported that the field enhancement exhibited saturation behavior rather than a monotonous increasing nature. This showed that the nanogap acts like a charged capacitor\cite{Seo2009} which is shown in the Fig.~\ref{fig6} (A), whose total induced charge is inversely proportional to the frequency of a light-induced alternating current of THz-frequency. FDTD analysis of the electric field distribution around 1.5 nm wide gap in 150 nm thick gold film at a frequency of 1.5 THz, as shown in Fig.~\ref{fig10} (A). As expected the horizontal electric field is strongly concentrated and is enhanced by a factor of 2000 inside the gap. Fig.~\ref{fig10} (B) shows the electric field enhancement in different gap size for various frequencies. The theoretical calculation of the modal expansion for the perfect electric conductor (PEC) model is shown by solid lines.

\subsection{The Finite-Difference Time-Domain (FDTD) Algorithm : A Computational Electromagnetism Approach}
The finite difference time domain (FDTD) method\cite{AllenTaflove2005} is used to solve Maxwell\textquotesingle s equations in the time domain and generally used for computational electrodynamics.
\begin{equation}
	\vec{\nabla}\boldsymbol{\cdot}\vec{D}=\rho
\end{equation}
\begin{equation}
	\vec{\nabla}\boldsymbol{\cdot}\vec{B}=0
\end{equation}
\begin{equation}
	\vec{\nabla}\times\vec{E}=-\frac{\partial\vec{B}}{\partial t}
\end{equation}
\begin{equation}
	\vec{\nabla}\times\vec{H}=\vec{J}+\frac{\partial\vec{D}}{\partial t}
\end{equation}
are the four Maxwell\textquotesingle s equations where, the electric field, magnetic field, electric displacement field, magnetic flux density, free electric charge density, and free current density are denoted by E, H, D, B, $ \rho $ and J, respectively.\cite{Jackson1998,Griffiths2013} These equations are solved numerically on a discrete grid in both space and time, and derivatives are handled with finite differences. No approximations or assumptions are made about the system, making this method highly versatile and accurate. It is a fully vectorial simulation method as it solves for all electric and magnetic field vector components. Being a time-domain method, FDTD can be used to calculate broadband results from a single simulation. FDTD method is typically used when the feature size is of the order of the wavelength. It is general, versatile, accurate, broadband and fasts which makes it the most reliable method in computational electrodynamics.
\begin{figure*}
	\centering
	\includegraphics[scale=0.58]{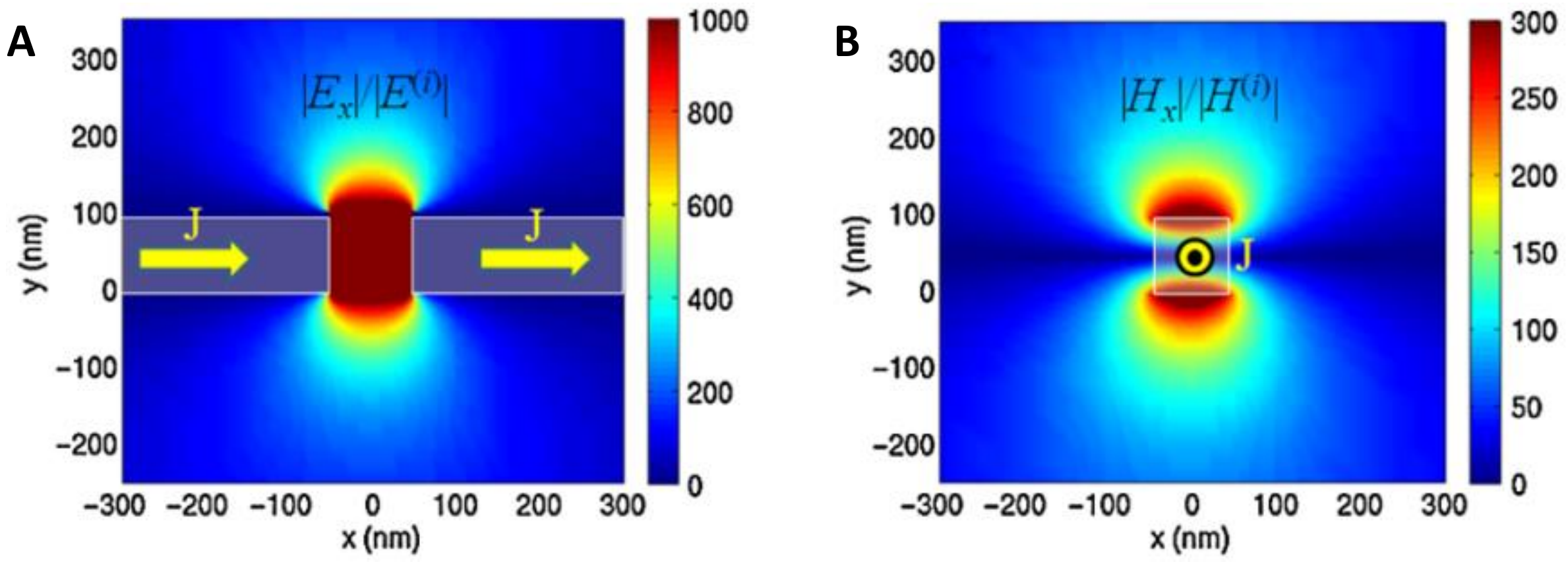}
	\caption{FDTD analysis of a gold nanogap and nanorod. \textbf{(A)} Horizontal electric field around 100 nm gold nanogap. \textbf{(B)} Simulated magnetic field around a gold nanorod. Yellow arrows represent the current flow. Reprinted with permission from Ref.~\cite{Koo2009}. \textcopyright~2009, American Physical Society.{\label{fig11}}}
\end{figure*}

FDTD algorithm solves Maxwell\textquotesingle s curl equations in non-magnetic materials.
\begin{equation}
	\frac{\partial\vec{D}}{\partial t}=\nabla\times\vec{H}
\end{equation}
where, $ \vec{D}=\epsilon\vec{E} $. Since the region is vacuum, $ \vec{J}=0 $.
\begin{equation}
	\vec{D}(\omega)=\epsilon_{0}\epsilon_{r}(\omega)\vec{E}(\omega)
\end{equation}
\begin{equation}
	\frac{\partial\vec{H}}{\partial t}=-\frac{1}{\mu_{0}}\nabla\times\vec{E}
\end{equation}
with H, E, and D describing the magnetic field, electric field and displacement field, respectively, while $ \epsilon_{0}(\omega) $ is a complex relative dielectric constant given by
\begin{equation}
	\epsilon_{r}(\omega)=n^{2}
\end{equation}
, where n denotes refractive index of the material. In 3-D, the six electromagnetic components of Maxwell\textquotesingle s equations are - $ E_{x}, E_{y}, E_{z} $ and $ H_{x}, H_{y}, H_{z} $. Assuming that in z-dimension, the structure is infinite and the fields are independent of z, such that
\begin{equation}
	\epsilon_{r}(\omega,x,y,z)=\epsilon_{r}(\omega,x,y)~,~and
\end{equation}
\begin{equation}
	\frac{\partial\vec{E}}{\partial z}=\frac{\partial\vec{H}}{\partial z}=0
\end{equation}
then, two independent sets of Maxwell\textquotesingle s equations will be created. Each set will contain three vector quantities which can only be solved in x-y plane. The equation containing the components - $ E_{x} $, $ E_{y} $, $ H_{z} $, is known as transverse electric (TE) equation, and the equation containing the components - $ H_{x} $, $ H_{y} $, $ E_{z} $, is known as transverse magnetic (TM) equation.\cite{AllenTaflove2005} Therefore, in TM case, Maxwell\textquotesingle s equations reduces to
\begin{equation}
	\frac{\partial D_{z}}{\partial t}=\frac{\partial H_{y}}{\partial x}-\frac{\partial H_{x}}{\partial y}
\end{equation}
\begin{equation}
	D_{z}(\omega)=\epsilon_{0}\epsilon_{r}(\omega)E_{z}(\omega)
\end{equation}
\begin{equation}
	\frac{\partial H_{x}}{\partial t}=-\frac{1}{\mu_{0}}\frac{\partial E_{z}}{\partial y}
\end{equation}
\begin{equation}
	\frac{\partial H_{y}}{\partial t}=-\frac{1}{\mu_{0}}\frac{\partial E_{z}}{\partial x}
\end{equation}
and for TE case, Maxwell\textquotesingle s equations reduces to
\begin{equation}
	\frac{\partial H_{z}}{\partial t}=-\frac{1}{\mu_{0}}\bigg[\frac{\partial E_{y}}{\partial x}-\frac{\partial E_{x}}{\partial y}\bigg]
\end{equation}
\begin{equation}
	D_{x}(\omega)=\epsilon_{0}\epsilon_{r}(\omega)E_{x}(\omega)
\end{equation}
\begin{equation}
	D_{y}(\omega)=\epsilon_{0}\epsilon_{r}(\omega)E_{y}(\omega)
\end{equation}
\begin{equation}
	\frac{\partial D_{x}}{\partial t}=\frac{\partial H_{z}}{\partial y}
\end{equation}
\begin{equation}
	\frac{\partial D_{y}}{\partial t}=\frac{\partial H_{z}}{\partial x}
\end{equation}
These equations are solved on a discrete, spatial and temporal grid. Each field component is solved at a slightly different location within the grid cell (or, Yee cell).\cite{AllenTaflove2005} 

With this method we can simulate the electric and magnetic fields around any nanostructures using commercially available softwares like, FDTD Solutions (Lumerical Inc., Canada), COMSOL, etc. Fig.~\ref{fig11} shows the FDTD analysis of gold nanogap (100 nm thick and wide gap) and nanorod, respectively.\cite{Koo2009}
\begin{figure*}
	\centering
	\includegraphics[scale=0.55]{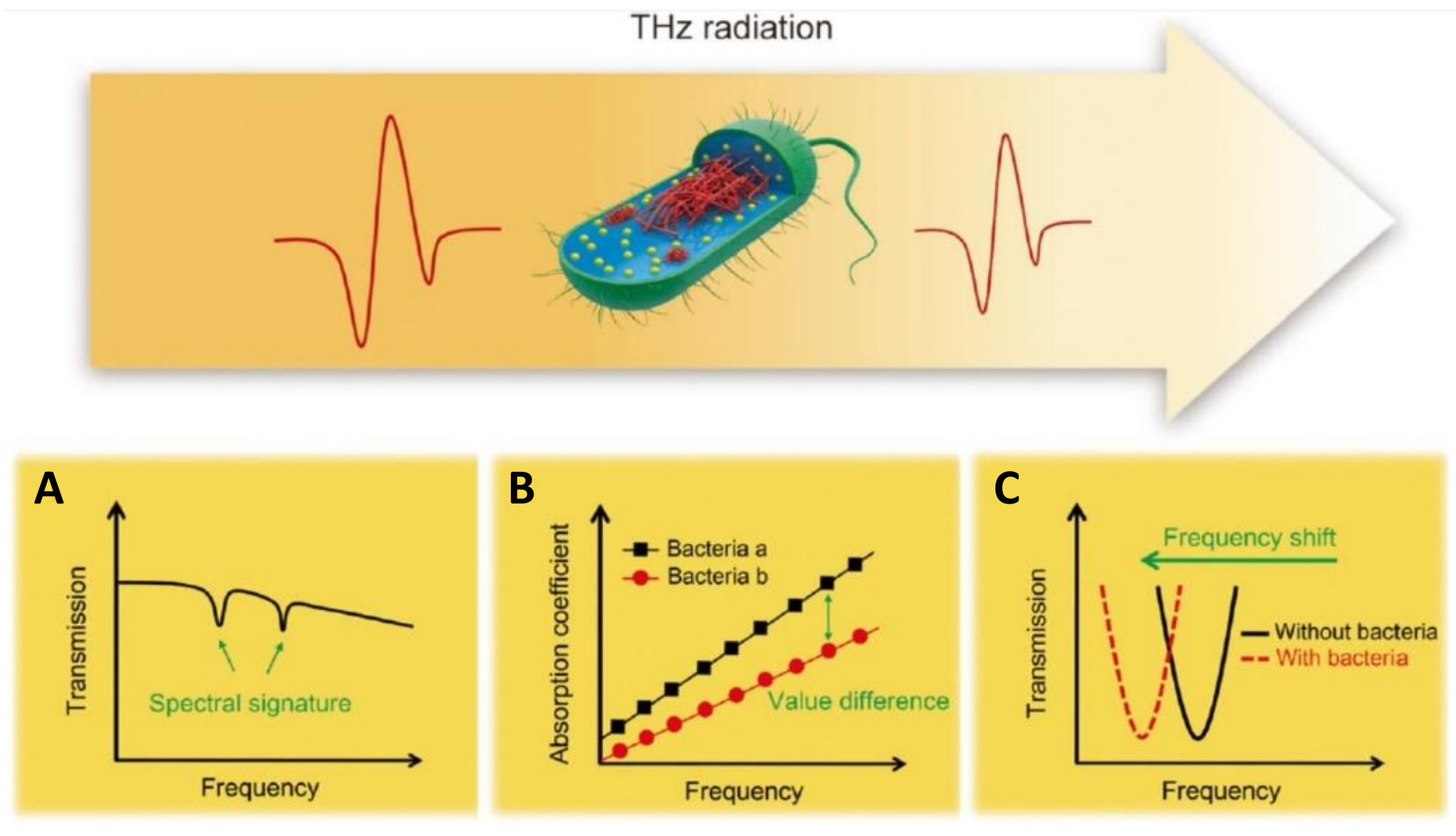}
	\caption{Spectral analysis for identification of biomaterials. \textbf{(A)} Distinct spectral signatures can be used to discriminate a bacterial cell, spore and intracellular metabolite. \textbf{(B)} Bacterial species can be differentiated from their differences in THz optical constants. \textbf{(C)} Resonant frequency shift analysis using THz nanoantenna biosensors can be used to study the presence of biomaterials. Adapted with permission from Ref.~\cite{Yang2016a}. \textcopyright~2016, Springer Nature.}
\end{figure*}
\begin{figure}
	\centering
	\includegraphics[scale=0.35]{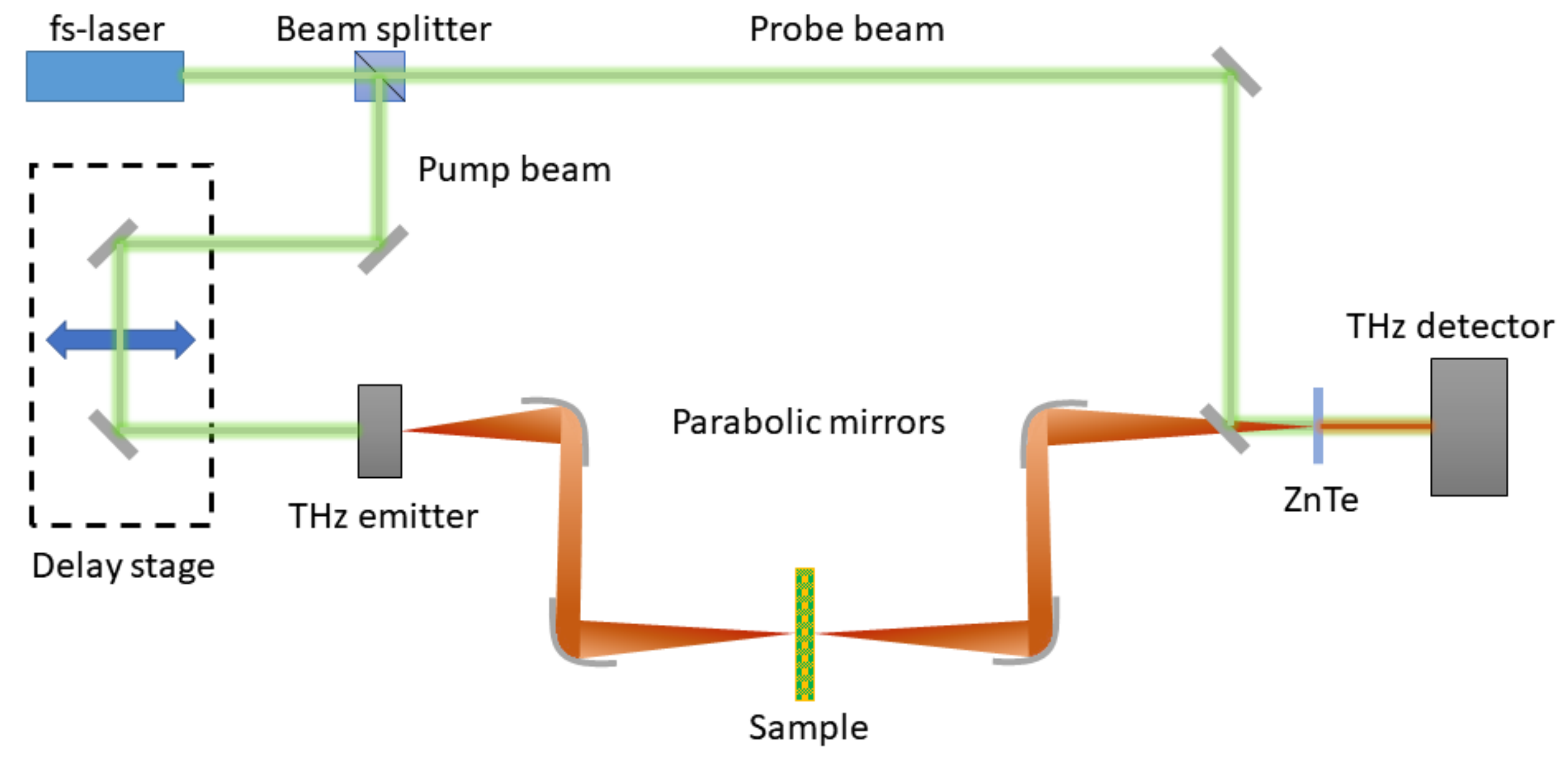}
	\caption{Conventional THz time-domain spectroscopy setup developed by Grischkowski et al. Adapted with permission from Ref.~\cite{Koo2009}. \textcopyright~2016, Springer Nature.{\label{fig13}}}
\end{figure}

\section{THz-TDS and Sensing of Molecules and Biomolecules}
In the late 1980s, Grischkowsky et al.\cite{Fattinger1988,Fattinger1989,Grischkowsky1990} introduced the terahertz time-domain spectroscopy technique (Fig.~\ref{fig13}). It is a spectroscopic technique in which short-THz pulses are used to probe the properties of matter. The system generally consisted of the femtosecond laser (fs-laser), which operates at a repetition rate of 100 MHz, producing a 100 fs laser pulse train. Using a beam splitter, the fs-pulse train is split into two beams: a pump beam and a probe beam. The pump beam is made to be incident on a THz emitter to emit THz pulses, which is collimated to the sample using a pair of parabolic mirrors. Concurrently, the probe beam is used in a time-gated manner for the detection of THz electric field which contains time-domain information of phase and amplitude. Coherently, the transmitted THz electric field is measured as a function of time to obtain a time-domain signal which is converted to noise-free frequency-domain signal using Fourier transformation function.

THz-TDS is a non-contact, non-invasive and label-free detection technique and is extensively applied in biomedical imaging. However, THz-TDS has proved to be less sensitive for detection of micro-organisms (like yeast, molds, and bacteria) of size equal to or less than $\lambda$/100, due to their transparency under THz frequencies. Recently, plasmonic nanoantennas\cite{Lee2015,Park2013,Park2014a} and metamaterials\cite{Lee2017,Park2014,Park2017} have proved to be effective devices for detection of micro-organisms, organic- and biomolecules. These devices show resonances with strong field enhancement across the gap in the THz frequency range, which is highly sensitive to changes in dielectric constant of the gap region. Therefore, these devices are used as biosensors for ultra-sensitive and label-free detection of micro-organisms, organic- and biomolecules. In this section, we discuss the two widely used label-free THz-TDS methods : (i) THz-TDS using molecular dynamic (MD) simulations, and (ii) THz sensing by nanoantennas and metamaterials.

\subsection{Computational THz-TDS using Molecular Dynamics (MD) Simulation}
MD simulation is a computational method used to study the dynamics of a system of molecules in the condensed phase. To accurately study the dynamics of a complex molecule, one needs to employ the quantum model to study the wave function of each sub-atomic particle. But MD employs classical Newtonian mechanics to study the dynamics of the system, which makes it less accurate. Specifically, it integrates Newton\textquotesingle s equation of motion in discrete time-steps. Due to high THz absorption by water, THz-TDS becomes challenging for water-based systems. Therefore, MD simulation is one of the preferred methods to study the water dynamics of a hydrated molecular system. In MD, water dynamics has a great significance in the structural arrangements of molecules, which occurs because the molecule alters the dynamics of the surrounding water molecules, adopting a quasi-coherent character, caused by reorganized, loose hydrogen-bond network.
\begin{figure}
	\centering 
	\includegraphics[scale=0.5]{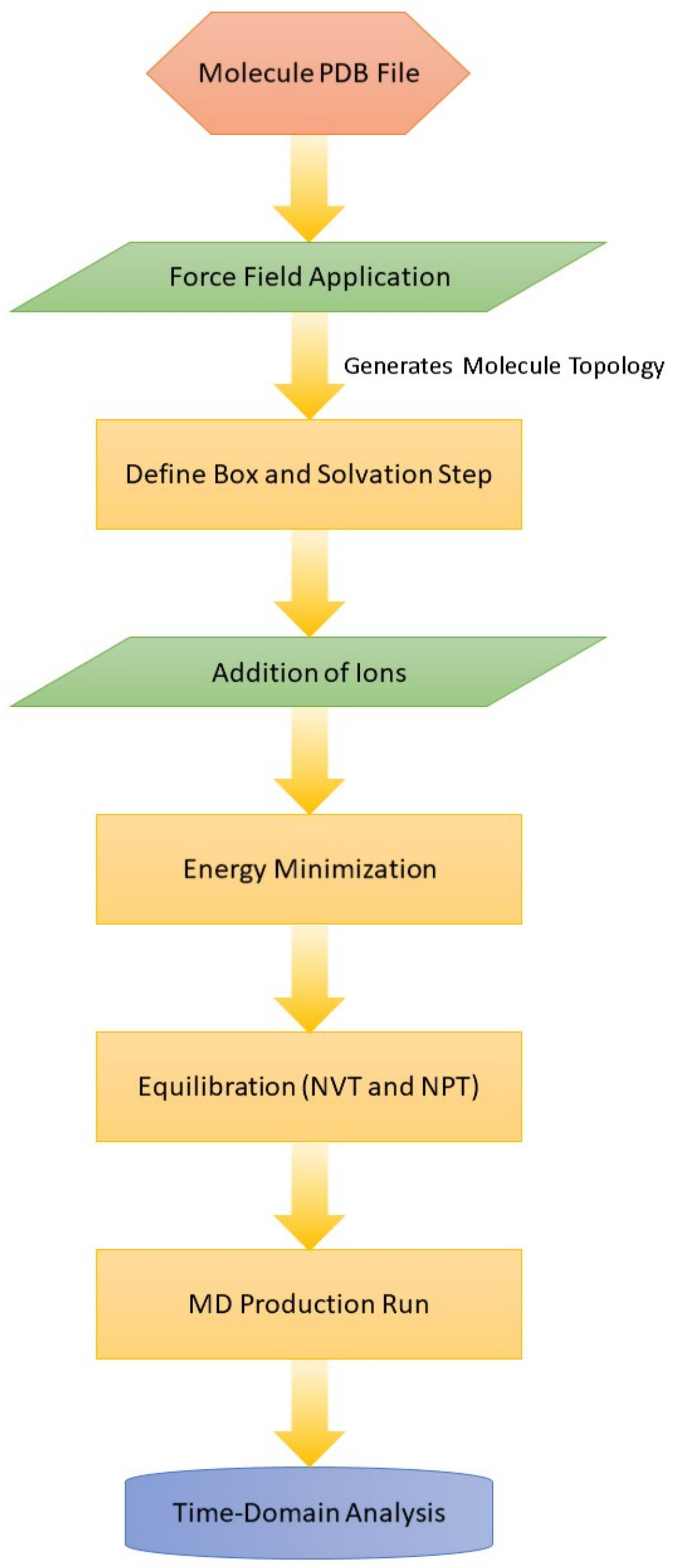}
	\caption{Flowchart of a general MD simulation process.{\label{fig14}}}
\end{figure}
\begin{figure}
	\centering 
	\includegraphics[scale=0.54]{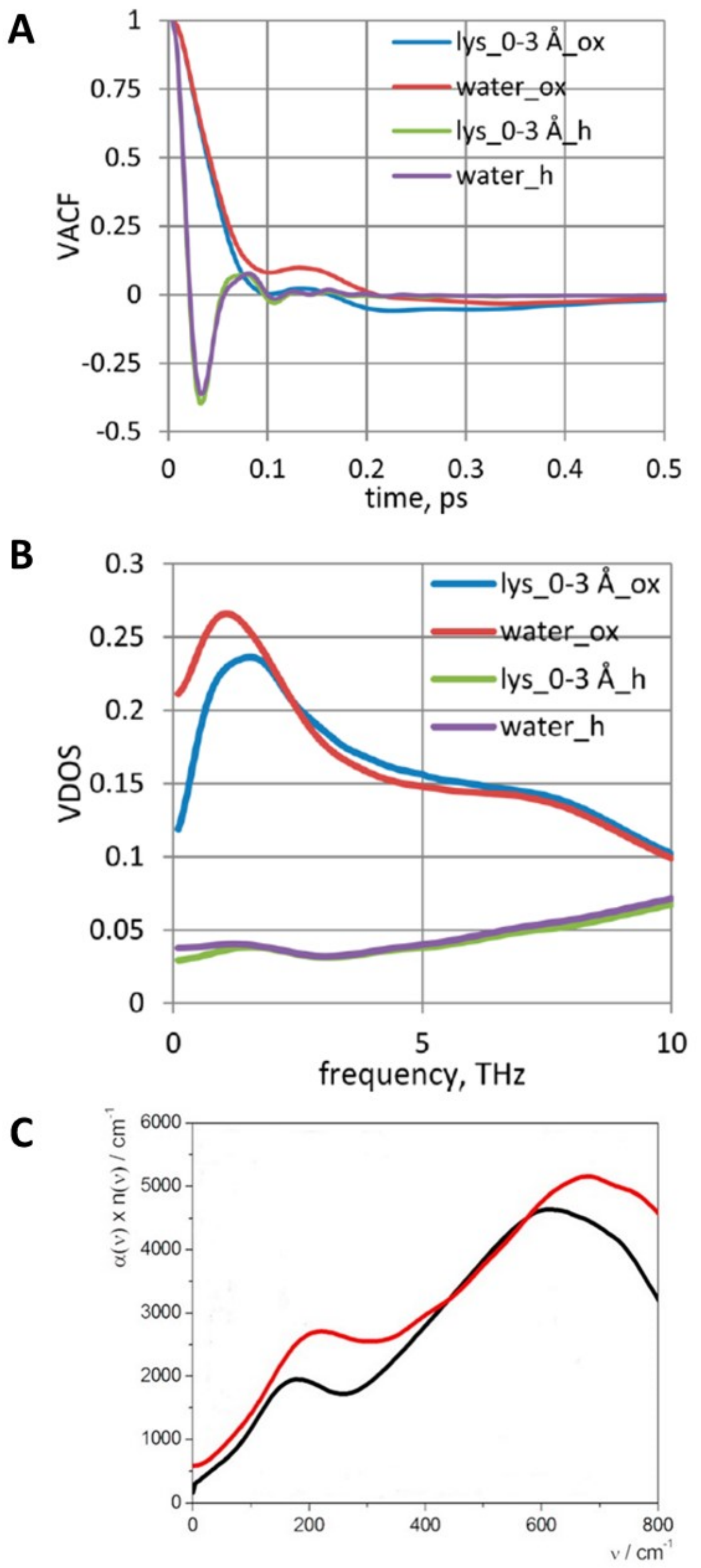}
	\caption{THz time-domain analysis obtained using MD simulations. \textbf{(A)} Velocity autocorrelation fuction (VACF) and \textbf{(B)} vibrational density of states of oxygen (ox) and hydrogen (h) in first solvation layer (3 \AA) of lysosome and bulk water. Reprinted with permission from Ref.\cite{Sushko2013}. \textcopyright~2013, American Chemical Society. \textbf{(C)} Absorption spectrum of water obtained experimentally (black line) and from MD simulation (red line). Adapted with permission from Ref.~\cite{Nibali2014}. \textcopyright~2014, American Chemical Society.{\label{fig15}}}
\end{figure}

A significant number of MD-based computational analysis have been performed to interpret the THz absorption spectra of protein solutions. Many studies showed the anomalous dynamics of water in the hydration shell due to the fact that surrounding water molecules were heterogeneously perturbed by the solute.\cite{Rocchi1998,Marchi2002,Sengupta2008,Sinha2008,Xu2012} Bandyopadhyay et al. have performed a significant amount of MD simulations to study the effects of hydration of protein molecules.\cite{Sinha2008,Chakraborty2007,Sinha2011,Bandyopadhyay2006,Bandyopadhyay2004,Pal2013,Pal2013a,Sinha2012,Sinha2012a} Various methods have been used to study the thickness of villin headpiece sub-domain HP-36 which includes MSD, velocity autocorrelation function analysis, H-bond time correlation function analysis and reorientational correlation function analysis of water molecules. According to the simulation results, the different helical segments had a heterogeneous influence on the water dynamics, which were limited to the first hydration layer.\cite{Sinha2008} Some similar results were obtained in which it was demonstrated a clear distinct power spectrum (VDOS) for water molecules bonded to different planes of antifreeze protein by hydrogen in the spectral domain of 1-4 THz.\cite{Xu2012} While studying the vibrational spectrum of water in the villin headpiece subdomain hydration layer, O\boldsymbol{$\cdot$}\boldsymbol{$\cdot$}\boldsymbol{$\cdot$}O\boldsymbol{$\cdot$}\boldsymbol{$\cdot$}\boldsymbol{$\cdot$}O bending mode showed a blue-shift in the first hydration layer, which is noticed for water molecules bonded to protein by hydrogen.\cite{Chakraborty2007} It also showed the possibility of the structural flexibility of protein. Later an atomistic MD simulation of hen egg-white lysosome solvated in explicit water molecules at room temperature was performed,\cite{Sinha2011} which reported that a few large-amplitude bistable motions exhibited by two coils controlled the overall flexibility of protein molecules. A series of MD simulations were performed to compare the experimental data of far-infrared spectroscopy used to study the dynamics of three aqueous peptides with varied helicity.\cite{Ding2011} Using the first principles, extensive studies on water, resolved in time and space, reported that the group motion of H-bonded molecules in the second solvation shell significantly contributed to the absorption at about 2.4 THz, also showing the presence of third-shell effects.\cite{Heyden2010a} Heyden et al. in their future publication\cite{Heyden2012} showed the capability of nonpolarizable water models in reproducing low-frequency, inter-molecular vibration of water since electronic polarization is dominated by the static molecular dipoles. The depth of hydration shell in the hydrated lysosome, BPTI, TRP-cage, and TRP-tail was estimated using MD simulations.\cite{Sushko2013}
\begin{figure}
	\centering
	\includegraphics[scale=0.46]{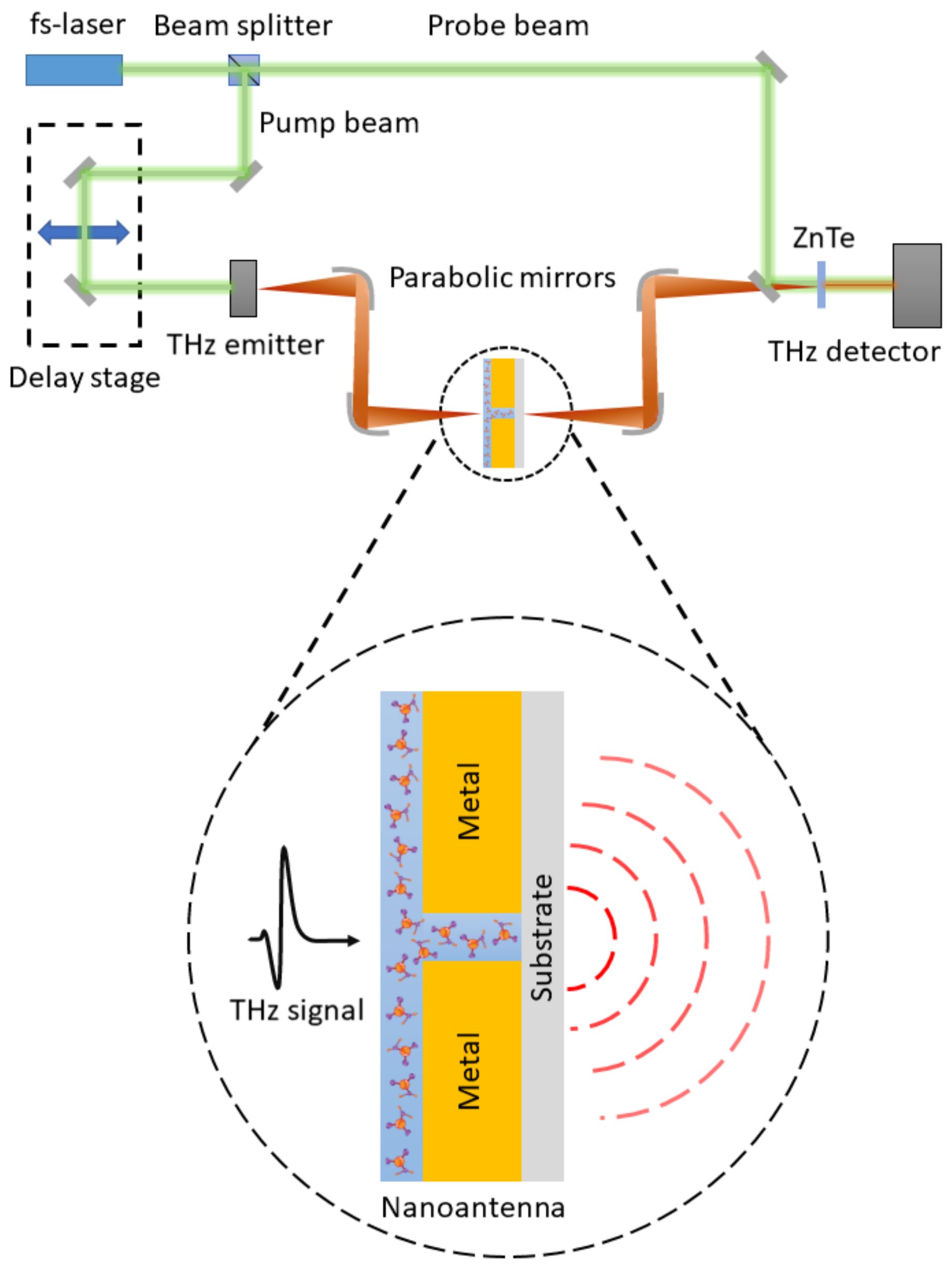}
	\caption{Application of THz nanoantenna in developing a highly-sensitive THz-TDS technique.{\label{fig16}}}
\end{figure}

To perform a basic MD simulation of a molecule,\cite{Allen2004} first and foremost a force field must be applied to characterize the molecular interaction of the molecules. In the next step, the molecule is positioned in a unit cell of desired shape and size followed by the addition of water molecules, known as solvation step. Then ions are added to neutralize the net charge of the system. After the addition of ions, the system is then relaxed through a process known as energy minimization. This process also ensures that the system is free from steric clashes or inappropriate geometry. For further simulation, the system needs to be brought to the desired temperature. After attaining the desired temperature, an adequate amount of pressure must be applied until it reaches a proper density. This whole process is known as equilibration and is conducted in two phases. The first phase is conducted to attain the desired temperature under NVT (constant Number of particles, Volume, and Temperature) or isothermal-isochoric ensemble. The second phase is conducted to stabilize both pressure and density of the system under NPT (constant Number of particles, Pressure, and Temperature) or isothermal-isobaric ensemble. After attaining the desired temperature and pressure, the system is finally ready for the MD simulation, known as MD production run. After performing the final MD-run, the system is ready to be analyzed. From MD simulation, the time-dependent properties are obtained via. correlation function, like velocity autocorrelation function (VACF). The time-domain properties are then converted to frequency-domain properties using Fourier transformation function, for example, the vibrational density of states (VDOS). The whole MD simulation steps are summed up in a flowchart (Fig.~\ref{fig14}).
\begin{figure}
	\centering
	\includegraphics[scale=0.63]{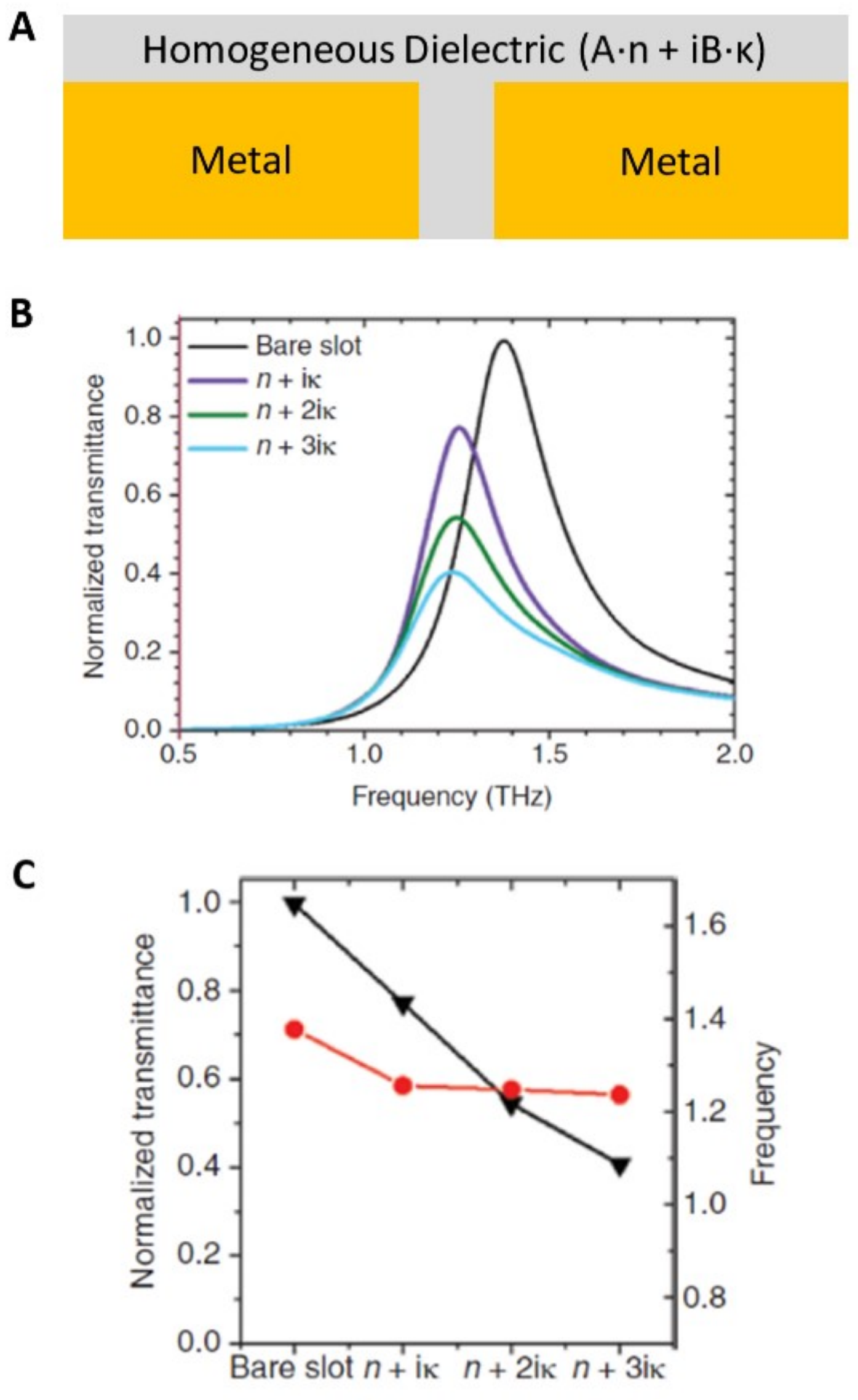}
	\caption{Interpretation of nanoantenna enhanced THz interactions of biomaterials using FDTD method. \textbf{(A)} Schematic of the cross section view of a metallic nanogap cladded with a absorptive homogeneous dielectric film with complex refractive index. \textbf{(B)} Normalized transmittance simulated by varying the imaginary part of the complex refractive index of the homogeneous dielectric film. \textbf{(C)} Transmittance peak values and resonance frequencies are plotted for various complex refractive indices. Adapted with permission from Ref.~\cite{Kang2018}. \textcopyright~2018, De Gruyter.{\label{fig17}}}
\end{figure}

Fig.~\ref{fig15} (A) shows the MD simulation generated VACF curves for oxygen and hydrogen atoms in bulk water and in the water within 3 \AA~of lysosome molecule (first hydration layer). The figure depicts two results: i) The hydrogen dynamics get uncorrelated faster (0.85 ps earlier) than oxygen dynamics, and ii) The dynamics of the oxygen atoms are much more restricted in the first hydration layer than in bulk water (as compared from their extremums) due to caging-effect. The corresponding VACF data was further Fourier transformed to get a set of VDOS data as shown in Fig.~\ref{fig15} (B). In the figure, at 1.1 THz a well-defined peak is shown by the oxygen atoms of bulk water signifying the bending motion of O -- O -- O atoms (known as triplets of H-bonded oxygen atoms). But in the case of lysosome-bounded water molecules, this peak is blue-shifted by 0.4 THz signifying that the H-bonds between O atoms gets stronger in the hydration layer. This peak has a lower amplitude compared to that of the oxygen atoms of bulk water, as the presence of lysosome does not have much influence on the vibrational mode of water molecules. The VDOS of hydrogen in both situations is very flat as they are not much influenced by the protein molecule. The absorption spectrum can be solved by applying Fourier transform on total dipole moment autocorrelation function of the system computed from MD simulation,\cite{Nibali2014} given by
\begin{equation}
	\alpha(\omega)=\frac{1}{4\pi\epsilon_{0}}\frac{2\pi\omega^{2}}{3Vk_{B}Tcn(\omega)}\int_{0}^{\infty}e^{i\omega t}\big<M(0)M(t)\big> dt
\end{equation}
here, V denotes the volume of the system, k$_{B}$ denotes the Boltzmann\textquotesingle s constant, T denotes the absolute temperature of the system, c denotes the speed of light and n($\omega$) denotes the frequency dependent refractive index. Figure~\ref{fig15} (C) is the absorption spectra of bulk water obtained experimentally (black line)\cite{Bertie1996} and by MD simulation (red line),\cite{Heyden2010} describing the inter-molecular hydrogen bond collective modes.
\begin{figure}
	\centering
	\includegraphics[scale=0.59]{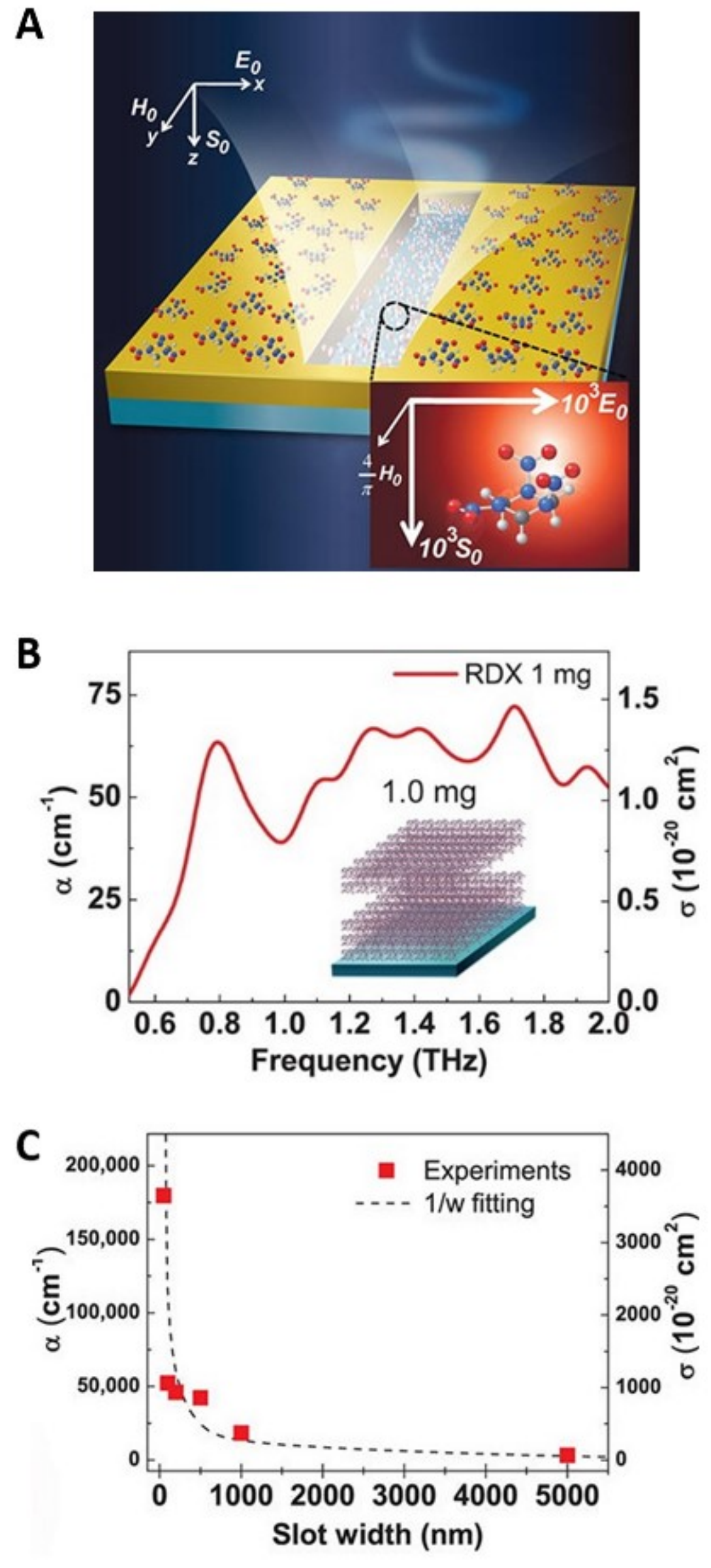}
	\caption{Effects of various gap-widths on THz-sensing of RDX molecules. \textbf{(A)} Schematic of the cross section of a THz single-slot nanoantenna containing RDX molecules. \textbf{(B)} Absorption spectrum of RDX molecules (1 mg) onto the bare quartz substrate. \textbf{(C)} Absorption coefficients peaks and cross section of RDX molecules in various nanoslot gap-widths \textit{w}. The black dashed line signifies the $ 1/w $ dependence. Reprinted with permission from Ref.~\cite{Park2013}. \textcopyright~2013, American Chemical Society.{\label{fig18}}}
\end{figure}
\begin{figure*}
	\centering
	\includegraphics[scale=0.7]{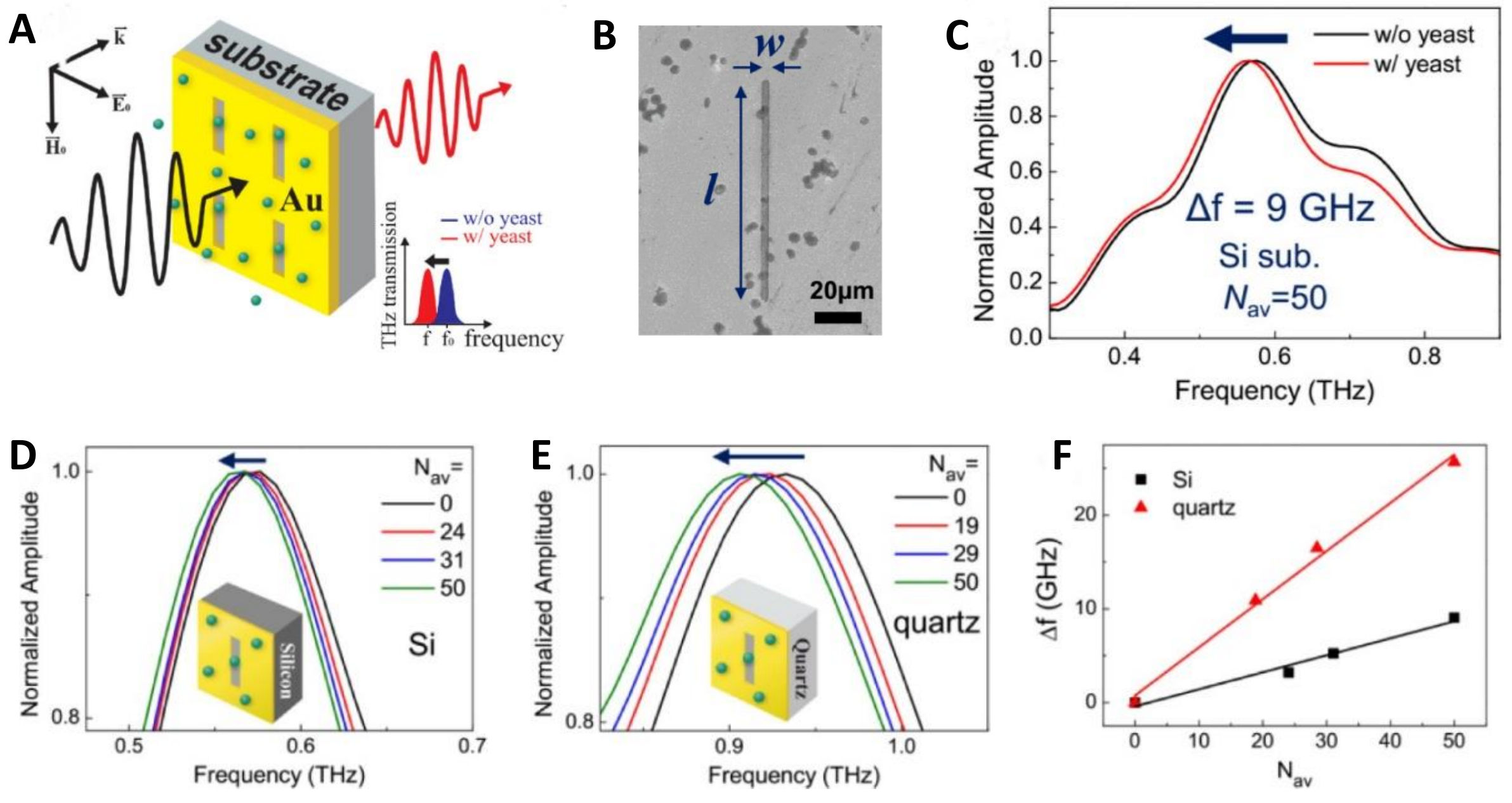}
	\caption{Effects of different substrates on yeast detection. \textbf{(A)} Schematic of THz sensing of yeast cells using a THz slot antenna. \textbf{(B)} SEM image of yeast cells deposited on 2 $\mu$m wide and 100 $\mu$m long slot antenna. \textbf{(C)} THz transmission amplitudes for THz slot antenna on Si substrate showing a red-shift in the resonant frequency of the slot antenna due to the presence of yeast cells. THz transmission amplitudes for THz slot antenna on \textbf{(D)} Si- and \textbf{(E)} quartz-substrates at different yeast cell concentration (N$_{av}$). \textbf{(F)} Shifts in resonance frequency of THz slot antenna on Si- and quartz-substrates with respect to N$_{av}$. Reprinted with permission from Ref.~\cite{Park2014}. \textcopyright~2014, Springer Nature.{\label{fig19}}}
\end{figure*}

\subsection{THz-Nanoantenna and Metamaterials Sensing of Micro-organisms, Organic- and Biomolecules}
THz nano-structures can be used for ultra-sensitive, label-free and non-invasive sensing of molecules and biomaterials. As discussed earlier in this section, the strong THz field enhancement along with resonances in the gap region of these nano-structures are sensitive to the changes in the dielectric constant of the insulating gap material. If the gap is filled with desired biomaterial sample, due to enhanced THz-interaction, the dielectric (i.e., the biomaterial sample) will show an enhanced THz-absorption therefore, the THz optical properties can be studied from the THz dielectric response. Thereby, making the necessary change in the THz-TDS setup (Fig.~\ref{fig16}), enables enhanced sensing of molecules and micro-organisms which were not sensed by conventional THz-TDS setup. Therefore, from the Fourier transformed spectra (frequency-domain amplitude and phase) obtained from the sample and reference transmitted signals, the complex optical properties ae computed from the following relation:
\begin{equation}
	A_{s}(\omega)=A_{r}(\omega)\cdot e^{-\big(\frac{d\cdot \alpha(\omega)}{2}\big)}\cdot e^{\big(i\frac{\omega}{c}n(\omega)d\big)}
\end{equation}
where, A$_{s}$ and A$_{r}$ are the amplitudes of sample and reference transmitted signals, respectively. The real parts of the absorption coefficient and the index of refraction are denoted by $\alpha$($\omega$) and n($\omega$), respectively, and thickness of the dielectric is denoted by \textit{d}. The real part of refractive index\cite{Yang2016a} is given by
\begin{equation}
	n(\omega)=1+\frac{(\phi_{r}-\phi_{s})c}{2\pi df}
\end{equation}
and applying Beer-Lambert law, absorption coefficient\cite{Kang2018} is given by
\begin{equation}
	\alpha(\omega)=-\frac{2}{d}\ln(T)=-\frac{2}{d}\ln\Bigg(\bigg(\frac{A_{s}(\omega)}{A_{r}(\omega)}\bigg)^{2}\Bigg)=\frac{4\pi fk}{c}
\end{equation}
where, $ \phi_{r} $ and $ \phi_{s} $ are the phases of sample and reference transmitted signals, respectively, T denotes the transmittance and f is the frequency of the input THz-signal.
\begin{figure*}
	\centering
	\includegraphics[scale=0.7]{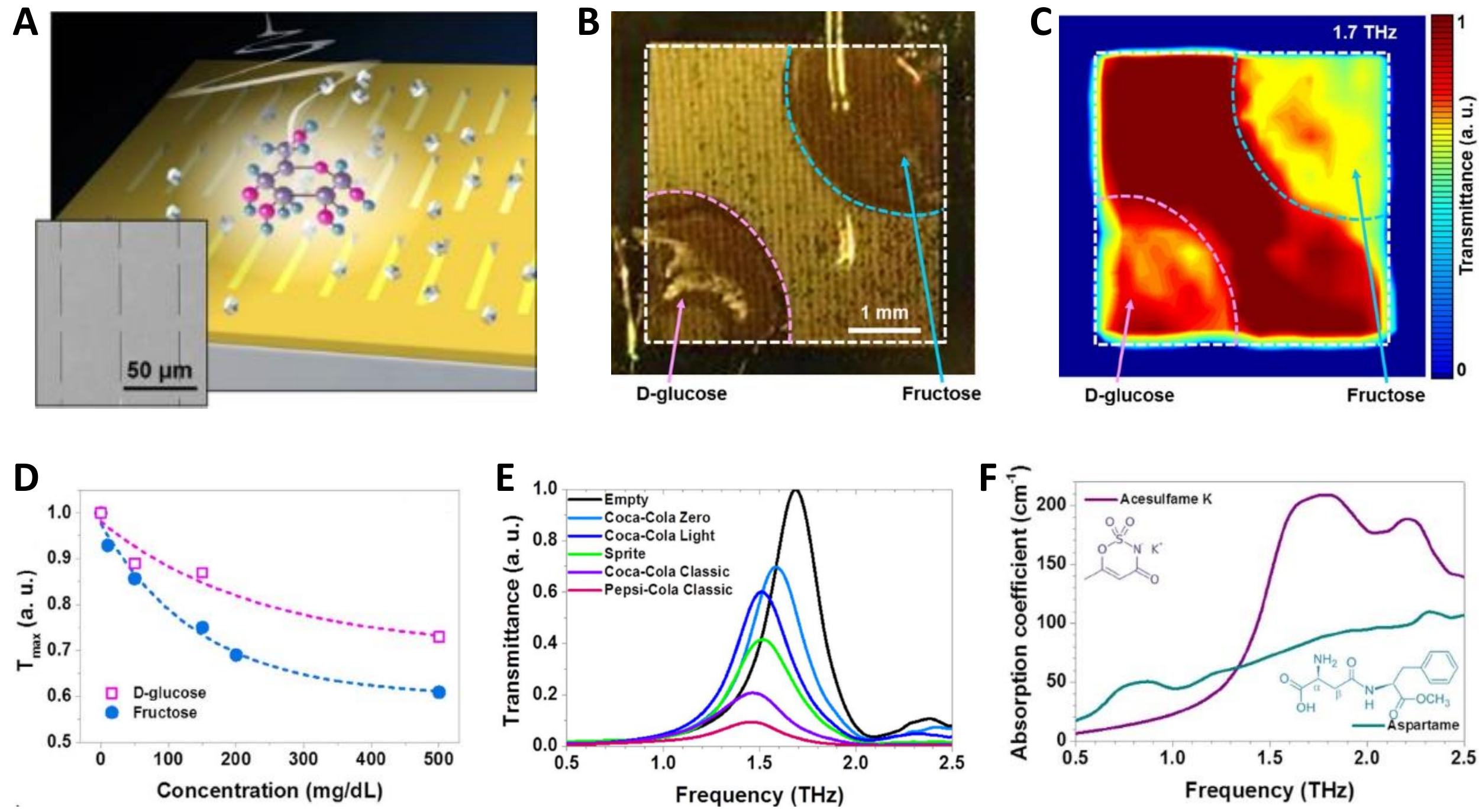}
	\caption{Highly-sensitive and selective nanoantenna. \textbf{(A)} Schematic of a THz nanoantenna for detection of sugar molecules. \textbf{(B)} Fructose antenna with stains of 250 mg/dL fructose (top-right) and D-glucose (bottom-left). \textbf{(C)} THz transmittance through stained fructose antenna illustrating the selective nature of nanoantenna. \textbf{(D)} THz transmittance peaks of fructose and D-glucose at different concentrations using fructose antenna. \textbf{(E)} THz transmittances of added sugars and artificial sweeteners in various branded beverages and dietary sodas using a fructose antenna. \textbf{(F)} THz absorption coefficients of low-concentrated artificial sweeteners: Acesulfame K and Aspartame. Reprinted with permission from Ref.~\cite{Lee2015}. \textcopyright~2015, Springer Nature.{\label{fig20}}}
\end{figure*}

Through FDTD simulation, an intense study of this biosensing process can be done. The biomaterial sample film is assumed to be a homogeneous dielectric film of a thickness proportional to the molecular concentration, is sandwiched between two gold films (Fig.~\ref{fig17} (A)). The dielectric film has a complex refractive index A$\cdot$n + i B$\cdot$$\kappa$, where A and B are constants whose values range from 1.0 - 3.0 over a frequency band. The absorptive dielectric film was inserted in the gold nanogap by applying an auxiliary differential equation. To study the electric field interaction in the absorptive dielectric film, a non-uniform mesh was applied over the whole film with the smallest step-size of 10 nm. To the nanoantenna, THz electromagnetic waves are made to be incident normally. As the THz waves pass through the dielectric film, the transmitted THz-electric field decays exponentially and is given by
\begin{equation}
	\frac{T_{s}(\omega)}{T_{r}(\omega)}=Ce^{-\kappa k(\omega)h}
\end{equation}
here, $T_{s}= (E_{s}(\omega))^{2}$ and $ T_{r}(\omega)=(E_{r})^{2} $ are the transmittances through dielectric and air nanogaps, respectively. In the above relation, C is known as the transmittance ratio at the air-dielectric interface, $ k=\frac{2\pi f}{c} $ is known as the incidence momentum and h is the thickness of the dielectric. Fig.~\ref{fig17} (B - C) shows the simulated transmittance for different complex refractive indices. It is evident from the figure, that the absorption is dependent on the imaginary part of the complex refractive index ($\kappa$), and the resonance frequency is solely dependent on the real part of the complex refractive index (n). However, the change in resonance frequency is quite small but stronger absorption can lead to an appreciable change. Therefore, the variation in the transmission spectra will provide evident proofs for identification of molecules and species contained in the biomaterial sample.
\begin{figure*}
	\centering
	\includegraphics[scale=0.7]{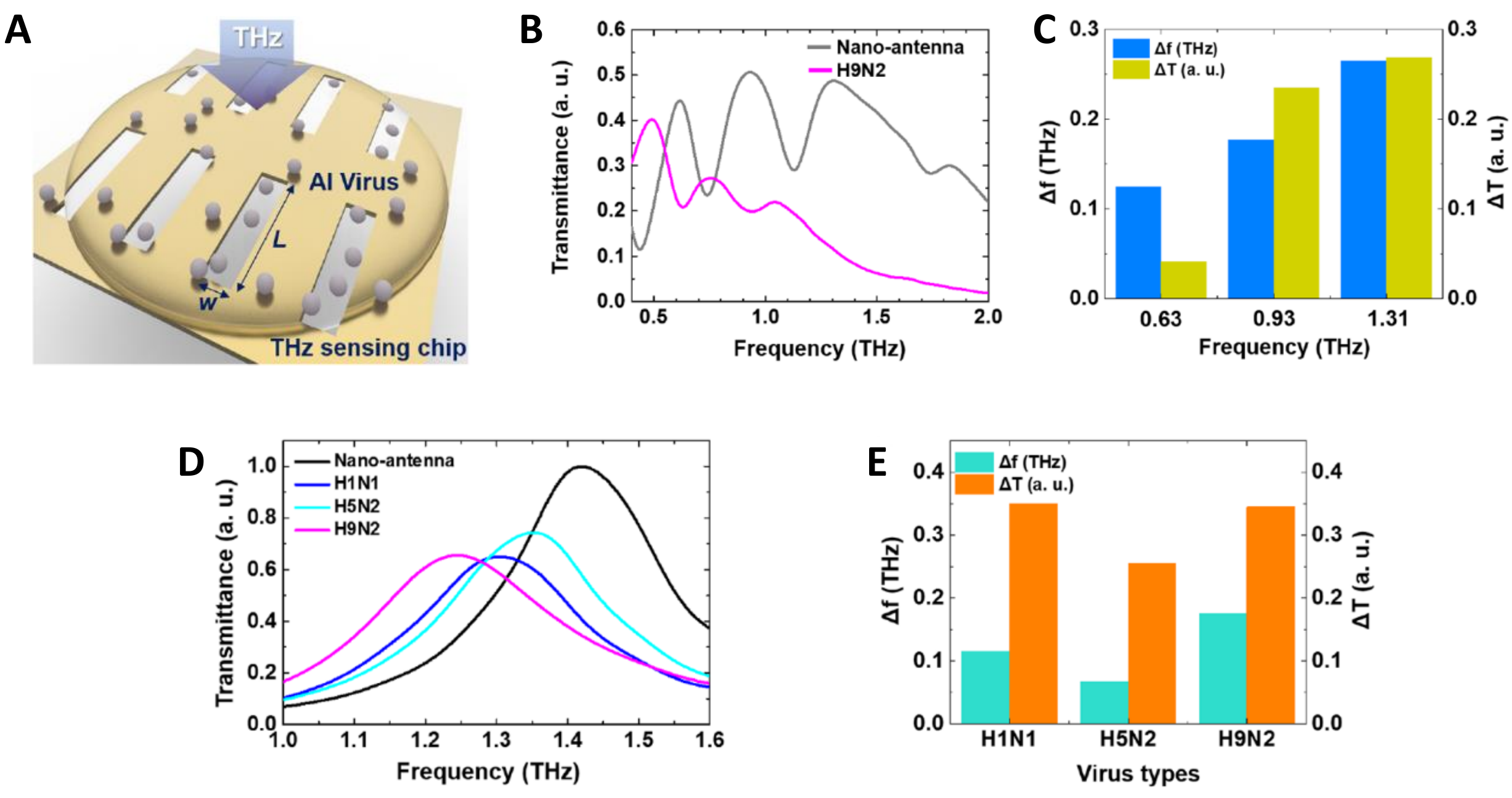}
	\caption{Detection of virus samples using THz nanoantenna sensing chip. \textbf{(A)} Schematic of a THz nano-slot antenna array sensing chip used to detect virus samples in liquid state. \textbf{(B)} Transmittance spectra measured using a multi-resonance nanoantenna, with and without H9N2. \textbf{(C)} Difference in transmitted intensity ($\Delta$T) and resonance frequency shifts from each fundamental resonance peak ($\Delta$f) for H9N2 virus sample. \textbf{(D)} THz transmission spectra through single-resonance nanoantenna, with and without virus samples (H1N1, H5N2 and H9N2). \textbf{(E)} Difference in transmitted intensity ($\Delta$T) and resonance frequency shifts from the fundamental resonance peak ($\Delta$f) of single-resonance nanoantenna for different virus samples. Reprinted with permission from Ref.~\cite{Lee2017}. \textcopyright~2017, Springer Nature.{\label{fig21}}}
\end{figure*}

To study the effects of various gap-widths on THz-sensing,\cite{Park2013} a single-slot THz nanoantennas of length 90 $\mu$m and slot-widths of 50 nm (50 nm thick), 100, 200, 500, 1000 and 5000 nm (100 nm thick) was fabricated on a thick quartz substrate and filled with 1 mg/ml RDX (1,3,5-trinitroperhydro-1,3,5-triazine) molecules. Initially, the THz-absorption ($\alpha$) of 1 mg RDX molecules placed over a bare quartz substrate was studied which is shown in Fig.~\ref{fig18} (B). Then the absorption of RDX molecules inside the various nano-slots was studied and the data obtained was plotted against the slot-widths as shown in the Fig.~\ref{fig18} (C). From Fig.~\ref{fig18} (B - C), it is evident that nanoantenna enhances the THz-absorption by a large factor of 10$^{3}$. 
\begin{figure*}
	\centering
	\includegraphics[scale=0.67]{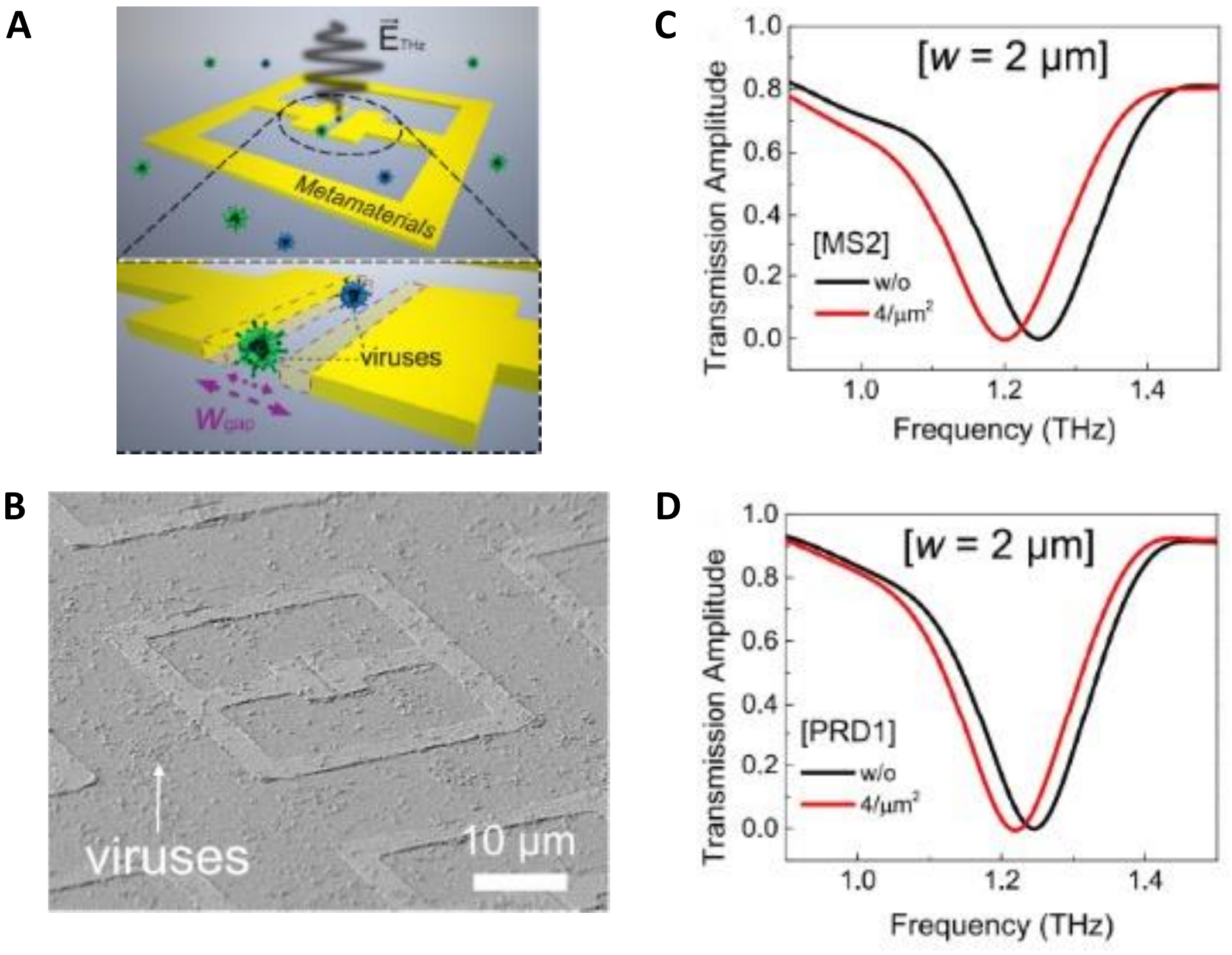}
	\caption{Detection of virus samples using THz nanogap metamaterial. \textbf{(A)} Schematic of THz nanogap metamaterial used to sense virus samples. \textbf{(B)} SEM image virus samples deposited on a THz nanogap metamaterial with 200 nm gap-width (\textit{w}). THz transmission spectra of THz metamaterial ($ w=2 \mu m $) containing \textbf{(C)} PRD1 and \textbf{(D)} MS2 virus samples. Noticeable red-shifts in resonance frequency are seen describing the sensitive detection of viruses using THz nanogap metamaterial. Reprinted with permission from Ref.~\cite{Park2017}. \textcopyright~2017, Optical Society of America.{\label{fig22}}}
\end{figure*}

The effects of different substrates on biomaterial detection\cite{Park2014} were studied using THz-antenna (length = 100 $\mu$m) with 10 $ \times $ 10 slot-array each with a slot-width of 2 $\mu$m and periodicity of 200 $\mu$m (Fig.~\ref{fig19} (A - B)), was lithographed using electron-beam on two different substrates, Si (undoped) and quartz, to sense yeast samples. Using FFT (Fast Fourier Transform) algorithm, the normalized THz-transmission amplitudes for THz slot-antenna, with (an average of 50 yeast molecules, or N$_{av}=50$) and without yeast sample was analyzed (Fig.~\ref{fig19} (C)). From the figure, an evident 9 GHz red-shift was observed in the resonance frequency. To compare the effects of different substrate, the normalized THz-transmission amplitudes were measured for Si- and quartz-substrate THz antennas as shown in Fig.~\ref{fig19} (D - E). Comparing the measured amplitudes, quartz showed stronger red-shifts compared to Si-substrate (Fig.~\ref{fig19} (F)), showing 1/$\epsilon_{eff}$ dependence of the resonant frequency-shift ($\Delta$f/f$_{0}$), where $ \epsilon_{eff}=n_{eff}^{2}$ is known as effective dielectric constant. Hence it can be concluded that by using low dielectric constant substrate, the sensitivity of a THz nanoantenna can be enhanced.

THz nanoantennas are also selective in nature.\cite{Lee2015} This selective detection was shown using a nanoantenna of length 35 $\mu$m having a resonance frequency of 1.7 THz. This antenna was specially designed to differentiate between fructose and D-glucose, and is also known as fructose antenna (Fig.~\ref{fig20} (B)). These molecules were discriminated from their measured transmittances using the fructose antenna as shown in Fig.~\ref{fig20} (C - D). Using the same nanoantenna, the sugars and low-concentrated artificial sweeteners (acesulfame K and aspartame) contained in beverages and dietary sodas, respectively, of popular brands were detected. The detection of these artificial sweeteners is important as these are recently found to be addictive and toxic in nature. The different transmittances of sugars contained in sweetened beverages of various brands in the THz range (0.5-2.5 THz) are shown in Fig.~\ref{fig20} (E). The low transmittances seen in the figure are due to the two artificially added sweeteners whose THz-absorptions is shown in Fig.~\ref{fig20} (F). Hence, nanoantennas can provide highly sensitive detection even at low molecular concentration.

Recently, Lee et al.\cite{Lee2017} studied the THz-transmittances of various virus samples (H1N1, H5N2, AND H9N2) using two THz nanoantennas: i) multi-resonance nanoantenna (with resonance frequencies of 0.63, 0.93 and 1.31 THz), and ii) single-resonance nanoantenna (with resonance frequency of 1.4 THz), which are shown in Fig.~\ref{fig21} (B - E). Similar work has been reported by Park et al.\cite{Park2017} in which they studied the transmission amplitude in both presence and absence of virus samples (PRD1 and MS2 viruses) using a THz nanogap metamaterial (Fig.~\ref{fig22}).

\section{Conclusion and Future Outlook}
THz waves are less penetrating and non-ionising electromagnetic waves. These waves are also known as 'sub-millimeter waves' and hence, is widely used in the fields of astronomy and spectroscopy. Since the description of electromagnetic interactions with metal is elucidated by using Maxwell\textquotesingle s equations, researchers are very much fascinated to study the electric field localization and other plasmonic effects in various metallic nanostructures. Numerous investigations reported that the electromagnetic interactions with metal film increase the mobility of charges in the film, tending it to move towards the metal edges, leading to the enhancement of the electric field inside different nanogap structures. To study the light-matter interactions, researchers are fascinated towards THz waves as they can squeeze through sub-nanometer metallic gaps (THz nanoantennas) and show non-linear optical responses. These nanoantennas are further integrated with different materials to study its novel plasmonic properties. Recently, successful fabrication of graphene-integrated plasmonic system\cite{Gao2014,Shi2014,Jiang2017,Zhao2017,Gu2013,Zhao2015} has boosted further investigations on their plasmonic properties because graphene is atomically thin and is electrically tunable.

As discussed earlier in this review, THz nanoantennas have played a crucial role as biosensors for ultra-sensitive detection of various biomolecules and biomaterials, which is our prime focus in this review. We have come across various literatures in which we have seen the development of nanoantennas, becoming more sensitive and selective. Due to its enhanced detection in low molecular concentrations, in future nanoantennas can be used to monitor blood sugar level.\cite{Lee2015} As an emerging future technology, nanoantennas can be used for early detection of cancerous tumours\cite{Conde2014,Huang2010} which will lead to the development of effective cancer treatment. Few recent works have shown that biosensing based on surface plasmon resonance (SPR) has proved to be effective in label-free tumor detection\cite{Abbas2011,Yanase2014,Helmerhorst2012,Dodson2015} and can be used to detect a single molecule of an early stage tumor. Currently, research and development are in progress in making a nanoantenna based SPR biosensor which will be able to detect cancer cells at an early stage.

\section{Acknowledgement}
The authors thanks Birla Institute of Technology, Mesra, Ranchi for providing research facilities and MHRD, government of India for support through TEQIP - III. The authors also thanks Pawan Kumar Dubey, Akriti Raj, Sameer Kumar Tiwari, Kamana Mishra, Priyanshi Srivastava, Dhruv Sood and Devotosh Ganguly for useful discussions and proof reading of the manuscript.
 
\section{Conflict of Interest}
The authors declares no conflict of interest.

\section{Keywords}


\begin{thebibliography}{248}%
	\makeatletter
	\providecommand \@ifxundefined [1]{%
		\@ifx{#1\undefined}
	}%
	\providecommand \@ifnum [1]{%
		\ifnum #1\expandafter \@firstoftwo
		\else \expandafter \@secondoftwo
		\fi
	}%
	\providecommand \@ifx [1]{%
		\ifx #1\expandafter \@firstoftwo
		\else \expandafter \@secondoftwo
		\fi
	}%
	\providecommand \natexlab [1]{#1}%
	\providecommand \enquote  [1]{``#1''}%
	\providecommand \bibnamefont  [1]{#1}%
	\providecommand \bibfnamefont [1]{#1}%
	\providecommand \citenamefont [1]{#1}%
	\providecommand \href@noop [0]{\@secondoftwo}%
	\providecommand \href [0]{\begingroup \@sanitize@url \@href}%
	\providecommand \@href[1]{\@@startlink{#1}\@@href}%
	\providecommand \@@href[1]{\endgroup#1\@@endlink}%
	\providecommand \@sanitize@url [0]{\catcode `\\12\catcode `\$12\catcode
		`\&12\catcode `\#12\catcode `\^12\catcode `\_12\catcode `\%12\relax}%
	\providecommand \@@startlink[1]{}%
	\providecommand \@@endlink[0]{}%
	\providecommand \url  [0]{\begingroup\@sanitize@url \@url }%
	\providecommand \@url [1]{\endgroup\@href {#1}{\urlprefix }}%
	\providecommand \urlprefix  [0]{URL }%
	\providecommand \Eprint [0]{\href }%
	\providecommand \doibase [0]{http://dx.doi.org/}%
	\providecommand \selectlanguage [0]{\@gobble}%
	\providecommand \bibinfo  [0]{\@secondoftwo}%
	\providecommand \bibfield  [0]{\@secondoftwo}%
	\providecommand \translation [1]{[#1]}%
	\providecommand \BibitemOpen [0]{}%
	\providecommand \bibitemStop [0]{}%
	\providecommand \bibitemNoStop [0]{.\EOS\space}%
	\providecommand \EOS [0]{\spacefactor3000\relax}%
	\providecommand \BibitemShut  [1]{\csname bibitem#1\endcsname}%
	\let\auto@bib@innerbib\@empty
	\bibitem [{\citenamefont {Bharadwaj}\ \emph {et~al.}(2009)\citenamefont
		{Bharadwaj}, \citenamefont {Deutsch},\ and\ \citenamefont
		{Novotny}}]{Bharadwaj2009}%
	\BibitemOpen
	\bibfield  {author} {\bibinfo {author} {\bibfnamefont {P.}~\bibnamefont
			{Bharadwaj}}, \bibinfo {author} {\bibfnamefont {B.}~\bibnamefont {Deutsch}},
		\ and\ \bibinfo {author} {\bibfnamefont {L.}~\bibnamefont {Novotny}},\
	}\bibfield  {title} {\enquote {\bibinfo {title} {Optical antennas},}\ }\href
	{\doibase 10.1364/aop.1.000438} {\bibfield  {journal} {\bibinfo  {journal}
			{Advances in Optics and Photonics}\ }\textbf {\bibinfo {volume} {1}},\
		\bibinfo {pages} {438} (\bibinfo {year} {2009})}\BibitemShut {NoStop}%
	\bibitem [{\citenamefont {Muhlschlegel}(2005)}]{Muhlschlegel2005}%
	\BibitemOpen
	\bibfield  {author} {\bibinfo {author} {\bibfnamefont {P.}~\bibnamefont
			{Muhlschlegel}},\ }\bibfield  {title} {\enquote {\bibinfo {title} {Resonant
				optical antennas},}\ }\href {\doibase 10.1126/science.1111886} {\bibfield
		{journal} {\bibinfo  {journal} {Science}\ }\textbf {\bibinfo {volume}
			{308}},\ \bibinfo {pages} {1607} (\bibinfo {year} {2005})}\BibitemShut
	{NoStop}%
	\bibitem [{\citenamefont {Alda}\ \emph {et~al.}(2005)\citenamefont {Alda},
		\citenamefont {Rico-Garc{\'{\i}}a}, \citenamefont {L{\'{o}}pez-Alonso},\ and\
		\citenamefont {Boreman}}]{Alda2005}%
	\BibitemOpen
	\bibfield  {author} {\bibinfo {author} {\bibfnamefont {J.}~\bibnamefont
			{Alda}}, \bibinfo {author} {\bibfnamefont {J.~M.}\ \bibnamefont
			{Rico-Garc{\'{\i}}a}}, \bibinfo {author} {\bibfnamefont {J.~M.}\ \bibnamefont
			{L{\'{o}}pez-Alonso}}, \ and\ \bibinfo {author} {\bibfnamefont
			{G.}~\bibnamefont {Boreman}},\ }\bibfield  {title} {\enquote {\bibinfo
			{title} {Optical antennas for nano-photonic applications},}\ }\href {\doibase
		10.1088/0957-4484/16/5/017} {\bibfield  {journal} {\bibinfo  {journal}
			{Nanotechnology}\ }\textbf {\bibinfo {volume} {16}},\ \bibinfo {pages} {S230}
		(\bibinfo {year} {2005})}\BibitemShut {NoStop}%
	\bibitem [{\citenamefont {Ross}\ and\ \citenamefont {Lee}(2009)}]{Ross2009}%
	\BibitemOpen
	\bibfield  {author} {\bibinfo {author} {\bibfnamefont {B.~M.}\ \bibnamefont
			{Ross}}\ and\ \bibinfo {author} {\bibfnamefont {L.~P.}\ \bibnamefont {Lee}},\
	}\bibfield  {title} {\enquote {\bibinfo {title} {Comparison of near- and
				far-field measures for plasmon resonance of metallic nanoparticles},}\ }\href
	{\doibase 10.1364/ol.34.000896} {\bibfield  {journal} {\bibinfo  {journal}
			{Optics Letters}\ }\textbf {\bibinfo {volume} {34}},\ \bibinfo {pages} {896}
		(\bibinfo {year} {2009})}\BibitemShut {NoStop}%
	\bibitem [{\citenamefont {Al{\`{u}}}\ and\ \citenamefont
		{Engheta}(2008)}]{Alu2008}%
	\BibitemOpen
	\bibfield  {author} {\bibinfo {author} {\bibfnamefont {A.}~\bibnamefont
			{Al{\`{u}}}}\ and\ \bibinfo {author} {\bibfnamefont {N.}~\bibnamefont
			{Engheta}},\ }\bibfield  {title} {\enquote {\bibinfo {title} {Input
				impedance, nanocircuit loading, and radiation tuning of optical
				nanoantennas},}\ }\href {\doibase 10.1103/physrevlett.101.043901} {\bibfield
		{journal} {\bibinfo  {journal} {Physical Review Letters}\ }\textbf {\bibinfo
			{volume} {101}} (\bibinfo {year} {2008}),\
		10.1103/physrevlett.101.043901}\BibitemShut {NoStop}%
	\bibitem [{\citenamefont {Novotny}(2007)}]{Novotny2007}%
	\BibitemOpen
	\bibfield  {author} {\bibinfo {author} {\bibfnamefont {L.}~\bibnamefont
			{Novotny}},\ }\bibfield  {title} {\enquote {\bibinfo {title} {Effective
				wavelength scaling for optical antennas},}\ }\href {\doibase
		10.1103/physrevlett.98.266802} {\bibfield  {journal} {\bibinfo  {journal}
			{Physical Review Letters}\ }\textbf {\bibinfo {volume} {98}} (\bibinfo {year}
		{2007}),\ 10.1103/physrevlett.98.266802}\BibitemShut {NoStop}%
	\bibitem [{\citenamefont {Kneipp}\ \emph {et~al.}(1997)\citenamefont {Kneipp},
		\citenamefont {Wang}, \citenamefont {Kneipp}, \citenamefont {Perelman},
		\citenamefont {Itzkan}, \citenamefont {Dasari},\ and\ \citenamefont
		{Feld}}]{Kneipp1997}%
	\BibitemOpen
	\bibfield  {author} {\bibinfo {author} {\bibfnamefont {K.}~\bibnamefont
			{Kneipp}}, \bibinfo {author} {\bibfnamefont {Y.}~\bibnamefont {Wang}},
		\bibinfo {author} {\bibfnamefont {H.}~\bibnamefont {Kneipp}}, \bibinfo
		{author} {\bibfnamefont {L.~T.}\ \bibnamefont {Perelman}}, \bibinfo {author}
		{\bibfnamefont {I.}~\bibnamefont {Itzkan}}, \bibinfo {author} {\bibfnamefont
			{R.~R.}\ \bibnamefont {Dasari}}, \ and\ \bibinfo {author} {\bibfnamefont
			{M.~S.}\ \bibnamefont {Feld}},\ }\bibfield  {title} {\enquote {\bibinfo
			{title} {Single molecule detection using surface-enhanced raman scattering
				({SERS})},}\ }\href {\doibase 10.1103/physrevlett.78.1667} {\bibfield
		{journal} {\bibinfo  {journal} {Physical Review Letters}\ }\textbf {\bibinfo
			{volume} {78}},\ \bibinfo {pages} {1667} (\bibinfo {year}
		{1997})}\BibitemShut {NoStop}%
	\bibitem [{\citenamefont {Nie}(1997)}]{Nie1997}%
	\BibitemOpen
	\bibfield  {author} {\bibinfo {author} {\bibfnamefont {S.}~\bibnamefont
			{Nie}},\ }\bibfield  {title} {\enquote {\bibinfo {title} {Probing single
				molecules and single nanoparticles by surface-enhanced raman scattering},}\
	}\href {\doibase 10.1126/science.275.5303.1102} {\bibfield  {journal}
		{\bibinfo  {journal} {Science}\ }\textbf {\bibinfo {volume} {275}},\ \bibinfo
		{pages} {1102} (\bibinfo {year} {1997})}\BibitemShut {NoStop}%
	\bibitem [{\citenamefont {Xu}\ \emph {et~al.}(1999)\citenamefont {Xu},
		\citenamefont {Bjerneld}, \citenamefont {Käll},\ and\ \citenamefont
		{Börjesson}}]{Xu1999}%
	\BibitemOpen
	\bibfield  {author} {\bibinfo {author} {\bibfnamefont {H.}~\bibnamefont
			{Xu}}, \bibinfo {author} {\bibfnamefont {E.~J.}\ \bibnamefont {Bjerneld}},
		\bibinfo {author} {\bibfnamefont {M.}~\bibnamefont {Käll}}, \ and\ \bibinfo
		{author} {\bibfnamefont {L.}~\bibnamefont {Börjesson}},\ }\bibfield  {title}
	{\enquote {\bibinfo {title} {Spectroscopy of single hemoglobin molecules by
				surface enhanced raman scattering},}\ }\href {\doibase
		10.1103/physrevlett.83.4357} {\bibfield  {journal} {\bibinfo  {journal}
			{Physical Review Letters}\ }\textbf {\bibinfo {volume} {83}},\ \bibinfo
		{pages} {4357} (\bibinfo {year} {1999})}\BibitemShut {NoStop}%
	\bibitem [{\citenamefont {Juan}\ \emph {et~al.}(2011)\citenamefont {Juan},
		\citenamefont {Righini},\ and\ \citenamefont {Quidant}}]{Juan2011}%
	\BibitemOpen
	\bibfield  {author} {\bibinfo {author} {\bibfnamefont {M.~L.}\ \bibnamefont
			{Juan}}, \bibinfo {author} {\bibfnamefont {M.}~\bibnamefont {Righini}}, \
		and\ \bibinfo {author} {\bibfnamefont {R.}~\bibnamefont {Quidant}},\
	}\bibfield  {title} {\enquote {\bibinfo {title} {Plasmon nano-optical
				tweezers},}\ }\href {\doibase 10.1038/nphoton.2011.56} {\bibfield  {journal}
		{\bibinfo  {journal} {Nature Photonics}\ }\textbf {\bibinfo {volume} {5}},\
		\bibinfo {pages} {349} (\bibinfo {year} {2011})}\BibitemShut {NoStop}%
	\bibitem [{\citenamefont {Marag{\`{o}}}\ \emph {et~al.}(2013)\citenamefont
		{Marag{\`{o}}}, \citenamefont {Jones}, \citenamefont {Gucciardi},
		\citenamefont {Volpe},\ and\ \citenamefont {Ferrari}}]{Marago2013}%
	\BibitemOpen
	\bibfield  {author} {\bibinfo {author} {\bibfnamefont {O.~M.}\ \bibnamefont
			{Marag{\`{o}}}}, \bibinfo {author} {\bibfnamefont {P.~H.}\ \bibnamefont
			{Jones}}, \bibinfo {author} {\bibfnamefont {P.~G.}\ \bibnamefont
			{Gucciardi}}, \bibinfo {author} {\bibfnamefont {G.}~\bibnamefont {Volpe}}, \
		and\ \bibinfo {author} {\bibfnamefont {A.~C.}\ \bibnamefont {Ferrari}},\
	}\bibfield  {title} {\enquote {\bibinfo {title} {Optical trapping and
				manipulation of nanostructures},}\ }\href {\doibase 10.1038/nnano.2013.208}
	{\bibfield  {journal} {\bibinfo  {journal} {Nature Nanotechnology}\ }\textbf
		{\bibinfo {volume} {8}},\ \bibinfo {pages} {807} (\bibinfo {year}
		{2013})}\BibitemShut {NoStop}%
	\bibitem [{\citenamefont {Ward}\ \emph {et~al.}(2010)\citenamefont {Ward},
		\citenamefont {Hüser}, \citenamefont {Pauly}, \citenamefont {Cuevas},\ and\
		\citenamefont {Natelson}}]{Ward2010}%
	\BibitemOpen
	\bibfield  {author} {\bibinfo {author} {\bibfnamefont {D.~R.}\ \bibnamefont
			{Ward}}, \bibinfo {author} {\bibfnamefont {F.}~\bibnamefont {Hüser}},
		\bibinfo {author} {\bibfnamefont {F.}~\bibnamefont {Pauly}}, \bibinfo
		{author} {\bibfnamefont {J.~C.}\ \bibnamefont {Cuevas}}, \ and\ \bibinfo
		{author} {\bibfnamefont {D.}~\bibnamefont {Natelson}},\ }\bibfield  {title}
	{\enquote {\bibinfo {title} {Optical rectification and field enhancement in a
				plasmonic nanogap},}\ }\href {\doibase 10.1038/nnano.2010.176} {\bibfield
		{journal} {\bibinfo  {journal} {Nature Nanotechnology}\ }\textbf {\bibinfo
			{volume} {5}},\ \bibinfo {pages} {732} (\bibinfo {year} {2010})}\BibitemShut
	{NoStop}%
	\bibitem [{\citenamefont {Kauranen}\ and\ \citenamefont
		{Zayats}(2012)}]{Kauranen2012}%
	\BibitemOpen
	\bibfield  {author} {\bibinfo {author} {\bibfnamefont {M.}~\bibnamefont
			{Kauranen}}\ and\ \bibinfo {author} {\bibfnamefont {A.~V.}\ \bibnamefont
			{Zayats}},\ }\bibfield  {title} {\enquote {\bibinfo {title} {Nonlinear
				plasmonics},}\ }\href {\doibase 10.1038/nphoton.2012.244} {\bibfield
		{journal} {\bibinfo  {journal} {Nature Photonics}\ }\textbf {\bibinfo
			{volume} {6}},\ \bibinfo {pages} {737} (\bibinfo {year} {2012})}\BibitemShut
	{NoStop}%
	\bibitem [{\citenamefont {Bahk}\ \emph {et~al.}(2014)\citenamefont {Bahk},
		\citenamefont {Ramakrishnan}, \citenamefont {Choi}, \citenamefont {Song},
		\citenamefont {Choi}, \citenamefont {Kim}, \citenamefont {Ahn}, \citenamefont
		{Kim},\ and\ \citenamefont {Planken}}]{Bahk2014}%
	\BibitemOpen
	\bibfield  {author} {\bibinfo {author} {\bibfnamefont {Y.-M.}\ \bibnamefont
			{Bahk}}, \bibinfo {author} {\bibfnamefont {G.}~\bibnamefont {Ramakrishnan}},
		\bibinfo {author} {\bibfnamefont {J.}~\bibnamefont {Choi}}, \bibinfo {author}
		{\bibfnamefont {H.}~\bibnamefont {Song}}, \bibinfo {author} {\bibfnamefont
			{G.}~\bibnamefont {Choi}}, \bibinfo {author} {\bibfnamefont {Y.~H.}\
			\bibnamefont {Kim}}, \bibinfo {author} {\bibfnamefont {K.~J.}\ \bibnamefont
			{Ahn}}, \bibinfo {author} {\bibfnamefont {D.-S.}\ \bibnamefont {Kim}}, \ and\
		\bibinfo {author} {\bibfnamefont {P.~C.~M.}\ \bibnamefont {Planken}},\
	}\bibfield  {title} {\enquote {\bibinfo {title} {Plasmon enhanced terahertz
				emission from single layer graphene},}\ }\href {\doibase 10.1021/nn5025237}
	{\bibfield  {journal} {\bibinfo  {journal} {{ACS} Nano}\ }\textbf {\bibinfo
			{volume} {8}},\ \bibinfo {pages} {9089} (\bibinfo {year} {2014})}\BibitemShut
	{NoStop}%
	\bibitem [{\citenamefont {Lassiter}\ \emph {et~al.}(2014)\citenamefont
		{Lassiter}, \citenamefont {Chen}, \citenamefont {Liu}, \citenamefont
		{Cirac{\`{\i}}}, \citenamefont {Hoang}, \citenamefont {Larouche},
		\citenamefont {Oh}, \citenamefont {Mikkelsen},\ and\ \citenamefont
		{Smith}}]{Lassiter2014}%
	\BibitemOpen
	\bibfield  {author} {\bibinfo {author} {\bibfnamefont {J.~B.}\ \bibnamefont
			{Lassiter}}, \bibinfo {author} {\bibfnamefont {X.}~\bibnamefont {Chen}},
		\bibinfo {author} {\bibfnamefont {X.}~\bibnamefont {Liu}}, \bibinfo {author}
		{\bibfnamefont {C.}~\bibnamefont {Cirac{\`{\i}}}}, \bibinfo {author}
		{\bibfnamefont {T.~B.}\ \bibnamefont {Hoang}}, \bibinfo {author}
		{\bibfnamefont {S.}~\bibnamefont {Larouche}}, \bibinfo {author}
		{\bibfnamefont {S.-H.}\ \bibnamefont {Oh}}, \bibinfo {author} {\bibfnamefont
			{M.~H.}\ \bibnamefont {Mikkelsen}}, \ and\ \bibinfo {author} {\bibfnamefont
			{D.~R.}\ \bibnamefont {Smith}},\ }\bibfield  {title} {\enquote {\bibinfo
			{title} {Third-harmonic generation enhancement by film-coupled plasmonic
				stripe resonators},}\ }\href {\doibase 10.1021/ph500276v} {\bibfield
		{journal} {\bibinfo  {journal} {{ACS} Photonics}\ }\textbf {\bibinfo {volume}
			{1}},\ \bibinfo {pages} {1212} (\bibinfo {year} {2014})}\BibitemShut
	{NoStop}%
	\bibitem [{\citenamefont {Bahk}\ \emph {et~al.}(2017)\citenamefont {Bahk},
		\citenamefont {Han}, \citenamefont {Rhie}, \citenamefont {Park},
		\citenamefont {Jeon}, \citenamefont {Park},\ and\ \citenamefont
		{Kim}}]{Bahk2017}%
	\BibitemOpen
	\bibfield  {author} {\bibinfo {author} {\bibfnamefont {Y.-M.}\ \bibnamefont
			{Bahk}}, \bibinfo {author} {\bibfnamefont {S.}~\bibnamefont {Han}}, \bibinfo
		{author} {\bibfnamefont {J.}~\bibnamefont {Rhie}}, \bibinfo {author}
		{\bibfnamefont {J.}~\bibnamefont {Park}}, \bibinfo {author} {\bibfnamefont
			{H.}~\bibnamefont {Jeon}}, \bibinfo {author} {\bibfnamefont {N.}~\bibnamefont
			{Park}}, \ and\ \bibinfo {author} {\bibfnamefont {D.-S.}\ \bibnamefont
			{Kim}},\ }\bibfield  {title} {\enquote {\bibinfo {title} {Ultimate terahertz
				field enhancement of single nanoslits},}\ }\href {\doibase
		10.1103/physrevb.95.075424} {\bibfield  {journal} {\bibinfo  {journal}
			{Physical Review B}\ }\textbf {\bibinfo {volume} {95}} (\bibinfo {year}
		{2017}),\ 10.1103/physrevb.95.075424}\BibitemShut {NoStop}%
	\bibitem [{\citenamefont {Bahk}\ \emph {et~al.}(2018)\citenamefont {Bahk},
		\citenamefont {Kim},\ and\ \citenamefont {Park}}]{Bahk2018}%
	\BibitemOpen
	\bibfield  {author} {\bibinfo {author} {\bibfnamefont {Y.-M.}\ \bibnamefont
			{Bahk}}, \bibinfo {author} {\bibfnamefont {D.-S.}\ \bibnamefont {Kim}}, \
		and\ \bibinfo {author} {\bibfnamefont {H.-R.}\ \bibnamefont {Park}},\
	}\bibfield  {title} {\enquote {\bibinfo {title} {Large-area metal gaps and
				their optical applications},}\ }\href {\doibase 10.1002/adom.201800426}
	{\bibfield  {journal} {\bibinfo  {journal} {Advanced Optical Materials}\
		}\textbf {\bibinfo {volume} {7}},\ \bibinfo {pages} {1800426} (\bibinfo
		{year} {2018})}\BibitemShut {NoStop}%
	\bibitem [{\citenamefont {Bonod}\ \emph {et~al.}(2008)\citenamefont {Bonod},
		\citenamefont {Popov}, \citenamefont {G{\'{e}}rard}, \citenamefont {Wenger},\
		and\ \citenamefont {Rigneault}}]{Bonod2008}%
	\BibitemOpen
	\bibfield  {author} {\bibinfo {author} {\bibfnamefont {N.}~\bibnamefont
			{Bonod}}, \bibinfo {author} {\bibfnamefont {E.}~\bibnamefont {Popov}},
		\bibinfo {author} {\bibfnamefont {D.}~\bibnamefont {G{\'{e}}rard}}, \bibinfo
		{author} {\bibfnamefont {J.}~\bibnamefont {Wenger}}, \ and\ \bibinfo {author}
		{\bibfnamefont {H.}~\bibnamefont {Rigneault}},\ }\bibfield  {title} {\enquote
		{\bibinfo {title} {Field enhancement in a circular aperture surrounded by a
				single channel groove},}\ }\href {\doibase 10.1364/oe.16.002276} {\bibfield
		{journal} {\bibinfo  {journal} {Optics Express}\ }\textbf {\bibinfo {volume}
			{16}},\ \bibinfo {pages} {2276} (\bibinfo {year} {2008})}\BibitemShut
	{NoStop}%
	\bibitem [{\citenamefont {Cao}\ and\ \citenamefont {Nahata}(2004)}]{Cao2004}%
	\BibitemOpen
	\bibfield  {author} {\bibinfo {author} {\bibfnamefont {H.}~\bibnamefont
			{Cao}}\ and\ \bibinfo {author} {\bibfnamefont {A.}~\bibnamefont {Nahata}},\
	}\bibfield  {title} {\enquote {\bibinfo {title} {Resonantly enhanced
				transmission of terahertz radiation through a periodic array of subwavelength
				apertures},}\ }\href {\doibase 10.1364/opex.12.001004} {\bibfield  {journal}
		{\bibinfo  {journal} {Optics Express}\ }\textbf {\bibinfo {volume} {12}},\
		\bibinfo {pages} {1004} (\bibinfo {year} {2004})}\BibitemShut {NoStop}%
	\bibitem [{\citenamefont {Chen}\ \emph {et~al.}(2006)\citenamefont {Chen},
		\citenamefont {Padilla}, \citenamefont {Zide}, \citenamefont {Gossard},
		\citenamefont {Taylor},\ and\ \citenamefont {Averitt}}]{Chen2006}%
	\BibitemOpen
	\bibfield  {author} {\bibinfo {author} {\bibfnamefont {H.-T.}\ \bibnamefont
			{Chen}}, \bibinfo {author} {\bibfnamefont {W.~J.}\ \bibnamefont {Padilla}},
		\bibinfo {author} {\bibfnamefont {J.~M.~O.}\ \bibnamefont {Zide}}, \bibinfo
		{author} {\bibfnamefont {A.~C.}\ \bibnamefont {Gossard}}, \bibinfo {author}
		{\bibfnamefont {A.~J.}\ \bibnamefont {Taylor}}, \ and\ \bibinfo {author}
		{\bibfnamefont {R.~D.}\ \bibnamefont {Averitt}},\ }\bibfield  {title}
	{\enquote {\bibinfo {title} {Active terahertz metamaterial devices},}\ }\href
	{\doibase 10.1038/nature05343} {\bibfield  {journal} {\bibinfo  {journal}
			{Nature}\ }\textbf {\bibinfo {volume} {444}},\ \bibinfo {pages} {597}
		(\bibinfo {year} {2006})}\BibitemShut {NoStop}%
	\bibitem [{\citenamefont {Chen}\ \emph {et~al.}(2014)\citenamefont {Chen},
		\citenamefont {Park}, \citenamefont {Lindquist}, \citenamefont {Shaver},
		\citenamefont {Pelton},\ and\ \citenamefont {Oh}}]{Chen2014}%
	\BibitemOpen
	\bibfield  {author} {\bibinfo {author} {\bibfnamefont {X.}~\bibnamefont
			{Chen}}, \bibinfo {author} {\bibfnamefont {H.-R.}\ \bibnamefont {Park}},
		\bibinfo {author} {\bibfnamefont {N.~C.}\ \bibnamefont {Lindquist}}, \bibinfo
		{author} {\bibfnamefont {J.}~\bibnamefont {Shaver}}, \bibinfo {author}
		{\bibfnamefont {M.}~\bibnamefont {Pelton}}, \ and\ \bibinfo {author}
		{\bibfnamefont {S.-H.}\ \bibnamefont {Oh}},\ }\bibfield  {title} {\enquote
		{\bibinfo {title} {Squeezing millimeter waves through a single,
				nanometer-wide, centimeter-long slit},}\ }\href {\doibase 10.1038/srep06722}
	{\bibfield  {journal} {\bibinfo  {journal} {Scientific Reports}\ }\textbf
		{\bibinfo {volume} {4}} (\bibinfo {year} {2014}),\
		10.1038/srep06722}\BibitemShut {NoStop}%
	\bibitem [{\citenamefont {Jeong}\ \emph {et~al.}(2014)\citenamefont {Jeong},
		\citenamefont {Rhie}, \citenamefont {Jeon}, \citenamefont {Hwang},\ and\
		\citenamefont {Kim}}]{Jeong2014}%
	\BibitemOpen
	\bibfield  {author} {\bibinfo {author} {\bibfnamefont {J.}~\bibnamefont
			{Jeong}}, \bibinfo {author} {\bibfnamefont {J.}~\bibnamefont {Rhie}},
		\bibinfo {author} {\bibfnamefont {W.}~\bibnamefont {Jeon}}, \bibinfo {author}
		{\bibfnamefont {C.~S.}\ \bibnamefont {Hwang}}, \ and\ \bibinfo {author}
		{\bibfnamefont {D.-S.}\ \bibnamefont {Kim}},\ }\bibfield  {title} {\enquote
		{\bibinfo {title} {High-throughput fabrication of infinitely long 10 nm slit
				arrays for terahertz applications},}\ }\href {\doibase
		10.1007/s10762-014-0135-3} {\bibfield  {journal} {\bibinfo  {journal}
			{Journal of Infrared, Millimeter, and Terahertz Waves}\ }\textbf {\bibinfo
			{volume} {36}},\ \bibinfo {pages} {262} (\bibinfo {year} {2014})}\BibitemShut
	{NoStop}%
	\bibitem [{\citenamefont {Kang}\ \emph {et~al.}(2018)\citenamefont {Kang},
		\citenamefont {Kim},\ and\ \citenamefont {Seo}}]{Kang2018}%
	\BibitemOpen
	\bibfield  {author} {\bibinfo {author} {\bibfnamefont {J.-H.}\ \bibnamefont
			{Kang}}, \bibinfo {author} {\bibfnamefont {D.-S.}\ \bibnamefont {Kim}}, \
		and\ \bibinfo {author} {\bibfnamefont {M.}~\bibnamefont {Seo}},\ }\bibfield
	{title} {\enquote {\bibinfo {title} {Terahertz wave interaction with metallic
				nanostructures},}\ }\href {\doibase 10.1515/nanoph-2017-0093} {\bibfield
		{journal} {\bibinfo  {journal} {Nanophotonics}\ }\textbf {\bibinfo {volume}
			{7}},\ \bibinfo {pages} {763} (\bibinfo {year} {2018})}\BibitemShut {NoStop}%
	\bibitem [{\citenamefont {Kim}\ \emph {et~al.}(2015)\citenamefont {Kim},
		\citenamefont {Kang}, \citenamefont {Park}, \citenamefont {Bahk},
		\citenamefont {Kim}, \citenamefont {Rhie}, \citenamefont {Jeon},
		\citenamefont {Rotermund},\ and\ \citenamefont {Kim}}]{Kim2015}%
	\BibitemOpen
	\bibfield  {author} {\bibinfo {author} {\bibfnamefont {J.-Y.}\ \bibnamefont
			{Kim}}, \bibinfo {author} {\bibfnamefont {B.~J.}\ \bibnamefont {Kang}},
		\bibinfo {author} {\bibfnamefont {J.}~\bibnamefont {Park}}, \bibinfo {author}
		{\bibfnamefont {Y.-M.}\ \bibnamefont {Bahk}}, \bibinfo {author}
		{\bibfnamefont {W.~T.}\ \bibnamefont {Kim}}, \bibinfo {author} {\bibfnamefont
			{J.}~\bibnamefont {Rhie}}, \bibinfo {author} {\bibfnamefont {H.}~\bibnamefont
			{Jeon}}, \bibinfo {author} {\bibfnamefont {F.}~\bibnamefont {Rotermund}}, \
		and\ \bibinfo {author} {\bibfnamefont {D.-S.}\ \bibnamefont {Kim}},\
	}\bibfield  {title} {\enquote {\bibinfo {title} {Terahertz quantum plasmonics
				of nanoslot antennas in nonlinear regime},}\ }\href {\doibase
		10.1021/acs.nanolett.5b02505} {\bibfield  {journal} {\bibinfo  {journal}
			{Nano Letters}\ }\textbf {\bibinfo {volume} {15}},\ \bibinfo {pages} {6683}
		(\bibinfo {year} {2015})}\BibitemShut {NoStop}%
	\bibitem [{\citenamefont {Kim}\ \emph {et~al.}(2017)\citenamefont {Kim},
		\citenamefont {In}, \citenamefont {Lee}, \citenamefont {Rhie}, \citenamefont
		{Jeong}, \citenamefont {Kim},\ and\ \citenamefont {Park}}]{Kim2017}%
	\BibitemOpen
	\bibfield  {author} {\bibinfo {author} {\bibfnamefont {N.}~\bibnamefont
			{Kim}}, \bibinfo {author} {\bibfnamefont {S.}~\bibnamefont {In}}, \bibinfo
		{author} {\bibfnamefont {D.}~\bibnamefont {Lee}}, \bibinfo {author}
		{\bibfnamefont {J.}~\bibnamefont {Rhie}}, \bibinfo {author} {\bibfnamefont
			{J.}~\bibnamefont {Jeong}}, \bibinfo {author} {\bibfnamefont {D.-S.}\
			\bibnamefont {Kim}}, \ and\ \bibinfo {author} {\bibfnamefont
			{N.}~\bibnamefont {Park}},\ }\bibfield  {title} {\enquote {\bibinfo {title}
			{Colossal terahertz field enhancement using split-ring resonators with a
				sub-10 nm gap},}\ }\href {\doibase 10.1021/acsphotonics.7b00627} {\bibfield
		{journal} {\bibinfo  {journal} {{ACS} Photonics}\ }\textbf {\bibinfo {volume}
			{5}},\ \bibinfo {pages} {278} (\bibinfo {year} {2017})}\BibitemShut {NoStop}%
	\bibitem [{\citenamefont {Lee}\ \emph {et~al.}(2007)\citenamefont {Lee},
		\citenamefont {Seo}, \citenamefont {Kang}, \citenamefont {Khim},
		\citenamefont {Jeoung},\ and\ \citenamefont {Kim}}]{Lee2007}%
	\BibitemOpen
	\bibfield  {author} {\bibinfo {author} {\bibfnamefont {J.~W.}\ \bibnamefont
			{Lee}}, \bibinfo {author} {\bibfnamefont {M.~A.}\ \bibnamefont {Seo}},
		\bibinfo {author} {\bibfnamefont {D.~H.}\ \bibnamefont {Kang}}, \bibinfo
		{author} {\bibfnamefont {K.~S.}\ \bibnamefont {Khim}}, \bibinfo {author}
		{\bibfnamefont {S.~C.}\ \bibnamefont {Jeoung}}, \ and\ \bibinfo {author}
		{\bibfnamefont {D.~S.}\ \bibnamefont {Kim}},\ }\bibfield  {title} {\enquote
		{\bibinfo {title} {Terahertz electromagnetic wave transmission through random
				arrays of single rectangular holes and slits in thin metallic sheets},}\
	}\href {\doibase 10.1103/physrevlett.99.137401} {\bibfield  {journal}
		{\bibinfo  {journal} {Physical Review Letters}\ }\textbf {\bibinfo {volume}
			{99}} (\bibinfo {year} {2007}),\ 10.1103/physrevlett.99.137401}\BibitemShut
	{NoStop}%
	\bibitem [{\citenamefont {Moser}\ \emph {et~al.}(2005)\citenamefont {Moser},
		\citenamefont {Casse}, \citenamefont {Wilhelmi},\ and\ \citenamefont
		{Saw}}]{Moser2005}%
	\BibitemOpen
	\bibfield  {author} {\bibinfo {author} {\bibfnamefont {H.~O.}\ \bibnamefont
			{Moser}}, \bibinfo {author} {\bibfnamefont {B.~D.~F.}\ \bibnamefont {Casse}},
		\bibinfo {author} {\bibfnamefont {O.}~\bibnamefont {Wilhelmi}}, \ and\
		\bibinfo {author} {\bibfnamefont {B.~T.}\ \bibnamefont {Saw}},\ }\bibfield
	{title} {\enquote {\bibinfo {title} {Terahertz response of a microfabricated
				rod{\textendash}split-ring-resonator electromagnetic metamaterial},}\ }\href
	{\doibase 10.1103/physrevlett.94.063901} {\bibfield  {journal} {\bibinfo
			{journal} {Physical Review Letters}\ }\textbf {\bibinfo {volume} {94}}
		(\bibinfo {year} {2005}),\ 10.1103/physrevlett.94.063901}\BibitemShut
	{NoStop}%
	\bibitem [{\citenamefont {Novitsky}\ \emph {et~al.}(2012)\citenamefont
		{Novitsky}, \citenamefont {Ivinskaya}, \citenamefont {Zalkovskij},
		\citenamefont {Malureanu}, \citenamefont {Jepsen},\ and\ \citenamefont
		{Lavrinenko}}]{Novitsky2012}%
	\BibitemOpen
	\bibfield  {author} {\bibinfo {author} {\bibfnamefont {A.}~\bibnamefont
			{Novitsky}}, \bibinfo {author} {\bibfnamefont {A.~M.}\ \bibnamefont
			{Ivinskaya}}, \bibinfo {author} {\bibfnamefont {M.}~\bibnamefont
			{Zalkovskij}}, \bibinfo {author} {\bibfnamefont {R.}~\bibnamefont
			{Malureanu}}, \bibinfo {author} {\bibfnamefont {P.~U.}\ \bibnamefont
			{Jepsen}}, \ and\ \bibinfo {author} {\bibfnamefont {A.~V.}\ \bibnamefont
			{Lavrinenko}},\ }\bibfield  {title} {\enquote {\bibinfo {title} {Non-resonant
				terahertz field enhancement in periodically arranged nanoslits},}\ }\href
	{\doibase 10.1063/1.4757024} {\bibfield  {journal} {\bibinfo  {journal}
			{Journal of Applied Physics}\ }\textbf {\bibinfo {volume} {112}},\ \bibinfo
		{pages} {074318} (\bibinfo {year} {2012})}\BibitemShut {NoStop}%
	\bibitem [{\citenamefont {Novitsky}\ \emph {et~al.}(2011)\citenamefont
		{Novitsky}, \citenamefont {Zalkovskij}, \citenamefont {Malureanu},\ and\
		\citenamefont {Lavrinenko}}]{Novitsky2011}%
	\BibitemOpen
	\bibfield  {author} {\bibinfo {author} {\bibfnamefont {A.}~\bibnamefont
			{Novitsky}}, \bibinfo {author} {\bibfnamefont {M.}~\bibnamefont
			{Zalkovskij}}, \bibinfo {author} {\bibfnamefont {R.}~\bibnamefont
			{Malureanu}}, \ and\ \bibinfo {author} {\bibfnamefont {A.}~\bibnamefont
			{Lavrinenko}},\ }\bibfield  {title} {\enquote {\bibinfo {title} {Microscopic
				model of the {THz} field enhancement in a metal nanoslit},}\ }\href {\doibase
		10.1016/j.optcom.2011.08.019} {\bibfield  {journal} {\bibinfo  {journal}
			{Optics Communications}\ }\textbf {\bibinfo {volume} {284}},\ \bibinfo
		{pages} {5495} (\bibinfo {year} {2011})}\BibitemShut {NoStop}%
	\bibitem [{\citenamefont {Park}\ \emph {et~al.}(2010)\citenamefont {Park},
		\citenamefont {Park}, \citenamefont {Kim}, \citenamefont {Kyoung},
		\citenamefont {Seo}, \citenamefont {Park}, \citenamefont {Ahn}, \citenamefont
		{Ahn},\ and\ \citenamefont {Kim}}]{Park2010}%
	\BibitemOpen
	\bibfield  {author} {\bibinfo {author} {\bibfnamefont {H.~R.}\ \bibnamefont
			{Park}}, \bibinfo {author} {\bibfnamefont {Y.~M.}\ \bibnamefont {Park}},
		\bibinfo {author} {\bibfnamefont {H.~S.}\ \bibnamefont {Kim}}, \bibinfo
		{author} {\bibfnamefont {J.~S.}\ \bibnamefont {Kyoung}}, \bibinfo {author}
		{\bibfnamefont {M.~A.}\ \bibnamefont {Seo}}, \bibinfo {author} {\bibfnamefont
			{D.~J.}\ \bibnamefont {Park}}, \bibinfo {author} {\bibfnamefont {Y.~H.}\
			\bibnamefont {Ahn}}, \bibinfo {author} {\bibfnamefont {K.~J.}\ \bibnamefont
			{Ahn}}, \ and\ \bibinfo {author} {\bibfnamefont {D.~S.}\ \bibnamefont
			{Kim}},\ }\bibfield  {title} {\enquote {\bibinfo {title} {Terahertz
				nanoresonators: Giant field enhancement and ultrabroadband performance},}\
	}\href {\doibase 10.1063/1.3368690} {\bibfield  {journal} {\bibinfo
			{journal} {Applied Physics Letters}\ }\textbf {\bibinfo {volume} {96}},\
		\bibinfo {pages} {121106} (\bibinfo {year} {2010})}\BibitemShut {NoStop}%
	\bibitem [{\citenamefont {Qu}\ \emph {et~al.}(2004)\citenamefont {Qu},
		\citenamefont {Grischkowsky},\ and\ \citenamefont {Zhang}}]{Qu2004a}%
	\BibitemOpen
	\bibfield  {author} {\bibinfo {author} {\bibfnamefont {D.}~\bibnamefont
			{Qu}}, \bibinfo {author} {\bibfnamefont {D.}~\bibnamefont {Grischkowsky}}, \
		and\ \bibinfo {author} {\bibfnamefont {W.}~\bibnamefont {Zhang}},\ }\bibfield
	{title} {\enquote {\bibinfo {title} {Terahertz transmission properties of
				thin, subwavelength metallic hole arrays},}\ }\href {\doibase
		10.1364/ol.29.000896} {\bibfield  {journal} {\bibinfo  {journal} {Optics
				Letters}\ }\textbf {\bibinfo {volume} {29}},\ \bibinfo {pages} {896}
		(\bibinfo {year} {2004})}\BibitemShut {NoStop}%
	\bibitem [{\citenamefont {Rivas}\ \emph {et~al.}(2003)\citenamefont {Rivas},
		\citenamefont {Schotsch}, \citenamefont {Bolivar},\ and\ \citenamefont
		{Kurz}}]{Rivas2003}%
	\BibitemOpen
	\bibfield  {author} {\bibinfo {author} {\bibfnamefont {J.~G.}\ \bibnamefont
			{Rivas}}, \bibinfo {author} {\bibfnamefont {C.}~\bibnamefont {Schotsch}},
		\bibinfo {author} {\bibfnamefont {P.~H.}\ \bibnamefont {Bolivar}}, \ and\
		\bibinfo {author} {\bibfnamefont {H.}~\bibnamefont {Kurz}},\ }\bibfield
	{title} {\enquote {\bibinfo {title} {Enhanced transmission of {THz} radiation
				through subwavelength holes},}\ }\href {\doibase 10.1103/physrevb.68.201306}
	{\bibfield  {journal} {\bibinfo  {journal} {Physical Review B}\ }\textbf
		{\bibinfo {volume} {68}} (\bibinfo {year} {2003}),\
		10.1103/physrevb.68.201306}\BibitemShut {NoStop}%
	\bibitem [{\citenamefont {Seo}\ \emph {et~al.}(2008)\citenamefont {Seo},
		\citenamefont {Adam}, \citenamefont {Kang}, \citenamefont {Lee},
		\citenamefont {Ahn}, \citenamefont {Park}, \citenamefont {Planken},\ and\
		\citenamefont {Kim}}]{Seo2008}%
	\BibitemOpen
	\bibfield  {author} {\bibinfo {author} {\bibfnamefont {M.~A.}\ \bibnamefont
			{Seo}}, \bibinfo {author} {\bibfnamefont {A.~J.~L.}\ \bibnamefont {Adam}},
		\bibinfo {author} {\bibfnamefont {J.~H.}\ \bibnamefont {Kang}}, \bibinfo
		{author} {\bibfnamefont {J.~W.}\ \bibnamefont {Lee}}, \bibinfo {author}
		{\bibfnamefont {K.~J.}\ \bibnamefont {Ahn}}, \bibinfo {author} {\bibfnamefont
			{Q.~H.}\ \bibnamefont {Park}}, \bibinfo {author} {\bibfnamefont {P.~C.~M.}\
			\bibnamefont {Planken}}, \ and\ \bibinfo {author} {\bibfnamefont {D.~S.}\
			\bibnamefont {Kim}},\ }\bibfield  {title} {\enquote {\bibinfo {title} {Near
				field imaging of terahertz focusing onto rectangular apertures},}\ }\href
	{\doibase 10.1364/oe.16.020484} {\bibfield  {journal} {\bibinfo  {journal}
			{Optics Express}\ }\textbf {\bibinfo {volume} {16}},\ \bibinfo {pages}
		{20484} (\bibinfo {year} {2008})}\BibitemShut {NoStop}%
	\bibitem [{\citenamefont {Seo}\ \emph {et~al.}(2009)\citenamefont {Seo},
		\citenamefont {Park}, \citenamefont {Koo}, \citenamefont {Park},
		\citenamefont {Kang}, \citenamefont {Suwal}, \citenamefont {Choi},
		\citenamefont {Planken}, \citenamefont {Park}, \citenamefont {Park},
		\citenamefont {Park},\ and\ \citenamefont {Kim}}]{Seo2009}%
	\BibitemOpen
	\bibfield  {author} {\bibinfo {author} {\bibfnamefont {M.~A.}\ \bibnamefont
			{Seo}}, \bibinfo {author} {\bibfnamefont {H.~R.}\ \bibnamefont {Park}},
		\bibinfo {author} {\bibfnamefont {S.~M.}\ \bibnamefont {Koo}}, \bibinfo
		{author} {\bibfnamefont {D.~J.}\ \bibnamefont {Park}}, \bibinfo {author}
		{\bibfnamefont {J.~H.}\ \bibnamefont {Kang}}, \bibinfo {author}
		{\bibfnamefont {O.~K.}\ \bibnamefont {Suwal}}, \bibinfo {author}
		{\bibfnamefont {S.~S.}\ \bibnamefont {Choi}}, \bibinfo {author}
		{\bibfnamefont {P.~C.~M.}\ \bibnamefont {Planken}}, \bibinfo {author}
		{\bibfnamefont {G.~S.}\ \bibnamefont {Park}}, \bibinfo {author}
		{\bibfnamefont {N.~K.}\ \bibnamefont {Park}}, \bibinfo {author}
		{\bibfnamefont {Q.~H.}\ \bibnamefont {Park}}, \ and\ \bibinfo {author}
		{\bibfnamefont {D.~S.}\ \bibnamefont {Kim}},\ }\bibfield  {title} {\enquote
		{\bibinfo {title} {Terahertz field enhancement by a metallic nano slit
				operating beyond the skin-depth limit},}\ }\href {\doibase
		10.1038/nphoton.2009.22} {\bibfield  {journal} {\bibinfo  {journal} {Nature
				Photonics}\ }\textbf {\bibinfo {volume} {3}},\ \bibinfo {pages} {152}
		(\bibinfo {year} {2009})}\BibitemShut {NoStop}%
	\bibitem [{\citenamefont {Shalaby}\ \emph {et~al.}(2011)\citenamefont
		{Shalaby}, \citenamefont {Merbold}, \citenamefont {Peccianti}, \citenamefont
		{Razzari}, \citenamefont {Sharma}, \citenamefont {Ozaki}, \citenamefont
		{Morandotti}, \citenamefont {Feurer}, \citenamefont {Weber}, \citenamefont
		{Heyderman}, \citenamefont {Patterson},\ and\ \citenamefont
		{Sigg}}]{Shalaby2011}%
	\BibitemOpen
	\bibfield  {author} {\bibinfo {author} {\bibfnamefont {M.}~\bibnamefont
			{Shalaby}}, \bibinfo {author} {\bibfnamefont {H.}~\bibnamefont {Merbold}},
		\bibinfo {author} {\bibfnamefont {M.}~\bibnamefont {Peccianti}}, \bibinfo
		{author} {\bibfnamefont {L.}~\bibnamefont {Razzari}}, \bibinfo {author}
		{\bibfnamefont {G.}~\bibnamefont {Sharma}}, \bibinfo {author} {\bibfnamefont
			{T.}~\bibnamefont {Ozaki}}, \bibinfo {author} {\bibfnamefont
			{R.}~\bibnamefont {Morandotti}}, \bibinfo {author} {\bibfnamefont
			{T.}~\bibnamefont {Feurer}}, \bibinfo {author} {\bibfnamefont
			{A.}~\bibnamefont {Weber}}, \bibinfo {author} {\bibfnamefont
			{L.}~\bibnamefont {Heyderman}}, \bibinfo {author} {\bibfnamefont
			{B.}~\bibnamefont {Patterson}}, \ and\ \bibinfo {author} {\bibfnamefont
			{H.}~\bibnamefont {Sigg}},\ }\bibfield  {title} {\enquote {\bibinfo {title}
			{Concurrent field enhancement and high transmission of {THz} radiation in
				nanoslit arrays},}\ }\href {\doibase 10.1063/1.3617476} {\bibfield  {journal}
		{\bibinfo  {journal} {Applied Physics Letters}\ }\textbf {\bibinfo {volume}
			{99}},\ \bibinfo {pages} {041110} (\bibinfo {year} {2011})}\BibitemShut
	{NoStop}%
	\bibitem [{\citenamefont {Toma}\ \emph {et~al.}(2014)\citenamefont {Toma},
		\citenamefont {Tuccio}, \citenamefont {Prato}, \citenamefont {Donato},
		\citenamefont {Perucchi}, \citenamefont {Pietro}, \citenamefont {Marras},
		\citenamefont {Liberale}, \citenamefont {Zaccaria}, \citenamefont {Angelis},
		\citenamefont {Manna}, \citenamefont {Lupi}, \citenamefont {Fabrizio},\ and\
		\citenamefont {Razzari}}]{Toma2014}%
	\BibitemOpen
	\bibfield  {author} {\bibinfo {author} {\bibfnamefont {A.}~\bibnamefont
			{Toma}}, \bibinfo {author} {\bibfnamefont {S.}~\bibnamefont {Tuccio}},
		\bibinfo {author} {\bibfnamefont {M.}~\bibnamefont {Prato}}, \bibinfo
		{author} {\bibfnamefont {F.~D.}\ \bibnamefont {Donato}}, \bibinfo {author}
		{\bibfnamefont {A.}~\bibnamefont {Perucchi}}, \bibinfo {author}
		{\bibfnamefont {P.~D.}\ \bibnamefont {Pietro}}, \bibinfo {author}
		{\bibfnamefont {S.}~\bibnamefont {Marras}}, \bibinfo {author} {\bibfnamefont
			{C.}~\bibnamefont {Liberale}}, \bibinfo {author} {\bibfnamefont {R.~P.}\
			\bibnamefont {Zaccaria}}, \bibinfo {author} {\bibfnamefont {F.~D.}\
			\bibnamefont {Angelis}}, \bibinfo {author} {\bibfnamefont {L.}~\bibnamefont
			{Manna}}, \bibinfo {author} {\bibfnamefont {S.}~\bibnamefont {Lupi}},
		\bibinfo {author} {\bibfnamefont {E.~D.}\ \bibnamefont {Fabrizio}}, \ and\
		\bibinfo {author} {\bibfnamefont {L.}~\bibnamefont {Razzari}},\ }\bibfield
	{title} {\enquote {\bibinfo {title} {Squeezing terahertz light into
				nanovolumes: Nanoantenna enhanced terahertz spectroscopy ({NETS}) of
				semiconductor quantum dots},}\ }\href {\doibase 10.1021/nl503705w} {\bibfield
		{journal} {\bibinfo  {journal} {Nano Letters}\ }\textbf {\bibinfo {volume}
			{15}},\ \bibinfo {pages} {386} (\bibinfo {year} {2014})}\BibitemShut
	{NoStop}%
	\bibitem [{\citenamefont {Tripathi}\ \emph {et~al.}(2016)\citenamefont
		{Tripathi}, \citenamefont {Bahk}, \citenamefont {Choi}, \citenamefont {Han},
		\citenamefont {Park},\ and\ \citenamefont {Kim}}]{Tripathi2016}%
	\BibitemOpen
	\bibfield  {author} {\bibinfo {author} {\bibfnamefont {L.-N.}\ \bibnamefont
			{Tripathi}}, \bibinfo {author} {\bibfnamefont {Y.-M.}\ \bibnamefont {Bahk}},
		\bibinfo {author} {\bibfnamefont {G.}~\bibnamefont {Choi}}, \bibinfo {author}
		{\bibfnamefont {S.}~\bibnamefont {Han}}, \bibinfo {author} {\bibfnamefont
			{N.}~\bibnamefont {Park}}, \ and\ \bibinfo {author} {\bibfnamefont {D.-S.}\
			\bibnamefont {Kim}},\ }\bibfield  {title} {\enquote {\bibinfo {title}
			{Terahertz transmission through rings of quantum dots-nanogap},}\ }\href
	{\doibase 10.7567/apex.9.032001} {\bibfield  {journal} {\bibinfo  {journal}
			{Applied Physics Express}\ }\textbf {\bibinfo {volume} {9}},\ \bibinfo
		{pages} {032001} (\bibinfo {year} {2016})}\BibitemShut {NoStop}%
	\bibitem [{\citenamefont {Werley}\ \emph {et~al.}(2012)\citenamefont {Werley},
		\citenamefont {Fan}, \citenamefont {Strikwerda}, \citenamefont {Teo},
		\citenamefont {Zhang}, \citenamefont {Averitt},\ and\ \citenamefont
		{Nelson}}]{Werley2012}%
	\BibitemOpen
	\bibfield  {author} {\bibinfo {author} {\bibfnamefont {C.~A.}\ \bibnamefont
			{Werley}}, \bibinfo {author} {\bibfnamefont {K.}~\bibnamefont {Fan}},
		\bibinfo {author} {\bibfnamefont {A.~C.}\ \bibnamefont {Strikwerda}},
		\bibinfo {author} {\bibfnamefont {S.~M.}\ \bibnamefont {Teo}}, \bibinfo
		{author} {\bibfnamefont {X.}~\bibnamefont {Zhang}}, \bibinfo {author}
		{\bibfnamefont {R.~D.}\ \bibnamefont {Averitt}}, \ and\ \bibinfo {author}
		{\bibfnamefont {K.~A.}\ \bibnamefont {Nelson}},\ }\bibfield  {title}
	{\enquote {\bibinfo {title} {Time-resolved imaging of near-fields in {THz}
				antennas and direct quantitative measurement of field enhancements},}\ }\href
	{\doibase 10.1364/oe.20.008551} {\bibfield  {journal} {\bibinfo  {journal}
			{Optics Express}\ }\textbf {\bibinfo {volume} {20}},\ \bibinfo {pages} {8551}
		(\bibinfo {year} {2012})}\BibitemShut {NoStop}%
	\bibitem [{\citenamefont {Park}\ \emph {et~al.}(2018)\citenamefont {Park},
		\citenamefont {Lee}, \citenamefont {Kang}, \citenamefont {Jeong},\ and\
		\citenamefont {Kim}}]{Park2018}%
	\BibitemOpen
	\bibfield  {author} {\bibinfo {author} {\bibfnamefont {W.}~\bibnamefont
			{Park}}, \bibinfo {author} {\bibfnamefont {Y.}~\bibnamefont {Lee}}, \bibinfo
		{author} {\bibfnamefont {T.}~\bibnamefont {Kang}}, \bibinfo {author}
		{\bibfnamefont {J.}~\bibnamefont {Jeong}}, \ and\ \bibinfo {author}
		{\bibfnamefont {D.-S.}\ \bibnamefont {Kim}},\ }\bibfield  {title} {\enquote
		{\bibinfo {title} {Terahertz-driven polymerization of resists in
				nanoantennas},}\ }\href {\doibase 10.1038/s41598-018-26214-w} {\bibfield
		{journal} {\bibinfo  {journal} {Scientific Reports}\ }\textbf {\bibinfo
			{volume} {8}} (\bibinfo {year} {2018}),\
		10.1038/s41598-018-26214-w}\BibitemShut {NoStop}%
	\bibitem [{\citenamefont {Park}\ \emph
		{et~al.}(2011{\natexlab{a}})\citenamefont {Park}, \citenamefont {Bahk},
		\citenamefont {Ahn}, \citenamefont {Park}, \citenamefont {Kim}, \citenamefont
		{Mart{\'{\i}}n-Moreno}, \citenamefont {Garc{\'{\i}}a-Vidal},\ and\
		\citenamefont {Bravo-Abad}}]{Park2011}%
	\BibitemOpen
	\bibfield  {author} {\bibinfo {author} {\bibfnamefont {H.-R.}\ \bibnamefont
			{Park}}, \bibinfo {author} {\bibfnamefont {Y.-M.}\ \bibnamefont {Bahk}},
		\bibinfo {author} {\bibfnamefont {K.~J.}\ \bibnamefont {Ahn}}, \bibinfo
		{author} {\bibfnamefont {Q.-H.}\ \bibnamefont {Park}}, \bibinfo {author}
		{\bibfnamefont {D.-S.}\ \bibnamefont {Kim}}, \bibinfo {author} {\bibfnamefont
			{L.}~\bibnamefont {Mart{\'{\i}}n-Moreno}}, \bibinfo {author} {\bibfnamefont
			{F.~J.}\ \bibnamefont {Garc{\'{\i}}a-Vidal}}, \ and\ \bibinfo {author}
		{\bibfnamefont {J.}~\bibnamefont {Bravo-Abad}},\ }\bibfield  {title}
	{\enquote {\bibinfo {title} {Controlling terahertz radiation with nanoscale
				metal barriers embedded in nano slot antennas},}\ }\href {\doibase
		10.1021/nn2031885} {\bibfield  {journal} {\bibinfo  {journal} {{ACS} Nano}\
		}\textbf {\bibinfo {volume} {5}},\ \bibinfo {pages} {8340} (\bibinfo {year}
		{2011}{\natexlab{a}})}\BibitemShut {NoStop}%
	\bibitem [{\citenamefont {Lindquist}\ \emph {et~al.}(2012)\citenamefont
		{Lindquist}, \citenamefont {Nagpal}, \citenamefont {McPeak}, \citenamefont
		{Norris},\ and\ \citenamefont {Oh}}]{Lindquist2012}%
	\BibitemOpen
	\bibfield  {author} {\bibinfo {author} {\bibfnamefont {N.~C.}\ \bibnamefont
			{Lindquist}}, \bibinfo {author} {\bibfnamefont {P.}~\bibnamefont {Nagpal}},
		\bibinfo {author} {\bibfnamefont {K.~M.}\ \bibnamefont {McPeak}}, \bibinfo
		{author} {\bibfnamefont {D.~J.}\ \bibnamefont {Norris}}, \ and\ \bibinfo
		{author} {\bibfnamefont {S.-H.}\ \bibnamefont {Oh}},\ }\bibfield  {title}
	{\enquote {\bibinfo {title} {Engineering metallic nanostructures for
				plasmonics and nanophotonics},}\ }\href {\doibase
		10.1088/0034-4885/75/3/036501} {\bibfield  {journal} {\bibinfo  {journal}
			{Reports on Progress in Physics}\ }\textbf {\bibinfo {volume} {75}},\
		\bibinfo {pages} {036501} (\bibinfo {year} {2012})}\BibitemShut {NoStop}%
	\bibitem [{\citenamefont {Hoffmann}\ \emph {et~al.}(2010)\citenamefont
		{Hoffmann}, \citenamefont {Monozon}, \citenamefont {Livshits}, \citenamefont
		{Rafailov},\ and\ \citenamefont {Turchinovich}}]{Hoffmann2010}%
	\BibitemOpen
	\bibfield  {author} {\bibinfo {author} {\bibfnamefont {M.~C.}\ \bibnamefont
			{Hoffmann}}, \bibinfo {author} {\bibfnamefont {B.~S.}\ \bibnamefont
			{Monozon}}, \bibinfo {author} {\bibfnamefont {D.}~\bibnamefont {Livshits}},
		\bibinfo {author} {\bibfnamefont {E.~U.}\ \bibnamefont {Rafailov}}, \ and\
		\bibinfo {author} {\bibfnamefont {D.}~\bibnamefont {Turchinovich}},\
	}\bibfield  {title} {\enquote {\bibinfo {title} {Terahertz electro-absorption
				effect enabling femtosecond all-optical switching in semiconductor quantum
				dots},}\ }\href {\doibase 10.1063/1.3515909} {\bibfield  {journal} {\bibinfo
			{journal} {Applied Physics Letters}\ }\textbf {\bibinfo {volume} {97}},\
		\bibinfo {pages} {231108} (\bibinfo {year} {2010})}\BibitemShut {NoStop}%
	\bibitem [{\citenamefont {Tanoto}\ \emph {et~al.}(2012)\citenamefont {Tanoto},
		\citenamefont {Teng}, \citenamefont {Wu}, \citenamefont {Sun}, \citenamefont
		{Chen}, \citenamefont {Maier}, \citenamefont {Wang}, \citenamefont {Chum},
		\citenamefont {Si}, \citenamefont {Danner},\ and\ \citenamefont
		{Chua}}]{Tanoto2012}%
	\BibitemOpen
	\bibfield  {author} {\bibinfo {author} {\bibfnamefont {H.}~\bibnamefont
			{Tanoto}}, \bibinfo {author} {\bibfnamefont {J.~H.}\ \bibnamefont {Teng}},
		\bibinfo {author} {\bibfnamefont {Q.~Y.}\ \bibnamefont {Wu}}, \bibinfo
		{author} {\bibfnamefont {M.}~\bibnamefont {Sun}}, \bibinfo {author}
		{\bibfnamefont {Z.~N.}\ \bibnamefont {Chen}}, \bibinfo {author}
		{\bibfnamefont {S.~A.}\ \bibnamefont {Maier}}, \bibinfo {author}
		{\bibfnamefont {B.}~\bibnamefont {Wang}}, \bibinfo {author} {\bibfnamefont
			{C.~C.}\ \bibnamefont {Chum}}, \bibinfo {author} {\bibfnamefont {G.~Y.}\
			\bibnamefont {Si}}, \bibinfo {author} {\bibfnamefont {A.~J.}\ \bibnamefont
			{Danner}}, \ and\ \bibinfo {author} {\bibfnamefont {S.~J.}\ \bibnamefont
			{Chua}},\ }\bibfield  {title} {\enquote {\bibinfo {title} {Greatly enhanced
				continuous-wave terahertz emission by nano-electrodes in a photoconductive
				photomixer},}\ }\href {\doibase 10.1038/nphoton.2011.322} {\bibfield
		{journal} {\bibinfo  {journal} {Nature Photonics}\ }\textbf {\bibinfo
			{volume} {6}},\ \bibinfo {pages} {121} (\bibinfo {year} {2012})}\BibitemShut
	{NoStop}%
	\bibitem [{\citenamefont {Ju}\ \emph {et~al.}(2011)\citenamefont {Ju},
		\citenamefont {Geng}, \citenamefont {Horng}, \citenamefont {Girit},
		\citenamefont {Martin}, \citenamefont {Hao}, \citenamefont {Bechtel},
		\citenamefont {Liang}, \citenamefont {Zettl}, \citenamefont {Shen},\ and\
		\citenamefont {Wang}}]{Ju2011}%
	\BibitemOpen
	\bibfield  {author} {\bibinfo {author} {\bibfnamefont {L.}~\bibnamefont
			{Ju}}, \bibinfo {author} {\bibfnamefont {B.}~\bibnamefont {Geng}}, \bibinfo
		{author} {\bibfnamefont {J.}~\bibnamefont {Horng}}, \bibinfo {author}
		{\bibfnamefont {C.}~\bibnamefont {Girit}}, \bibinfo {author} {\bibfnamefont
			{M.}~\bibnamefont {Martin}}, \bibinfo {author} {\bibfnamefont
			{Z.}~\bibnamefont {Hao}}, \bibinfo {author} {\bibfnamefont {H.~A.}\
			\bibnamefont {Bechtel}}, \bibinfo {author} {\bibfnamefont {X.}~\bibnamefont
			{Liang}}, \bibinfo {author} {\bibfnamefont {A.}~\bibnamefont {Zettl}},
		\bibinfo {author} {\bibfnamefont {Y.~R.}\ \bibnamefont {Shen}}, \ and\
		\bibinfo {author} {\bibfnamefont {F.}~\bibnamefont {Wang}},\ }\bibfield
	{title} {\enquote {\bibinfo {title} {Graphene plasmonics for tunable
				terahertz metamaterials},}\ }\href {\doibase 10.1038/nnano.2011.146}
	{\bibfield  {journal} {\bibinfo  {journal} {Nature Nanotechnology}\ }\textbf
		{\bibinfo {volume} {6}},\ \bibinfo {pages} {630} (\bibinfo {year}
		{2011})}\BibitemShut {NoStop}%
	\bibitem [{\citenamefont {Shur}(2010)}]{Shur2010}%
	\BibitemOpen
	\bibfield  {author} {\bibinfo {author} {\bibfnamefont {M.}~\bibnamefont
			{Shur}},\ }\bibfield  {title} {\enquote {\bibinfo {title} {Plasma wave
				terahertz electronics},}\ }\href {\doibase 10.1049/el.2010.8457} {\bibfield
		{journal} {\bibinfo  {journal} {Electronics Letters}\ }\textbf {\bibinfo
			{volume} {46}},\ \bibinfo {pages} {S18} (\bibinfo {year} {2010})}\BibitemShut
	{NoStop}%
	\bibitem [{\citenamefont {Upadhyaya}\ \emph {et~al.}(2008)\citenamefont
		{Upadhyaya}, \citenamefont {Pramanik},\ and\ \citenamefont
		{Bandyopadhyay}}]{Upadhyaya2008}%
	\BibitemOpen
	\bibfield  {author} {\bibinfo {author} {\bibfnamefont {P.}~\bibnamefont
			{Upadhyaya}}, \bibinfo {author} {\bibfnamefont {S.}~\bibnamefont {Pramanik}},
		\ and\ \bibinfo {author} {\bibfnamefont {S.}~\bibnamefont {Bandyopadhyay}},\
	}\bibfield  {title} {\enquote {\bibinfo {title} {Optical transitions in a
				quantum wire with spin-orbit interaction and its applications in terahertz
				electronics: Beyond zeroth-order theory},}\ }\href {\doibase
		10.1103/physrevb.77.155439} {\bibfield  {journal} {\bibinfo  {journal}
			{Physical Review B}\ }\textbf {\bibinfo {volume} {77}} (\bibinfo {year}
		{2008}),\ 10.1103/physrevb.77.155439}\BibitemShut {NoStop}%
	\bibitem [{\citenamefont {Bahramipanah}\ \emph {et~al.}(2014)\citenamefont
		{Bahramipanah}, \citenamefont {Abrishamian}, \citenamefont {Mirtaheri},\ and\
		\citenamefont {Liu}}]{Bahramipanah2014}%
	\BibitemOpen
	\bibfield  {author} {\bibinfo {author} {\bibfnamefont {M.}~\bibnamefont
			{Bahramipanah}}, \bibinfo {author} {\bibfnamefont {M.~S.}\ \bibnamefont
			{Abrishamian}}, \bibinfo {author} {\bibfnamefont {S.~A.}\ \bibnamefont
			{Mirtaheri}}, \ and\ \bibinfo {author} {\bibfnamefont {J.-M.}\ \bibnamefont
			{Liu}},\ }\bibfield  {title} {\enquote {\bibinfo {title} {Ultracompact
				plasmonic loop{\textendash}stub notch filter and sensor},}\ }\href {\doibase
		10.1016/j.snb.2013.12.084} {\bibfield  {journal} {\bibinfo  {journal}
			{Sensors and Actuators B: Chemical}\ }\textbf {\bibinfo {volume} {194}},\
		\bibinfo {pages} {311} (\bibinfo {year} {2014})}\BibitemShut {NoStop}%
	\bibitem [{\citenamefont {Mittleman}(2013)}]{Mittleman2013}%
	\BibitemOpen
	\bibfield  {author} {\bibinfo {author} {\bibfnamefont {D.~M.}\ \bibnamefont
			{Mittleman}},\ }\bibfield  {title} {\enquote {\bibinfo {title} {Frontiers in
				terahertz sources and plasmonics},}\ }\href {\doibase
		10.1038/nphoton.2013.235} {\bibfield  {journal} {\bibinfo  {journal} {Nature
				Photonics}\ }\textbf {\bibinfo {volume} {7}},\ \bibinfo {pages} {666}
		(\bibinfo {year} {2013})}\BibitemShut {NoStop}%
	\bibitem [{\citenamefont {Williams}\ \emph {et~al.}(2008)\citenamefont
		{Williams}, \citenamefont {Andrews}, \citenamefont {Maier}, \citenamefont
		{Fern{\'{a}}ndez-Dom{\'{\i}}nguez}, \citenamefont {Mart{\'{\i}}n-Moreno},\
		and\ \citenamefont {Garc{\'{\i}}a-Vidal}}]{Williams2008}%
	\BibitemOpen
	\bibfield  {author} {\bibinfo {author} {\bibfnamefont {C.~R.}\ \bibnamefont
			{Williams}}, \bibinfo {author} {\bibfnamefont {S.~R.}\ \bibnamefont
			{Andrews}}, \bibinfo {author} {\bibfnamefont {S.~A.}\ \bibnamefont {Maier}},
		\bibinfo {author} {\bibfnamefont {A.~I.}\ \bibnamefont
			{Fern{\'{a}}ndez-Dom{\'{\i}}nguez}}, \bibinfo {author} {\bibfnamefont
			{L.}~\bibnamefont {Mart{\'{\i}}n-Moreno}}, \ and\ \bibinfo {author}
		{\bibfnamefont {F.~J.}\ \bibnamefont {Garc{\'{\i}}a-Vidal}},\ }\bibfield
	{title} {\enquote {\bibinfo {title} {Highly confined guiding of terahertz
				surface plasmon polaritons on structured metal surfaces},}\ }\href {\doibase
		10.1038/nphoton.2007.301} {\bibfield  {journal} {\bibinfo  {journal} {Nature
				Photonics}\ }\textbf {\bibinfo {volume} {2}},\ \bibinfo {pages} {175}
		(\bibinfo {year} {2008})}\BibitemShut {NoStop}%
	\bibitem [{\citenamefont {Fitzgerald}\ \emph {et~al.}(2002)\citenamefont
		{Fitzgerald}, \citenamefont {Berry}, \citenamefont {Zinovev}, \citenamefont
		{Walker}, \citenamefont {Smith},\ and\ \citenamefont
		{Chamberlain}}]{Fitzgerald2002}%
	\BibitemOpen
	\bibfield  {author} {\bibinfo {author} {\bibfnamefont {A.~J.}\ \bibnamefont
			{Fitzgerald}}, \bibinfo {author} {\bibfnamefont {E.}~\bibnamefont {Berry}},
		\bibinfo {author} {\bibfnamefont {N.~N.}\ \bibnamefont {Zinovev}}, \bibinfo
		{author} {\bibfnamefont {G.~C.}\ \bibnamefont {Walker}}, \bibinfo {author}
		{\bibfnamefont {M.~A.}\ \bibnamefont {Smith}}, \ and\ \bibinfo {author}
		{\bibfnamefont {J.~M.}\ \bibnamefont {Chamberlain}},\ }\bibfield  {title}
	{\enquote {\bibinfo {title} {An introduction to medical imaging with coherent
				terahertz frequency radiation},}\ }\href {\doibase
		10.1088/0031-9155/47/7/201} {\bibfield  {journal} {\bibinfo  {journal}
			{Physics in Medicine and Biology}\ }\textbf {\bibinfo {volume} {47}},\
		\bibinfo {pages} {R67} (\bibinfo {year} {2002})}\BibitemShut {NoStop}%
	\bibitem [{\citenamefont {Parrott}\ \emph {et~al.}(2011)\citenamefont
		{Parrott}, \citenamefont {Sun},\ and\ \citenamefont
		{Pickwell-MacPherson}}]{Parrott2011}%
	\BibitemOpen
	\bibfield  {author} {\bibinfo {author} {\bibfnamefont {E.~P.~J.}\
			\bibnamefont {Parrott}}, \bibinfo {author} {\bibfnamefont {Y.}~\bibnamefont
			{Sun}}, \ and\ \bibinfo {author} {\bibfnamefont {E.}~\bibnamefont
			{Pickwell-MacPherson}},\ }\bibfield  {title} {\enquote {\bibinfo {title}
			{Terahertz spectroscopy: Its future role in medical diagnoses},}\ }\href
	{\doibase 10.1016/j.molstruc.2011.05.048} {\bibfield  {journal} {\bibinfo
			{journal} {Journal of Molecular Structure}\ }\textbf {\bibinfo {volume}
			{1006}},\ \bibinfo {pages} {66} (\bibinfo {year} {2011})}\BibitemShut
	{NoStop}%
	\bibitem [{\citenamefont {Handley}\ \emph {et~al.}(2002)\citenamefont
		{Handley}, \citenamefont {Fitzgerald}, \citenamefont {Berry},\ and\
		\citenamefont {Boyle}}]{Handley2002}%
	\BibitemOpen
	\bibfield  {author} {\bibinfo {author} {\bibfnamefont {J.~W.}\ \bibnamefont
			{Handley}}, \bibinfo {author} {\bibfnamefont {A.~J.}\ \bibnamefont
			{Fitzgerald}}, \bibinfo {author} {\bibfnamefont {E.}~\bibnamefont {Berry}}, \
		and\ \bibinfo {author} {\bibfnamefont {R.~D.}\ \bibnamefont {Boyle}},\
	}\bibfield  {title} {\enquote {\bibinfo {title} {Wavelet compression in
				medical terahertz pulsed imaging},}\ }\href {\doibase
		10.1088/0031-9155/47/21/328} {\bibfield  {journal} {\bibinfo  {journal}
			{Physics in Medicine and Biology}\ }\textbf {\bibinfo {volume} {47}},\
		\bibinfo {pages} {3885} (\bibinfo {year} {2002})}\BibitemShut {NoStop}%
	\bibitem [{\citenamefont {Tu}\ \emph {et~al.}(2016)\citenamefont {Tu},
		\citenamefont {Zhong}, \citenamefont {Shen},\ and\ \citenamefont
		{Incecik}}]{Tu2016}%
	\BibitemOpen
	\bibfield  {author} {\bibinfo {author} {\bibfnamefont {W.}~\bibnamefont
			{Tu}}, \bibinfo {author} {\bibfnamefont {S.}~\bibnamefont {Zhong}}, \bibinfo
		{author} {\bibfnamefont {Y.}~\bibnamefont {Shen}}, \ and\ \bibinfo {author}
		{\bibfnamefont {A.}~\bibnamefont {Incecik}},\ }\bibfield  {title} {\enquote
		{\bibinfo {title} {Nondestructive testing of marine protective coatings using
				terahertz waves with stationary wavelet transform},}\ }\href {\doibase
		10.1016/j.oceaneng.2015.11.028} {\bibfield  {journal} {\bibinfo  {journal}
			{Ocean Engineering}\ }\textbf {\bibinfo {volume} {111}},\ \bibinfo {pages}
		{582} (\bibinfo {year} {2016})}\BibitemShut {NoStop}%
	\bibitem [{\citenamefont {Li}\ and\ \citenamefont {Pi}(2015)}]{Li2015}%
	\BibitemOpen
	\bibfield  {author} {\bibinfo {author} {\bibfnamefont {J.}~\bibnamefont
			{Li}}\ and\ \bibinfo {author} {\bibfnamefont {Y.}~\bibnamefont {Pi}},\
	}\bibfield  {title} {\enquote {\bibinfo {title} {Target detection for
				terahertz radar networks based on micro-doppler signatures},}\ }\href
	{\doibase 10.1504/ijsnet.2015.067861} {\bibfield  {journal} {\bibinfo
			{journal} {International Journal of Sensor Networks}\ }\textbf {\bibinfo
			{volume} {17}},\ \bibinfo {pages} {115} (\bibinfo {year} {2015})}\BibitemShut
	{NoStop}%
	\bibitem [{\citenamefont {Semashkin}\ and\ \citenamefont
		{Artyushkina}(2015)}]{Semashkin2015}%
	\BibitemOpen
	\bibfield  {author} {\bibinfo {author} {\bibfnamefont {E.~N.}\ \bibnamefont
			{Semashkin}}\ and\ \bibinfo {author} {\bibfnamefont {T.~V.}\ \bibnamefont
			{Artyushkina}},\ }\bibfield  {title} {\enquote {\bibinfo {title} {Operating
				range and all-weather capability of terahertz (01{\hspace{0.167em}}{THz}) and
				gigahertz (3{\textendash}333{\hspace{0.167em}}{GHz}) radars on horizontal and
				oblique tracks},}\ }\href {\doibase 10.1364/jot.82.000430} {\bibfield
		{journal} {\bibinfo  {journal} {Journal of Optical Technology}\ }\textbf
		{\bibinfo {volume} {82}},\ \bibinfo {pages} {430} (\bibinfo {year}
		{2015})}\BibitemShut {NoStop}%
	\bibitem [{\citenamefont {Zimdars}\ \emph {et~al.}(2006)\citenamefont
		{Zimdars}, \citenamefont {White}, \citenamefont {Stuk}, \citenamefont
		{Chernovsky}, \citenamefont {Fichter},\ and\ \citenamefont
		{Williamson}}]{Zimdars2006}%
	\BibitemOpen
	\bibfield  {author} {\bibinfo {author} {\bibfnamefont {D.}~\bibnamefont
			{Zimdars}}, \bibinfo {author} {\bibfnamefont {J.~S.}\ \bibnamefont {White}},
		\bibinfo {author} {\bibfnamefont {G.}~\bibnamefont {Stuk}}, \bibinfo {author}
		{\bibfnamefont {A.}~\bibnamefont {Chernovsky}}, \bibinfo {author}
		{\bibfnamefont {G.}~\bibnamefont {Fichter}}, \ and\ \bibinfo {author}
		{\bibfnamefont {S.}~\bibnamefont {Williamson}},\ }\bibfield  {title}
	{\enquote {\bibinfo {title} {Security and non destructive evaluation
				application of high speed time domain terahertz imaging},}\ }in\ \href
	{\doibase 10.1109/cleo.2006.4627807} {\emph {\bibinfo {booktitle} {2006
				Conference on Lasers and Electro-Optics and 2006 Quantum Electronics and
				Laser Science Conference}}}\ (\bibinfo  {publisher} {{IEEE}},\ \bibinfo
	{year} {2006})\BibitemShut {NoStop}%
	\bibitem [{\citenamefont {Han}\ \emph {et~al.}(2016)\citenamefont {Han},
		\citenamefont {Kim}, \citenamefont {Lee}, \citenamefont {Lee}, \citenamefont
		{Ko}, \citenamefont {Lee}, \citenamefont {Moon}, \citenamefont {Lee},\ and\
		\citenamefont {Park}}]{Han2016}%
	\BibitemOpen
	\bibfield  {author} {\bibinfo {author} {\bibfnamefont {S.-P.}\ \bibnamefont
			{Han}}, \bibinfo {author} {\bibfnamefont {N.}~\bibnamefont {Kim}}, \bibinfo
		{author} {\bibfnamefont {W.-H.}\ \bibnamefont {Lee}}, \bibinfo {author}
		{\bibfnamefont {E.~S.}\ \bibnamefont {Lee}}, \bibinfo {author} {\bibfnamefont
			{H.}~\bibnamefont {Ko}}, \bibinfo {author} {\bibfnamefont {I.-M.}\
			\bibnamefont {Lee}}, \bibinfo {author} {\bibfnamefont {K.}~\bibnamefont
			{Moon}}, \bibinfo {author} {\bibfnamefont {D.~H.}\ \bibnamefont {Lee}}, \
		and\ \bibinfo {author} {\bibfnamefont {K.~H.}\ \bibnamefont {Park}},\
	}\bibfield  {title} {\enquote {\bibinfo {title} {Real-time imaging of moving
				living objects using a compact terahertz scanner},}\ }\href {\doibase
		10.7567/apex.9.022501} {\bibfield  {journal} {\bibinfo  {journal} {Applied
				Physics Express}\ }\textbf {\bibinfo {volume} {9}},\ \bibinfo {pages}
		{022501} (\bibinfo {year} {2016})}\BibitemShut {NoStop}%
	\bibitem [{\citenamefont {Fukunaga}\ and\ \citenamefont
		{Hosako}(2010)}]{Fukunaga2010}%
	\BibitemOpen
	\bibfield  {author} {\bibinfo {author} {\bibfnamefont {K.}~\bibnamefont
			{Fukunaga}}\ and\ \bibinfo {author} {\bibfnamefont {I.}~\bibnamefont
			{Hosako}},\ }\bibfield  {title} {\enquote {\bibinfo {title} {Innovative
				non-invasive analysis techniques for cultural heritage using terahertz
				technology},}\ }\href {\doibase 10.1016/j.crhy.2010.05.004} {\bibfield
		{journal} {\bibinfo  {journal} {Comptes Rendus Physique}\ }\textbf {\bibinfo
			{volume} {11}},\ \bibinfo {pages} {519} (\bibinfo {year} {2010})}\BibitemShut
	{NoStop}%
	\bibitem [{\citenamefont {Manceau}\ \emph {et~al.}(2008)\citenamefont
		{Manceau}, \citenamefont {Nevin}, \citenamefont {Fotakis},\ and\
		\citenamefont {Tzortzakis}}]{Manceau2008}%
	\BibitemOpen
	\bibfield  {author} {\bibinfo {author} {\bibfnamefont {J.-M.}\ \bibnamefont
			{Manceau}}, \bibinfo {author} {\bibfnamefont {A.}~\bibnamefont {Nevin}},
		\bibinfo {author} {\bibfnamefont {C.}~\bibnamefont {Fotakis}}, \ and\
		\bibinfo {author} {\bibfnamefont {S.}~\bibnamefont {Tzortzakis}},\ }\bibfield
	{title} {\enquote {\bibinfo {title} {Terahertz time domain spectroscopy for
				the analysis of cultural heritage related materials},}\ }\href {\doibase
		10.1007/s00340-008-2933-6} {\bibfield  {journal} {\bibinfo  {journal}
			{Applied Physics B}\ }\textbf {\bibinfo {volume} {90}},\ \bibinfo {pages}
		{365} (\bibinfo {year} {2008})}\BibitemShut {NoStop}%
	\bibitem [{\citenamefont {Lee}\ \emph {et~al.}(2006{\natexlab{a}})\citenamefont
		{Lee}, \citenamefont {Seo}, \citenamefont {Park}, \citenamefont {Jeoung},
		\citenamefont {Park}, \citenamefont {Lienau},\ and\ \citenamefont
		{Kim}}]{Lee2006a}%
	\BibitemOpen
	\bibfield  {author} {\bibinfo {author} {\bibfnamefont {J.~W.}\ \bibnamefont
			{Lee}}, \bibinfo {author} {\bibfnamefont {M.~A.}\ \bibnamefont {Seo}},
		\bibinfo {author} {\bibfnamefont {D.~J.}\ \bibnamefont {Park}}, \bibinfo
		{author} {\bibfnamefont {S.~C.}\ \bibnamefont {Jeoung}}, \bibinfo {author}
		{\bibfnamefont {Q.~H.}\ \bibnamefont {Park}}, \bibinfo {author}
		{\bibfnamefont {C.}~\bibnamefont {Lienau}}, \ and\ \bibinfo {author}
		{\bibfnamefont {D.~S.}\ \bibnamefont {Kim}},\ }\bibfield  {title} {\enquote
		{\bibinfo {title} {Terahertz transparency at fabry-perot resonances of
				periodic slit arrays in a metal plate: experiment and theory},}\ }\href
	{\doibase 10.1364/oe.14.012637} {\bibfield  {journal} {\bibinfo  {journal}
			{Optics Express}\ }\textbf {\bibinfo {volume} {14}},\ \bibinfo {pages}
		{12637} (\bibinfo {year} {2006}{\natexlab{a}})}\BibitemShut {NoStop}%
	\bibitem [{\citenamefont {Libon}\ \emph {et~al.}(2000)\citenamefont {Libon},
		\citenamefont {Baumgärtner}, \citenamefont {Hempel}, \citenamefont {Hecker},
		\citenamefont {Feldmann}, \citenamefont {Koch},\ and\ \citenamefont
		{Dawson}}]{Libon2000}%
	\BibitemOpen
	\bibfield  {author} {\bibinfo {author} {\bibfnamefont {I.~H.}\ \bibnamefont
			{Libon}}, \bibinfo {author} {\bibfnamefont {S.}~\bibnamefont {Baumgärtner}},
		\bibinfo {author} {\bibfnamefont {M.}~\bibnamefont {Hempel}}, \bibinfo
		{author} {\bibfnamefont {N.~E.}\ \bibnamefont {Hecker}}, \bibinfo {author}
		{\bibfnamefont {J.}~\bibnamefont {Feldmann}}, \bibinfo {author}
		{\bibfnamefont {M.}~\bibnamefont {Koch}}, \ and\ \bibinfo {author}
		{\bibfnamefont {P.}~\bibnamefont {Dawson}},\ }\bibfield  {title} {\enquote
		{\bibinfo {title} {An optically controllable terahertz filter},}\ }\href
	{\doibase 10.1063/1.126484} {\bibfield  {journal} {\bibinfo  {journal}
			{Applied Physics Letters}\ }\textbf {\bibinfo {volume} {76}},\ \bibinfo
		{pages} {2821} (\bibinfo {year} {2000})}\BibitemShut {NoStop}%
	\bibitem [{\citenamefont {Mendis}\ \emph {et~al.}(2010)\citenamefont {Mendis},
		\citenamefont {Nag}, \citenamefont {Chen},\ and\ \citenamefont
		{Mittleman}}]{Mendis2010}%
	\BibitemOpen
	\bibfield  {author} {\bibinfo {author} {\bibfnamefont {R.}~\bibnamefont
			{Mendis}}, \bibinfo {author} {\bibfnamefont {A.}~\bibnamefont {Nag}},
		\bibinfo {author} {\bibfnamefont {F.}~\bibnamefont {Chen}}, \ and\ \bibinfo
		{author} {\bibfnamefont {D.~M.}\ \bibnamefont {Mittleman}},\ }\bibfield
	{title} {\enquote {\bibinfo {title} {A tunable universal terahertz filter
				using artificial dielectrics based on parallel-plate waveguides},}\ }\href
	{\doibase 10.1063/1.3495994} {\bibfield  {journal} {\bibinfo  {journal}
			{Applied Physics Letters}\ }\textbf {\bibinfo {volume} {97}},\ \bibinfo
		{pages} {131106} (\bibinfo {year} {2010})}\BibitemShut {NoStop}%
	\bibitem [{\citenamefont {Kaliteevski}\ \emph {et~al.}(2008)\citenamefont
		{Kaliteevski}, \citenamefont {Brand}, \citenamefont {Garvie-Cook},
		\citenamefont {Abram},\ and\ \citenamefont {Chamberlain}}]{Kaliteevski2008}%
	\BibitemOpen
	\bibfield  {author} {\bibinfo {author} {\bibfnamefont {M.~A.}\ \bibnamefont
			{Kaliteevski}}, \bibinfo {author} {\bibfnamefont {S.}~\bibnamefont {Brand}},
		\bibinfo {author} {\bibfnamefont {J.}~\bibnamefont {Garvie-Cook}}, \bibinfo
		{author} {\bibfnamefont {R.~A.}\ \bibnamefont {Abram}}, \ and\ \bibinfo
		{author} {\bibfnamefont {J.~M.}\ \bibnamefont {Chamberlain}},\ }\bibfield
	{title} {\enquote {\bibinfo {title} {Terahertz filter based on refractive
				properties of metallic photonic crystal},}\ }\href {\doibase
		10.1364/oe.16.007330} {\bibfield  {journal} {\bibinfo  {journal} {Optics
				Express}\ }\textbf {\bibinfo {volume} {16}},\ \bibinfo {pages} {7330}
		(\bibinfo {year} {2008})}\BibitemShut {NoStop}%
	\bibitem [{\citenamefont {Lan}\ \emph {et~al.}(2014)\citenamefont {Lan},
		\citenamefont {Yang}, \citenamefont {Qi}, \citenamefont {Gao},\ and\
		\citenamefont {Shi}}]{Lan2014}%
	\BibitemOpen
	\bibfield  {author} {\bibinfo {author} {\bibfnamefont {F.}~\bibnamefont
			{Lan}}, \bibinfo {author} {\bibfnamefont {Z.}~\bibnamefont {Yang}}, \bibinfo
		{author} {\bibfnamefont {L.}~\bibnamefont {Qi}}, \bibinfo {author}
		{\bibfnamefont {X.}~\bibnamefont {Gao}}, \ and\ \bibinfo {author}
		{\bibfnamefont {Z.}~\bibnamefont {Shi}},\ }\bibfield  {title} {\enquote
		{\bibinfo {title} {Terahertz dual-resonance bandpass filter using bilayer
				reformative complementary metamaterial structures},}\ }\href {\doibase
		10.1364/ol.39.001709} {\bibfield  {journal} {\bibinfo  {journal} {Optics
				Letters}\ }\textbf {\bibinfo {volume} {39}},\ \bibinfo {pages} {1709}
		(\bibinfo {year} {2014})}\BibitemShut {NoStop}%
	\bibitem [{\citenamefont {Heshmat}\ \emph {et~al.}(2012)\citenamefont
		{Heshmat}, \citenamefont {Pahlevaninezhad}, \citenamefont {Yuanjie},
		\citenamefont {Masnadi-Shirazi}, \citenamefont {Lewis}, \citenamefont
		{Tiedje}, \citenamefont {Gordon},\ and\ \citenamefont
		{Darcie}}]{Heshmat2012}%
	\BibitemOpen
	\bibfield  {author} {\bibinfo {author} {\bibfnamefont {B.}~\bibnamefont
			{Heshmat}}, \bibinfo {author} {\bibfnamefont {H.}~\bibnamefont
			{Pahlevaninezhad}}, \bibinfo {author} {\bibfnamefont {P.}~\bibnamefont
			{Yuanjie}}, \bibinfo {author} {\bibfnamefont {M.}~\bibnamefont
			{Masnadi-Shirazi}}, \bibinfo {author} {\bibfnamefont {R.~B.}\ \bibnamefont
			{Lewis}}, \bibinfo {author} {\bibfnamefont {T.}~\bibnamefont {Tiedje}},
		\bibinfo {author} {\bibfnamefont {R.}~\bibnamefont {Gordon}}, \ and\ \bibinfo
		{author} {\bibfnamefont {T.~E.}\ \bibnamefont {Darcie}},\ }\bibfield  {title}
	{\enquote {\bibinfo {title} {Nanoplasmonic terahertz photoconductive switch
				on {GaAs}},}\ }\href {\doibase 10.1021/nl303314a} {\bibfield  {journal}
		{\bibinfo  {journal} {Nano Letters}\ }\textbf {\bibinfo {volume} {12}},\
		\bibinfo {pages} {6255} (\bibinfo {year} {2012})}\BibitemShut {NoStop}%
	\bibitem [{\citenamefont {Rahm}\ \emph {et~al.}(2012)\citenamefont {Rahm},
		\citenamefont {Li},\ and\ \citenamefont {Padilla}}]{Rahm2012}%
	\BibitemOpen
	\bibfield  {author} {\bibinfo {author} {\bibfnamefont {M.}~\bibnamefont
			{Rahm}}, \bibinfo {author} {\bibfnamefont {J.-S.}\ \bibnamefont {Li}}, \ and\
		\bibinfo {author} {\bibfnamefont {W.~J.}\ \bibnamefont {Padilla}},\
	}\bibfield  {title} {\enquote {\bibinfo {title} {{THz} wave modulators: A
				brief review on different modulation techniques},}\ }\href {\doibase
		10.1007/s10762-012-9946-2} {\bibfield  {journal} {\bibinfo  {journal}
			{Journal of Infrared, Millimeter, and Terahertz Waves}\ }\textbf {\bibinfo
			{volume} {34}},\ \bibinfo {pages} {1} (\bibinfo {year} {2012})}\BibitemShut
	{NoStop}%
	\bibitem [{\citenamefont {Choi}\ \emph
		{et~al.}(2011{\natexlab{a}})\citenamefont {Choi}, \citenamefont {Kyoung},
		\citenamefont {Kim}, \citenamefont {Park}, \citenamefont {Park},
		\citenamefont {Kim}, \citenamefont {Ahn}, \citenamefont {Rotermund},
		\citenamefont {Kim}, \citenamefont {Ahn},\ and\ \citenamefont
		{Kim}}]{Choi2011}%
	\BibitemOpen
	\bibfield  {author} {\bibinfo {author} {\bibfnamefont {S.~B.}\ \bibnamefont
			{Choi}}, \bibinfo {author} {\bibfnamefont {J.~S.}\ \bibnamefont {Kyoung}},
		\bibinfo {author} {\bibfnamefont {H.~S.}\ \bibnamefont {Kim}}, \bibinfo
		{author} {\bibfnamefont {H.~R.}\ \bibnamefont {Park}}, \bibinfo {author}
		{\bibfnamefont {D.~J.}\ \bibnamefont {Park}}, \bibinfo {author}
		{\bibfnamefont {B.-J.}\ \bibnamefont {Kim}}, \bibinfo {author} {\bibfnamefont
			{Y.~H.}\ \bibnamefont {Ahn}}, \bibinfo {author} {\bibfnamefont
			{F.}~\bibnamefont {Rotermund}}, \bibinfo {author} {\bibfnamefont {H.-T.}\
			\bibnamefont {Kim}}, \bibinfo {author} {\bibfnamefont {K.~J.}\ \bibnamefont
			{Ahn}}, \ and\ \bibinfo {author} {\bibfnamefont {D.~S.}\ \bibnamefont
			{Kim}},\ }\bibfield  {title} {\enquote {\bibinfo {title} {Nanopattern enabled
				terahertz all-optical switching on vanadium dioxide thin film},}\ }\href
	{\doibase 10.1063/1.3553504} {\bibfield  {journal} {\bibinfo  {journal}
			{Applied Physics Letters}\ }\textbf {\bibinfo {volume} {98}},\ \bibinfo
		{pages} {071105} (\bibinfo {year} {2011}{\natexlab{a}})}\BibitemShut
	{NoStop}%
	\bibitem [{\citenamefont {Unlu}\ and\ \citenamefont
		{Jarrahi}(2014)}]{Unlu2014}%
	\BibitemOpen
	\bibfield  {author} {\bibinfo {author} {\bibfnamefont {M.}~\bibnamefont
			{Unlu}}\ and\ \bibinfo {author} {\bibfnamefont {M.}~\bibnamefont {Jarrahi}},\
	}\bibfield  {title} {\enquote {\bibinfo {title} {Miniature multi-contact
				{MEMS} switch for broadband terahertz modulation},}\ }\href {\doibase
		10.1364/oe.22.032245} {\bibfield  {journal} {\bibinfo  {journal} {Optics
				Express}\ }\textbf {\bibinfo {volume} {22}},\ \bibinfo {pages} {32245}
		(\bibinfo {year} {2014})}\BibitemShut {NoStop}%
	\bibitem [{\citenamefont {Li}\ \emph {et~al.}(2007)\citenamefont {Li},
		\citenamefont {He},\ and\ \citenamefont {Hong}}]{Li2007}%
	\BibitemOpen
	\bibfield  {author} {\bibinfo {author} {\bibfnamefont {J.}~\bibnamefont
			{Li}}, \bibinfo {author} {\bibfnamefont {J.}~\bibnamefont {He}}, \ and\
		\bibinfo {author} {\bibfnamefont {Z.}~\bibnamefont {Hong}},\ }\bibfield
	{title} {\enquote {\bibinfo {title} {Terahertz wave switch based on silicon
				photonic crystals},}\ }\href {\doibase 10.1364/ao.46.005034} {\bibfield
		{journal} {\bibinfo  {journal} {Applied Optics}\ }\textbf {\bibinfo {volume}
			{46}},\ \bibinfo {pages} {5034} (\bibinfo {year} {2007})}\BibitemShut
	{NoStop}%
	\bibitem [{\citenamefont {Gao}\ \emph {et~al.}(2014)\citenamefont {Gao},
		\citenamefont {Shu}, \citenamefont {Reichel}, \citenamefont {Nickel},
		\citenamefont {He}, \citenamefont {Shi}, \citenamefont {Vajtai},
		\citenamefont {Ajayan}, \citenamefont {Kono}, \citenamefont {Mittleman},\
		and\ \citenamefont {Xu}}]{Gao2014}%
	\BibitemOpen
	\bibfield  {author} {\bibinfo {author} {\bibfnamefont {W.}~\bibnamefont
			{Gao}}, \bibinfo {author} {\bibfnamefont {J.}~\bibnamefont {Shu}}, \bibinfo
		{author} {\bibfnamefont {K.}~\bibnamefont {Reichel}}, \bibinfo {author}
		{\bibfnamefont {D.~V.}\ \bibnamefont {Nickel}}, \bibinfo {author}
		{\bibfnamefont {X.}~\bibnamefont {He}}, \bibinfo {author} {\bibfnamefont
			{G.}~\bibnamefont {Shi}}, \bibinfo {author} {\bibfnamefont {R.}~\bibnamefont
			{Vajtai}}, \bibinfo {author} {\bibfnamefont {P.~M.}\ \bibnamefont {Ajayan}},
		\bibinfo {author} {\bibfnamefont {J.}~\bibnamefont {Kono}}, \bibinfo {author}
		{\bibfnamefont {D.~M.}\ \bibnamefont {Mittleman}}, \ and\ \bibinfo {author}
		{\bibfnamefont {Q.}~\bibnamefont {Xu}},\ }\bibfield  {title} {\enquote
		{\bibinfo {title} {High-contrast terahertz wave modulation by gated graphene
				enhanced by extraordinary transmission through ring apertures},}\ }\href
	{\doibase 10.1021/nl4041274} {\bibfield  {journal} {\bibinfo  {journal} {Nano
				Letters}\ }\textbf {\bibinfo {volume} {14}},\ \bibinfo {pages} {1242}
		(\bibinfo {year} {2014})}\BibitemShut {NoStop}%
	\bibitem [{\citenamefont {Liu}\ \emph {et~al.}(2013)\citenamefont {Liu},
		\citenamefont {He}, \citenamefont {Tian}, \citenamefont {Li},\ and\
		\citenamefont {Liu}}]{Liu2013}%
	\BibitemOpen
	\bibfield  {author} {\bibinfo {author} {\bibfnamefont {G.}~\bibnamefont
			{Liu}}, \bibinfo {author} {\bibfnamefont {M.}~\bibnamefont {He}}, \bibinfo
		{author} {\bibfnamefont {Z.}~\bibnamefont {Tian}}, \bibinfo {author}
		{\bibfnamefont {J.}~\bibnamefont {Li}}, \ and\ \bibinfo {author}
		{\bibfnamefont {J.}~\bibnamefont {Liu}},\ }\bibfield  {title} {\enquote
		{\bibinfo {title} {Terahertz surface plasmon sensor for distinguishing
				gasolines},}\ }\href {\doibase 10.1364/ao.52.005695} {\bibfield  {journal}
		{\bibinfo  {journal} {Applied Optics}\ }\textbf {\bibinfo {volume} {52}},\
		\bibinfo {pages} {5695} (\bibinfo {year} {2013})}\BibitemShut {NoStop}%
	\bibitem [{\citenamefont {Astley}\ \emph {et~al.}(2012)\citenamefont {Astley},
		\citenamefont {Reichel}, \citenamefont {Mendis},\ and\ \citenamefont
		{Mittleman}}]{Astley2012}%
	\BibitemOpen
	\bibfield  {author} {\bibinfo {author} {\bibfnamefont {V.}~\bibnamefont
			{Astley}}, \bibinfo {author} {\bibfnamefont {K.}~\bibnamefont {Reichel}},
		\bibinfo {author} {\bibfnamefont {R.}~\bibnamefont {Mendis}}, \ and\ \bibinfo
		{author} {\bibfnamefont {D.~M.}\ \bibnamefont {Mittleman}},\ }\bibfield
	{title} {\enquote {\bibinfo {title} {Terahertz microfluidic sensing using a
				parallel-plate waveguide sensor},}\ }\href {\doibase 10.3791/4304} {\bibfield
		{journal} {\bibinfo  {journal} {Journal of Visualized Experiments}\ }
		(\bibinfo {year} {2012}),\ 10.3791/4304}\BibitemShut {NoStop}%
	\bibitem [{\citenamefont {Alves}\ \emph {et~al.}(2012)\citenamefont {Alves},
		\citenamefont {Grbovic}, \citenamefont {Kearney},\ and\ \citenamefont
		{Karunasiri}}]{Alves2012}%
	\BibitemOpen
	\bibfield  {author} {\bibinfo {author} {\bibfnamefont {F.}~\bibnamefont
			{Alves}}, \bibinfo {author} {\bibfnamefont {D.}~\bibnamefont {Grbovic}},
		\bibinfo {author} {\bibfnamefont {B.}~\bibnamefont {Kearney}}, \ and\
		\bibinfo {author} {\bibfnamefont {G.}~\bibnamefont {Karunasiri}},\ }\bibfield
	{title} {\enquote {\bibinfo {title} {Microelectromechanical systems
				bimaterial terahertz sensor with integrated metamaterial absorber},}\ }\href
	{\doibase 10.1364/ol.37.001886} {\bibfield  {journal} {\bibinfo  {journal}
			{Optics Letters}\ }\textbf {\bibinfo {volume} {37}},\ \bibinfo {pages} {1886}
		(\bibinfo {year} {2012})}\BibitemShut {NoStop}%
	\bibitem [{\citenamefont {Hassani}\ and\ \citenamefont
		{Skorobogatiy}(2008)}]{Hassani2008}%
	\BibitemOpen
	\bibfield  {author} {\bibinfo {author} {\bibfnamefont {A.}~\bibnamefont
			{Hassani}}\ and\ \bibinfo {author} {\bibfnamefont {M.}~\bibnamefont
			{Skorobogatiy}},\ }\bibfield  {title} {\enquote {\bibinfo {title} {Surface
				plasmon resonance-like integrated sensor at terahertz frequencies for gaseous
				analytes},}\ }\href {\doibase 10.1364/oe.16.020206} {\bibfield  {journal}
		{\bibinfo  {journal} {Optics Express}\ }\textbf {\bibinfo {volume} {16}},\
		\bibinfo {pages} {20206} (\bibinfo {year} {2008})}\BibitemShut {NoStop}%
	\bibitem [{\citenamefont {Miyamaru}\ \emph {et~al.}(2006)\citenamefont
		{Miyamaru}, \citenamefont {Hayashi}, \citenamefont {Otani}, \citenamefont
		{Kawase}, \citenamefont {Ogawa}, \citenamefont {Yoshida},\ and\ \citenamefont
		{Kato}}]{Miyamaru2006}%
	\BibitemOpen
	\bibfield  {author} {\bibinfo {author} {\bibfnamefont {F.}~\bibnamefont
			{Miyamaru}}, \bibinfo {author} {\bibfnamefont {S.}~\bibnamefont {Hayashi}},
		\bibinfo {author} {\bibfnamefont {C.}~\bibnamefont {Otani}}, \bibinfo
		{author} {\bibfnamefont {K.}~\bibnamefont {Kawase}}, \bibinfo {author}
		{\bibfnamefont {Y.}~\bibnamefont {Ogawa}}, \bibinfo {author} {\bibfnamefont
			{H.}~\bibnamefont {Yoshida}}, \ and\ \bibinfo {author} {\bibfnamefont
			{E.}~\bibnamefont {Kato}},\ }\bibfield  {title} {\enquote {\bibinfo {title}
			{Terahertz surface-wave resonant sensor with a metal hole array},}\ }\href
	{\doibase 10.1364/ol.31.001118} {\bibfield  {journal} {\bibinfo  {journal}
			{Optics Letters}\ }\textbf {\bibinfo {volume} {31}},\ \bibinfo {pages} {1118}
		(\bibinfo {year} {2006})}\BibitemShut {NoStop}%
	\bibitem [{\citenamefont {Xu}(2016)}]{Xu2016}%
	\BibitemOpen
	\bibfield  {author} {\bibinfo {author} {\bibfnamefont {K.}~\bibnamefont
			{Xu}},\ }\bibfield  {title} {\enquote {\bibinfo {title} {Integrated silicon
				directly modulated light source using p-well in standard {CMOS}
				technology},}\ }\href {\doibase 10.1109/jsen.2016.2582840} {\bibfield
		{journal} {\bibinfo  {journal} {{IEEE} Sensors Journal}\ }\textbf {\bibinfo
			{volume} {16}},\ \bibinfo {pages} {6184} (\bibinfo {year}
		{2016})}\BibitemShut {NoStop}%
	\bibitem [{\citenamefont {Lee}(2008)}]{Lee2008}%
	\BibitemOpen
	\bibfield  {author} {\bibinfo {author} {\bibfnamefont {Y.-S.}\ \bibnamefont
			{Lee}},\ }\href
	{https://www.ebook.de/de/product/7469349/yun_shik_lee_principles_of_terahertz_science_and_technology.html}
	{\emph {\bibinfo {title} {Principles of Terahertz Science and Technology}}}\
	(\bibinfo  {publisher} {SPRINGER NATURE},\ \bibinfo {year}
	{2008})\BibitemShut {NoStop}%
	\bibitem [{\citenamefont {Baxter}\ and\ \citenamefont
		{Guglietta}(2011)}]{Baxter2011}%
	\BibitemOpen
	\bibfield  {author} {\bibinfo {author} {\bibfnamefont {J.~B.}\ \bibnamefont
			{Baxter}}\ and\ \bibinfo {author} {\bibfnamefont {G.~W.}\ \bibnamefont
			{Guglietta}},\ }\bibfield  {title} {\enquote {\bibinfo {title} {Terahertz
				spectroscopy},}\ }\href {\doibase 10.1021/ac200907z} {\bibfield  {journal}
		{\bibinfo  {journal} {Analytical Chemistry}\ }\textbf {\bibinfo {volume}
			{83}},\ \bibinfo {pages} {4342} (\bibinfo {year} {2011})}\BibitemShut
	{NoStop}%
	\bibitem [{\citenamefont {Xi-Cheng~Zhang}(2009)}]{Xi-ChengZhang2009}%
	\BibitemOpen
	\bibfield  {author} {\bibinfo {author} {\bibfnamefont {J.~X.}\ \bibnamefont
			{Xi-Cheng~Zhang}},\ }\href
	{https://www.ebook.de/de/product/8616879/xi_cheng_zhang_jingzhou_xu_introduction_to_thz_wave_photonics.html}
	{\emph {\bibinfo {title} {Introduction to THz Wave Photonics}}}\ (\bibinfo
	{publisher} {Springer-Verlag GmbH},\ \bibinfo {year} {2009})\BibitemShut
	{NoStop}%
	\bibitem [{\citenamefont {Dexheimer}(2007)}]{Dexheimer2007}%
	\BibitemOpen
	\bibinfo {editor} {\bibfnamefont {S.}~\bibnamefont {Dexheimer}},\ ed.,\ \href
	{\doibase 10.1201/9781420007701} {\emph {\bibinfo {title} {Terahertz
				Spectroscopy}}}\ (\bibinfo  {publisher} {{CRC} Press},\ \bibinfo {year}
	{2007})\BibitemShut {NoStop}%
	\bibitem [{\citenamefont {Nagel}\ \emph {et~al.}(2006)\citenamefont {Nagel},
		\citenamefont {Först},\ and\ \citenamefont {Kurz}}]{Nagel2006}%
	\BibitemOpen
	\bibfield  {author} {\bibinfo {author} {\bibfnamefont {M.}~\bibnamefont
			{Nagel}}, \bibinfo {author} {\bibfnamefont {M.}~\bibnamefont {Först}}, \
		and\ \bibinfo {author} {\bibfnamefont {H.}~\bibnamefont {Kurz}},\ }\bibfield
	{title} {\enquote {\bibinfo {title} {{THz} biosensing devices: fundamentals
				and technology},}\ }\href {\doibase 10.1088/0953-8984/18/18/s07} {\bibfield
		{journal} {\bibinfo  {journal} {Journal of Physics: Condensed Matter}\
		}\textbf {\bibinfo {volume} {18}},\ \bibinfo {pages} {S601} (\bibinfo {year}
		{2006})}\BibitemShut {NoStop}%
	\bibitem [{\citenamefont {Kar}\ \emph {et~al.}(2018)\citenamefont {Kar},
		\citenamefont {Nguyen}, \citenamefont {Mohapatra}, \citenamefont {Lee},\ and\
		\citenamefont {Sood}}]{Kar2018}%
	\BibitemOpen
	\bibfield  {author} {\bibinfo {author} {\bibfnamefont {S.}~\bibnamefont
			{Kar}}, \bibinfo {author} {\bibfnamefont {V.~L.}\ \bibnamefont {Nguyen}},
		\bibinfo {author} {\bibfnamefont {D.~R.}\ \bibnamefont {Mohapatra}}, \bibinfo
		{author} {\bibfnamefont {Y.~H.}\ \bibnamefont {Lee}}, \ and\ \bibinfo
		{author} {\bibfnamefont {A.~K.}\ \bibnamefont {Sood}},\ }\bibfield  {title}
	{\enquote {\bibinfo {title} {Ultrafast spectral photoresponse of bilayer
				graphene: Optical pump{\textendash}terahertz probe spectroscopy},}\ }\href
	{\doibase 10.1021/acsnano.7b08555} {\bibfield  {journal} {\bibinfo  {journal}
			{{ACS} Nano}\ }\textbf {\bibinfo {volume} {12}},\ \bibinfo {pages} {1785}
		(\bibinfo {year} {2018})}\BibitemShut {NoStop}%
	\bibitem [{\citenamefont {Fischer}\ \emph {et~al.}(2002)\citenamefont
		{Fischer}, \citenamefont {Walther},\ and\ \citenamefont
		{Jepsen}}]{Fischer2002}%
	\BibitemOpen
	\bibfield  {author} {\bibinfo {author} {\bibfnamefont {B.~M.}\ \bibnamefont
			{Fischer}}, \bibinfo {author} {\bibfnamefont {M.}~\bibnamefont {Walther}}, \
		and\ \bibinfo {author} {\bibfnamefont {P.~U.}\ \bibnamefont {Jepsen}},\
	}\bibfield  {title} {\enquote {\bibinfo {title} {Far-infrared vibrational
				modes of {DNA} components studied by terahertz time-domain spectroscopy},}\
	}\href {\doibase 10.1088/0031-9155/47/21/319} {\bibfield  {journal} {\bibinfo
			{journal} {Physics in Medicine and Biology}\ }\textbf {\bibinfo {volume}
			{47}},\ \bibinfo {pages} {3807} (\bibinfo {year} {2002})}\BibitemShut
	{NoStop}%
	\bibitem [{\citenamefont {Markelz}\ \emph {et~al.}(2002)\citenamefont
		{Markelz}, \citenamefont {Whitmire}, \citenamefont {Hillebrecht},\ and\
		\citenamefont {Birge}}]{Markelz2002}%
	\BibitemOpen
	\bibfield  {author} {\bibinfo {author} {\bibfnamefont {A.}~\bibnamefont
			{Markelz}}, \bibinfo {author} {\bibfnamefont {S.}~\bibnamefont {Whitmire}},
		\bibinfo {author} {\bibfnamefont {J.}~\bibnamefont {Hillebrecht}}, \ and\
		\bibinfo {author} {\bibfnamefont {R.}~\bibnamefont {Birge}},\ }\bibfield
	{title} {\enquote {\bibinfo {title} {{THz} time domain spectroscopy of
				biomolecular conformational modes},}\ }\href {\doibase
		10.1088/0031-9155/47/21/318} {\bibfield  {journal} {\bibinfo  {journal}
			{Physics in Medicine and Biology}\ }\textbf {\bibinfo {volume} {47}},\
		\bibinfo {pages} {3797} (\bibinfo {year} {2002})}\BibitemShut {NoStop}%
	\bibitem [{\citenamefont {Arora}\ \emph {et~al.}(2012)\citenamefont {Arora},
		\citenamefont {Luong}, \citenamefont {Krüger}, \citenamefont {Kim},
		\citenamefont {Nam}, \citenamefont {Manz},\ and\ \citenamefont
		{Havenith}}]{Arora2012}%
	\BibitemOpen
	\bibfield  {author} {\bibinfo {author} {\bibfnamefont {A.}~\bibnamefont
			{Arora}}, \bibinfo {author} {\bibfnamefont {T.~Q.}\ \bibnamefont {Luong}},
		\bibinfo {author} {\bibfnamefont {M.}~\bibnamefont {Krüger}}, \bibinfo
		{author} {\bibfnamefont {Y.~J.}\ \bibnamefont {Kim}}, \bibinfo {author}
		{\bibfnamefont {C.-H.}\ \bibnamefont {Nam}}, \bibinfo {author} {\bibfnamefont
			{A.}~\bibnamefont {Manz}}, \ and\ \bibinfo {author} {\bibfnamefont
			{M.}~\bibnamefont {Havenith}},\ }\bibfield  {title} {\enquote {\bibinfo
			{title} {Terahertz-time domain spectroscopy for the detection of {PCR}
				amplified {DNA} in aqueous solution},}\ }\href {\doibase 10.1039/c2an15820e}
	{\bibfield  {journal} {\bibinfo  {journal} {The Analyst}\ }\textbf {\bibinfo
			{volume} {137}},\ \bibinfo {pages} {575} (\bibinfo {year}
		{2012})}\BibitemShut {NoStop}%
	\bibitem [{\citenamefont {Yang}\ \emph
		{et~al.}(2016{\natexlab{a}})\citenamefont {Yang}, \citenamefont {Wei},
		\citenamefont {Yan}, \citenamefont {Liu}, \citenamefont {Yu}, \citenamefont
		{Zhang}, \citenamefont {Yang}, \citenamefont {Zhu}, \citenamefont {Huang},
		\citenamefont {Cui},\ and\ \citenamefont {Fu}}]{Yang2016}%
	\BibitemOpen
	\bibfield  {author} {\bibinfo {author} {\bibfnamefont {X.}~\bibnamefont
			{Yang}}, \bibinfo {author} {\bibfnamefont {D.}~\bibnamefont {Wei}}, \bibinfo
		{author} {\bibfnamefont {S.}~\bibnamefont {Yan}}, \bibinfo {author}
		{\bibfnamefont {Y.}~\bibnamefont {Liu}}, \bibinfo {author} {\bibfnamefont
			{S.}~\bibnamefont {Yu}}, \bibinfo {author} {\bibfnamefont {M.}~\bibnamefont
			{Zhang}}, \bibinfo {author} {\bibfnamefont {Z.}~\bibnamefont {Yang}},
		\bibinfo {author} {\bibfnamefont {X.}~\bibnamefont {Zhu}}, \bibinfo {author}
		{\bibfnamefont {Q.}~\bibnamefont {Huang}}, \bibinfo {author} {\bibfnamefont
			{H.-L.}\ \bibnamefont {Cui}}, \ and\ \bibinfo {author} {\bibfnamefont
			{W.}~\bibnamefont {Fu}},\ }\bibfield  {title} {\enquote {\bibinfo {title}
			{Rapid and label-free detection and assessment of bacteria by terahertz
				time-domain spectroscopy},}\ }\href {\doibase 10.1002/jbio.201500270}
	{\bibfield  {journal} {\bibinfo  {journal} {Journal of Biophotonics}\
		}\textbf {\bibinfo {volume} {9}},\ \bibinfo {pages} {1050} (\bibinfo {year}
		{2016}{\natexlab{a}})}\BibitemShut {NoStop}%
	\bibitem [{\citenamefont {Yang}\ \emph
		{et~al.}(2016{\natexlab{b}})\citenamefont {Yang}, \citenamefont {Yang},
		\citenamefont {Luo},\ and\ \citenamefont {Fu}}]{Yang2016a}%
	\BibitemOpen
	\bibfield  {author} {\bibinfo {author} {\bibfnamefont {X.}~\bibnamefont
			{Yang}}, \bibinfo {author} {\bibfnamefont {K.}~\bibnamefont {Yang}}, \bibinfo
		{author} {\bibfnamefont {Y.}~\bibnamefont {Luo}}, \ and\ \bibinfo {author}
		{\bibfnamefont {W.}~\bibnamefont {Fu}},\ }\bibfield  {title} {\enquote
		{\bibinfo {title} {Terahertz spectroscopy for bacterial detection:
				opportunities and challenges},}\ }\href {\doibase 10.1007/s00253-016-7569-6}
	{\bibfield  {journal} {\bibinfo  {journal} {Applied Microbiology and
				Biotechnology}\ }\textbf {\bibinfo {volume} {100}},\ \bibinfo {pages} {5289}
		(\bibinfo {year} {2016}{\natexlab{b}})}\BibitemShut {NoStop}%
	\bibitem [{\citenamefont {Lee}\ \emph {et~al.}(2015)\citenamefont {Lee},
		\citenamefont {Kang}, \citenamefont {Lee}, \citenamefont {Kim}, \citenamefont
		{Kim}, \citenamefont {Kim}, \citenamefont {Lee}, \citenamefont {Son},
		\citenamefont {Park},\ and\ \citenamefont {Seo}}]{Lee2015}%
	\BibitemOpen
	\bibfield  {author} {\bibinfo {author} {\bibfnamefont {D.-K.}\ \bibnamefont
			{Lee}}, \bibinfo {author} {\bibfnamefont {J.-H.}\ \bibnamefont {Kang}},
		\bibinfo {author} {\bibfnamefont {J.-S.}\ \bibnamefont {Lee}}, \bibinfo
		{author} {\bibfnamefont {H.-S.}\ \bibnamefont {Kim}}, \bibinfo {author}
		{\bibfnamefont {C.}~\bibnamefont {Kim}}, \bibinfo {author} {\bibfnamefont
			{J.~H.}\ \bibnamefont {Kim}}, \bibinfo {author} {\bibfnamefont
			{T.}~\bibnamefont {Lee}}, \bibinfo {author} {\bibfnamefont {J.-H.}\
			\bibnamefont {Son}}, \bibinfo {author} {\bibfnamefont {Q.-H.}\ \bibnamefont
			{Park}}, \ and\ \bibinfo {author} {\bibfnamefont {M.}~\bibnamefont {Seo}},\
	}\bibfield  {title} {\enquote {\bibinfo {title} {Highly sensitive and
				selective sugar detection by terahertz nano-antennas},}\ }\href {\doibase
		10.1038/srep15459} {\bibfield  {journal} {\bibinfo  {journal} {Scientific
				Reports}\ }\textbf {\bibinfo {volume} {5}} (\bibinfo {year} {2015}),\
		10.1038/srep15459}\BibitemShut {NoStop}%
	\bibitem [{\citenamefont {Park}\ \emph {et~al.}(2013)\citenamefont {Park},
		\citenamefont {Ahn}, \citenamefont {Han}, \citenamefont {Bahk}, \citenamefont
		{Park},\ and\ \citenamefont {Kim}}]{Park2013}%
	\BibitemOpen
	\bibfield  {author} {\bibinfo {author} {\bibfnamefont {H.-R.}\ \bibnamefont
			{Park}}, \bibinfo {author} {\bibfnamefont {K.~J.}\ \bibnamefont {Ahn}},
		\bibinfo {author} {\bibfnamefont {S.}~\bibnamefont {Han}}, \bibinfo {author}
		{\bibfnamefont {Y.-M.}\ \bibnamefont {Bahk}}, \bibinfo {author}
		{\bibfnamefont {N.}~\bibnamefont {Park}}, \ and\ \bibinfo {author}
		{\bibfnamefont {D.-S.}\ \bibnamefont {Kim}},\ }\bibfield  {title} {\enquote
		{\bibinfo {title} {Colossal absorption of molecules inside single terahertz
				nanoantennas},}\ }\href {\doibase 10.1021/nl400374z} {\bibfield  {journal}
		{\bibinfo  {journal} {Nano Letters}\ }\textbf {\bibinfo {volume} {13}},\
		\bibinfo {pages} {1782} (\bibinfo {year} {2013})}\BibitemShut {NoStop}%
	\bibitem [{\citenamefont {Limaj}\ \emph {et~al.}(2016)\citenamefont {Limaj},
		\citenamefont {Etezadi}, \citenamefont {Wittenberg}, \citenamefont {Rodrigo},
		\citenamefont {Yoo}, \citenamefont {Oh},\ and\ \citenamefont
		{Altug}}]{Limaj2016}%
	\BibitemOpen
	\bibfield  {author} {\bibinfo {author} {\bibfnamefont {O.}~\bibnamefont
			{Limaj}}, \bibinfo {author} {\bibfnamefont {D.}~\bibnamefont {Etezadi}},
		\bibinfo {author} {\bibfnamefont {N.~J.}\ \bibnamefont {Wittenberg}},
		\bibinfo {author} {\bibfnamefont {D.}~\bibnamefont {Rodrigo}}, \bibinfo
		{author} {\bibfnamefont {D.}~\bibnamefont {Yoo}}, \bibinfo {author}
		{\bibfnamefont {S.-H.}\ \bibnamefont {Oh}}, \ and\ \bibinfo {author}
		{\bibfnamefont {H.}~\bibnamefont {Altug}},\ }\bibfield  {title} {\enquote
		{\bibinfo {title} {Infrared plasmonic biosensor for real-time and label-free
				monitoring of lipid membranes},}\ }\href {\doibase
		10.1021/acs.nanolett.5b05316} {\bibfield  {journal} {\bibinfo  {journal}
			{Nano Letters}\ }\textbf {\bibinfo {volume} {16}},\ \bibinfo {pages} {1502}
		(\bibinfo {year} {2016})}\BibitemShut {NoStop}%
	\bibitem [{\citenamefont {Rodrigo}\ \emph {et~al.}(2015)\citenamefont
		{Rodrigo}, \citenamefont {Limaj}, \citenamefont {Janner}, \citenamefont
		{Etezadi}, \citenamefont {de~Abajo}, \citenamefont {Pruneri},\ and\
		\citenamefont {Altug}}]{Rodrigo2015}%
	\BibitemOpen
	\bibfield  {author} {\bibinfo {author} {\bibfnamefont {D.}~\bibnamefont
			{Rodrigo}}, \bibinfo {author} {\bibfnamefont {O.}~\bibnamefont {Limaj}},
		\bibinfo {author} {\bibfnamefont {D.}~\bibnamefont {Janner}}, \bibinfo
		{author} {\bibfnamefont {D.}~\bibnamefont {Etezadi}}, \bibinfo {author}
		{\bibfnamefont {F.~J.~G.}\ \bibnamefont {de~Abajo}}, \bibinfo {author}
		{\bibfnamefont {V.}~\bibnamefont {Pruneri}}, \ and\ \bibinfo {author}
		{\bibfnamefont {H.}~\bibnamefont {Altug}},\ }\bibfield  {title} {\enquote
		{\bibinfo {title} {Mid-infrared plasmonic biosensing with graphene},}\ }\href
	{\doibase 10.1126/science.aab2051} {\bibfield  {journal} {\bibinfo  {journal}
			{Science}\ }\textbf {\bibinfo {volume} {349}},\ \bibinfo {pages} {165}
		(\bibinfo {year} {2015})}\BibitemShut {NoStop}%
	\bibitem [{\citenamefont {Lee}\ \emph {et~al.}(2017)\citenamefont {Lee},
		\citenamefont {Kang}, \citenamefont {Kwon}, \citenamefont {Lee},
		\citenamefont {Lee}, \citenamefont {Woo}, \citenamefont {Kim}, \citenamefont
		{Song}, \citenamefont {Park},\ and\ \citenamefont {Seo}}]{Lee2017}%
	\BibitemOpen
	\bibfield  {author} {\bibinfo {author} {\bibfnamefont {D.-K.}\ \bibnamefont
			{Lee}}, \bibinfo {author} {\bibfnamefont {J.-H.}\ \bibnamefont {Kang}},
		\bibinfo {author} {\bibfnamefont {J.}~\bibnamefont {Kwon}}, \bibinfo {author}
		{\bibfnamefont {J.-S.}\ \bibnamefont {Lee}}, \bibinfo {author} {\bibfnamefont
			{S.}~\bibnamefont {Lee}}, \bibinfo {author} {\bibfnamefont {D.~H.}\
			\bibnamefont {Woo}}, \bibinfo {author} {\bibfnamefont {J.~H.}\ \bibnamefont
			{Kim}}, \bibinfo {author} {\bibfnamefont {C.-S.}\ \bibnamefont {Song}},
		\bibinfo {author} {\bibfnamefont {Q.-H.}\ \bibnamefont {Park}}, \ and\
		\bibinfo {author} {\bibfnamefont {M.}~\bibnamefont {Seo}},\ }\bibfield
	{title} {\enquote {\bibinfo {title} {Nano metamaterials for ultrasensitive
				terahertz biosensing},}\ }\href {\doibase 10.1038/s41598-017-08508-7}
	{\bibfield  {journal} {\bibinfo  {journal} {Scientific Reports}\ }\textbf
		{\bibinfo {volume} {7}} (\bibinfo {year} {2017}),\
		10.1038/s41598-017-08508-7}\BibitemShut {NoStop}%
	\bibitem [{\citenamefont {Park}\ \emph
		{et~al.}(2014{\natexlab{a}})\citenamefont {Park}, \citenamefont {Hong},
		\citenamefont {Choi}, \citenamefont {Kim}, \citenamefont {Park},
		\citenamefont {Han}, \citenamefont {Park}, \citenamefont {Lee}, \citenamefont
		{Kim},\ and\ \citenamefont {Ahn}}]{Park2014}%
	\BibitemOpen
	\bibfield  {author} {\bibinfo {author} {\bibfnamefont {S.~J.}\ \bibnamefont
			{Park}}, \bibinfo {author} {\bibfnamefont {J.~T.}\ \bibnamefont {Hong}},
		\bibinfo {author} {\bibfnamefont {S.~J.}\ \bibnamefont {Choi}}, \bibinfo
		{author} {\bibfnamefont {H.~S.}\ \bibnamefont {Kim}}, \bibinfo {author}
		{\bibfnamefont {W.~K.}\ \bibnamefont {Park}}, \bibinfo {author}
		{\bibfnamefont {S.~T.}\ \bibnamefont {Han}}, \bibinfo {author} {\bibfnamefont
			{J.~Y.}\ \bibnamefont {Park}}, \bibinfo {author} {\bibfnamefont
			{S.}~\bibnamefont {Lee}}, \bibinfo {author} {\bibfnamefont {D.~S.}\
			\bibnamefont {Kim}}, \ and\ \bibinfo {author} {\bibfnamefont {Y.~H.}\
			\bibnamefont {Ahn}},\ }\bibfield  {title} {\enquote {\bibinfo {title}
			{Detection of microorganisms using terahertz metamaterials},}\ }\href
	{\doibase 10.1038/srep04988} {\bibfield  {journal} {\bibinfo  {journal}
			{Scientific Reports}\ }\textbf {\bibinfo {volume} {4}} (\bibinfo {year}
		{2014}{\natexlab{a}}),\ 10.1038/srep04988}\BibitemShut {NoStop}%
	\bibitem [{\citenamefont {Park}\ \emph {et~al.}(2017)\citenamefont {Park},
		\citenamefont {Cha}, \citenamefont {Shin},\ and\ \citenamefont
		{Ahn}}]{Park2017}%
	\BibitemOpen
	\bibfield  {author} {\bibinfo {author} {\bibfnamefont {S.~J.}\ \bibnamefont
			{Park}}, \bibinfo {author} {\bibfnamefont {S.~H.}\ \bibnamefont {Cha}},
		\bibinfo {author} {\bibfnamefont {G.~A.}\ \bibnamefont {Shin}}, \ and\
		\bibinfo {author} {\bibfnamefont {Y.~H.}\ \bibnamefont {Ahn}},\ }\bibfield
	{title} {\enquote {\bibinfo {title} {Sensing viruses using terahertz nano-gap
				metamaterials},}\ }\href {\doibase 10.1364/boe.8.003551} {\bibfield
		{journal} {\bibinfo  {journal} {Biomedical Optics Express}\ }\textbf
		{\bibinfo {volume} {8}},\ \bibinfo {pages} {3551} (\bibinfo {year}
		{2017})}\BibitemShut {NoStop}%
	\bibitem [{\citenamefont {Chen}\ \emph {et~al.}(2013)\citenamefont {Chen},
		\citenamefont {Park}, \citenamefont {Pelton}, \citenamefont {Piao},
		\citenamefont {Lindquist}, \citenamefont {Im}, \citenamefont {Kim},
		\citenamefont {Ahn}, \citenamefont {Ahn}, \citenamefont {Park}, \citenamefont
		{Kim},\ and\ \citenamefont {Oh}}]{Chen2013}%
	\BibitemOpen
	\bibfield  {author} {\bibinfo {author} {\bibfnamefont {X.}~\bibnamefont
			{Chen}}, \bibinfo {author} {\bibfnamefont {H.-R.}\ \bibnamefont {Park}},
		\bibinfo {author} {\bibfnamefont {M.}~\bibnamefont {Pelton}}, \bibinfo
		{author} {\bibfnamefont {X.}~\bibnamefont {Piao}}, \bibinfo {author}
		{\bibfnamefont {N.~C.}\ \bibnamefont {Lindquist}}, \bibinfo {author}
		{\bibfnamefont {H.}~\bibnamefont {Im}}, \bibinfo {author} {\bibfnamefont
			{Y.~J.}\ \bibnamefont {Kim}}, \bibinfo {author} {\bibfnamefont {J.~S.}\
			\bibnamefont {Ahn}}, \bibinfo {author} {\bibfnamefont {K.~J.}\ \bibnamefont
			{Ahn}}, \bibinfo {author} {\bibfnamefont {N.}~\bibnamefont {Park}}, \bibinfo
		{author} {\bibfnamefont {D.-S.}\ \bibnamefont {Kim}}, \ and\ \bibinfo
		{author} {\bibfnamefont {S.-H.}\ \bibnamefont {Oh}},\ }\bibfield  {title}
	{\enquote {\bibinfo {title} {Atomic layer lithography of wafer-scale nanogap
				arrays for extreme confinement of electromagnetic waves},}\ }\href {\doibase
		10.1038/ncomms3361} {\bibfield  {journal} {\bibinfo  {journal} {Nature
				Communications}\ }\textbf {\bibinfo {volume} {4}} (\bibinfo {year} {2013}),\
		10.1038/ncomms3361}\BibitemShut {NoStop}%
	\bibitem [{\citenamefont {Lee}\ \emph {et~al.}(2006{\natexlab{b}})\citenamefont
		{Lee}, \citenamefont {Seo}, \citenamefont {Park}, \citenamefont {Kim},
		\citenamefont {Jeoung}, \citenamefont {Lienau}, \citenamefont {Park},\ and\
		\citenamefont {Planken}}]{Lee2006}%
	\BibitemOpen
	\bibfield  {author} {\bibinfo {author} {\bibfnamefont {J.}~\bibnamefont
			{Lee}}, \bibinfo {author} {\bibfnamefont {M.}~\bibnamefont {Seo}}, \bibinfo
		{author} {\bibfnamefont {D.}~\bibnamefont {Park}}, \bibinfo {author}
		{\bibfnamefont {D.}~\bibnamefont {Kim}}, \bibinfo {author} {\bibfnamefont
			{S.}~\bibnamefont {Jeoung}}, \bibinfo {author} {\bibfnamefont
			{C.}~\bibnamefont {Lienau}}, \bibinfo {author} {\bibfnamefont {Q.-H.}\
			\bibnamefont {Park}}, \ and\ \bibinfo {author} {\bibfnamefont
			{P.}~\bibnamefont {Planken}},\ }\bibfield  {title} {\enquote {\bibinfo
			{title} {Shape resonance omni-directional terahertz filters with near-unity
				transmittance},}\ }\href {\doibase 10.1364/oe.14.001253} {\bibfield
		{journal} {\bibinfo  {journal} {Optics Express}\ }\textbf {\bibinfo {volume}
			{14}},\ \bibinfo {pages} {1253} (\bibinfo {year}
		{2006}{\natexlab{b}})}\BibitemShut {NoStop}%
	\bibitem [{\citenamefont {Lee}\ \emph {et~al.}(2006{\natexlab{c}})\citenamefont
		{Lee}, \citenamefont {Seo}, \citenamefont {Kim}, \citenamefont {Jeoung},
		\citenamefont {Lienau}, \citenamefont {Kang},\ and\ \citenamefont
		{Park}}]{Lee2006b}%
	\BibitemOpen
	\bibfield  {author} {\bibinfo {author} {\bibfnamefont {J.~W.}\ \bibnamefont
			{Lee}}, \bibinfo {author} {\bibfnamefont {M.~A.}\ \bibnamefont {Seo}},
		\bibinfo {author} {\bibfnamefont {D.~S.}\ \bibnamefont {Kim}}, \bibinfo
		{author} {\bibfnamefont {S.~C.}\ \bibnamefont {Jeoung}}, \bibinfo {author}
		{\bibfnamefont {C.}~\bibnamefont {Lienau}}, \bibinfo {author} {\bibfnamefont
			{J.~H.}\ \bibnamefont {Kang}}, \ and\ \bibinfo {author} {\bibfnamefont
			{Q.-H.}\ \bibnamefont {Park}},\ }\bibfield  {title} {\enquote {\bibinfo
			{title} {Fabry{\textendash}perot effects in {THz} time-domain spectroscopy of
				plasmonic band-gap structures},}\ }\href {\doibase 10.1063/1.2174104}
	{\bibfield  {journal} {\bibinfo  {journal} {Applied Physics Letters}\
		}\textbf {\bibinfo {volume} {88}},\ \bibinfo {pages} {071114} (\bibinfo
		{year} {2006}{\natexlab{c}})}\BibitemShut {NoStop}%
	\bibitem [{\citenamefont {Ward}\ \emph {et~al.}(2006)\citenamefont {Ward},
		\citenamefont {Statz},\ and\ \citenamefont {Nelson}}]{Ward2006}%
	\BibitemOpen
	\bibfield  {author} {\bibinfo {author} {\bibfnamefont {D.}~\bibnamefont
			{Ward}}, \bibinfo {author} {\bibfnamefont {E.}~\bibnamefont {Statz}}, \ and\
		\bibinfo {author} {\bibfnamefont {K.}~\bibnamefont {Nelson}},\ }\bibfield
	{title} {\enquote {\bibinfo {title} {Fabrication of polaritonic structures in
				{LiNbO}3 and {LiTaO}3 using femtosecond laser machining},}\ }\href {\doibase
		10.1007/s00339-006-3721-y} {\bibfield  {journal} {\bibinfo  {journal}
			{Applied Physics A}\ }\textbf {\bibinfo {volume} {86}},\ \bibinfo {pages}
		{49} (\bibinfo {year} {2006})}\BibitemShut {NoStop}%
	\bibitem [{\citenamefont {Enkrich}\ \emph {et~al.}(2005)\citenamefont
		{Enkrich}, \citenamefont {P{\'{e}}rez-Willard}, \citenamefont {Gerthsen},
		\citenamefont {Zhou}, \citenamefont {Koschny}, \citenamefont {Soukoulis},
		\citenamefont {Wegener},\ and\ \citenamefont {Linden}}]{Enkrich2005}%
	\BibitemOpen
	\bibfield  {author} {\bibinfo {author} {\bibfnamefont {C.}~\bibnamefont
			{Enkrich}}, \bibinfo {author} {\bibfnamefont {F.}~\bibnamefont
			{P{\'{e}}rez-Willard}}, \bibinfo {author} {\bibfnamefont {D.}~\bibnamefont
			{Gerthsen}}, \bibinfo {author} {\bibfnamefont {J.~F.}\ \bibnamefont {Zhou}},
		\bibinfo {author} {\bibfnamefont {T.}~\bibnamefont {Koschny}}, \bibinfo
		{author} {\bibfnamefont {C.~M.}\ \bibnamefont {Soukoulis}}, \bibinfo {author}
		{\bibfnamefont {M.}~\bibnamefont {Wegener}}, \ and\ \bibinfo {author}
		{\bibfnamefont {S.}~\bibnamefont {Linden}},\ }\bibfield  {title} {\enquote
		{\bibinfo {title} {Focused-ion-beam nanofabrication of near-infrared magnetic
				metamaterials},}\ }\href {\doibase 10.1002/adma.200500804} {\bibfield
		{journal} {\bibinfo  {journal} {Advanced Materials}\ }\textbf {\bibinfo
			{volume} {17}},\ \bibinfo {pages} {2547} (\bibinfo {year}
		{2005})}\BibitemShut {NoStop}%
	\bibitem [{\citenamefont {Ocelic}\ and\ \citenamefont
		{Hillenbrand}(2004)}]{Ocelic2004}%
	\BibitemOpen
	\bibfield  {author} {\bibinfo {author} {\bibfnamefont {N.}~\bibnamefont
			{Ocelic}}\ and\ \bibinfo {author} {\bibfnamefont {R.}~\bibnamefont
			{Hillenbrand}},\ }\bibfield  {title} {\enquote {\bibinfo {title}
			{Subwavelength-scale tailoring of surface phonon polaritons by focused
				ion-beam implantation},}\ }\href {\doibase 10.1038/nmat1194} {\bibfield
		{journal} {\bibinfo  {journal} {Nature Materials}\ }\textbf {\bibinfo
			{volume} {3}},\ \bibinfo {pages} {606} (\bibinfo {year} {2004})}\BibitemShut
	{NoStop}%
	\bibitem [{\citenamefont {Ghaemi}\ \emph {et~al.}(1998)\citenamefont {Ghaemi},
		\citenamefont {Thio}, \citenamefont {Grupp}, \citenamefont {Ebbesen},\ and\
		\citenamefont {Lezec}}]{Ghaemi1998}%
	\BibitemOpen
	\bibfield  {author} {\bibinfo {author} {\bibfnamefont {H.~F.}\ \bibnamefont
			{Ghaemi}}, \bibinfo {author} {\bibfnamefont {T.}~\bibnamefont {Thio}},
		\bibinfo {author} {\bibfnamefont {D.~E.}\ \bibnamefont {Grupp}}, \bibinfo
		{author} {\bibfnamefont {T.~W.}\ \bibnamefont {Ebbesen}}, \ and\ \bibinfo
		{author} {\bibfnamefont {H.~J.}\ \bibnamefont {Lezec}},\ }\bibfield  {title}
	{\enquote {\bibinfo {title} {Surface plasmons enhance optical transmission
				through subwavelength holes},}\ }\href {\doibase 10.1103/physrevb.58.6779}
	{\bibfield  {journal} {\bibinfo  {journal} {Physical Review B}\ }\textbf
		{\bibinfo {volume} {58}},\ \bibinfo {pages} {6779} (\bibinfo {year}
		{1998})}\BibitemShut {NoStop}%
	\bibitem [{\citenamefont {Ebbesen}\ \emph {et~al.}(1998)\citenamefont
		{Ebbesen}, \citenamefont {Lezec}, \citenamefont {Ghaemi}, \citenamefont
		{Thio},\ and\ \citenamefont {Wolff}}]{Ebbesen1998}%
	\BibitemOpen
	\bibfield  {author} {\bibinfo {author} {\bibfnamefont {T.~W.}\ \bibnamefont
			{Ebbesen}}, \bibinfo {author} {\bibfnamefont {H.~J.}\ \bibnamefont {Lezec}},
		\bibinfo {author} {\bibfnamefont {H.~F.}\ \bibnamefont {Ghaemi}}, \bibinfo
		{author} {\bibfnamefont {T.}~\bibnamefont {Thio}}, \ and\ \bibinfo {author}
		{\bibfnamefont {P.~A.}\ \bibnamefont {Wolff}},\ }\bibfield  {title} {\enquote
		{\bibinfo {title} {Extraordinary optical transmission through sub-wavelength
				hole arrays},}\ }\href {\doibase 10.1038/35570} {\bibfield  {journal}
		{\bibinfo  {journal} {Nature}\ }\textbf {\bibinfo {volume} {391}},\ \bibinfo
		{pages} {667} (\bibinfo {year} {1998})}\BibitemShut {NoStop}%
	\bibitem [{\citenamefont {Park}\ \emph
		{et~al.}(2011{\natexlab{b}})\citenamefont {Park}, \citenamefont {Bahk},
		\citenamefont {Choe}, \citenamefont {Han}, \citenamefont {Choi},
		\citenamefont {Ahn}, \citenamefont {Park}, \citenamefont {Park},\ and\
		\citenamefont {Kim}}]{Park2011a}%
	\BibitemOpen
	\bibfield  {author} {\bibinfo {author} {\bibfnamefont {H.-R.}\ \bibnamefont
			{Park}}, \bibinfo {author} {\bibfnamefont {Y.-M.}\ \bibnamefont {Bahk}},
		\bibinfo {author} {\bibfnamefont {J.~H.}\ \bibnamefont {Choe}}, \bibinfo
		{author} {\bibfnamefont {S.}~\bibnamefont {Han}}, \bibinfo {author}
		{\bibfnamefont {S.~S.}\ \bibnamefont {Choi}}, \bibinfo {author}
		{\bibfnamefont {K.~J.}\ \bibnamefont {Ahn}}, \bibinfo {author} {\bibfnamefont
			{N.}~\bibnamefont {Park}}, \bibinfo {author} {\bibfnamefont {Q.-H.}\
			\bibnamefont {Park}}, \ and\ \bibinfo {author} {\bibfnamefont {D.-S.}\
			\bibnamefont {Kim}},\ }\bibfield  {title} {\enquote {\bibinfo {title}
			{Terahertz pinch harmonics enabled by single nano rods},}\ }\href {\doibase
		10.1364/oe.19.024775} {\bibfield  {journal} {\bibinfo  {journal} {Optics
				Express}\ }\textbf {\bibinfo {volume} {19}},\ \bibinfo {pages} {24775}
		(\bibinfo {year} {2011}{\natexlab{b}})}\BibitemShut {NoStop}%
	\bibitem [{\citenamefont {Bell}\ \emph {et~al.}(2009)\citenamefont {Bell},
		\citenamefont {Lemme}, \citenamefont {Stern}, \citenamefont {Williams},\ and\
		\citenamefont {Marcus}}]{Bell2009}%
	\BibitemOpen
	\bibfield  {author} {\bibinfo {author} {\bibfnamefont {D.~C.}\ \bibnamefont
			{Bell}}, \bibinfo {author} {\bibfnamefont {M.~C.}\ \bibnamefont {Lemme}},
		\bibinfo {author} {\bibfnamefont {L.~A.}\ \bibnamefont {Stern}}, \bibinfo
		{author} {\bibfnamefont {J.~R.}\ \bibnamefont {Williams}}, \ and\ \bibinfo
		{author} {\bibfnamefont {C.~M.}\ \bibnamefont {Marcus}},\ }\bibfield  {title}
	{\enquote {\bibinfo {title} {Precision cutting and patterning of graphene
				with helium ions},}\ }\href {\doibase 10.1088/0957-4484/20/45/455301}
	{\bibfield  {journal} {\bibinfo  {journal} {Nanotechnology}\ }\textbf
		{\bibinfo {volume} {20}},\ \bibinfo {pages} {455301} (\bibinfo {year}
		{2009})}\BibitemShut {NoStop}%
	\bibitem [{\citenamefont {Chen}\ \emph {et~al.}(2008)\citenamefont {Chen},
		\citenamefont {O{\textquotesingle}Hara}, \citenamefont {Azad}, \citenamefont
		{Taylor}, \citenamefont {Averitt}, \citenamefont {Shrekenhamer},\ and\
		\citenamefont {Padilla}}]{Chen2008}%
	\BibitemOpen
	\bibfield  {author} {\bibinfo {author} {\bibfnamefont {H.-T.}\ \bibnamefont
			{Chen}}, \bibinfo {author} {\bibfnamefont {J.~F.}\ \bibnamefont
			{O{\textquotesingle}Hara}}, \bibinfo {author} {\bibfnamefont {A.~K.}\
			\bibnamefont {Azad}}, \bibinfo {author} {\bibfnamefont {A.~J.}\ \bibnamefont
			{Taylor}}, \bibinfo {author} {\bibfnamefont {R.~D.}\ \bibnamefont {Averitt}},
		\bibinfo {author} {\bibfnamefont {D.~B.}\ \bibnamefont {Shrekenhamer}}, \
		and\ \bibinfo {author} {\bibfnamefont {W.~J.}\ \bibnamefont {Padilla}},\
	}\bibfield  {title} {\enquote {\bibinfo {title} {Experimental demonstration
				of frequency-agile terahertz metamaterials},}\ }\href {\doibase
		10.1038/nphoton.2008.52} {\bibfield  {journal} {\bibinfo  {journal} {Nature
				Photonics}\ }\textbf {\bibinfo {volume} {2}},\ \bibinfo {pages} {295}
		(\bibinfo {year} {2008})}\BibitemShut {NoStop}%
	\bibitem [{\citenamefont {Singh}\ \emph {et~al.}(2009)\citenamefont {Singh},
		\citenamefont {Plum}, \citenamefont {Menzel}, \citenamefont {Rockstuhl},
		\citenamefont {Azad}, \citenamefont {Cheville}, \citenamefont {Lederer},
		\citenamefont {Zhang},\ and\ \citenamefont {Zheludev}}]{Singh2009}%
	\BibitemOpen
	\bibfield  {author} {\bibinfo {author} {\bibfnamefont {R.}~\bibnamefont
			{Singh}}, \bibinfo {author} {\bibfnamefont {E.}~\bibnamefont {Plum}},
		\bibinfo {author} {\bibfnamefont {C.}~\bibnamefont {Menzel}}, \bibinfo
		{author} {\bibfnamefont {C.}~\bibnamefont {Rockstuhl}}, \bibinfo {author}
		{\bibfnamefont {A.~K.}\ \bibnamefont {Azad}}, \bibinfo {author}
		{\bibfnamefont {R.~A.}\ \bibnamefont {Cheville}}, \bibinfo {author}
		{\bibfnamefont {F.}~\bibnamefont {Lederer}}, \bibinfo {author} {\bibfnamefont
			{W.}~\bibnamefont {Zhang}}, \ and\ \bibinfo {author} {\bibfnamefont {N.~I.}\
			\bibnamefont {Zheludev}},\ }\bibfield  {title} {\enquote {\bibinfo {title}
			{Terahertz metamaterial with asymmetric transmission},}\ }\href {\doibase
		10.1103/physrevb.80.153104} {\bibfield  {journal} {\bibinfo  {journal}
			{Physical Review B}\ }\textbf {\bibinfo {volume} {80}} (\bibinfo {year}
		{2009}),\ 10.1103/physrevb.80.153104}\BibitemShut {NoStop}%
	\bibitem [{\citenamefont {Choi}\ \emph
		{et~al.}(2011{\natexlab{b}})\citenamefont {Choi}, \citenamefont {Lee},
		\citenamefont {Kim}, \citenamefont {Kang}, \citenamefont {Shin},
		\citenamefont {Kwak}, \citenamefont {Kang}, \citenamefont {Lee},
		\citenamefont {Park},\ and\ \citenamefont {Min}}]{Choi2011a}%
	\BibitemOpen
	\bibfield  {author} {\bibinfo {author} {\bibfnamefont {M.}~\bibnamefont
			{Choi}}, \bibinfo {author} {\bibfnamefont {S.~H.}\ \bibnamefont {Lee}},
		\bibinfo {author} {\bibfnamefont {Y.}~\bibnamefont {Kim}}, \bibinfo {author}
		{\bibfnamefont {S.~B.}\ \bibnamefont {Kang}}, \bibinfo {author}
		{\bibfnamefont {J.}~\bibnamefont {Shin}}, \bibinfo {author} {\bibfnamefont
			{M.~H.}\ \bibnamefont {Kwak}}, \bibinfo {author} {\bibfnamefont {K.-Y.}\
			\bibnamefont {Kang}}, \bibinfo {author} {\bibfnamefont {Y.-H.}\ \bibnamefont
			{Lee}}, \bibinfo {author} {\bibfnamefont {N.}~\bibnamefont {Park}}, \ and\
		\bibinfo {author} {\bibfnamefont {B.}~\bibnamefont {Min}},\ }\bibfield
	{title} {\enquote {\bibinfo {title} {A terahertz metamaterial with
				unnaturally high refractive index},}\ }\href {\doibase 10.1038/nature09776}
	{\bibfield  {journal} {\bibinfo  {journal} {Nature}\ }\textbf {\bibinfo
			{volume} {470}},\ \bibinfo {pages} {369} (\bibinfo {year}
		{2011}{\natexlab{b}})}\BibitemShut {NoStop}%
	\bibitem [{\citenamefont {Willson}\ \emph {et~al.}(1997)\citenamefont
		{Willson}, \citenamefont {Dammel},\ and\ \citenamefont
		{Reiser}}]{Willson1997}%
	\BibitemOpen
	\bibfield  {author} {\bibinfo {author} {\bibfnamefont {C.~G.}\ \bibnamefont
			{Willson}}, \bibinfo {author} {\bibfnamefont {R.~R.}\ \bibnamefont {Dammel}},
		\ and\ \bibinfo {author} {\bibfnamefont {A.}~\bibnamefont {Reiser}},\
	}\bibfield  {title} {\enquote {\bibinfo {title} {Photoresist materials: a
				historical perspective},}\ }in\ \href {\doibase 10.1117/12.275921} {\emph
		{\bibinfo {booktitle} {Metrology, Inspection, and Process Control for
				Microlithography {XI}}}}\ (\bibinfo  {publisher} {{SPIE}},\ \bibinfo {year}
	{1997})\BibitemShut {NoStop}%
	\bibitem [{\citenamefont {Lin}\ \emph {et~al.}(2004)\citenamefont {Lin},
		\citenamefont {Chen}, \citenamefont {Schneider}, \citenamefont {Yao},
		\citenamefont {Shi}, \citenamefont {Sharkawy},\ and\ \citenamefont
		{Prather}}]{Lin2004}%
	\BibitemOpen
	\bibfield  {author} {\bibinfo {author} {\bibfnamefont {C.}~\bibnamefont
			{Lin}}, \bibinfo {author} {\bibfnamefont {C.}~\bibnamefont {Chen}}, \bibinfo
		{author} {\bibfnamefont {G.~J.}\ \bibnamefont {Schneider}}, \bibinfo {author}
		{\bibfnamefont {P.}~\bibnamefont {Yao}}, \bibinfo {author} {\bibfnamefont
			{S.}~\bibnamefont {Shi}}, \bibinfo {author} {\bibfnamefont {A.}~\bibnamefont
			{Sharkawy}}, \ and\ \bibinfo {author} {\bibfnamefont {D.~W.}\ \bibnamefont
			{Prather}},\ }\bibfield  {title} {\enquote {\bibinfo {title} {Wavelength
				scale terahertz two-dimensional photonic crystal waveguides},}\ }\href
	{\doibase 10.1364/opex.12.005723} {\bibfield  {journal} {\bibinfo  {journal}
			{Optics Express}\ }\textbf {\bibinfo {volume} {12}},\ \bibinfo {pages} {5723}
		(\bibinfo {year} {2004})}\BibitemShut {NoStop}%
	\bibitem [{\citenamefont {Park}\ \emph {et~al.}(2015)\citenamefont {Park},
		\citenamefont {Chen}, \citenamefont {Nguyen}, \citenamefont {Peraire},\ and\
		\citenamefont {Oh}}]{Park2015}%
	\BibitemOpen
	\bibfield  {author} {\bibinfo {author} {\bibfnamefont {H.-R.}\ \bibnamefont
			{Park}}, \bibinfo {author} {\bibfnamefont {X.}~\bibnamefont {Chen}}, \bibinfo
		{author} {\bibfnamefont {N.-C.}\ \bibnamefont {Nguyen}}, \bibinfo {author}
		{\bibfnamefont {J.}~\bibnamefont {Peraire}}, \ and\ \bibinfo {author}
		{\bibfnamefont {S.-H.}\ \bibnamefont {Oh}},\ }\bibfield  {title} {\enquote
		{\bibinfo {title} {Nanogap-enhanced terahertz sensing of 1 nm thick
				($\lambda$/106) dielectric films},}\ }\href {\doibase 10.1021/ph500464j}
	{\bibfield  {journal} {\bibinfo  {journal} {{ACS} Photonics}\ }\textbf
		{\bibinfo {volume} {2}},\ \bibinfo {pages} {417} (\bibinfo {year}
		{2015})}\BibitemShut {NoStop}%
	\bibitem [{\citenamefont {Fursina}\ \emph {et~al.}(2008)\citenamefont
		{Fursina}, \citenamefont {Lee}, \citenamefont {Sofin}, \citenamefont
		{Shvets},\ and\ \citenamefont {Natelson}}]{Fursina2008}%
	\BibitemOpen
	\bibfield  {author} {\bibinfo {author} {\bibfnamefont {A.}~\bibnamefont
			{Fursina}}, \bibinfo {author} {\bibfnamefont {S.}~\bibnamefont {Lee}},
		\bibinfo {author} {\bibfnamefont {R.~G.~S.}\ \bibnamefont {Sofin}}, \bibinfo
		{author} {\bibfnamefont {I.~V.}\ \bibnamefont {Shvets}}, \ and\ \bibinfo
		{author} {\bibfnamefont {D.}~\bibnamefont {Natelson}},\ }\bibfield  {title}
	{\enquote {\bibinfo {title} {Nanogaps with very large aspect ratios for
				electrical measurements},}\ }\href {\doibase 10.1063/1.2895644} {\bibfield
		{journal} {\bibinfo  {journal} {Applied Physics Letters}\ }\textbf {\bibinfo
			{volume} {92}},\ \bibinfo {pages} {113102} (\bibinfo {year}
		{2008})}\BibitemShut {NoStop}%
	\bibitem [{\citenamefont {Vieu}\ \emph {et~al.}(2000)\citenamefont {Vieu},
		\citenamefont {Carcenac}, \citenamefont {P{\'{e}}pin}, \citenamefont {Chen},
		\citenamefont {Mejias}, \citenamefont {Lebib}, \citenamefont
		{Manin-Ferlazzo}, \citenamefont {Couraud},\ and\ \citenamefont
		{Launois}}]{Vieu2000}%
	\BibitemOpen
	\bibfield  {author} {\bibinfo {author} {\bibfnamefont {C.}~\bibnamefont
			{Vieu}}, \bibinfo {author} {\bibfnamefont {F.}~\bibnamefont {Carcenac}},
		\bibinfo {author} {\bibfnamefont {A.}~\bibnamefont {P{\'{e}}pin}}, \bibinfo
		{author} {\bibfnamefont {Y.}~\bibnamefont {Chen}}, \bibinfo {author}
		{\bibfnamefont {M.}~\bibnamefont {Mejias}}, \bibinfo {author} {\bibfnamefont
			{A.}~\bibnamefont {Lebib}}, \bibinfo {author} {\bibfnamefont
			{L.}~\bibnamefont {Manin-Ferlazzo}}, \bibinfo {author} {\bibfnamefont
			{L.}~\bibnamefont {Couraud}}, \ and\ \bibinfo {author} {\bibfnamefont
			{H.}~\bibnamefont {Launois}},\ }\bibfield  {title} {\enquote {\bibinfo
			{title} {Electron beam lithography: resolution limits and applications},}\
	}\href {\doibase 10.1016/s0169-4332(00)00352-4} {\bibfield  {journal}
		{\bibinfo  {journal} {Applied Surface Science}\ }\textbf {\bibinfo {volume}
			{164}},\ \bibinfo {pages} {111} (\bibinfo {year} {2000})}\BibitemShut
	{NoStop}%
	\bibitem [{\citenamefont {Broers}\ \emph {et~al.}(1996)\citenamefont {Broers},
		\citenamefont {Hoole},\ and\ \citenamefont {Ryan}}]{Broers1996}%
	\BibitemOpen
	\bibfield  {author} {\bibinfo {author} {\bibfnamefont {A.}~\bibnamefont
			{Broers}}, \bibinfo {author} {\bibfnamefont {A.}~\bibnamefont {Hoole}}, \
		and\ \bibinfo {author} {\bibfnamefont {J.}~\bibnamefont {Ryan}},\ }\bibfield
	{title} {\enquote {\bibinfo {title} {Electron beam
				lithography{\textemdash}resolution limits},}\ }\href {\doibase
		10.1016/0167-9317(95)00368-1} {\bibfield  {journal} {\bibinfo  {journal}
			{Microelectronic Engineering}\ }\textbf {\bibinfo {volume} {32}},\ \bibinfo
		{pages} {131} (\bibinfo {year} {1996})}\BibitemShut {NoStop}%
	\bibitem [{\citenamefont {Chen}(2015)}]{Chen2015}%
	\BibitemOpen
	\bibfield  {author} {\bibinfo {author} {\bibfnamefont {Y.}~\bibnamefont
			{Chen}},\ }\bibfield  {title} {\enquote {\bibinfo {title} {Nanofabrication by
				electron beam lithography and its applications: A review},}\ }\href {\doibase
		10.1016/j.mee.2015.02.042} {\bibfield  {journal} {\bibinfo  {journal}
			{Microelectronic Engineering}\ }\textbf {\bibinfo {volume} {135}},\ \bibinfo
		{pages} {57} (\bibinfo {year} {2015})}\BibitemShut {NoStop}%
	\bibitem [{\citenamefont {Tseng}\ \emph {et~al.}(2003)\citenamefont {Tseng},
		\citenamefont {Chen}, \citenamefont {Chen},\ and\ \citenamefont
		{Ma}}]{Tseng2003}%
	\BibitemOpen
	\bibfield  {author} {\bibinfo {author} {\bibfnamefont {A.}~\bibnamefont
			{Tseng}}, \bibinfo {author} {\bibfnamefont {K.}~\bibnamefont {Chen}},
		\bibinfo {author} {\bibfnamefont {C.}~\bibnamefont {Chen}}, \ and\ \bibinfo
		{author} {\bibfnamefont {K.}~\bibnamefont {Ma}},\ }\bibfield  {title}
	{\enquote {\bibinfo {title} {Electron beam lithography in nanoscale
				fabrication: recent development},}\ }\href {\doibase
		10.1109/tepm.2003.817714} {\bibfield  {journal} {\bibinfo  {journal} {{IEEE}
				Transactions on Electronics Packaging Manufacturing}\ }\textbf {\bibinfo
			{volume} {26}},\ \bibinfo {pages} {141} (\bibinfo {year} {2003})}\BibitemShut
	{NoStop}%
	\bibitem [{\citenamefont {Hulteen}\ and\ \citenamefont
		{Duyne}(1995)}]{Hulteen1995}%
	\BibitemOpen
	\bibfield  {author} {\bibinfo {author} {\bibfnamefont {J.~C.}\ \bibnamefont
			{Hulteen}}\ and\ \bibinfo {author} {\bibfnamefont {R.~P.~V.}\ \bibnamefont
			{Duyne}},\ }\bibfield  {title} {\enquote {\bibinfo {title} {Nanosphere
				lithography: A materials general fabrication process for periodic particle
				array surfaces},}\ }\href {\doibase 10.1116/1.579726} {\bibfield  {journal}
		{\bibinfo  {journal} {Journal of Vacuum Science {\&} Technology A: Vacuum,
				Surfaces, and Films}\ }\textbf {\bibinfo {volume} {13}},\ \bibinfo {pages}
		{1553} (\bibinfo {year} {1995})}\BibitemShut {NoStop}%
	\bibitem [{\citenamefont {Vettiger}(1989)}]{Vettiger1989}%
	\BibitemOpen
	\bibfield  {author} {\bibinfo {author} {\bibfnamefont {P.}~\bibnamefont
			{Vettiger}},\ }\bibfield  {title} {\enquote {\bibinfo {title} {Nanometer
				sidewall lithography by resist silylation},}\ }\href {\doibase
		10.1116/1.584452} {\bibfield  {journal} {\bibinfo  {journal} {Journal of
				Vacuum Science {\&} Technology B: Microelectronics and Nanometer Structures}\
		}\textbf {\bibinfo {volume} {7}},\ \bibinfo {pages} {1756} (\bibinfo {year}
		{1989})}\BibitemShut {NoStop}%
	\bibitem [{\citenamefont {Chou}\ \emph {et~al.}(1996)\citenamefont {Chou},
		\citenamefont {Krauss},\ and\ \citenamefont {Renstrom}}]{Chou1996}%
	\BibitemOpen
	\bibfield  {author} {\bibinfo {author} {\bibfnamefont {S.~Y.}\ \bibnamefont
			{Chou}}, \bibinfo {author} {\bibfnamefont {P.~R.}\ \bibnamefont {Krauss}}, \
		and\ \bibinfo {author} {\bibfnamefont {P.~J.}\ \bibnamefont {Renstrom}},\
	}\bibfield  {title} {\enquote {\bibinfo {title} {Imprint lithography with
				25-nanometer resolution},}\ }\href {\doibase 10.1126/science.272.5258.85}
	{\bibfield  {journal} {\bibinfo  {journal} {Science}\ }\textbf {\bibinfo
			{volume} {272}},\ \bibinfo {pages} {85} (\bibinfo {year} {1996})}\BibitemShut
	{NoStop}%
	\bibitem [{\citenamefont {Chou}(1996)}]{Chou1996a}%
	\BibitemOpen
	\bibfield  {author} {\bibinfo {author} {\bibfnamefont {S.~Y.}\ \bibnamefont
			{Chou}},\ }\bibfield  {title} {\enquote {\bibinfo {title} {Nanoimprint
				lithography},}\ }\href {\doibase 10.1116/1.588605} {\bibfield  {journal}
		{\bibinfo  {journal} {Journal of Vacuum Science {\&} Technology B:
				Microelectronics and Nanometer Structures}\ }\textbf {\bibinfo {volume}
			{14}},\ \bibinfo {pages} {4129} (\bibinfo {year} {1996})}\BibitemShut
	{NoStop}%
	\bibitem [{\citenamefont {Austin}\ \emph {et~al.}(2004)\citenamefont {Austin},
		\citenamefont {Ge}, \citenamefont {Wu}, \citenamefont {Li}, \citenamefont
		{Yu}, \citenamefont {Wasserman}, \citenamefont {Lyon},\ and\ \citenamefont
		{Chou}}]{Austin2004}%
	\BibitemOpen
	\bibfield  {author} {\bibinfo {author} {\bibfnamefont {M.~D.}\ \bibnamefont
			{Austin}}, \bibinfo {author} {\bibfnamefont {H.}~\bibnamefont {Ge}}, \bibinfo
		{author} {\bibfnamefont {W.}~\bibnamefont {Wu}}, \bibinfo {author}
		{\bibfnamefont {M.}~\bibnamefont {Li}}, \bibinfo {author} {\bibfnamefont
			{Z.}~\bibnamefont {Yu}}, \bibinfo {author} {\bibfnamefont {D.}~\bibnamefont
			{Wasserman}}, \bibinfo {author} {\bibfnamefont {S.~A.}\ \bibnamefont {Lyon}},
		\ and\ \bibinfo {author} {\bibfnamefont {S.~Y.}\ \bibnamefont {Chou}},\
	}\bibfield  {title} {\enquote {\bibinfo {title} {Fabrication of 5nm linewidth
				and 14nm pitch features by nanoimprint lithography},}\ }\href {\doibase
		10.1063/1.1766071} {\bibfield  {journal} {\bibinfo  {journal} {Applied
				Physics Letters}\ }\textbf {\bibinfo {volume} {84}},\ \bibinfo {pages} {5299}
		(\bibinfo {year} {2004})}\BibitemShut {NoStop}%
	\bibitem [{\citenamefont {Guo}(2007)}]{Guo2007}%
	\BibitemOpen
	\bibfield  {author} {\bibinfo {author} {\bibfnamefont {L.}~\bibnamefont
			{Guo}},\ }\bibfield  {title} {\enquote {\bibinfo {title} {Nanoimprint
				lithography: Methods and material requirements},}\ }\href {\doibase
		10.1002/adma.200600882} {\bibfield  {journal} {\bibinfo  {journal} {Advanced
				Materials}\ }\textbf {\bibinfo {volume} {19}},\ \bibinfo {pages} {495}
		(\bibinfo {year} {2007})}\BibitemShut {NoStop}%
	\bibitem [{\citenamefont {Guo}\ \emph {et~al.}(1997)\citenamefont {Guo},
		\citenamefont {Krauss},\ and\ \citenamefont {Chou}}]{Guo1997}%
	\BibitemOpen
	\bibfield  {author} {\bibinfo {author} {\bibfnamefont {L.}~\bibnamefont
			{Guo}}, \bibinfo {author} {\bibfnamefont {P.~R.}\ \bibnamefont {Krauss}}, \
		and\ \bibinfo {author} {\bibfnamefont {S.~Y.}\ \bibnamefont {Chou}},\
	}\bibfield  {title} {\enquote {\bibinfo {title} {Nanoscale silicon field
				effect transistors fabricated using imprint lithography},}\ }\href {\doibase
		10.1063/1.119426} {\bibfield  {journal} {\bibinfo  {journal} {Applied Physics
				Letters}\ }\textbf {\bibinfo {volume} {71}},\ \bibinfo {pages} {1881}
		(\bibinfo {year} {1997})}\BibitemShut {NoStop}%
	\bibitem [{\citenamefont {Ambhire}\ \emph {et~al.}(2018)\citenamefont
		{Ambhire}, \citenamefont {Palkhivala}, \citenamefont {Agrawal}, \citenamefont
		{Gupta}, \citenamefont {Rana}, \citenamefont {Mehta}, \citenamefont
		{Ghindani}, \citenamefont {Bhattacharya}, \citenamefont {Achanta},\ and\
		\citenamefont {Prabhu}}]{Ambhire2018}%
	\BibitemOpen
	\bibfield  {author} {\bibinfo {author} {\bibfnamefont {S.~C.}\ \bibnamefont
			{Ambhire}}, \bibinfo {author} {\bibfnamefont {S.}~\bibnamefont {Palkhivala}},
		\bibinfo {author} {\bibfnamefont {A.}~\bibnamefont {Agrawal}}, \bibinfo
		{author} {\bibfnamefont {A.}~\bibnamefont {Gupta}}, \bibinfo {author}
		{\bibfnamefont {G.}~\bibnamefont {Rana}}, \bibinfo {author} {\bibfnamefont
			{R.}~\bibnamefont {Mehta}}, \bibinfo {author} {\bibfnamefont
			{D.}~\bibnamefont {Ghindani}}, \bibinfo {author} {\bibfnamefont
			{A.}~\bibnamefont {Bhattacharya}}, \bibinfo {author} {\bibfnamefont {V.~G.}\
			\bibnamefont {Achanta}}, \ and\ \bibinfo {author} {\bibfnamefont {S.~S.}\
			\bibnamefont {Prabhu}},\ }\bibfield  {title} {\enquote {\bibinfo {title}
			{{\textquotedblleft}pattern and peel{\textquotedblright} method for
				fabricating mechanically tunable terahertz metasurface on an elastomeric
				substrate},}\ }\href {\doibase 10.1364/ome.8.003382} {\bibfield  {journal}
		{\bibinfo  {journal} {Optical Materials Express}\ }\textbf {\bibinfo {volume}
			{8}},\ \bibinfo {pages} {3382} (\bibinfo {year} {2018})}\BibitemShut
	{NoStop}%
	\bibitem [{\citenamefont {Feuillet-Palma}\ \emph {et~al.}(2013)\citenamefont
		{Feuillet-Palma}, \citenamefont {Todorov}, \citenamefont {Vasanelli},\ and\
		\citenamefont {Sirtori}}]{Feuillet-Palma2013}%
	\BibitemOpen
	\bibfield  {author} {\bibinfo {author} {\bibfnamefont {C.}~\bibnamefont
			{Feuillet-Palma}}, \bibinfo {author} {\bibfnamefont {Y.}~\bibnamefont
			{Todorov}}, \bibinfo {author} {\bibfnamefont {A.}~\bibnamefont {Vasanelli}},
		\ and\ \bibinfo {author} {\bibfnamefont {C.}~\bibnamefont {Sirtori}},\
	}\bibfield  {title} {\enquote {\bibinfo {title} {Strong near field
				enhancement in {THz} nano-antenna arrays},}\ }\href {\doibase
		10.1038/srep01361} {\bibfield  {journal} {\bibinfo  {journal} {Scientific
				Reports}\ }\textbf {\bibinfo {volume} {3}} (\bibinfo {year} {2013}),\
		10.1038/srep01361}\BibitemShut {NoStop}%
	\bibitem [{\citenamefont {Suwal}\ \emph {et~al.}(2017)\citenamefont {Suwal},
		\citenamefont {Rhie}, \citenamefont {Kim},\ and\ \citenamefont
		{Kim}}]{Suwal2017}%
	\BibitemOpen
	\bibfield  {author} {\bibinfo {author} {\bibfnamefont {O.~K.}\ \bibnamefont
			{Suwal}}, \bibinfo {author} {\bibfnamefont {J.}~\bibnamefont {Rhie}},
		\bibinfo {author} {\bibfnamefont {N.}~\bibnamefont {Kim}}, \ and\ \bibinfo
		{author} {\bibfnamefont {D.-S.}\ \bibnamefont {Kim}},\ }\bibfield  {title}
	{\enquote {\bibinfo {title} {Nonresonant 104 terahertz field enhancement with
				5-nm slits},}\ }\href {\doibase 10.1038/srep45638} {\bibfield  {journal}
		{\bibinfo  {journal} {Scientific Reports}\ }\textbf {\bibinfo {volume} {7}}
		(\bibinfo {year} {2017}),\ 10.1038/srep45638}\BibitemShut {NoStop}%
	\bibitem [{\citenamefont {Park}\ \emph {et~al.}(2016)\citenamefont {Park},
		\citenamefont {Rhie}, \citenamefont {Kim}, \citenamefont {Hong},\ and\
		\citenamefont {Kim}}]{Park2016}%
	\BibitemOpen
	\bibfield  {author} {\bibinfo {author} {\bibfnamefont {W.}~\bibnamefont
			{Park}}, \bibinfo {author} {\bibfnamefont {J.}~\bibnamefont {Rhie}}, \bibinfo
		{author} {\bibfnamefont {N.~Y.}\ \bibnamefont {Kim}}, \bibinfo {author}
		{\bibfnamefont {S.}~\bibnamefont {Hong}}, \ and\ \bibinfo {author}
		{\bibfnamefont {D.-S.}\ \bibnamefont {Kim}},\ }\bibfield  {title} {\enquote
		{\bibinfo {title} {Sub-10{\hspace{0.167em}}nm feature chromium photomasks for
				contact lithography patterning of square metal ring arrays},}\ }\href
	{\doibase 10.1038/srep23823} {\bibfield  {journal} {\bibinfo  {journal}
			{Scientific Reports}\ }\textbf {\bibinfo {volume} {6}} (\bibinfo {year}
		{2016}),\ 10.1038/srep23823}\BibitemShut {NoStop}%
	\bibitem [{\citenamefont {Im}\ \emph {et~al.}(2010)\citenamefont {Im},
		\citenamefont {Bantz}, \citenamefont {Lindquist}, \citenamefont {Haynes},\
		and\ \citenamefont {Oh}}]{Im2010}%
	\BibitemOpen
	\bibfield  {author} {\bibinfo {author} {\bibfnamefont {H.}~\bibnamefont
			{Im}}, \bibinfo {author} {\bibfnamefont {K.~C.}\ \bibnamefont {Bantz}},
		\bibinfo {author} {\bibfnamefont {N.~C.}\ \bibnamefont {Lindquist}}, \bibinfo
		{author} {\bibfnamefont {C.~L.}\ \bibnamefont {Haynes}}, \ and\ \bibinfo
		{author} {\bibfnamefont {S.-H.}\ \bibnamefont {Oh}},\ }\bibfield  {title}
	{\enquote {\bibinfo {title} {Vertically oriented sub-10-nm plasmonic nanogap
				arrays},}\ }\href {\doibase 10.1021/nl1012085} {\bibfield  {journal}
		{\bibinfo  {journal} {Nano Letters}\ }\textbf {\bibinfo {volume} {10}},\
		\bibinfo {pages} {2231} (\bibinfo {year} {2010})}\BibitemShut {NoStop}%
	\bibitem [{\citenamefont {Miyazaki}\ and\ \citenamefont
		{Kurokawa}(2006)}]{Miyazaki2006}%
	\BibitemOpen
	\bibfield  {author} {\bibinfo {author} {\bibfnamefont {H.~T.}\ \bibnamefont
			{Miyazaki}}\ and\ \bibinfo {author} {\bibfnamefont {Y.}~\bibnamefont
			{Kurokawa}},\ }\bibfield  {title} {\enquote {\bibinfo {title} {Squeezing
				visible light waves into a 3-nm-thick and 55-nm-long plasmon cavity},}\
	}\href {\doibase 10.1103/physrevlett.96.097401} {\bibfield  {journal}
		{\bibinfo  {journal} {Physical Review Letters}\ }\textbf {\bibinfo {volume}
			{96}} (\bibinfo {year} {2006}),\ 10.1103/physrevlett.96.097401}\BibitemShut
	{NoStop}%
	\bibitem [{\citenamefont {Zhu}\ \emph {et~al.}(2011)\citenamefont {Zhu},
		\citenamefont {Banaee}, \citenamefont {Wang}, \citenamefont {Chu},\ and\
		\citenamefont {Crozier}}]{Zhu2011}%
	\BibitemOpen
	\bibfield  {author} {\bibinfo {author} {\bibfnamefont {W.}~\bibnamefont
			{Zhu}}, \bibinfo {author} {\bibfnamefont {M.~G.}\ \bibnamefont {Banaee}},
		\bibinfo {author} {\bibfnamefont {D.}~\bibnamefont {Wang}}, \bibinfo {author}
		{\bibfnamefont {Y.}~\bibnamefont {Chu}}, \ and\ \bibinfo {author}
		{\bibfnamefont {K.~B.}\ \bibnamefont {Crozier}},\ }\bibfield  {title}
	{\enquote {\bibinfo {title} {Lithographically fabricated optical antennas
				with gaps well below 10 nm},}\ }\href {\doibase 10.1002/smll.201100371}
	{\bibfield  {journal} {\bibinfo  {journal} {Small}\ }\textbf {\bibinfo
			{volume} {7}},\ \bibinfo {pages} {1761} (\bibinfo {year} {2011})}\BibitemShut
	{NoStop}%
	\bibitem [{\citenamefont {Fan}\ \emph {et~al.}(2005)\citenamefont {Fan},
		\citenamefont {Zhang}, \citenamefont {Malloy},\ and\ \citenamefont
		{Brueck}}]{Fan2005}%
	\BibitemOpen
	\bibfield  {author} {\bibinfo {author} {\bibfnamefont {W.}~\bibnamefont
			{Fan}}, \bibinfo {author} {\bibfnamefont {S.}~\bibnamefont {Zhang}}, \bibinfo
		{author} {\bibfnamefont {K.~J.}\ \bibnamefont {Malloy}}, \ and\ \bibinfo
		{author} {\bibfnamefont {S.~R.~J.}\ \bibnamefont {Brueck}},\ }\bibfield
	{title} {\enquote {\bibinfo {title} {Enhanced mid-infrared transmission
				through nanoscale metallic coaxial-aperture arrays},}\ }\href {\doibase
		10.1364/opex.13.004406} {\bibfield  {journal} {\bibinfo  {journal} {Optics
				Express}\ }\textbf {\bibinfo {volume} {13}},\ \bibinfo {pages} {4406}
		(\bibinfo {year} {2005})}\BibitemShut {NoStop}%
	\bibitem [{\citenamefont {Ji}\ \emph {et~al.}(2017)\citenamefont {Ji},
		\citenamefont {Cheney}, \citenamefont {Zhang}, \citenamefont {Song},
		\citenamefont {Gao}, \citenamefont {Zeng}, \citenamefont {Hu}, \citenamefont
		{Jiang}, \citenamefont {Yu},\ and\ \citenamefont {Gan}}]{Ji2017}%
	\BibitemOpen
	\bibfield  {author} {\bibinfo {author} {\bibfnamefont {D.}~\bibnamefont
			{Ji}}, \bibinfo {author} {\bibfnamefont {A.}~\bibnamefont {Cheney}}, \bibinfo
		{author} {\bibfnamefont {N.}~\bibnamefont {Zhang}}, \bibinfo {author}
		{\bibfnamefont {H.}~\bibnamefont {Song}}, \bibinfo {author} {\bibfnamefont
			{J.}~\bibnamefont {Gao}}, \bibinfo {author} {\bibfnamefont {X.}~\bibnamefont
			{Zeng}}, \bibinfo {author} {\bibfnamefont {H.}~\bibnamefont {Hu}}, \bibinfo
		{author} {\bibfnamefont {S.}~\bibnamefont {Jiang}}, \bibinfo {author}
		{\bibfnamefont {Z.}~\bibnamefont {Yu}}, \ and\ \bibinfo {author}
		{\bibfnamefont {Q.}~\bibnamefont {Gan}},\ }\bibfield  {title} {\enquote
		{\bibinfo {title} {Efficient mid-infrared light confinement within sub-5-nm
				gaps for extreme field enhancement},}\ }\href {\doibase
		10.1002/adom.201700223} {\bibfield  {journal} {\bibinfo  {journal} {Advanced
				Optical Materials}\ }\textbf {\bibinfo {volume} {5}},\ \bibinfo {pages}
		{1700223} (\bibinfo {year} {2017})}\BibitemShut {NoStop}%
	\bibitem [{\citenamefont {Ding}\ \emph {et~al.}(2015)\citenamefont {Ding},
		\citenamefont {Sigle}, \citenamefont {Zhang}, \citenamefont {Mertens},
		\citenamefont {de~Nijs},\ and\ \citenamefont {Baumberg}}]{Ding2015}%
	\BibitemOpen
	\bibfield  {author} {\bibinfo {author} {\bibfnamefont {T.}~\bibnamefont
			{Ding}}, \bibinfo {author} {\bibfnamefont {D.}~\bibnamefont {Sigle}},
		\bibinfo {author} {\bibfnamefont {L.}~\bibnamefont {Zhang}}, \bibinfo
		{author} {\bibfnamefont {J.}~\bibnamefont {Mertens}}, \bibinfo {author}
		{\bibfnamefont {B.}~\bibnamefont {de~Nijs}}, \ and\ \bibinfo {author}
		{\bibfnamefont {J.}~\bibnamefont {Baumberg}},\ }\bibfield  {title} {\enquote
		{\bibinfo {title} {Controllable tuning plasmonic coupling with nanoscale
				oxidation},}\ }\href {\doibase 10.1021/acsnano.5b01283} {\bibfield  {journal}
		{\bibinfo  {journal} {{ACS} Nano}\ }\textbf {\bibinfo {volume} {9}},\
		\bibinfo {pages} {6110} (\bibinfo {year} {2015})}\BibitemShut {NoStop}%
	\bibitem [{\citenamefont {Tripathi}\ \emph {et~al.}(2015)\citenamefont
		{Tripathi}, \citenamefont {Kang}, \citenamefont {Bahk}, \citenamefont {Han},
		\citenamefont {Choi}, \citenamefont {Rhie}, \citenamefont {Jeong},\ and\
		\citenamefont {Kim}}]{Tripathi2015}%
	\BibitemOpen
	\bibfield  {author} {\bibinfo {author} {\bibfnamefont {L.~N.}\ \bibnamefont
			{Tripathi}}, \bibinfo {author} {\bibfnamefont {T.}~\bibnamefont {Kang}},
		\bibinfo {author} {\bibfnamefont {Y.-M.}\ \bibnamefont {Bahk}}, \bibinfo
		{author} {\bibfnamefont {S.}~\bibnamefont {Han}}, \bibinfo {author}
		{\bibfnamefont {G.}~\bibnamefont {Choi}}, \bibinfo {author} {\bibfnamefont
			{J.}~\bibnamefont {Rhie}}, \bibinfo {author} {\bibfnamefont {J.}~\bibnamefont
			{Jeong}}, \ and\ \bibinfo {author} {\bibfnamefont {D.-S.}\ \bibnamefont
			{Kim}},\ }\bibfield  {title} {\enquote {\bibinfo {title} {Quantum
				dots-nanogap metamaterials fabrication by self-assembly lithography and
				photoluminescence studies},}\ }\href {\doibase 10.1364/oe.23.014937}
	{\bibfield  {journal} {\bibinfo  {journal} {Optics Express}\ }\textbf
		{\bibinfo {volume} {23}},\ \bibinfo {pages} {14937} (\bibinfo {year}
		{2015})}\BibitemShut {NoStop}%
	\bibitem [{\citenamefont {Kang}\ \emph
		{et~al.}(2009{\natexlab{a}})\citenamefont {Kang}, \citenamefont {Choe},
		\citenamefont {Kim},\ and\ \citenamefont {Park}}]{Kang2009}%
	\BibitemOpen
	\bibfield  {author} {\bibinfo {author} {\bibfnamefont {J.~H.}\ \bibnamefont
			{Kang}}, \bibinfo {author} {\bibfnamefont {J.-H.}\ \bibnamefont {Choe}},
		\bibinfo {author} {\bibfnamefont {D.~S.}\ \bibnamefont {Kim}}, \ and\
		\bibinfo {author} {\bibfnamefont {Q.-H.}\ \bibnamefont {Park}},\ }\bibfield
	{title} {\enquote {\bibinfo {title} {Substrate effect on aperture resonances
				in a thin metal film},}\ }\href {\doibase 10.1364/oe.17.015652} {\bibfield
		{journal} {\bibinfo  {journal} {Optics Express}\ }\textbf {\bibinfo {volume}
			{17}},\ \bibinfo {pages} {15652} (\bibinfo {year}
		{2009}{\natexlab{a}})}\BibitemShut {NoStop}%
	\bibitem [{\citenamefont {Choe}\ \emph {et~al.}(2012)\citenamefont {Choe},
		\citenamefont {Kang}, \citenamefont {Kim},\ and\ \citenamefont
		{Park}}]{Choe2012}%
	\BibitemOpen
	\bibfield  {author} {\bibinfo {author} {\bibfnamefont {J.-H.}\ \bibnamefont
			{Choe}}, \bibinfo {author} {\bibfnamefont {J.-H.}\ \bibnamefont {Kang}},
		\bibinfo {author} {\bibfnamefont {D.-S.}\ \bibnamefont {Kim}}, \ and\
		\bibinfo {author} {\bibfnamefont {Q.-H.}\ \bibnamefont {Park}},\ }\bibfield
	{title} {\enquote {\bibinfo {title} {Slot antenna as a bound charge
				oscillator},}\ }\href {\doibase 10.1364/oe.20.006521} {\bibfield  {journal}
		{\bibinfo  {journal} {Optics Express}\ }\textbf {\bibinfo {volume} {20}},\
		\bibinfo {pages} {6521} (\bibinfo {year} {2012})}\BibitemShut {NoStop}%
	\bibitem [{\citenamefont {Garc{\'{\i}}a-Vidal}\ \emph
		{et~al.}(2006)\citenamefont {Garc{\'{\i}}a-Vidal}, \citenamefont
		{Mart{\'{\i}}n-Moreno}, \citenamefont {Moreno}, \citenamefont {Kumar},\ and\
		\citenamefont {Gordon}}]{Garcia-Vidal2006}%
	\BibitemOpen
	\bibfield  {author} {\bibinfo {author} {\bibfnamefont {F.~J.}\ \bibnamefont
			{Garc{\'{\i}}a-Vidal}}, \bibinfo {author} {\bibfnamefont {L.}~\bibnamefont
			{Mart{\'{\i}}n-Moreno}}, \bibinfo {author} {\bibfnamefont {E.}~\bibnamefont
			{Moreno}}, \bibinfo {author} {\bibfnamefont {L.~K.~S.}\ \bibnamefont
			{Kumar}}, \ and\ \bibinfo {author} {\bibfnamefont {R.}~\bibnamefont
			{Gordon}},\ }\bibfield  {title} {\enquote {\bibinfo {title} {Transmission of
				light through a single rectangular hole in a real metal},}\ }\href {\doibase
		10.1103/physrevb.74.153411} {\bibfield  {journal} {\bibinfo  {journal}
			{Physical Review B}\ }\textbf {\bibinfo {volume} {74}} (\bibinfo {year}
		{2006}),\ 10.1103/physrevb.74.153411}\BibitemShut {NoStop}%
	\bibitem [{\citenamefont {Park}\ \emph {et~al.}(2012)\citenamefont {Park},
		\citenamefont {Choi}, \citenamefont {Oh},\ and\ \citenamefont
		{Jeong}}]{Park2012}%
	\BibitemOpen
	\bibfield  {author} {\bibinfo {author} {\bibfnamefont {S.-G.}\ \bibnamefont
			{Park}}, \bibinfo {author} {\bibfnamefont {Y.}~\bibnamefont {Choi}}, \bibinfo
		{author} {\bibfnamefont {Y.-J.}\ \bibnamefont {Oh}}, \ and\ \bibinfo {author}
		{\bibfnamefont {K.-H.}\ \bibnamefont {Jeong}},\ }\bibfield  {title} {\enquote
		{\bibinfo {title} {Terahertz photoconductive antenna with metal
				nanoislands},}\ }\href {\doibase 10.1364/oe.20.025530} {\bibfield  {journal}
		{\bibinfo  {journal} {Optics Express}\ }\textbf {\bibinfo {volume} {20}},\
		\bibinfo {pages} {25530} (\bibinfo {year} {2012})}\BibitemShut {NoStop}%
	\bibitem [{\citenamefont {Yang}\ \emph {et~al.}(2011)\citenamefont {Yang},
		\citenamefont {Singh},\ and\ \citenamefont {Zhang}}]{Yang2011}%
	\BibitemOpen
	\bibfield  {author} {\bibinfo {author} {\bibfnamefont {Y.}~\bibnamefont
			{Yang}}, \bibinfo {author} {\bibfnamefont {R.}~\bibnamefont {Singh}}, \ and\
		\bibinfo {author} {\bibfnamefont {W.}~\bibnamefont {Zhang}},\ }\bibfield
	{title} {\enquote {\bibinfo {title} {Anomalous terahertz transmission in
				bow-tie plasmonic antenna apertures},}\ }\href {\doibase
		10.1364/ol.36.002901} {\bibfield  {journal} {\bibinfo  {journal} {Optics
				Letters}\ }\textbf {\bibinfo {volume} {36}},\ \bibinfo {pages} {2901}
		(\bibinfo {year} {2011})}\BibitemShut {NoStop}%
	\bibitem [{\citenamefont {Maksymov}\ \emph {et~al.}(2012)\citenamefont
		{Maksymov}, \citenamefont {Miroshnichenko},\ and\ \citenamefont
		{Kivshar}}]{Maksymov2012}%
	\BibitemOpen
	\bibfield  {author} {\bibinfo {author} {\bibfnamefont {I.~S.}\ \bibnamefont
			{Maksymov}}, \bibinfo {author} {\bibfnamefont {A.~E.}\ \bibnamefont
			{Miroshnichenko}}, \ and\ \bibinfo {author} {\bibfnamefont {Y.~S.}\
			\bibnamefont {Kivshar}},\ }\bibfield  {title} {\enquote {\bibinfo {title}
			{Actively tunable bistable optical yagi-uda nanoantenna},}\ }\href {\doibase
		10.1364/oe.20.008929} {\bibfield  {journal} {\bibinfo  {journal} {Optics
				Express}\ }\textbf {\bibinfo {volume} {20}},\ \bibinfo {pages} {8929}
		(\bibinfo {year} {2012})}\BibitemShut {NoStop}%
	\bibitem [{\citenamefont {Taminiau}\ \emph {et~al.}(2008)\citenamefont
		{Taminiau}, \citenamefont {Stefani},\ and\ \citenamefont {van
			Hulst}}]{Taminiau2008}%
	\BibitemOpen
	\bibfield  {author} {\bibinfo {author} {\bibfnamefont {T.~H.}\ \bibnamefont
			{Taminiau}}, \bibinfo {author} {\bibfnamefont {F.~D.}\ \bibnamefont
			{Stefani}}, \ and\ \bibinfo {author} {\bibfnamefont {N.~F.}\ \bibnamefont
			{van Hulst}},\ }\bibfield  {title} {\enquote {\bibinfo {title} {Enhanced
				directional excitation and emission of single emitters by a nano-optical
				yagi-uda antenna},}\ }\href {\doibase 10.1364/oe.16.010858} {\bibfield
		{journal} {\bibinfo  {journal} {Optics Express}\ }\textbf {\bibinfo {volume}
			{16}},\ \bibinfo {pages} {10858} (\bibinfo {year} {2008})}\BibitemShut
	{NoStop}%
	\bibitem [{\citenamefont {Pelosi}\ and\ \citenamefont
		{Selleri}(2017)}]{Pelosi2017}%
	\BibitemOpen
	\bibfield  {author} {\bibinfo {author} {\bibfnamefont {G.}~\bibnamefont
			{Pelosi}}\ and\ \bibinfo {author} {\bibfnamefont {S.}~\bibnamefont
			{Selleri}},\ }\bibfield  {title} {\enquote {\bibinfo {title}
			{Babinet{\textquotesingle}s principle in electromagnetics: Why does a slot
				radiate like a dipole? [historical corner]},}\ }\href {\doibase
		10.1109/map.2017.2658345} {\bibfield  {journal} {\bibinfo  {journal} {{IEEE}
				Antennas and Propagation Magazine}\ }\textbf {\bibinfo {volume} {59}},\
		\bibinfo {pages} {144} (\bibinfo {year} {2017})}\BibitemShut {NoStop}%
	\bibitem [{\citenamefont {Booker}(1946)}]{Booker1946}%
	\BibitemOpen
	\bibfield  {author} {\bibinfo {author} {\bibfnamefont {H.}~\bibnamefont
			{Booker}},\ }\bibfield  {title} {\enquote {\bibinfo {title} {Slot aerials and
				their relation to complementary wire aerials (babinet{\textquotesingle}s
				principle)},}\ }\href {\doibase 10.1049/ji-3a-1.1946.0150} {\bibfield
		{journal} {\bibinfo  {journal} {Journal of the Institution of Electrical
				Engineers - Part {IIIA}: Radiolocation}\ }\textbf {\bibinfo {volume} {93}},\
		\bibinfo {pages} {620} (\bibinfo {year} {1946})}\BibitemShut {NoStop}%
	\bibitem [{\citenamefont {Mushiake}(2004)}]{Mushiake2004}%
	\BibitemOpen
	\bibfield  {author} {\bibinfo {author} {\bibfnamefont {Y.}~\bibnamefont
			{Mushiake}},\ }\bibfield  {title} {\enquote {\bibinfo {title} {A report on
				japanese development of antennas: from the yagi-uda antenna to
				self-complementary antennas},}\ }\href {\doibase 10.1109/map.2004.1373999}
	{\bibfield  {journal} {\bibinfo  {journal} {{IEEE} Antennas and Propagation
				Magazine}\ }\textbf {\bibinfo {volume} {46}},\ \bibinfo {pages} {47}
		(\bibinfo {year} {2004})}\BibitemShut {NoStop}%
	\bibitem [{\citenamefont {Born}\ and\ \citenamefont {Wolf}(1981)}]{Born1981}%
	\BibitemOpen
	\bibfield  {author} {\bibinfo {author} {\bibfnamefont {M.}~\bibnamefont
			{Born}}\ and\ \bibinfo {author} {\bibfnamefont {E.}~\bibnamefont {Wolf}},\
	}\href
	{https://www.amazon.com/Principles-Optics-Electromagnetic-Propagation-Interference/dp/0080264824?SubscriptionId=AKIAIOBINVZYXZQZ2U3A&tag=chimbori05-20&linkCode=xm2&camp=2025&creative=165953&creativeASIN=0080264824}
	{\emph {\bibinfo {title} {Principles of Optics: Electromagnetic Theory of
				Propagation Interference and Diffraction of Light}}}\ (\bibinfo  {publisher}
	{Pergamon Pr},\ \bibinfo {year} {1981})\BibitemShut {NoStop}%
	\bibitem [{\citenamefont {Koo}\ \emph {et~al.}(2009)\citenamefont {Koo},
		\citenamefont {Kumar}, \citenamefont {Shin}, \citenamefont {Kim},\ and\
		\citenamefont {Park}}]{Koo2009}%
	\BibitemOpen
	\bibfield  {author} {\bibinfo {author} {\bibfnamefont {S.}~\bibnamefont
			{Koo}}, \bibinfo {author} {\bibfnamefont {M.~S.}\ \bibnamefont {Kumar}},
		\bibinfo {author} {\bibfnamefont {J.}~\bibnamefont {Shin}}, \bibinfo {author}
		{\bibfnamefont {D.}~\bibnamefont {Kim}}, \ and\ \bibinfo {author}
		{\bibfnamefont {N.}~\bibnamefont {Park}},\ }\bibfield  {title} {\enquote
		{\bibinfo {title} {Extraordinary magnetic field enhancement with metallic
				nanowire: Role of surface impedance in babinet's principle for sub-skin-depth
				regime},}\ }\href {\doibase 10.1103/physrevlett.103.263901} {\bibfield
		{journal} {\bibinfo  {journal} {Physical Review Letters}\ }\textbf {\bibinfo
			{volume} {103}} (\bibinfo {year} {2009}),\
		10.1103/physrevlett.103.263901}\BibitemShut {NoStop}%
	\bibitem [{\citenamefont {Razzari}\ \emph {et~al.}(2012)\citenamefont
		{Razzari}, \citenamefont {Toma}, \citenamefont {Clerici}, \citenamefont
		{Shalaby}, \citenamefont {Das}, \citenamefont {Liberale}, \citenamefont
		{Chirumamilla}, \citenamefont {Zaccaria}, \citenamefont {Angelis},
		\citenamefont {Peccianti}, \citenamefont {Morandotti},\ and\ \citenamefont
		{Fabrizio}}]{Razzari2012}%
	\BibitemOpen
	\bibfield  {author} {\bibinfo {author} {\bibfnamefont {L.}~\bibnamefont
			{Razzari}}, \bibinfo {author} {\bibfnamefont {A.}~\bibnamefont {Toma}},
		\bibinfo {author} {\bibfnamefont {M.}~\bibnamefont {Clerici}}, \bibinfo
		{author} {\bibfnamefont {M.}~\bibnamefont {Shalaby}}, \bibinfo {author}
		{\bibfnamefont {G.}~\bibnamefont {Das}}, \bibinfo {author} {\bibfnamefont
			{C.}~\bibnamefont {Liberale}}, \bibinfo {author} {\bibfnamefont
			{M.}~\bibnamefont {Chirumamilla}}, \bibinfo {author} {\bibfnamefont {R.~P.}\
			\bibnamefont {Zaccaria}}, \bibinfo {author} {\bibfnamefont {F.~D.}\
			\bibnamefont {Angelis}}, \bibinfo {author} {\bibfnamefont {M.}~\bibnamefont
			{Peccianti}}, \bibinfo {author} {\bibfnamefont {R.}~\bibnamefont
			{Morandotti}}, \ and\ \bibinfo {author} {\bibfnamefont {E.~D.}\ \bibnamefont
			{Fabrizio}},\ }\bibfield  {title} {\enquote {\bibinfo {title} {Terahertz
				dipole nanoantenna arrays: Resonance characteristics},}\ }\href {\doibase
		10.1007/s11468-012-9439-0} {\bibfield  {journal} {\bibinfo  {journal}
			{Plasmonics}\ }\textbf {\bibinfo {volume} {8}},\ \bibinfo {pages} {133}
		(\bibinfo {year} {2012})}\BibitemShut {NoStop}%
	\bibitem [{\citenamefont {Jeong}\ \emph {et~al.}(2013)\citenamefont {Jeong},
		\citenamefont {Paul}, \citenamefont {Kim}, \citenamefont {Yee}, \citenamefont
		{Kim},\ and\ \citenamefont {Lee}}]{Jeong2013}%
	\BibitemOpen
	\bibfield  {author} {\bibinfo {author} {\bibfnamefont {Y.-G.}\ \bibnamefont
			{Jeong}}, \bibinfo {author} {\bibfnamefont {M.~J.}\ \bibnamefont {Paul}},
		\bibinfo {author} {\bibfnamefont {S.-H.}\ \bibnamefont {Kim}}, \bibinfo
		{author} {\bibfnamefont {K.-J.}\ \bibnamefont {Yee}}, \bibinfo {author}
		{\bibfnamefont {D.-S.}\ \bibnamefont {Kim}}, \ and\ \bibinfo {author}
		{\bibfnamefont {Y.-S.}\ \bibnamefont {Lee}},\ }\bibfield  {title} {\enquote
		{\bibinfo {title} {Large enhancement of nonlinear terahertz absorption in
				intrinsic {GaAs} by plasmonic nano antennas},}\ }\href {\doibase
		10.1063/1.4826272} {\bibfield  {journal} {\bibinfo  {journal} {Applied
				Physics Letters}\ }\textbf {\bibinfo {volume} {103}},\ \bibinfo {pages}
		{171109} (\bibinfo {year} {2013})}\BibitemShut {NoStop}%
	\bibitem [{\citenamefont {Shu}\ \emph {et~al.}(2011)\citenamefont {Shu},
		\citenamefont {Qiu}, \citenamefont {Astley}, \citenamefont {Nickel},
		\citenamefont {Mittleman},\ and\ \citenamefont {Xu}}]{Shu2011}%
	\BibitemOpen
	\bibfield  {author} {\bibinfo {author} {\bibfnamefont {J.}~\bibnamefont
			{Shu}}, \bibinfo {author} {\bibfnamefont {C.}~\bibnamefont {Qiu}}, \bibinfo
		{author} {\bibfnamefont {V.}~\bibnamefont {Astley}}, \bibinfo {author}
		{\bibfnamefont {D.}~\bibnamefont {Nickel}}, \bibinfo {author} {\bibfnamefont
			{D.~M.}\ \bibnamefont {Mittleman}}, \ and\ \bibinfo {author} {\bibfnamefont
			{Q.}~\bibnamefont {Xu}},\ }\bibfield  {title} {\enquote {\bibinfo {title}
			{High-contrast terahertz modulator based on extraordinary transmission
				through a ring aperture},}\ }\href {\doibase 10.1364/oe.19.026666} {\bibfield
		{journal} {\bibinfo  {journal} {Optics Express}\ }\textbf {\bibinfo {volume}
			{19}},\ \bibinfo {pages} {26666} (\bibinfo {year} {2011})}\BibitemShut
	{NoStop}%
	\bibitem [{\citenamefont {Merbold}\ \emph {et~al.}(2011)\citenamefont
		{Merbold}, \citenamefont {Bitzer},\ and\ \citenamefont
		{Feurer}}]{Merbold2011}%
	\BibitemOpen
	\bibfield  {author} {\bibinfo {author} {\bibfnamefont {H.}~\bibnamefont
			{Merbold}}, \bibinfo {author} {\bibfnamefont {A.}~\bibnamefont {Bitzer}}, \
		and\ \bibinfo {author} {\bibfnamefont {T.}~\bibnamefont {Feurer}},\
	}\bibfield  {title} {\enquote {\bibinfo {title} {Second harmonic generation
				based on strong field enhancement in nanostructured {THz} materials},}\
	}\href {\doibase 10.1364/oe.19.007262} {\bibfield  {journal} {\bibinfo
			{journal} {Optics Express}\ }\textbf {\bibinfo {volume} {19}},\ \bibinfo
		{pages} {7262} (\bibinfo {year} {2011})}\BibitemShut {NoStop}%
	\bibitem [{\citenamefont {Dai}\ and\ \citenamefont {Jiang}(2009)}]{Dai2009}%
	\BibitemOpen
	\bibfield  {author} {\bibinfo {author} {\bibfnamefont {L.}~\bibnamefont
			{Dai}}\ and\ \bibinfo {author} {\bibfnamefont {C.}~\bibnamefont {Jiang}},\
	}\bibfield  {title} {\enquote {\bibinfo {title} {Anomalous near-perfect extraordinary optical absorption on subwavelength thin metal film grating.},}\ }\href {\doibase 10.1364/oe.17.020502} {\bibfield  {journal}
		{\bibinfo  {journal} {Optics Express}\ }\textbf {\bibinfo {volume} {17}},\
		\bibinfo {pages} {20502} (\bibinfo {year} {2009})}\BibitemShut {NoStop}%
	\bibitem [{\citenamefont {Ögüt}\ \emph {et~al.}(2011)\citenamefont {Ögüt},
		\citenamefont {Vogelgesang}, \citenamefont {Sigle}, \citenamefont {Talebi},
		\citenamefont {Koch},\ and\ \citenamefont {van Aken}}]{Oeguet2011}%
	\BibitemOpen
	\bibfield  {author} {\bibinfo {author} {\bibfnamefont {B.}~\bibnamefont
			{Ögüt}}, \bibinfo {author} {\bibfnamefont {R.}~\bibnamefont {Vogelgesang}},
		\bibinfo {author} {\bibfnamefont {W.}~\bibnamefont {Sigle}}, \bibinfo
		{author} {\bibfnamefont {N.}~\bibnamefont {Talebi}}, \bibinfo {author}
		{\bibfnamefont {C.~T.}\ \bibnamefont {Koch}}, \ and\ \bibinfo {author}
		{\bibfnamefont {P.~A.}\ \bibnamefont {van Aken}},\ }\bibfield  {title}
	{\enquote {\bibinfo {title} {Hybridized metal slit eigenmodes as an
				illustration of babinet's principle},}\ }\href {\doibase 10.1021/nn2022414}
	{\bibfield  {journal} {\bibinfo  {journal} {{ACS} Nano}\ }\textbf {\bibinfo
			{volume} {5}},\ \bibinfo {pages} {6701} (\bibinfo {year} {2011})}\BibitemShut
	{NoStop}%
	\bibitem [{\citenamefont {Yang}\ \emph {et~al.}(2010)\citenamefont {Yang},
		\citenamefont {Cao},\ and\ \citenamefont {Zhou}}]{Yang2010}%
	\BibitemOpen
	\bibfield  {author} {\bibinfo {author} {\bibfnamefont {J.}~\bibnamefont
			{Yang}}, \bibinfo {author} {\bibfnamefont {Q.}~\bibnamefont {Cao}}, \ and\
		\bibinfo {author} {\bibfnamefont {C.}~\bibnamefont {Zhou}},\ }\bibfield
	{title} {\enquote {\bibinfo {title} {Theory for terahertz plasmons of
				metallic nanowires with sub-skin-depth diameters},}\ }\href {\doibase
		10.1364/oe.18.018550} {\bibfield  {journal} {\bibinfo  {journal} {Optics
				Express}\ }\textbf {\bibinfo {volume} {18}},\ \bibinfo {pages} {18550}
		(\bibinfo {year} {2010})}\BibitemShut {NoStop}%
	\bibitem [{\citenamefont {Hu}\ \emph {et~al.}(2013)\citenamefont {Hu},
		\citenamefont {Wang}, \citenamefont {Feng}, \citenamefont {Ye}, \citenamefont
		{Sun}, \citenamefont {Kan}, \citenamefont {Klar},\ and\ \citenamefont
		{Zhang}}]{Hu2013}%
	\BibitemOpen
	\bibfield  {author} {\bibinfo {author} {\bibfnamefont {D.}~\bibnamefont
			{Hu}}, \bibinfo {author} {\bibfnamefont {X.}~\bibnamefont {Wang}}, \bibinfo
		{author} {\bibfnamefont {S.}~\bibnamefont {Feng}}, \bibinfo {author}
		{\bibfnamefont {J.}~\bibnamefont {Ye}}, \bibinfo {author} {\bibfnamefont
			{W.}~\bibnamefont {Sun}}, \bibinfo {author} {\bibfnamefont {Q.}~\bibnamefont
			{Kan}}, \bibinfo {author} {\bibfnamefont {P.~J.}\ \bibnamefont {Klar}}, \
		and\ \bibinfo {author} {\bibfnamefont {Y.}~\bibnamefont {Zhang}},\ }\bibfield
	{title} {\enquote {\bibinfo {title} {Ultrathin terahertz planar elements},}\
	}\href {\doibase 10.1002/adom.201200044} {\bibfield  {journal} {\bibinfo
			{journal} {Advanced Optical Materials}\ }\textbf {\bibinfo {volume} {1}},\
		\bibinfo {pages} {186} (\bibinfo {year} {2013})}\BibitemShut {NoStop}%
	\bibitem [{\citenamefont {Iwaszczuk}\ \emph {et~al.}(2012)\citenamefont
		{Iwaszczuk}, \citenamefont {Andryieuski}, \citenamefont {Lavrinenko},
		\citenamefont {Zhang},\ and\ \citenamefont {Jepsen}}]{Iwaszczuk2012}%
	\BibitemOpen
	\bibfield  {author} {\bibinfo {author} {\bibfnamefont {K.}~\bibnamefont
			{Iwaszczuk}}, \bibinfo {author} {\bibfnamefont {A.}~\bibnamefont
			{Andryieuski}}, \bibinfo {author} {\bibfnamefont {A.}~\bibnamefont
			{Lavrinenko}}, \bibinfo {author} {\bibfnamefont {X.-C.}\ \bibnamefont
			{Zhang}}, \ and\ \bibinfo {author} {\bibfnamefont {P.~U.}\ \bibnamefont
			{Jepsen}},\ }\bibfield  {title} {\enquote {\bibinfo {title} {Terahertz field
				enhancement to the {MV}/cm regime in a tapered parallel plate waveguide},}\
	}\href {\doibase 10.1364/oe.20.008344} {\bibfield  {journal} {\bibinfo
			{journal} {Optics Express}\ }\textbf {\bibinfo {volume} {20}},\ \bibinfo
		{pages} {8344} (\bibinfo {year} {2012})}\BibitemShut {NoStop}%
	\bibitem [{\citenamefont {Bulgarevich}\ \emph {et~al.}(2012)\citenamefont
		{Bulgarevich}, \citenamefont {Watanabe},\ and\ \citenamefont
		{Shiwa}}]{Bulgarevich2012}%
	\BibitemOpen
	\bibfield  {author} {\bibinfo {author} {\bibfnamefont {D.~S.}\ \bibnamefont
			{Bulgarevich}}, \bibinfo {author} {\bibfnamefont {M.}~\bibnamefont
			{Watanabe}}, \ and\ \bibinfo {author} {\bibfnamefont {M.}~\bibnamefont
			{Shiwa}},\ }\bibfield  {title} {\enquote {\bibinfo {title} {Single
				sub-wavelength aperture with greatly enhanced transmission},}\ }\href
	{\doibase 10.1088/1367-2630/14/5/053001} {\bibfield  {journal} {\bibinfo
			{journal} {New Journal of Physics}\ }\textbf {\bibinfo {volume} {14}},\
		\bibinfo {pages} {053001} (\bibinfo {year} {2012})}\BibitemShut {NoStop}%
	\bibitem [{\citenamefont {Gadalla}\ \emph {et~al.}(2014)\citenamefont
		{Gadalla}, \citenamefont {Abdel-Rahman},\ and\ \citenamefont
		{Shamim}}]{Gadalla2014}%
	\BibitemOpen
	\bibfield  {author} {\bibinfo {author} {\bibfnamefont {M.~N.}\ \bibnamefont
			{Gadalla}}, \bibinfo {author} {\bibfnamefont {M.}~\bibnamefont
			{Abdel-Rahman}}, \ and\ \bibinfo {author} {\bibfnamefont {A.}~\bibnamefont
			{Shamim}},\ }\bibfield  {title} {\enquote {\bibinfo {title} {Design,
				optimization and fabrication of a 28.3{\hspace{0.25em}}{THz} nano-rectenna
				for infrared detection and rectification},}\ }\href {\doibase
		10.1038/srep04270} {\bibfield  {journal} {\bibinfo  {journal} {Scientific
				Reports}\ }\textbf {\bibinfo {volume} {4}} (\bibinfo {year} {2014}),\
		10.1038/srep04270}\BibitemShut {NoStop}%
	\bibitem [{\citenamefont {Fan}\ \emph {et~al.}(2016)\citenamefont {Fan},
		\citenamefont {Xu}, \citenamefont {Wang},\ and\ \citenamefont
		{Chang}}]{Fan2016}%
	\BibitemOpen
	\bibfield  {author} {\bibinfo {author} {\bibfnamefont {F.}~\bibnamefont
			{Fan}}, \bibinfo {author} {\bibfnamefont {S.-T.}\ \bibnamefont {Xu}},
		\bibinfo {author} {\bibfnamefont {X.-H.}\ \bibnamefont {Wang}}, \ and\
		\bibinfo {author} {\bibfnamefont {S.-J.}\ \bibnamefont {Chang}},\ }\bibfield
	{title} {\enquote {\bibinfo {title} {Terahertz polarization converter and
				one-way transmission based on double-layer magneto-plasmonics of magnetized
				{InSb}},}\ }\href {\doibase 10.1364/oe.24.026431} {\bibfield  {journal}
		{\bibinfo  {journal} {Optics Express}\ }\textbf {\bibinfo {volume} {24}},\
		\bibinfo {pages} {26431} (\bibinfo {year} {2016})}\BibitemShut {NoStop}%
	\bibitem [{\citenamefont {Low}\ and\ \citenamefont {Avouris}(2014)}]{Low2014}%
	\BibitemOpen
	\bibfield  {author} {\bibinfo {author} {\bibfnamefont {T.}~\bibnamefont
			{Low}}\ and\ \bibinfo {author} {\bibfnamefont {P.}~\bibnamefont {Avouris}},\
	}\bibfield  {title} {\enquote {\bibinfo {title} {Graphene plasmonics for
				terahertz to mid-infrared applications},}\ }\href {\doibase
		10.1021/nn406627u} {\bibfield  {journal} {\bibinfo  {journal} {{ACS} Nano}\
		}\textbf {\bibinfo {volume} {8}},\ \bibinfo {pages} {1086} (\bibinfo {year}
		{2014})}\BibitemShut {NoStop}%
	\bibitem [{\citenamefont {Fan}\ \emph {et~al.}(2013)\citenamefont {Fan},
		\citenamefont {Gu}, \citenamefont {Chen}, \citenamefont {Wang},\ and\
		\citenamefont {Chang}}]{Fan2013}%
	\BibitemOpen
	\bibfield  {author} {\bibinfo {author} {\bibfnamefont {F.}~\bibnamefont
			{Fan}}, \bibinfo {author} {\bibfnamefont {W.-H.}\ \bibnamefont {Gu}},
		\bibinfo {author} {\bibfnamefont {S.}~\bibnamefont {Chen}}, \bibinfo {author}
		{\bibfnamefont {X.-H.}\ \bibnamefont {Wang}}, \ and\ \bibinfo {author}
		{\bibfnamefont {S.-J.}\ \bibnamefont {Chang}},\ }\bibfield  {title} {\enquote
		{\bibinfo {title} {State conversion based on terahertz plasmonics with
				vanadium dioxide coating controlled by optical pumping},}\ }\href {\doibase
		10.1364/ol.38.001582} {\bibfield  {journal} {\bibinfo  {journal} {Optics
				Letters}\ }\textbf {\bibinfo {volume} {38}},\ \bibinfo {pages} {1582}
		(\bibinfo {year} {2013})}\BibitemShut {NoStop}%
	\bibitem [{\citenamefont {Panaretos}\ and\ \citenamefont
		{Werner}(2016)}]{Panaretos2016}%
	\BibitemOpen
	\bibfield  {author} {\bibinfo {author} {\bibfnamefont {A.~H.}\ \bibnamefont
			{Panaretos}}\ and\ \bibinfo {author} {\bibfnamefont {D.~H.}\ \bibnamefont
			{Werner}},\ }\bibfield  {title} {\enquote {\bibinfo {title} {Spoof plasmon
				radiation using sinusoidally modulated corrugated reactance surfaces},}\
	}\href {\doibase 10.1364/oe.24.002443} {\bibfield  {journal} {\bibinfo
			{journal} {Optics Express}\ }\textbf {\bibinfo {volume} {24}},\ \bibinfo
		{pages} {2443} (\bibinfo {year} {2016})}\BibitemShut {NoStop}%
	\bibitem [{\citenamefont {Yu}\ \emph {et~al.}(2010)\citenamefont {Yu},
		\citenamefont {Wang}, \citenamefont {Kats}, \citenamefont {Fan},
		\citenamefont {Khanna}, \citenamefont {Li}, \citenamefont {Davies},
		\citenamefont {Linfield},\ and\ \citenamefont {Capasso}}]{Yu2010}%
	\BibitemOpen
	\bibfield  {author} {\bibinfo {author} {\bibfnamefont {N.}~\bibnamefont
			{Yu}}, \bibinfo {author} {\bibfnamefont {Q.~J.}\ \bibnamefont {Wang}},
		\bibinfo {author} {\bibfnamefont {M.~A.}\ \bibnamefont {Kats}}, \bibinfo
		{author} {\bibfnamefont {J.~A.}\ \bibnamefont {Fan}}, \bibinfo {author}
		{\bibfnamefont {S.~P.}\ \bibnamefont {Khanna}}, \bibinfo {author}
		{\bibfnamefont {L.}~\bibnamefont {Li}}, \bibinfo {author} {\bibfnamefont
			{A.~G.}\ \bibnamefont {Davies}}, \bibinfo {author} {\bibfnamefont {E.~H.}\
			\bibnamefont {Linfield}}, \ and\ \bibinfo {author} {\bibfnamefont
			{F.}~\bibnamefont {Capasso}},\ }\bibfield  {title} {\enquote {\bibinfo
			{title} {Designer spoof surface plasmon structures collimate terahertz laser
				beams},}\ }\href {\doibase 10.1038/nmat2822} {\bibfield  {journal} {\bibinfo
			{journal} {Nature Materials}\ }\textbf {\bibinfo {volume} {9}},\ \bibinfo
		{pages} {730} (\bibinfo {year} {2010})}\BibitemShut {NoStop}%
	\bibitem [{\citenamefont {Ishikawa}\ \emph {et~al.}(2009)\citenamefont
		{Ishikawa}, \citenamefont {Zhang}, \citenamefont {Genov}, \citenamefont
		{Bartal},\ and\ \citenamefont {Zhang}}]{Ishikawa2009}%
	\BibitemOpen
	\bibfield  {author} {\bibinfo {author} {\bibfnamefont {A.}~\bibnamefont
			{Ishikawa}}, \bibinfo {author} {\bibfnamefont {S.}~\bibnamefont {Zhang}},
		\bibinfo {author} {\bibfnamefont {D.~A.}\ \bibnamefont {Genov}}, \bibinfo
		{author} {\bibfnamefont {G.}~\bibnamefont {Bartal}}, \ and\ \bibinfo {author}
		{\bibfnamefont {X.}~\bibnamefont {Zhang}},\ }\bibfield  {title} {\enquote
		{\bibinfo {title} {Deep subwavelength terahertz waveguides using gap magnetic
				plasmon},}\ }\href {\doibase 10.1103/physrevlett.102.043904} {\bibfield
		{journal} {\bibinfo  {journal} {Physical Review Letters}\ }\textbf {\bibinfo
			{volume} {102}} (\bibinfo {year} {2009}),\
		10.1103/physrevlett.102.043904}\BibitemShut {NoStop}%
	\bibitem [{\citenamefont {Chern}(2008)}]{Chern2008}%
	\BibitemOpen
	\bibfield  {author} {\bibinfo {author} {\bibfnamefont {R.-L.}\ \bibnamefont
			{Chern}},\ }\bibfield  {title} {\enquote {\bibinfo {title} {Magnetic and
				surface plasmon resonances for periodic lattices of plasmonic split-ring
				resonators},}\ }\href {\doibase 10.1103/physrevb.78.085116} {\bibfield
		{journal} {\bibinfo  {journal} {Physical Review B}\ }\textbf {\bibinfo
			{volume} {78}} (\bibinfo {year} {2008}),\
		10.1103/physrevb.78.085116}\BibitemShut {NoStop}%
	\bibitem [{\citenamefont {Martin-Cano}\ \emph {et~al.}(2011)\citenamefont
		{Martin-Cano}, \citenamefont {Quevedo-Teruel}, \citenamefont {Moreno},
		\citenamefont {Martin-Moreno},\ and\ \citenamefont
		{Garcia-Vidal}}]{Martin-Cano2011}%
	\BibitemOpen
	\bibfield  {author} {\bibinfo {author} {\bibfnamefont {D.}~\bibnamefont
			{Martin-Cano}}, \bibinfo {author} {\bibfnamefont {O.}~\bibnamefont
			{Quevedo-Teruel}}, \bibinfo {author} {\bibfnamefont {E.}~\bibnamefont
			{Moreno}}, \bibinfo {author} {\bibfnamefont {L.}~\bibnamefont
			{Martin-Moreno}}, \ and\ \bibinfo {author} {\bibfnamefont {F.~J.}\
			\bibnamefont {Garcia-Vidal}},\ }\bibfield  {title} {\enquote {\bibinfo
			{title} {Waveguided spoof surface plasmons with deep-subwavelength lateral
				confinement},}\ }\href {\doibase 10.1364/ol.36.004635} {\bibfield  {journal}
		{\bibinfo  {journal} {Optics Letters}\ }\textbf {\bibinfo {volume} {36}},\
		\bibinfo {pages} {4635} (\bibinfo {year} {2011})}\BibitemShut {NoStop}%
	\bibitem [{\citenamefont {Lee}\ \emph {et~al.}(2013)\citenamefont {Lee},
		\citenamefont {Choi}, \citenamefont {Kim}, \citenamefont {Choi},\ and\
		\citenamefont {Min}}]{Lee2013}%
	\BibitemOpen
	\bibfield  {author} {\bibinfo {author} {\bibfnamefont {S.~H.}\ \bibnamefont
			{Lee}}, \bibinfo {author} {\bibfnamefont {J.}~\bibnamefont {Choi}}, \bibinfo
		{author} {\bibfnamefont {H.-D.}\ \bibnamefont {Kim}}, \bibinfo {author}
		{\bibfnamefont {H.}~\bibnamefont {Choi}}, \ and\ \bibinfo {author}
		{\bibfnamefont {B.}~\bibnamefont {Min}},\ }\bibfield  {title} {\enquote
		{\bibinfo {title} {Ultrafast refractive index control of a terahertz graphene
				metamaterial},}\ }\href {\doibase 10.1038/srep02135} {\bibfield  {journal}
		{\bibinfo  {journal} {Scientific Reports}\ }\textbf {\bibinfo {volume} {3}}
		(\bibinfo {year} {2013}),\ 10.1038/srep02135}\BibitemShut {NoStop}%
	\bibitem [{\citenamefont {Wang}\ \emph {et~al.}(2015)\citenamefont {Wang},
		\citenamefont {Gu}, \citenamefont {Gong}, \citenamefont {Qiu},\ and\
		\citenamefont {Hong}}]{Wang2015}%
	\BibitemOpen
	\bibfield  {author} {\bibinfo {author} {\bibfnamefont {D.}~\bibnamefont
			{Wang}}, \bibinfo {author} {\bibfnamefont {Y.}~\bibnamefont {Gu}}, \bibinfo
		{author} {\bibfnamefont {Y.}~\bibnamefont {Gong}}, \bibinfo {author}
		{\bibfnamefont {C.-W.}\ \bibnamefont {Qiu}}, \ and\ \bibinfo {author}
		{\bibfnamefont {M.}~\bibnamefont {Hong}},\ }\bibfield  {title} {\enquote
		{\bibinfo {title} {An ultrathin terahertz quarter-wave plate using planar
				babinet-inverted metasurface},}\ }\href {\doibase 10.1364/oe.23.011114}
	{\bibfield  {journal} {\bibinfo  {journal} {Optics Express}\ }\textbf
		{\bibinfo {volume} {23}},\ \bibinfo {pages} {11114} (\bibinfo {year}
		{2015})}\BibitemShut {NoStop}%
	\bibitem [{\citenamefont {Shi}\ \emph {et~al.}(2014)\citenamefont {Shi},
		\citenamefont {Zeng}, \citenamefont {Han}, \citenamefont {Hong},
		\citenamefont {Tsai}, \citenamefont {Jung}, \citenamefont {Zettl},
		\citenamefont {Crommie},\ and\ \citenamefont {Wang}}]{Shi2014}%
	\BibitemOpen
	\bibfield  {author} {\bibinfo {author} {\bibfnamefont {S.-F.}\ \bibnamefont
			{Shi}}, \bibinfo {author} {\bibfnamefont {B.}~\bibnamefont {Zeng}}, \bibinfo
		{author} {\bibfnamefont {H.-L.}\ \bibnamefont {Han}}, \bibinfo {author}
		{\bibfnamefont {X.}~\bibnamefont {Hong}}, \bibinfo {author} {\bibfnamefont
			{H.-Z.}\ \bibnamefont {Tsai}}, \bibinfo {author} {\bibfnamefont {H.~S.}\
			\bibnamefont {Jung}}, \bibinfo {author} {\bibfnamefont {A.}~\bibnamefont
			{Zettl}}, \bibinfo {author} {\bibfnamefont {M.~F.}\ \bibnamefont {Crommie}},
		\ and\ \bibinfo {author} {\bibfnamefont {F.}~\bibnamefont {Wang}},\
	}\bibfield  {title} {\enquote {\bibinfo {title} {Optimizing broadband
				terahertz modulation with hybrid graphene/metasurface structures},}\ }\href
	{\doibase 10.1021/nl503670d} {\bibfield  {journal} {\bibinfo  {journal} {Nano
				Letters}\ }\textbf {\bibinfo {volume} {15}},\ \bibinfo {pages} {372}
		(\bibinfo {year} {2014})}\BibitemShut {NoStop}%
	\bibitem [{\citenamefont {Takano}\ \emph {et~al.}(2010)\citenamefont {Takano},
		\citenamefont {Shibuya}, \citenamefont {Akiyama}, \citenamefont {Nagashima},
		\citenamefont {Miyamaru},\ and\ \citenamefont {Hangyo}}]{Takano2010}%
	\BibitemOpen
	\bibfield  {author} {\bibinfo {author} {\bibfnamefont {K.}~\bibnamefont
			{Takano}}, \bibinfo {author} {\bibfnamefont {K.}~\bibnamefont {Shibuya}},
		\bibinfo {author} {\bibfnamefont {K.}~\bibnamefont {Akiyama}}, \bibinfo
		{author} {\bibfnamefont {T.}~\bibnamefont {Nagashima}}, \bibinfo {author}
		{\bibfnamefont {F.}~\bibnamefont {Miyamaru}}, \ and\ \bibinfo {author}
		{\bibfnamefont {M.}~\bibnamefont {Hangyo}},\ }\bibfield  {title} {\enquote
		{\bibinfo {title} {A metal-to-insulator transition in cut-wire-grid
				metamaterials in the terahertz region},}\ }\href {\doibase 10.1063/1.3284958}
	{\bibfield  {journal} {\bibinfo  {journal} {Journal of Applied Physics}\
		}\textbf {\bibinfo {volume} {107}},\ \bibinfo {pages} {024907} (\bibinfo
		{year} {2010})}\BibitemShut {NoStop}%
	\bibitem [{\citenamefont {Sommerfeld}(2004)}]{Sommerfeld2004}%
	\BibitemOpen
	\bibfield  {author} {\bibinfo {author} {\bibfnamefont {A.}~\bibnamefont
			{Sommerfeld}},\ }\bibfield  {title} {\enquote {\bibinfo {title} {Mathematical
				theory of diffraction},}\ }in\ \href {\doibase 10.1007/978-0-8176-8196-8_2}
	{\emph {\bibinfo {booktitle} {Mathematical Theory of Diffraction}}}\
	(\bibinfo  {publisher} {Birkhäuser Boston},\ \bibinfo {year} {2004})\ pp.\
	\bibinfo {pages} {9--68}\BibitemShut {NoStop}%
	\bibitem [{\citenamefont {Sheppard}\ \emph {et~al.}(2013)\citenamefont
		{Sheppard}, \citenamefont {Lin},\ and\ \citenamefont {Kou}}]{Sheppard2013}%
	\BibitemOpen
	\bibfield  {author} {\bibinfo {author} {\bibfnamefont {C.~J.~R.}\
			\bibnamefont {Sheppard}}, \bibinfo {author} {\bibfnamefont {J.}~\bibnamefont
			{Lin}}, \ and\ \bibinfo {author} {\bibfnamefont {S.~S.}\ \bibnamefont
			{Kou}},\ }\bibfield  {title} {\enquote {\bibinfo {title}
			{Rayleigh{\textendash}sommerfeld diffraction formula in k space},}\ }\href
	{\doibase 10.1364/josaa.30.001180} {\bibfield  {journal} {\bibinfo  {journal}
			{Journal of the Optical Society of America A}\ }\textbf {\bibinfo {volume}
			{30}},\ \bibinfo {pages} {1180} (\bibinfo {year} {2013})}\BibitemShut
	{NoStop}%
	\bibitem [{\citenamefont {Marathay}\ and\ \citenamefont
		{McCalmont}(2004)}]{Marathay2004}%
	\BibitemOpen
	\bibfield  {author} {\bibinfo {author} {\bibfnamefont {A.~S.}\ \bibnamefont
			{Marathay}}\ and\ \bibinfo {author} {\bibfnamefont {J.~F.}\ \bibnamefont
			{McCalmont}},\ }\bibfield  {title} {\enquote {\bibinfo {title} {On the usual
				approximation used in the rayleigh{\textendash}sommerfeld diffraction
				theory},}\ }\href {\doibase 10.1364/josaa.21.000510} {\bibfield  {journal}
		{\bibinfo  {journal} {Journal of the Optical Society of America A}\ }\textbf
		{\bibinfo {volume} {21}},\ \bibinfo {pages} {510} (\bibinfo {year}
		{2004})}\BibitemShut {NoStop}%
	\bibitem [{\citenamefont {Bethe}(1944)}]{Bethe1944}%
	\BibitemOpen
	\bibfield  {author} {\bibinfo {author} {\bibfnamefont {H.~A.}\ \bibnamefont
			{Bethe}},\ }\bibfield  {title} {\enquote {\bibinfo {title} {Theory of
				diffraction by small holes},}\ }\href {\doibase 10.1103/physrev.66.163}
	{\bibfield  {journal} {\bibinfo  {journal} {Physical Review}\ }\textbf
		{\bibinfo {volume} {66}},\ \bibinfo {pages} {163} (\bibinfo {year}
		{1944})}\BibitemShut {NoStop}%
	\bibitem [{\citenamefont {Bouwkamp}\ and\ \citenamefont
		{Casimir}(1954)}]{Bouwkamp1954}%
	\BibitemOpen
	\bibfield  {author} {\bibinfo {author} {\bibfnamefont {C.}~\bibnamefont
			{Bouwkamp}}\ and\ \bibinfo {author} {\bibfnamefont {H.}~\bibnamefont
			{Casimir}},\ }\bibfield  {title} {\enquote {\bibinfo {title} {On multipole
				expansions in the theory of electromagnetic radiation},}\ }\href {\doibase
		10.1016/s0031-8914(54)80068-1} {\bibfield  {journal} {\bibinfo  {journal}
			{Physica}\ }\textbf {\bibinfo {volume} {20}},\ \bibinfo {pages} {539}
		(\bibinfo {year} {1954})}\BibitemShut {NoStop}%
	\bibitem [{\citenamefont {Bouwkamp}(1954)}]{Bouwkamp1954a}%
	\BibitemOpen
	\bibfield  {author} {\bibinfo {author} {\bibfnamefont {C.~J.}\ \bibnamefont
			{Bouwkamp}},\ }\bibfield  {title} {\enquote {\bibinfo {title} {Diffraction
				theory},}\ }\href {\doibase 10.1088/0034-4885/17/1/302} {\bibfield  {journal}
		{\bibinfo  {journal} {Reports on Progress in Physics}\ }\textbf {\bibinfo
			{volume} {17}},\ \bibinfo {pages} {35} (\bibinfo {year} {1954})}\BibitemShut
	{NoStop}%
	\bibitem [{\citenamefont {Ulrich}(1967)}]{Ulrich1967}%
	\BibitemOpen
	\bibfield  {author} {\bibinfo {author} {\bibfnamefont {R.}~\bibnamefont
			{Ulrich}},\ }\bibfield  {title} {\enquote {\bibinfo {title} {Far-infrared
				properties of metallic mesh and its complementary structure},}\ }\href
	{\doibase 10.1016/0020-0891(67)90028-0} {\bibfield  {journal} {\bibinfo
			{journal} {Infrared Physics}\ }\textbf {\bibinfo {volume} {7}},\ \bibinfo
		{pages} {37} (\bibinfo {year} {1967})}\BibitemShut {NoStop}%
	\bibitem [{\citenamefont {Kim}\ \emph {et~al.}(1999)\citenamefont {Kim},
		\citenamefont {Thio}, \citenamefont {Ebbesen}, \citenamefont {Grupp},\ and\
		\citenamefont {Lezec}}]{Kim1999}%
	\BibitemOpen
	\bibfield  {author} {\bibinfo {author} {\bibfnamefont {T.~J.}\ \bibnamefont
			{Kim}}, \bibinfo {author} {\bibfnamefont {T.}~\bibnamefont {Thio}}, \bibinfo
		{author} {\bibfnamefont {T.~W.}\ \bibnamefont {Ebbesen}}, \bibinfo {author}
		{\bibfnamefont {D.~E.}\ \bibnamefont {Grupp}}, \ and\ \bibinfo {author}
		{\bibfnamefont {H.~J.}\ \bibnamefont {Lezec}},\ }\bibfield  {title} {\enquote
		{\bibinfo {title} {Control of optical transmission through metals perforated
				with subwavelength hole arrays},}\ }\href {\doibase 10.1364/ol.24.000256}
	{\bibfield  {journal} {\bibinfo  {journal} {Optics Letters}\ }\textbf
		{\bibinfo {volume} {24}},\ \bibinfo {pages} {256} (\bibinfo {year}
		{1999})}\BibitemShut {NoStop}%
	\bibitem [{\citenamefont {Kim}\ \emph {et~al.}(2003)\citenamefont {Kim},
		\citenamefont {Hohng}, \citenamefont {Malyarchuk}, \citenamefont {Yoon},
		\citenamefont {Ahn}, \citenamefont {Yee}, \citenamefont {Park}, \citenamefont
		{Kim}, \citenamefont {Park},\ and\ \citenamefont {Lienau}}]{Kim2003}%
	\BibitemOpen
	\bibfield  {author} {\bibinfo {author} {\bibfnamefont {D.~S.}\ \bibnamefont
			{Kim}}, \bibinfo {author} {\bibfnamefont {S.~C.}\ \bibnamefont {Hohng}},
		\bibinfo {author} {\bibfnamefont {V.}~\bibnamefont {Malyarchuk}}, \bibinfo
		{author} {\bibfnamefont {Y.~C.}\ \bibnamefont {Yoon}}, \bibinfo {author}
		{\bibfnamefont {Y.~H.}\ \bibnamefont {Ahn}}, \bibinfo {author} {\bibfnamefont
			{K.~J.}\ \bibnamefont {Yee}}, \bibinfo {author} {\bibfnamefont {J.~W.}\
			\bibnamefont {Park}}, \bibinfo {author} {\bibfnamefont {J.}~\bibnamefont
			{Kim}}, \bibinfo {author} {\bibfnamefont {Q.~H.}\ \bibnamefont {Park}}, \
		and\ \bibinfo {author} {\bibfnamefont {C.}~\bibnamefont {Lienau}},\
	}\bibfield  {title} {\enquote {\bibinfo {title} {Microscopic origin of
				surface-plasmon radiation in plasmonic band-gap nanostructures},}\ }\href
	{\doibase 10.1103/physrevlett.91.143901} {\bibfield  {journal} {\bibinfo
			{journal} {Physical Review Letters}\ }\textbf {\bibinfo {volume} {91}}
		(\bibinfo {year} {2003}),\ 10.1103/physrevlett.91.143901}\BibitemShut
	{NoStop}%
	\bibitem [{\citenamefont {Minhas}\ \emph {et~al.}(2002)\citenamefont {Minhas},
		\citenamefont {Fan}, \citenamefont {Agi}, \citenamefont {Brueck},\ and\
		\citenamefont {Malloy}}]{Minhas2002}%
	\BibitemOpen
	\bibfield  {author} {\bibinfo {author} {\bibfnamefont {B.~K.}\ \bibnamefont
			{Minhas}}, \bibinfo {author} {\bibfnamefont {W.}~\bibnamefont {Fan}},
		\bibinfo {author} {\bibfnamefont {K.}~\bibnamefont {Agi}}, \bibinfo {author}
		{\bibfnamefont {S.~R.~J.}\ \bibnamefont {Brueck}}, \ and\ \bibinfo {author}
		{\bibfnamefont {K.~J.}\ \bibnamefont {Malloy}},\ }\bibfield  {title}
	{\enquote {\bibinfo {title} {Metallic inductive and capacitive grids: theory
				and experiment},}\ }\href {\doibase 10.1364/josaa.19.001352} {\bibfield
		{journal} {\bibinfo  {journal} {Journal of the Optical Society of America A}\
		}\textbf {\bibinfo {volume} {19}},\ \bibinfo {pages} {1352} (\bibinfo {year}
		{2002})}\BibitemShut {NoStop}%
	\bibitem [{\citenamefont {Naweed}\ \emph {et~al.}(2003)\citenamefont {Naweed},
		\citenamefont {Baumann}, \citenamefont {William A.~Bailey}, \citenamefont
		{Karakashian},\ and\ \citenamefont {Goodhue}}]{Naweed2003}%
	\BibitemOpen
	\bibfield  {author} {\bibinfo {author} {\bibfnamefont {A.}~\bibnamefont
			{Naweed}}, \bibinfo {author} {\bibfnamefont {F.}~\bibnamefont {Baumann}},
		\bibinfo {author} {\bibfnamefont {J.}~\bibnamefont {William A.~Bailey}},
		\bibinfo {author} {\bibfnamefont {A.~S.}\ \bibnamefont {Karakashian}}, \ and\
		\bibinfo {author} {\bibfnamefont {W.~D.}\ \bibnamefont {Goodhue}},\
	}\bibfield  {title} {\enquote {\bibinfo {title} {Evidence for radiative
				damping in surface-plasmon-mediated light transmission through perforated
				conducting films},}\ }\href {\doibase 10.1364/josab.20.002534} {\bibfield
		{journal} {\bibinfo  {journal} {Journal of the Optical Society of America B}\
		}\textbf {\bibinfo {volume} {20}},\ \bibinfo {pages} {2534} (\bibinfo {year}
		{2003})}\BibitemShut {NoStop}%
	\bibitem [{\citenamefont {Qu}\ and\ \citenamefont
		{Grischkowsky}(2004)}]{Qu2004}%
	\BibitemOpen
	\bibfield  {author} {\bibinfo {author} {\bibfnamefont {D.}~\bibnamefont
			{Qu}}\ and\ \bibinfo {author} {\bibfnamefont {D.}~\bibnamefont
			{Grischkowsky}},\ }\bibfield  {title} {\enquote {\bibinfo {title}
			{Observation of a new type of~{THz} resonance of surface plasmons propagating
				on metal-film hole arrays},}\ }\href {\doibase 10.1103/physrevlett.93.196804}
	{\bibfield  {journal} {\bibinfo  {journal} {Physical Review Letters}\
		}\textbf {\bibinfo {volume} {93}} (\bibinfo {year} {2004}),\
		10.1103/physrevlett.93.196804}\BibitemShut {NoStop}%
	\bibitem [{\citenamefont {Kang}\ \emph
		{et~al.}(2009{\natexlab{b}})\citenamefont {Kang}, \citenamefont {Kim},\ and\
		\citenamefont {Park}}]{Kang2009a}%
	\BibitemOpen
	\bibfield  {author} {\bibinfo {author} {\bibfnamefont {J.~H.}\ \bibnamefont
			{Kang}}, \bibinfo {author} {\bibfnamefont {D.~S.}\ \bibnamefont {Kim}}, \
		and\ \bibinfo {author} {\bibfnamefont {Q.-H.}\ \bibnamefont {Park}},\
	}\bibfield  {title} {\enquote {\bibinfo {title} {Local capacitor model for
				plasmonic electric field enhancement},}\ }\href {\doibase
		10.1103/physrevlett.102.093906} {\bibfield  {journal} {\bibinfo  {journal}
			{Physical Review Letters}\ }\textbf {\bibinfo {volume} {102}} (\bibinfo
		{year} {2009}{\natexlab{b}}),\ 10.1103/physrevlett.102.093906}\BibitemShut
	{NoStop}%
	\bibitem [{\citenamefont {Gordon}\ and\ \citenamefont
		{Brolo}(2005)}]{Gordon2005}%
	\BibitemOpen
	\bibfield  {author} {\bibinfo {author} {\bibfnamefont {R.}~\bibnamefont
			{Gordon}}\ and\ \bibinfo {author} {\bibfnamefont {A.~G.}\ \bibnamefont
			{Brolo}},\ }\bibfield  {title} {\enquote {\bibinfo {title} {Increased cut-off
				wavelength for a subwavelength hole in a real metal},}\ }\href {\doibase
		10.1364/opex.13.001933} {\bibfield  {journal} {\bibinfo  {journal} {Optics
				Express}\ }\textbf {\bibinfo {volume} {13}},\ \bibinfo {pages} {1933}
		(\bibinfo {year} {2005})}\BibitemShut {NoStop}%
	\bibitem [{\citenamefont {Bravo-Abad}\ \emph {et~al.}(2007)\citenamefont
		{Bravo-Abad}, \citenamefont {Fern{\'{a}}ndez-Dom{\'{\i}}nguez}, \citenamefont
		{Garc{\'{\i}}a-Vidal},\ and\ \citenamefont
		{Mart{\'{\i}}n-Moreno}}]{Bravo-Abad2007}%
	\BibitemOpen
	\bibfield  {author} {\bibinfo {author} {\bibfnamefont {J.}~\bibnamefont
			{Bravo-Abad}}, \bibinfo {author} {\bibfnamefont {A.~I.}\ \bibnamefont
			{Fern{\'{a}}ndez-Dom{\'{\i}}nguez}}, \bibinfo {author} {\bibfnamefont
			{F.~J.}\ \bibnamefont {Garc{\'{\i}}a-Vidal}}, \ and\ \bibinfo {author}
		{\bibfnamefont {L.}~\bibnamefont {Mart{\'{\i}}n-Moreno}},\ }\bibfield
	{title} {\enquote {\bibinfo {title} {Theory of extraordinary transmission of
				light through quasiperiodic arrays of subwavelength holes},}\ }\href
	{\doibase 10.1103/physrevlett.99.203905} {\bibfield  {journal} {\bibinfo
			{journal} {Physical Review Letters}\ }\textbf {\bibinfo {volume} {99}}
		(\bibinfo {year} {2007}),\ 10.1103/physrevlett.99.203905}\BibitemShut
	{NoStop}%
	\bibitem [{\citenamefont {Takakura}(2001)}]{Takakura2001}%
	\BibitemOpen
	\bibfield  {author} {\bibinfo {author} {\bibfnamefont {Y.}~\bibnamefont
			{Takakura}},\ }\bibfield  {title} {\enquote {\bibinfo {title} {Optical
				resonance in a narrow slit in a thick metallic screen},}\ }\href {\doibase
		10.1103/physrevlett.86.5601} {\bibfield  {journal} {\bibinfo  {journal}
			{Physical Review Letters}\ }\textbf {\bibinfo {volume} {86}},\ \bibinfo
		{pages} {5601} (\bibinfo {year} {2001})}\BibitemShut {NoStop}%
	\bibitem [{\citenamefont {Delgado}\ and\ \citenamefont
		{Marqu{\'{e}}s}(2011)}]{Delgado2011}%
	\BibitemOpen
	\bibfield  {author} {\bibinfo {author} {\bibfnamefont {V.}~\bibnamefont
			{Delgado}}\ and\ \bibinfo {author} {\bibfnamefont {R.}~\bibnamefont
			{Marqu{\'{e}}s}},\ }\bibfield  {title} {\enquote {\bibinfo {title} {Surface
				impedance model for extraordinary transmission in 1d metallic and dielectric
				screens},}\ }\href {\doibase 10.1364/oe.19.025290} {\bibfield  {journal}
		{\bibinfo  {journal} {Optics Express}\ }\textbf {\bibinfo {volume} {19}},\
		\bibinfo {pages} {25290} (\bibinfo {year} {2011})}\BibitemShut {NoStop}%
	\bibitem [{\citenamefont {Galindo}\ and\ \citenamefont
		{Wu}(1966)}]{Galindo1966}%
	\BibitemOpen
	\bibfield  {author} {\bibinfo {author} {\bibfnamefont {V.}~\bibnamefont
			{Galindo}}\ and\ \bibinfo {author} {\bibfnamefont {C.}~\bibnamefont {Wu}},\
	}\bibfield  {title} {\enquote {\bibinfo {title} {Numerical solutions for an
				infinite phased array of rectangular waveguides with thick walls},}\ }\href
	{\doibase 10.1109/tap.1966.1138653} {\bibfield  {journal} {\bibinfo
			{journal} {{IEEE} Transactions on Antennas and Propagation}\ }\textbf
		{\bibinfo {volume} {14}},\ \bibinfo {pages} {149} (\bibinfo {year}
		{1966})}\BibitemShut {NoStop}%
	\bibitem [{\citenamefont {Sheng}\ \emph {et~al.}(1982)\citenamefont {Sheng},
		\citenamefont {Stepleman},\ and\ \citenamefont {Sanda}}]{Sheng1982}%
	\BibitemOpen
	\bibfield  {author} {\bibinfo {author} {\bibfnamefont {P.}~\bibnamefont
			{Sheng}}, \bibinfo {author} {\bibfnamefont {R.~S.}\ \bibnamefont
			{Stepleman}}, \ and\ \bibinfo {author} {\bibfnamefont {P.~N.}\ \bibnamefont
			{Sanda}},\ }\bibfield  {title} {\enquote {\bibinfo {title} {Exact
				eigenfunctions for square-wave gratings: Application to diffraction and
				surface-plasmon calculations},}\ }\href {\doibase 10.1103/physrevb.26.2907}
	{\bibfield  {journal} {\bibinfo  {journal} {Physical Review B}\ }\textbf
		{\bibinfo {volume} {26}},\ \bibinfo {pages} {2907} (\bibinfo {year}
		{1982})}\BibitemShut {NoStop}%
	\bibitem [{\citenamefont {Garcia-Vidal}\ \emph {et~al.}(2010)\citenamefont
		{Garcia-Vidal}, \citenamefont {Martin-Moreno}, \citenamefont {Ebbesen},\ and\
		\citenamefont {Kuipers}}]{Garcia-Vidal2010}%
	\BibitemOpen
	\bibfield  {author} {\bibinfo {author} {\bibfnamefont {F.~J.}\ \bibnamefont
			{Garcia-Vidal}}, \bibinfo {author} {\bibfnamefont {L.}~\bibnamefont
			{Martin-Moreno}}, \bibinfo {author} {\bibfnamefont {T.~W.}\ \bibnamefont
			{Ebbesen}}, \ and\ \bibinfo {author} {\bibfnamefont {L.}~\bibnamefont
			{Kuipers}},\ }\bibfield  {title} {\enquote {\bibinfo {title} {Light passing
				through subwavelength apertures},}\ }\href {\doibase
		10.1103/revmodphys.82.729} {\bibfield  {journal} {\bibinfo  {journal}
			{Reviews of Modern Physics}\ }\textbf {\bibinfo {volume} {82}},\ \bibinfo
		{pages} {729} (\bibinfo {year} {2010})}\BibitemShut {NoStop}%
	\bibitem [{\citenamefont {Liu}\ and\ \citenamefont {Lalanne}(2008)}]{Liu2008}%
	\BibitemOpen
	\bibfield  {author} {\bibinfo {author} {\bibfnamefont {H.}~\bibnamefont
			{Liu}}\ and\ \bibinfo {author} {\bibfnamefont {P.}~\bibnamefont {Lalanne}},\
	}\bibfield  {title} {\enquote {\bibinfo {title} {Microscopic theory of the
				extraordinary optical transmission},}\ }\href {\doibase 10.1038/nature06762}
	{\bibfield  {journal} {\bibinfo  {journal} {Nature}\ }\textbf {\bibinfo
			{volume} {452}},\ \bibinfo {pages} {728} (\bibinfo {year}
		{2008})}\BibitemShut {NoStop}%
	\bibitem [{\citenamefont {He}(2009)}]{He2009}%
	\BibitemOpen
	\bibfield  {author} {\bibinfo {author} {\bibfnamefont {X.-Y.}\ \bibnamefont
			{He}},\ }\bibfield  {title} {\enquote {\bibinfo {title} {Numerical analysis
				of the propagation properties of subwavelength semiconductor slit in the
				terahertz region},}\ }\href {\doibase 10.1364/oe.17.015359} {\bibfield
		{journal} {\bibinfo  {journal} {Optics Express}\ }\textbf {\bibinfo {volume}
			{17}},\ \bibinfo {pages} {15359} (\bibinfo {year} {2009})}\BibitemShut
	{NoStop}%
	\bibitem [{\citenamefont {Bell}\ \emph {et~al.}(1995)\citenamefont {Bell},
		\citenamefont {Pendry}, \citenamefont {Moreno},\ and\ \citenamefont
		{Ward}}]{Bell1995}%
	\BibitemOpen
	\bibfield  {author} {\bibinfo {author} {\bibfnamefont {P.}~\bibnamefont
			{Bell}}, \bibinfo {author} {\bibfnamefont {J.}~\bibnamefont {Pendry}},
		\bibinfo {author} {\bibfnamefont {L.}~\bibnamefont {Moreno}}, \ and\ \bibinfo
		{author} {\bibfnamefont {A.}~\bibnamefont {Ward}},\ }\bibfield  {title}
	{\enquote {\bibinfo {title} {A program for calculating photonic band
				structures and transmission coefficients of complex structures},}\ }\href
	{\doibase 10.1016/0010-4655(94)00131-k} {\bibfield  {journal} {\bibinfo
			{journal} {Computer Physics Communications}\ }\textbf {\bibinfo {volume}
			{85}},\ \bibinfo {pages} {306} (\bibinfo {year} {1995})}\BibitemShut
	{NoStop}%
	\bibitem [{\citenamefont {Li}(1997)}]{Li1997}%
	\BibitemOpen
	\bibfield  {author} {\bibinfo {author} {\bibfnamefont {L.}~\bibnamefont
			{Li}},\ }\bibfield  {title} {\enquote {\bibinfo {title} {New formulation of
				the fourier modal method for crossed surface-relief gratings},}\ }\href
	{\doibase 10.1364/josaa.14.002758} {\bibfield  {journal} {\bibinfo  {journal}
			{Journal of the Optical Society of America A}\ }\textbf {\bibinfo {volume}
			{14}},\ \bibinfo {pages} {2758} (\bibinfo {year} {1997})}\BibitemShut
	{NoStop}%
	\bibitem [{\citenamefont {Salomon}\ \emph {et~al.}(2001)\citenamefont
		{Salomon}, \citenamefont {Grillot}, \citenamefont {Zayats},\ and\
		\citenamefont {de~Fornel}}]{Salomon2001}%
	\BibitemOpen
	\bibfield  {author} {\bibinfo {author} {\bibfnamefont {L.}~\bibnamefont
			{Salomon}}, \bibinfo {author} {\bibfnamefont {F.}~\bibnamefont {Grillot}},
		\bibinfo {author} {\bibfnamefont {A.~V.}\ \bibnamefont {Zayats}}, \ and\
		\bibinfo {author} {\bibfnamefont {F.}~\bibnamefont {de~Fornel}},\ }\bibfield
	{title} {\enquote {\bibinfo {title} {Near-field distribution of optical
				transmission of periodic subwavelength holes in a metal film},}\ }\href
	{\doibase 10.1103/physrevlett.86.1110} {\bibfield  {journal} {\bibinfo
			{journal} {Physical Review Letters}\ }\textbf {\bibinfo {volume} {86}},\
		\bibinfo {pages} {1110} (\bibinfo {year} {2001})}\BibitemShut {NoStop}%
	\bibitem [{\citenamefont {Zuloaga}\ \emph {et~al.}(2009)\citenamefont
		{Zuloaga}, \citenamefont {Prodan},\ and\ \citenamefont
		{Nordlander}}]{Zuloaga2009}%
	\BibitemOpen
	\bibfield  {author} {\bibinfo {author} {\bibfnamefont {J.}~\bibnamefont
			{Zuloaga}}, \bibinfo {author} {\bibfnamefont {E.}~\bibnamefont {Prodan}}, \
		and\ \bibinfo {author} {\bibfnamefont {P.}~\bibnamefont {Nordlander}},\
	}\bibfield  {title} {\enquote {\bibinfo {title} {Quantum description of the
				plasmon resonances of a nanoparticle dimer},}\ }\href {\doibase
		10.1021/nl803811g} {\bibfield  {journal} {\bibinfo  {journal} {Nano Letters}\
		}\textbf {\bibinfo {volume} {9}},\ \bibinfo {pages} {887} (\bibinfo {year}
		{2009})}\BibitemShut {NoStop}%
	\bibitem [{\citenamefont {Esteban}\ \emph {et~al.}(2012)\citenamefont
		{Esteban}, \citenamefont {Borisov}, \citenamefont {Nordlander},\ and\
		\citenamefont {Aizpurua}}]{Esteban2012}%
	\BibitemOpen
	\bibfield  {author} {\bibinfo {author} {\bibfnamefont {R.}~\bibnamefont
			{Esteban}}, \bibinfo {author} {\bibfnamefont {A.~G.}\ \bibnamefont
			{Borisov}}, \bibinfo {author} {\bibfnamefont {P.}~\bibnamefont {Nordlander}},
		\ and\ \bibinfo {author} {\bibfnamefont {J.}~\bibnamefont {Aizpurua}},\
	}\bibfield  {title} {\enquote {\bibinfo {title} {Bridging quantum and
				classical plasmonics with a quantum-corrected model},}\ }\href {\doibase
		10.1038/ncomms1806} {\bibfield  {journal} {\bibinfo  {journal} {Nature
				Communications}\ }\textbf {\bibinfo {volume} {3}} (\bibinfo {year} {2012}),\
		10.1038/ncomms1806}\BibitemShut {NoStop}%
	\bibitem [{\citenamefont {Marinica}\ \emph {et~al.}(2012)\citenamefont
		{Marinica}, \citenamefont {Kazansky}, \citenamefont {Nordlander},
		\citenamefont {Aizpurua},\ and\ \citenamefont {Borisov}}]{Marinica2012}%
	\BibitemOpen
	\bibfield  {author} {\bibinfo {author} {\bibfnamefont {D.}~\bibnamefont
			{Marinica}}, \bibinfo {author} {\bibfnamefont {A.}~\bibnamefont {Kazansky}},
		\bibinfo {author} {\bibfnamefont {P.}~\bibnamefont {Nordlander}}, \bibinfo
		{author} {\bibfnamefont {J.}~\bibnamefont {Aizpurua}}, \ and\ \bibinfo
		{author} {\bibfnamefont {A.~G.}\ \bibnamefont {Borisov}},\ }\bibfield
	{title} {\enquote {\bibinfo {title} {Quantum plasmonics: Nonlinear effects in
				the field enhancement of a plasmonic nanoparticle dimer},}\ }\href {\doibase
		10.1021/nl300269c} {\bibfield  {journal} {\bibinfo  {journal} {Nano Letters}\
		}\textbf {\bibinfo {volume} {12}},\ \bibinfo {pages} {1333} (\bibinfo {year}
		{2012})}\BibitemShut {NoStop}%
	\bibitem [{\citenamefont {Ciraci}\ \emph {et~al.}(2012)\citenamefont {Ciraci},
		\citenamefont {Hill}, \citenamefont {Mock}, \citenamefont {Urzhumov},
		\citenamefont {Fernandez-Dominguez}, \citenamefont {Maier}, \citenamefont
		{Pendry}, \citenamefont {Chilkoti},\ and\ \citenamefont
		{Smith}}]{Ciraci2012}%
	\BibitemOpen
	\bibfield  {author} {\bibinfo {author} {\bibfnamefont {C.}~\bibnamefont
			{Ciraci}}, \bibinfo {author} {\bibfnamefont {R.~T.}\ \bibnamefont {Hill}},
		\bibinfo {author} {\bibfnamefont {J.~J.}\ \bibnamefont {Mock}}, \bibinfo
		{author} {\bibfnamefont {Y.}~\bibnamefont {Urzhumov}}, \bibinfo {author}
		{\bibfnamefont {A.~I.}\ \bibnamefont {Fernandez-Dominguez}}, \bibinfo
		{author} {\bibfnamefont {S.~A.}\ \bibnamefont {Maier}}, \bibinfo {author}
		{\bibfnamefont {J.~B.}\ \bibnamefont {Pendry}}, \bibinfo {author}
		{\bibfnamefont {A.}~\bibnamefont {Chilkoti}}, \ and\ \bibinfo {author}
		{\bibfnamefont {D.~R.}\ \bibnamefont {Smith}},\ }\bibfield  {title} {\enquote
		{\bibinfo {title} {Probing the ultimate limits of plasmonic enhancement},}\
	}\href {\doibase 10.1126/science.1224823} {\bibfield  {journal} {\bibinfo
			{journal} {Science}\ }\textbf {\bibinfo {volume} {337}},\ \bibinfo {pages}
		{1072} (\bibinfo {year} {2012})}\BibitemShut {NoStop}%
	\bibitem [{\citenamefont {Tame}\ \emph {et~al.}(2013)\citenamefont {Tame},
		\citenamefont {McEnery}, \citenamefont {Özdemir}, \citenamefont {Lee},
		\citenamefont {Maier},\ and\ \citenamefont {Kim}}]{Tame2013}%
	\BibitemOpen
	\bibfield  {author} {\bibinfo {author} {\bibfnamefont {M.~S.}\ \bibnamefont
			{Tame}}, \bibinfo {author} {\bibfnamefont {K.~R.}\ \bibnamefont {McEnery}},
		\bibinfo {author} {\bibfnamefont {{\c{S}}.~K.}\ \bibnamefont {Özdemir}},
		\bibinfo {author} {\bibfnamefont {J.}~\bibnamefont {Lee}}, \bibinfo {author}
		{\bibfnamefont {S.~A.}\ \bibnamefont {Maier}}, \ and\ \bibinfo {author}
		{\bibfnamefont {M.~S.}\ \bibnamefont {Kim}},\ }\bibfield  {title} {\enquote
		{\bibinfo {title} {Quantum plasmonics},}\ }\href {\doibase 10.1038/nphys2615}
	{\bibfield  {journal} {\bibinfo  {journal} {Nature Physics}\ }\textbf
		{\bibinfo {volume} {9}},\ \bibinfo {pages} {329} (\bibinfo {year}
		{2013})}\BibitemShut {NoStop}%
	\bibitem [{\citenamefont {Tan}\ \emph {et~al.}(2014)\citenamefont {Tan},
		\citenamefont {Wu}, \citenamefont {Yang}, \citenamefont {Bai}, \citenamefont
		{Bosman},\ and\ \citenamefont {Nijhuis}}]{Tan2014}%
	\BibitemOpen
	\bibfield  {author} {\bibinfo {author} {\bibfnamefont {S.~F.}\ \bibnamefont
			{Tan}}, \bibinfo {author} {\bibfnamefont {L.}~\bibnamefont {Wu}}, \bibinfo
		{author} {\bibfnamefont {J.~K.~W.}\ \bibnamefont {Yang}}, \bibinfo {author}
		{\bibfnamefont {P.}~\bibnamefont {Bai}}, \bibinfo {author} {\bibfnamefont
			{M.}~\bibnamefont {Bosman}}, \ and\ \bibinfo {author} {\bibfnamefont {C.~A.}\
			\bibnamefont {Nijhuis}},\ }\bibfield  {title} {\enquote {\bibinfo {title}
			{Quantum plasmon resonances controlled by molecular tunnel junctions},}\
	}\href {\doibase 10.1126/science.1248797} {\bibfield  {journal} {\bibinfo
			{journal} {Science}\ }\textbf {\bibinfo {volume} {343}},\ \bibinfo {pages}
		{1496} (\bibinfo {year} {2014})}\BibitemShut {NoStop}%
	\bibitem [{\citenamefont {Savage}\ \emph {et~al.}(2012)\citenamefont {Savage},
		\citenamefont {Hawkeye}, \citenamefont {Esteban}, \citenamefont {Borisov},
		\citenamefont {Aizpurua},\ and\ \citenamefont {Baumberg}}]{Savage2012}%
	\BibitemOpen
	\bibfield  {author} {\bibinfo {author} {\bibfnamefont {K.~J.}\ \bibnamefont
			{Savage}}, \bibinfo {author} {\bibfnamefont {M.~M.}\ \bibnamefont {Hawkeye}},
		\bibinfo {author} {\bibfnamefont {R.}~\bibnamefont {Esteban}}, \bibinfo
		{author} {\bibfnamefont {A.~G.}\ \bibnamefont {Borisov}}, \bibinfo {author}
		{\bibfnamefont {J.}~\bibnamefont {Aizpurua}}, \ and\ \bibinfo {author}
		{\bibfnamefont {J.~J.}\ \bibnamefont {Baumberg}},\ }\bibfield  {title}
	{\enquote {\bibinfo {title} {Revealing the quantum regime in tunnelling
				plasmonics},}\ }\href {\doibase 10.1038/nature11653} {\bibfield  {journal}
		{\bibinfo  {journal} {Nature}\ }\textbf {\bibinfo {volume} {491}},\ \bibinfo
		{pages} {574} (\bibinfo {year} {2012})}\BibitemShut {NoStop}%
	\bibitem [{\citenamefont {Scholl}\ \emph {et~al.}(2013)\citenamefont {Scholl},
		\citenamefont {Garc{\'{\i}}a-Etxarri}, \citenamefont {Koh},\ and\
		\citenamefont {Dionne}}]{Scholl2013}%
	\BibitemOpen
	\bibfield  {author} {\bibinfo {author} {\bibfnamefont {J.~A.}\ \bibnamefont
			{Scholl}}, \bibinfo {author} {\bibfnamefont {A.}~\bibnamefont
			{Garc{\'{\i}}a-Etxarri}}, \bibinfo {author} {\bibfnamefont {A.~L.}\
			\bibnamefont {Koh}}, \ and\ \bibinfo {author} {\bibfnamefont {J.~A.}\
			\bibnamefont {Dionne}},\ }\bibfield  {title} {\enquote {\bibinfo {title}
			{Observation of quantum tunneling between two plasmonic nanoparticles},}\
	}\href {\doibase 10.1021/nl304078v} {\bibfield  {journal} {\bibinfo
			{journal} {Nano Letters}\ }\textbf {\bibinfo {volume} {13}},\ \bibinfo
		{pages} {564} (\bibinfo {year} {2013})}\BibitemShut {NoStop}%
	\bibitem [{\citenamefont {Zhu}\ and\ \citenamefont {Crozier}(2014)}]{Zhu2014}%
	\BibitemOpen
	\bibfield  {author} {\bibinfo {author} {\bibfnamefont {W.}~\bibnamefont
			{Zhu}}\ and\ \bibinfo {author} {\bibfnamefont {K.~B.}\ \bibnamefont
			{Crozier}},\ }\bibfield  {title} {\enquote {\bibinfo {title} {Quantum
				mechanical limit to plasmonic enhancement as observed by surface-enhanced
				raman scattering},}\ }\href {\doibase 10.1038/ncomms6228} {\bibfield
		{journal} {\bibinfo  {journal} {Nature Communications}\ }\textbf {\bibinfo
			{volume} {5}} (\bibinfo {year} {2014}),\ 10.1038/ncomms6228}\BibitemShut
	{NoStop}%
	\bibitem [{\citenamefont {Zhu}\ \emph {et~al.}(2016)\citenamefont {Zhu},
		\citenamefont {Esteban}, \citenamefont {Borisov}, \citenamefont {Baumberg},
		\citenamefont {Nordlander}, \citenamefont {Lezec}, \citenamefont {Aizpurua},\
		and\ \citenamefont {Crozier}}]{Zhu2016}%
	\BibitemOpen
	\bibfield  {author} {\bibinfo {author} {\bibfnamefont {W.}~\bibnamefont
			{Zhu}}, \bibinfo {author} {\bibfnamefont {R.}~\bibnamefont {Esteban}},
		\bibinfo {author} {\bibfnamefont {A.~G.}\ \bibnamefont {Borisov}}, \bibinfo
		{author} {\bibfnamefont {J.~J.}\ \bibnamefont {Baumberg}}, \bibinfo {author}
		{\bibfnamefont {P.}~\bibnamefont {Nordlander}}, \bibinfo {author}
		{\bibfnamefont {H.~J.}\ \bibnamefont {Lezec}}, \bibinfo {author}
		{\bibfnamefont {J.}~\bibnamefont {Aizpurua}}, \ and\ \bibinfo {author}
		{\bibfnamefont {K.~B.}\ \bibnamefont {Crozier}},\ }\bibfield  {title}
	{\enquote {\bibinfo {title} {Quantum mechanical effects in plasmonic
				structures with subnanometre gaps},}\ }\href {\doibase 10.1038/ncomms11495}
	{\bibfield  {journal} {\bibinfo  {journal} {Nature Communications}\ }\textbf
		{\bibinfo {volume} {7}} (\bibinfo {year} {2016}),\
		10.1038/ncomms11495}\BibitemShut {NoStop}%
	\bibitem [{\citenamefont {de~Nijs}\ \emph {et~al.}(2017)\citenamefont
		{de~Nijs}, \citenamefont {Benz}, \citenamefont {Barrow}, \citenamefont
		{Sigle}, \citenamefont {Chikkaraddy}, \citenamefont {Palma}, \citenamefont
		{Carnegie}, \citenamefont {Kamp}, \citenamefont {Sundararaman}, \citenamefont
		{Narang}, \citenamefont {Scherman},\ and\ \citenamefont
		{Baumberg}}]{Nijs2017}%
	\BibitemOpen
	\bibfield  {author} {\bibinfo {author} {\bibfnamefont {B.}~\bibnamefont
			{de~Nijs}}, \bibinfo {author} {\bibfnamefont {F.}~\bibnamefont {Benz}},
		\bibinfo {author} {\bibfnamefont {S.~J.}\ \bibnamefont {Barrow}}, \bibinfo
		{author} {\bibfnamefont {D.~O.}\ \bibnamefont {Sigle}}, \bibinfo {author}
		{\bibfnamefont {R.}~\bibnamefont {Chikkaraddy}}, \bibinfo {author}
		{\bibfnamefont {A.}~\bibnamefont {Palma}}, \bibinfo {author} {\bibfnamefont
			{C.}~\bibnamefont {Carnegie}}, \bibinfo {author} {\bibfnamefont
			{M.}~\bibnamefont {Kamp}}, \bibinfo {author} {\bibfnamefont {R.}~\bibnamefont
			{Sundararaman}}, \bibinfo {author} {\bibfnamefont {P.}~\bibnamefont
			{Narang}}, \bibinfo {author} {\bibfnamefont {O.~A.}\ \bibnamefont
			{Scherman}}, \ and\ \bibinfo {author} {\bibfnamefont {J.~J.}\ \bibnamefont
			{Baumberg}},\ }\bibfield  {title} {\enquote {\bibinfo {title} {Plasmonic
				tunnel junctions for single-molecule redox chemistry},}\ }\href {\doibase
		10.1038/s41467-017-00819-7} {\bibfield  {journal} {\bibinfo  {journal}
			{Nature Communications}\ }\textbf {\bibinfo {volume} {8}} (\bibinfo {year}
		{2017}),\ 10.1038/s41467-017-00819-7}\BibitemShut {NoStop}%
	\bibitem [{\citenamefont {Cristofolini}\ \emph {et~al.}(2012)\citenamefont
		{Cristofolini}, \citenamefont {Christmann}, \citenamefont {Tsintzos},
		\citenamefont {Deligeorgis}, \citenamefont {Konstantinidis}, \citenamefont
		{Hatzopoulos}, \citenamefont {Savvidis},\ and\ \citenamefont
		{Baumberg}}]{Cristofolini2012}%
	\BibitemOpen
	\bibfield  {author} {\bibinfo {author} {\bibfnamefont {P.}~\bibnamefont
			{Cristofolini}}, \bibinfo {author} {\bibfnamefont {G.}~\bibnamefont
			{Christmann}}, \bibinfo {author} {\bibfnamefont {S.~I.}\ \bibnamefont
			{Tsintzos}}, \bibinfo {author} {\bibfnamefont {G.}~\bibnamefont
			{Deligeorgis}}, \bibinfo {author} {\bibfnamefont {G.}~\bibnamefont
			{Konstantinidis}}, \bibinfo {author} {\bibfnamefont {Z.}~\bibnamefont
			{Hatzopoulos}}, \bibinfo {author} {\bibfnamefont {P.~G.}\ \bibnamefont
			{Savvidis}}, \ and\ \bibinfo {author} {\bibfnamefont {J.~J.}\ \bibnamefont
			{Baumberg}},\ }\bibfield  {title} {\enquote {\bibinfo {title} {Coupling
				quantum tunneling with cavity photons},}\ }\href {\doibase
		10.1126/science.1219010} {\bibfield  {journal} {\bibinfo  {journal}
			{Science}\ }\textbf {\bibinfo {volume} {336}},\ \bibinfo {pages} {704}
		(\bibinfo {year} {2012})}\BibitemShut {NoStop}%
	\bibitem [{\citenamefont {Mar}\ \emph {et~al.}(2011)\citenamefont {Mar},
		\citenamefont {Xu}, \citenamefont {Baumberg}, \citenamefont {Irvine},
		\citenamefont {Stanley},\ and\ \citenamefont {Williams}}]{Mar2011}%
	\BibitemOpen
	\bibfield  {author} {\bibinfo {author} {\bibfnamefont {J.~D.}\ \bibnamefont
			{Mar}}, \bibinfo {author} {\bibfnamefont {X.~L.}\ \bibnamefont {Xu}},
		\bibinfo {author} {\bibfnamefont {J.~J.}\ \bibnamefont {Baumberg}}, \bibinfo
		{author} {\bibfnamefont {A.~C.}\ \bibnamefont {Irvine}}, \bibinfo {author}
		{\bibfnamefont {C.}~\bibnamefont {Stanley}}, \ and\ \bibinfo {author}
		{\bibfnamefont {D.~A.}\ \bibnamefont {Williams}},\ }\bibfield  {title}
	{\enquote {\bibinfo {title} {Voltage-controlled electron tunneling from a
				single self-assembled quantum dot embedded in a
				two-dimensional-electron-gas-based photovoltaic cell},}\ }\href {\doibase
		10.1063/1.3633216} {\bibfield  {journal} {\bibinfo  {journal} {Journal of
				Applied Physics}\ }\textbf {\bibinfo {volume} {110}},\ \bibinfo {pages}
		{053110} (\bibinfo {year} {2011})}\BibitemShut {NoStop}%
	\bibitem [{\citenamefont {Cazier}\ \emph {et~al.}(2016)\citenamefont {Cazier},
		\citenamefont {Buret}, \citenamefont {Uskov}, \citenamefont {Markey},
		\citenamefont {Arocas}, \citenamefont {Francs},\ and\ \citenamefont
		{Bouhelier}}]{Cazier2016}%
	\BibitemOpen
	\bibfield  {author} {\bibinfo {author} {\bibfnamefont {N.}~\bibnamefont
			{Cazier}}, \bibinfo {author} {\bibfnamefont {M.}~\bibnamefont {Buret}},
		\bibinfo {author} {\bibfnamefont {A.~V.}\ \bibnamefont {Uskov}}, \bibinfo
		{author} {\bibfnamefont {L.}~\bibnamefont {Markey}}, \bibinfo {author}
		{\bibfnamefont {J.}~\bibnamefont {Arocas}}, \bibinfo {author} {\bibfnamefont
			{G.~C.~D.}\ \bibnamefont {Francs}}, \ and\ \bibinfo {author} {\bibfnamefont
			{A.}~\bibnamefont {Bouhelier}},\ }\bibfield  {title} {\enquote {\bibinfo
			{title} {Electrical excitation of waveguided surface plasmons by a
				light-emitting tunneling optical gap antenna},}\ }\href {\doibase
		10.1364/oe.24.003873} {\bibfield  {journal} {\bibinfo  {journal} {Optics
				Express}\ }\textbf {\bibinfo {volume} {24}},\ \bibinfo {pages} {3873}
		(\bibinfo {year} {2016})}\BibitemShut {NoStop}%
	\bibitem [{\citenamefont {Uskov}\ \emph {et~al.}(2016)\citenamefont {Uskov},
		\citenamefont {Khurgin}, \citenamefont {Protsenko}, \citenamefont
		{Smetanin},\ and\ \citenamefont {Bouhelier}}]{Uskov2016}%
	\BibitemOpen
	\bibfield  {author} {\bibinfo {author} {\bibfnamefont {A.~V.}\ \bibnamefont
			{Uskov}}, \bibinfo {author} {\bibfnamefont {J.~B.}\ \bibnamefont {Khurgin}},
		\bibinfo {author} {\bibfnamefont {I.~E.}\ \bibnamefont {Protsenko}}, \bibinfo
		{author} {\bibfnamefont {I.~V.}\ \bibnamefont {Smetanin}}, \ and\ \bibinfo
		{author} {\bibfnamefont {A.}~\bibnamefont {Bouhelier}},\ }\bibfield  {title}
	{\enquote {\bibinfo {title} {Excitation of plasmonic nanoantennas by
				nonresonant and resonant electron tunnelling},}\ }\href {\doibase
		10.1039/c6nr01931e} {\bibfield  {journal} {\bibinfo  {journal} {Nanoscale}\
		}\textbf {\bibinfo {volume} {8}},\ \bibinfo {pages} {14573} (\bibinfo {year}
		{2016})}\BibitemShut {NoStop}%
	\bibitem [{\citenamefont {Stolz}\ \emph {et~al.}(2014)\citenamefont {Stolz},
		\citenamefont {Berthelot}, \citenamefont {Mennemanteuil}, \citenamefont {des
			Francs}, \citenamefont {Markey}, \citenamefont {Meunier},\ and\ \citenamefont
		{Bouhelier}}]{Stolz2014}%
	\BibitemOpen
	\bibfield  {author} {\bibinfo {author} {\bibfnamefont {A.}~\bibnamefont
			{Stolz}}, \bibinfo {author} {\bibfnamefont {J.}~\bibnamefont {Berthelot}},
		\bibinfo {author} {\bibfnamefont {M.-M.}\ \bibnamefont {Mennemanteuil}},
		\bibinfo {author} {\bibfnamefont {G.~C.}\ \bibnamefont {des Francs}},
		\bibinfo {author} {\bibfnamefont {L.}~\bibnamefont {Markey}}, \bibinfo
		{author} {\bibfnamefont {V.}~\bibnamefont {Meunier}}, \ and\ \bibinfo
		{author} {\bibfnamefont {A.}~\bibnamefont {Bouhelier}},\ }\bibfield  {title}
	{\enquote {\bibinfo {title} {Nonlinear photon-assisted tunneling transport in
				optical gap antennas},}\ }\href {\doibase 10.1021/nl404707t} {\bibfield
		{journal} {\bibinfo  {journal} {Nano Letters}\ }\textbf {\bibinfo {volume}
			{14}},\ \bibinfo {pages} {2330} (\bibinfo {year} {2014})}\BibitemShut
	{NoStop}%
	\bibitem [{\citenamefont {Kravtsov}\ \emph {et~al.}(2014)\citenamefont
		{Kravtsov}, \citenamefont {Berweger}, \citenamefont {Atkin},\ and\
		\citenamefont {Raschke}}]{Kravtsov2014}%
	\BibitemOpen
	\bibfield  {author} {\bibinfo {author} {\bibfnamefont {V.}~\bibnamefont
			{Kravtsov}}, \bibinfo {author} {\bibfnamefont {S.}~\bibnamefont {Berweger}},
		\bibinfo {author} {\bibfnamefont {J.~M.}\ \bibnamefont {Atkin}}, \ and\
		\bibinfo {author} {\bibfnamefont {M.~B.}\ \bibnamefont {Raschke}},\
	}\bibfield  {title} {\enquote {\bibinfo {title} {Control of plasmon emission
				and dynamics at the transition from classical to quantum coupling},}\ }\href
	{\doibase 10.1021/nl502297t} {\bibfield  {journal} {\bibinfo  {journal} {Nano
				Letters}\ }\textbf {\bibinfo {volume} {14}},\ \bibinfo {pages} {5270}
		(\bibinfo {year} {2014})}\BibitemShut {NoStop}%
	\bibitem [{\citenamefont {Allen~Taflove}(2005)}]{AllenTaflove2005}%
	\BibitemOpen
	\bibfield  {author} {\bibinfo {author} {\bibfnamefont {S.~C.~H.}\
			\bibnamefont {Allen~Taflove}},\ }\href
	{https://www.ebook.de/de/product/3580196/allen_taflove_susan_c_hagness_computational_electrodynamics_the_finite_difference_time_domain_method.html}
	{\emph {\bibinfo {title} {Computational Electrodynamics: The
				Finite-Difference Time-Domain Method}}}\ (\bibinfo  {publisher} {ARTECH HOUSE
		INC},\ \bibinfo {year} {2005})\BibitemShut {NoStop}%
	\bibitem [{\citenamefont {Jackson}(1998)}]{Jackson1998}%
	\BibitemOpen
	\bibfield  {author} {\bibinfo {author} {\bibfnamefont {J.~D.}\ \bibnamefont
			{Jackson}},\ }\href@noop {} {\emph {\bibinfo {title} {Classical
				electrodynamics, Third Edition}}}\ (\bibinfo  {publisher} {John Wiley \&
		Sons},\ \bibinfo {year} {1998})\BibitemShut {NoStop}%
	\bibitem [{\citenamefont {Griffiths}(2013)}]{Griffiths2013}%
	\BibitemOpen
	\bibfield  {author} {\bibinfo {author} {\bibfnamefont {D.~J.}\ \bibnamefont
			{Griffiths}},\ }\href {https://cds.cern.ch/record/1492149} {\emph {\bibinfo
			{title} {{Introduction to electrodynamics; 4th ed.}}}}\ (\bibinfo
	{publisher} {Pearson},\ \bibinfo {address} {Boston, MA},\ \bibinfo {year}
	{2013})\ \bibinfo {note} {re-published by Cambridge University Press in
		2017}\BibitemShut {NoStop}%
	\bibitem [{\citenamefont {Fattinger}\ and\ \citenamefont
		{Grischkowsky}(1988)}]{Fattinger1988}%
	\BibitemOpen
	\bibfield  {author} {\bibinfo {author} {\bibfnamefont {C.}~\bibnamefont
			{Fattinger}}\ and\ \bibinfo {author} {\bibfnamefont {D.}~\bibnamefont
			{Grischkowsky}},\ }\bibfield  {title} {\enquote {\bibinfo {title} {Point
				source terahertz optics},}\ }\href {\doibase 10.1063/1.99971} {\bibfield
		{journal} {\bibinfo  {journal} {Applied Physics Letters}\ }\textbf {\bibinfo
			{volume} {53}},\ \bibinfo {pages} {1480} (\bibinfo {year}
		{1988})}\BibitemShut {NoStop}%
	\bibitem [{\citenamefont {Fattinger}\ and\ \citenamefont
		{Grischkowsky}(1989)}]{Fattinger1989}%
	\BibitemOpen
	\bibfield  {author} {\bibinfo {author} {\bibfnamefont {C.}~\bibnamefont
			{Fattinger}}\ and\ \bibinfo {author} {\bibfnamefont {D.}~\bibnamefont
			{Grischkowsky}},\ }\bibfield  {title} {\enquote {\bibinfo {title} {Terahertz
				beams},}\ }\href {\doibase 10.1063/1.100958} {\bibfield  {journal} {\bibinfo
			{journal} {Applied Physics Letters}\ }\textbf {\bibinfo {volume} {54}},\
		\bibinfo {pages} {490} (\bibinfo {year} {1989})}\BibitemShut {NoStop}%
	\bibitem [{\citenamefont {Grischkowsky}\ \emph {et~al.}(1990)\citenamefont
		{Grischkowsky}, \citenamefont {Keiding}, \citenamefont {van Exter},\ and\
		\citenamefont {Fattinger}}]{Grischkowsky1990}%
	\BibitemOpen
	\bibfield  {author} {\bibinfo {author} {\bibfnamefont {D.}~\bibnamefont
			{Grischkowsky}}, \bibinfo {author} {\bibfnamefont {S.}~\bibnamefont
			{Keiding}}, \bibinfo {author} {\bibfnamefont {M.}~\bibnamefont {van Exter}},
		\ and\ \bibinfo {author} {\bibfnamefont {C.}~\bibnamefont {Fattinger}},\
	}\bibfield  {title} {\enquote {\bibinfo {title} {Far-infrared time-domain
				spectroscopy with terahertz beams of dielectrics and semiconductors},}\
	}\href {\doibase 10.1364/josab.7.002006} {\bibfield  {journal} {\bibinfo
			{journal} {Journal of the Optical Society of America B}\ }\textbf {\bibinfo
			{volume} {7}},\ \bibinfo {pages} {2006} (\bibinfo {year} {1990})}\BibitemShut
	{NoStop}%
	\bibitem [{\citenamefont {Park}\ \emph
		{et~al.}(2014{\natexlab{b}})\citenamefont {Park}, \citenamefont {Son},
		\citenamefont {Choi}, \citenamefont {Kim},\ and\ \citenamefont
		{Ahn}}]{Park2014a}%
	\BibitemOpen
	\bibfield  {author} {\bibinfo {author} {\bibfnamefont {S.~J.}\ \bibnamefont
			{Park}}, \bibinfo {author} {\bibfnamefont {B.~H.}\ \bibnamefont {Son}},
		\bibinfo {author} {\bibfnamefont {S.~J.}\ \bibnamefont {Choi}}, \bibinfo
		{author} {\bibfnamefont {H.~S.}\ \bibnamefont {Kim}}, \ and\ \bibinfo
		{author} {\bibfnamefont {Y.~H.}\ \bibnamefont {Ahn}},\ }\bibfield  {title}
	{\enquote {\bibinfo {title} {Sensitive detection of yeast using terahertz
				slot antennas},}\ }\href {\doibase 10.1364/oe.22.030467} {\bibfield
		{journal} {\bibinfo  {journal} {Optics Express}\ }\textbf {\bibinfo {volume}
			{22}},\ \bibinfo {pages} {30467} (\bibinfo {year}
		{2014}{\natexlab{b}})}\BibitemShut {NoStop}%
	\bibitem [{\citenamefont {Sushko}\ \emph {et~al.}(2013)\citenamefont {Sushko},
		\citenamefont {Dubrovka},\ and\ \citenamefont {Donnan}}]{Sushko2013}%
	\BibitemOpen
	\bibfield  {author} {\bibinfo {author} {\bibfnamefont {O.}~\bibnamefont
			{Sushko}}, \bibinfo {author} {\bibfnamefont {R.}~\bibnamefont {Dubrovka}}, \
		and\ \bibinfo {author} {\bibfnamefont {R.~S.}\ \bibnamefont {Donnan}},\
	}\bibfield  {title} {\enquote {\bibinfo {title} {Terahertz spectral domain
				computational analysis of hydration shell of proteins with increasingly
				complex tertiary structure},}\ }\href {\doibase 10.1021/jp407580y} {\bibfield
		{journal} {\bibinfo  {journal} {The Journal of Physical Chemistry B}\
		}\textbf {\bibinfo {volume} {117}},\ \bibinfo {pages} {16486} (\bibinfo
		{year} {2013})}\BibitemShut {NoStop}%
	\bibitem [{\citenamefont {Nibali}\ and\ \citenamefont
		{Havenith}(2014)}]{Nibali2014}%
	\BibitemOpen
	\bibfield  {author} {\bibinfo {author} {\bibfnamefont {V.~C.}\ \bibnamefont
			{Nibali}}\ and\ \bibinfo {author} {\bibfnamefont {M.}~\bibnamefont
			{Havenith}},\ }\bibfield  {title} {\enquote {\bibinfo {title} {New insights
				into the role of water in biological function: Studying solvated biomolecules
				using terahertz absorption spectroscopy in conjunction with molecular
				dynamics simulations},}\ }\href {\doibase 10.1021/ja504441h} {\bibfield
		{journal} {\bibinfo  {journal} {Journal of the American Chemical Society}\
		}\textbf {\bibinfo {volume} {136}},\ \bibinfo {pages} {12800} (\bibinfo
		{year} {2014})}\BibitemShut {NoStop}%
	\bibitem [{\citenamefont {Rocchi}\ \emph {et~al.}(1998)\citenamefont {Rocchi},
		\citenamefont {Bizzarri},\ and\ \citenamefont {Cannistraro}}]{Rocchi1998}%
	\BibitemOpen
	\bibfield  {author} {\bibinfo {author} {\bibfnamefont {C.}~\bibnamefont
			{Rocchi}}, \bibinfo {author} {\bibfnamefont {A.~R.}\ \bibnamefont
			{Bizzarri}}, \ and\ \bibinfo {author} {\bibfnamefont {S.}~\bibnamefont
			{Cannistraro}},\ }\bibfield  {title} {\enquote {\bibinfo {title} {Water
				dynamical anomalies evidenced by molecular-dynamics simulations at the
				solvent-protein interface},}\ }\href {\doibase 10.1103/physreve.57.3315}
	{\bibfield  {journal} {\bibinfo  {journal} {Physical Review E}\ }\textbf
		{\bibinfo {volume} {57}},\ \bibinfo {pages} {3315} (\bibinfo {year}
		{1998})}\BibitemShut {NoStop}%
	\bibitem [{\citenamefont {Marchi}\ \emph {et~al.}(2002)\citenamefont {Marchi},
		\citenamefont {Sterpone},\ and\ \citenamefont {Ceccarelli}}]{Marchi2002}%
	\BibitemOpen
	\bibfield  {author} {\bibinfo {author} {\bibfnamefont {M.}~\bibnamefont
			{Marchi}}, \bibinfo {author} {\bibfnamefont {F.}~\bibnamefont {Sterpone}}, \
		and\ \bibinfo {author} {\bibfnamefont {M.}~\bibnamefont {Ceccarelli}},\
	}\bibfield  {title} {\enquote {\bibinfo {title} {Water rotational relaxation
				and diffusion in hydrated lysozyme},}\ }\href {\doibase 10.1021/ja025905m}
	{\bibfield  {journal} {\bibinfo  {journal} {Journal of the American Chemical
				Society}\ }\textbf {\bibinfo {volume} {124}},\ \bibinfo {pages} {6787}
		(\bibinfo {year} {2002})}\BibitemShut {NoStop}%
	\bibitem [{\citenamefont {Sengupta}\ \emph {et~al.}(2008)\citenamefont
		{Sengupta}, \citenamefont {Jaud},\ and\ \citenamefont
		{Tobias}}]{Sengupta2008}%
	\BibitemOpen
	\bibfield  {author} {\bibinfo {author} {\bibfnamefont {N.}~\bibnamefont
			{Sengupta}}, \bibinfo {author} {\bibfnamefont {S.}~\bibnamefont {Jaud}}, \
		and\ \bibinfo {author} {\bibfnamefont {D.~J.}\ \bibnamefont {Tobias}},\
	}\bibfield  {title} {\enquote {\bibinfo {title} {Hydration dynamics in a
				partially denatured ensemble of the globular protein human
				$\alpha$-lactalbumin investigated with molecular dynamics simulations},}\
	}\href {\doibase 10.1529/biophysj.108.136531} {\bibfield  {journal} {\bibinfo
			{journal} {Biophysical Journal}\ }\textbf {\bibinfo {volume} {95}},\
		\bibinfo {pages} {5257} (\bibinfo {year} {2008})}\BibitemShut {NoStop}%
	\bibitem [{\citenamefont {Sinha}\ \emph {et~al.}(2008)\citenamefont {Sinha},
		\citenamefont {Chakraborty},\ and\ \citenamefont
		{Bandyopadhyay}}]{Sinha2008}%
	\BibitemOpen
	\bibfield  {author} {\bibinfo {author} {\bibfnamefont {S.~K.}\ \bibnamefont
			{Sinha}}, \bibinfo {author} {\bibfnamefont {S.}~\bibnamefont {Chakraborty}},
		\ and\ \bibinfo {author} {\bibfnamefont {S.}~\bibnamefont {Bandyopadhyay}},\
	}\bibfield  {title} {\enquote {\bibinfo {title} {Thickness of the hydration
				layer of a protein from molecular dynamics simulation},}\ }\href {\doibase
		10.1021/jp8000724} {\bibfield  {journal} {\bibinfo  {journal} {The Journal of
				Physical Chemistry B}\ }\textbf {\bibinfo {volume} {112}},\ \bibinfo {pages}
		{8203} (\bibinfo {year} {2008})}\BibitemShut {NoStop}%
	\bibitem [{\citenamefont {Xu}\ \emph {et~al.}(2012)\citenamefont {Xu},
		\citenamefont {Gnanasekaran},\ and\ \citenamefont {Leitner}}]{Xu2012}%
	\BibitemOpen
	\bibfield  {author} {\bibinfo {author} {\bibfnamefont {Y.}~\bibnamefont
			{Xu}}, \bibinfo {author} {\bibfnamefont {R.}~\bibnamefont {Gnanasekaran}}, \
		and\ \bibinfo {author} {\bibfnamefont {D.~M.}\ \bibnamefont {Leitner}},\
	}\bibfield  {title} {\enquote {\bibinfo {title} {Analysis of water and
				hydrogen bond dynamics at the surface of an antifreeze protein},}\ }\href
	{\doibase 10.1155/2012/125071} {\bibfield  {journal} {\bibinfo  {journal}
			{Journal of Atomic, Molecular, and Optical Physics}\ }\textbf {\bibinfo
			{volume} {2012}},\ \bibinfo {pages} {1} (\bibinfo {year} {2012})}\BibitemShut
	{NoStop}%
	\bibitem [{\citenamefont {Chakraborty}\ \emph {et~al.}(2007)\citenamefont
		{Chakraborty}, \citenamefont {Sinha},\ and\ \citenamefont
		{Bandyopadhyay}}]{Chakraborty2007}%
	\BibitemOpen
	\bibfield  {author} {\bibinfo {author} {\bibfnamefont {S.}~\bibnamefont
			{Chakraborty}}, \bibinfo {author} {\bibfnamefont {S.~K.}\ \bibnamefont
			{Sinha}}, \ and\ \bibinfo {author} {\bibfnamefont {S.}~\bibnamefont
			{Bandyopadhyay}},\ }\bibfield  {title} {\enquote {\bibinfo {title}
			{Low-frequency vibrational spectrum of water in the hydration layer of a
				protein:~ a molecular dynamics simulation study},}\ }\href {\doibase
		10.1021/jp0746401} {\bibfield  {journal} {\bibinfo  {journal} {The Journal of
				Physical Chemistry B}\ }\textbf {\bibinfo {volume} {111}},\ \bibinfo {pages}
		{13626} (\bibinfo {year} {2007})}\BibitemShut {NoStop}%
	\bibitem [{\citenamefont {Sinha}\ and\ \citenamefont
		{Bandyopadhyay}(2011)}]{Sinha2011}%
	\BibitemOpen
	\bibfield  {author} {\bibinfo {author} {\bibfnamefont {S.~K.}\ \bibnamefont
			{Sinha}}\ and\ \bibinfo {author} {\bibfnamefont {S.}~\bibnamefont
			{Bandyopadhyay}},\ }\bibfield  {title} {\enquote {\bibinfo {title}
			{Differential flexibility of the secondary structures of lysozyme and the
				structure and ordering of surrounding water molecules},}\ }\href {\doibase
		10.1063/1.3560442} {\bibfield  {journal} {\bibinfo  {journal} {The Journal of
				Chemical Physics}\ }\textbf {\bibinfo {volume} {134}},\ \bibinfo {pages}
		{115101} (\bibinfo {year} {2011})}\BibitemShut {NoStop}%
	\bibitem [{\citenamefont {Bandyopadhyay}\ \emph {et~al.}(2006)\citenamefont
		{Bandyopadhyay}, \citenamefont {Chakraborty},\ and\ \citenamefont
		{Bagchi}}]{Bandyopadhyay2006}%
	\BibitemOpen
	\bibfield  {author} {\bibinfo {author} {\bibfnamefont {S.}~\bibnamefont
			{Bandyopadhyay}}, \bibinfo {author} {\bibfnamefont {S.}~\bibnamefont
			{Chakraborty}}, \ and\ \bibinfo {author} {\bibfnamefont {B.}~\bibnamefont
			{Bagchi}},\ }\bibfield  {title} {\enquote {\bibinfo {title} {Exploration of
				the secondary structure specific differential solvation dynamics between the
				native and molten globule states of the protein {HP}-36},}\ }\href {\doibase
		10.1021/jp0633547} {\bibfield  {journal} {\bibinfo  {journal} {The Journal of
				Physical Chemistry B}\ }\textbf {\bibinfo {volume} {110}},\ \bibinfo {pages}
		{20629} (\bibinfo {year} {2006})}\BibitemShut {NoStop}%
	\bibitem [{\citenamefont {Bandyopadhyay}\ \emph {et~al.}(2004)\citenamefont
		{Bandyopadhyay}, \citenamefont {Chakraborty}, \citenamefont
		{Balasubramanian}, \citenamefont {Pal},\ and\ \citenamefont
		{Bagchi}}]{Bandyopadhyay2004}%
	\BibitemOpen
	\bibfield  {author} {\bibinfo {author} {\bibfnamefont {S.}~\bibnamefont
			{Bandyopadhyay}}, \bibinfo {author} {\bibfnamefont {S.}~\bibnamefont
			{Chakraborty}}, \bibinfo {author} {\bibfnamefont {S.}~\bibnamefont
			{Balasubramanian}}, \bibinfo {author} {\bibfnamefont {S.}~\bibnamefont
			{Pal}}, \ and\ \bibinfo {author} {\bibfnamefont {B.}~\bibnamefont {Bagchi}},\
	}\bibfield  {title} {\enquote {\bibinfo {title} {Atomistic simulation study
				of the coupled motion of amino acid residues and water molecules around
				protein {HP}-36: Fluctuations at and around the active sites},}\ }\href
	{\doibase 10.1021/jp048532f} {\bibfield  {journal} {\bibinfo  {journal} {The
				Journal of Physical Chemistry B}\ }\textbf {\bibinfo {volume} {108}},\
		\bibinfo {pages} {12608} (\bibinfo {year} {2004})}\BibitemShut {NoStop}%
	\bibitem [{\citenamefont {Pal}\ and\ \citenamefont
		{Bandyopadhyay}(2013{\natexlab{a}})}]{Pal2013}%
	\BibitemOpen
	\bibfield  {author} {\bibinfo {author} {\bibfnamefont {S.}~\bibnamefont
			{Pal}}\ and\ \bibinfo {author} {\bibfnamefont {S.}~\bibnamefont
			{Bandyopadhyay}},\ }\bibfield  {title} {\enquote {\bibinfo {title} {Effects
				of protein conformational flexibilities and electrostatic interactions on the
				low-frequency vibrational spectrum of hydration water},}\ }\href {\doibase
		10.1021/jp402662v} {\bibfield  {journal} {\bibinfo  {journal} {The Journal of
				Physical Chemistry B}\ }\textbf {\bibinfo {volume} {117}},\ \bibinfo {pages}
		{5848} (\bibinfo {year} {2013}{\natexlab{a}})}\BibitemShut {NoStop}%
	\bibitem [{\citenamefont {Pal}\ and\ \citenamefont
		{Bandyopadhyay}(2013{\natexlab{b}})}]{Pal2013a}%
	\BibitemOpen
	\bibfield  {author} {\bibinfo {author} {\bibfnamefont {S.}~\bibnamefont
			{Pal}}\ and\ \bibinfo {author} {\bibfnamefont {S.}~\bibnamefont
			{Bandyopadhyay}},\ }\bibfield  {title} {\enquote {\bibinfo {title}
			{Importance of protein conformational motions and electrostatic anchoring
				sites on the dynamics and hydrogen bond properties of hydration water},}\
	}\href {\doibase 10.1021/la303959m} {\bibfield  {journal} {\bibinfo
			{journal} {Langmuir}\ }\textbf {\bibinfo {volume} {29}},\ \bibinfo {pages}
		{1162} (\bibinfo {year} {2013}{\natexlab{b}})}\BibitemShut {NoStop}%
	\bibitem [{\citenamefont {Sinha}\ and\ \citenamefont
		{Bandyopadhyay}(2012{\natexlab{a}})}]{Sinha2012}%
	\BibitemOpen
	\bibfield  {author} {\bibinfo {author} {\bibfnamefont {S.~K.}\ \bibnamefont
			{Sinha}}\ and\ \bibinfo {author} {\bibfnamefont {S.}~\bibnamefont
			{Bandyopadhyay}},\ }\bibfield  {title} {\enquote {\bibinfo {title} {Polar
				solvation dynamics of lysozyme from molecular dynamics studies},}\ }\href
	{\doibase 10.1063/1.4712036} {\bibfield  {journal} {\bibinfo  {journal} {The
				Journal of Chemical Physics}\ }\textbf {\bibinfo {volume} {136}},\ \bibinfo
		{pages} {185102} (\bibinfo {year} {2012}{\natexlab{a}})}\BibitemShut
	{NoStop}%
	\bibitem [{\citenamefont {Sinha}\ and\ \citenamefont
		{Bandyopadhyay}(2012{\natexlab{b}})}]{Sinha2012a}%
	\BibitemOpen
	\bibfield  {author} {\bibinfo {author} {\bibfnamefont {S.~K.}\ \bibnamefont
			{Sinha}}\ and\ \bibinfo {author} {\bibfnamefont {S.}~\bibnamefont
			{Bandyopadhyay}},\ }\bibfield  {title} {\enquote {\bibinfo {title} {Local
				heterogeneous dynamics of water around lysozyme: a computer simulation
				study},}\ }\href {\doibase 10.1039/c1cp22575h} {\bibfield  {journal}
		{\bibinfo  {journal} {Phys. Chem. Chem. Phys.}\ }\textbf {\bibinfo {volume}
			{14}},\ \bibinfo {pages} {899} (\bibinfo {year}
		{2012}{\natexlab{b}})}\BibitemShut {NoStop}%
	\bibitem [{\citenamefont {Ding}\ \emph {et~al.}(2011)\citenamefont {Ding},
		\citenamefont {Huber}, \citenamefont {Middelberg},\ and\ \citenamefont
		{Falconer}}]{Ding2011}%
	\BibitemOpen
	\bibfield  {author} {\bibinfo {author} {\bibfnamefont {T.}~\bibnamefont
			{Ding}}, \bibinfo {author} {\bibfnamefont {T.}~\bibnamefont {Huber}},
		\bibinfo {author} {\bibfnamefont {A.~P.}\ \bibnamefont {Middelberg}}, \ and\
		\bibinfo {author} {\bibfnamefont {R.~J.}\ \bibnamefont {Falconer}},\
	}\bibfield  {title} {\enquote {\bibinfo {title} {Characterization of
				low-frequency modes in aqueous peptides using far-infrared spectroscopy and
				molecular dynamics simulation},}\ }\href {\doibase 10.1021/jp200553d}
	{\bibfield  {journal} {\bibinfo  {journal} {The Journal of Physical Chemistry
				A}\ }\textbf {\bibinfo {volume} {115}},\ \bibinfo {pages} {11559} (\bibinfo
		{year} {2011})}\BibitemShut {NoStop}%
	\bibitem [{\citenamefont {Heyden}\ and\ \citenamefont
		{Havenith}(2010)}]{Heyden2010a}%
	\BibitemOpen
	\bibfield  {author} {\bibinfo {author} {\bibfnamefont {M.}~\bibnamefont
			{Heyden}}\ and\ \bibinfo {author} {\bibfnamefont {M.}~\bibnamefont
			{Havenith}},\ }\bibfield  {title} {\enquote {\bibinfo {title} {Combining
				{THz} spectroscopy and {MD} simulations to study protein-hydration
				coupling},}\ }\href {\doibase 10.1016/j.ymeth.2010.05.007} {\bibfield
		{journal} {\bibinfo  {journal} {Methods}\ }\textbf {\bibinfo {volume} {52}},\
		\bibinfo {pages} {74} (\bibinfo {year} {2010})}\BibitemShut {NoStop}%
	\bibitem [{\citenamefont {Heyden}\ \emph {et~al.}(2012)\citenamefont {Heyden},
		\citenamefont {Sun}, \citenamefont {Forbert}, \citenamefont {Mathias},
		\citenamefont {Havenith},\ and\ \citenamefont {Marx}}]{Heyden2012}%
	\BibitemOpen
	\bibfield  {author} {\bibinfo {author} {\bibfnamefont {M.}~\bibnamefont
			{Heyden}}, \bibinfo {author} {\bibfnamefont {J.}~\bibnamefont {Sun}},
		\bibinfo {author} {\bibfnamefont {H.}~\bibnamefont {Forbert}}, \bibinfo
		{author} {\bibfnamefont {G.}~\bibnamefont {Mathias}}, \bibinfo {author}
		{\bibfnamefont {M.}~\bibnamefont {Havenith}}, \ and\ \bibinfo {author}
		{\bibfnamefont {D.}~\bibnamefont {Marx}},\ }\bibfield  {title} {\enquote
		{\bibinfo {title} {Understanding the origins of dipolar couplings and
				correlated motion in the vibrational spectrum of water},}\ }\href {\doibase
		10.1021/jz300748s} {\bibfield  {journal} {\bibinfo  {journal} {The Journal of
				Physical Chemistry Letters}\ }\textbf {\bibinfo {volume} {3}},\ \bibinfo
		{pages} {2135} (\bibinfo {year} {2012})}\BibitemShut {NoStop}%
	\bibitem [{\citenamefont {Allen}\ \emph {et~al.}(2004)\citenamefont {Allen}
		\emph {et~al.}}]{Allen2004}%
	\BibitemOpen
	\bibfield  {author} {\bibinfo {author} {\bibfnamefont {M.~P.}\ \bibnamefont
			{Allen}} \emph {et~al.},\ }\bibfield  {title} {\enquote {\bibinfo {title}
			{Introduction to molecular dynamics simulation},}\ }\href@noop {} {\bibfield
		{journal} {\bibinfo  {journal} {Computational soft matter: from synthetic
				polymers to proteins}\ }\textbf {\bibinfo {volume} {23}},\ \bibinfo {pages}
		{1} (\bibinfo {year} {2004})}\BibitemShut {NoStop}%
	\bibitem [{\citenamefont {Bertie}\ and\ \citenamefont
		{Lan}(1996)}]{Bertie1996}%
	\BibitemOpen
	\bibfield  {author} {\bibinfo {author} {\bibfnamefont {J.~E.}\ \bibnamefont
			{Bertie}}\ and\ \bibinfo {author} {\bibfnamefont {Z.}~\bibnamefont {Lan}},\
	}\bibfield  {title} {\enquote {\bibinfo {title} {Infrared intensities of
				liquids {XX}: The intensity of the {OH} stretching band of liquid water
				revisited, and the best current values of the optical constants of h2o(l) at
				25{\textdegree}c between 15,000 and 1 cm-1},}\ }\href {\doibase
		10.1366/0003702963905385} {\bibfield  {journal} {\bibinfo  {journal} {Applied
				Spectroscopy}\ }\textbf {\bibinfo {volume} {50}},\ \bibinfo {pages} {1047}
		(\bibinfo {year} {1996})}\BibitemShut {NoStop}%
	\bibitem [{\citenamefont {Heyden}\ \emph {et~al.}(2010)\citenamefont {Heyden},
		\citenamefont {Sun}, \citenamefont {Funkner}, \citenamefont {Mathias},
		\citenamefont {Forbert}, \citenamefont {Havenith},\ and\ \citenamefont
		{Marx}}]{Heyden2010}%
	\BibitemOpen
	\bibfield  {author} {\bibinfo {author} {\bibfnamefont {M.}~\bibnamefont
			{Heyden}}, \bibinfo {author} {\bibfnamefont {J.}~\bibnamefont {Sun}},
		\bibinfo {author} {\bibfnamefont {S.}~\bibnamefont {Funkner}}, \bibinfo
		{author} {\bibfnamefont {G.}~\bibnamefont {Mathias}}, \bibinfo {author}
		{\bibfnamefont {H.}~\bibnamefont {Forbert}}, \bibinfo {author} {\bibfnamefont
			{M.}~\bibnamefont {Havenith}}, \ and\ \bibinfo {author} {\bibfnamefont
			{D.}~\bibnamefont {Marx}},\ }\bibfield  {title} {\enquote {\bibinfo {title}
			{Dissecting the {THz} spectrum of liquid water from first principles via
				correlations in time and space},}\ }\href {\doibase 10.1073/pnas.0914885107}
	{\bibfield  {journal} {\bibinfo  {journal} {Proceedings of the National
				Academy of Sciences}\ }\textbf {\bibinfo {volume} {107}},\ \bibinfo {pages}
		{12068} (\bibinfo {year} {2010})}\BibitemShut {NoStop}%
	\bibitem [{\citenamefont {Jiang}\ \emph {et~al.}(2017)\citenamefont {Jiang},
		\citenamefont {Xiao}, \citenamefont {Wang}, \citenamefont {Liu},
		\citenamefont {Wang},\ and\ \citenamefont {Liu}}]{Jiang2017}%
	\BibitemOpen
	\bibfield  {author} {\bibinfo {author} {\bibfnamefont {J.}~\bibnamefont
			{Jiang}}, \bibinfo {author} {\bibfnamefont {M.}~\bibnamefont {Xiao}},
		\bibinfo {author} {\bibfnamefont {S.}~\bibnamefont {Wang}}, \bibinfo {author}
		{\bibfnamefont {K.}~\bibnamefont {Liu}}, \bibinfo {author} {\bibfnamefont
			{X.}~\bibnamefont {Wang}}, \ and\ \bibinfo {author} {\bibfnamefont
			{T.}~\bibnamefont {Liu}},\ }\bibfield  {title} {\enquote {\bibinfo {title}
			{Polarized low-coherence interferometer based on a matrix {CCD} and
				birefringence crystal with a two-dimensional angle},}\ }\href {\doibase
		10.1364/oe.25.015977} {\bibfield  {journal} {\bibinfo  {journal} {Optics
				Express}\ }\textbf {\bibinfo {volume} {25}},\ \bibinfo {pages} {15977}
		(\bibinfo {year} {2017})}\BibitemShut {NoStop}%
	\bibitem [{\citenamefont {liang Zhao}\ \emph {et~al.}(2017)\citenamefont {liang
			Zhao}, \citenamefont {jun Ren}, \citenamefont {Liu}, \citenamefont {peng
			Xin}, \citenamefont {bo~Bai},\ and\ \citenamefont {quan Yao}}]{Zhao2017}%
	\BibitemOpen
	\bibfield  {author} {\bibinfo {author} {\bibfnamefont {H.}~\bibnamefont
			{liang Zhao}}, \bibinfo {author} {\bibfnamefont {G.}~\bibnamefont {jun Ren}},
		\bibinfo {author} {\bibfnamefont {F.}~\bibnamefont {Liu}}, \bibinfo {author}
		{\bibfnamefont {H.}~\bibnamefont {peng Xin}}, \bibinfo {author}
		{\bibfnamefont {Y.}~\bibnamefont {bo~Bai}}, \ and\ \bibinfo {author}
		{\bibfnamefont {J.}~\bibnamefont {quan Yao}},\ }\bibfield  {title} {\enquote
		{\bibinfo {title} {Tunable terahertz source via liquid crystal grating coated
				with electron beam excited graphene: A theoretical analysis},}\ }\href
	{\doibase 10.1016/j.optcom.2016.12.063} {\bibfield  {journal} {\bibinfo
			{journal} {Optics Communications}\ }\textbf {\bibinfo {volume} {390}},\
		\bibinfo {pages} {137} (\bibinfo {year} {2017})}\BibitemShut {NoStop}%
	\bibitem [{\citenamefont {Gu}\ \emph {et~al.}(2013)\citenamefont {Gu},
		\citenamefont {Lin},\ and\ \citenamefont {Liu}}]{Gu2013}%
	\BibitemOpen
	\bibfield  {author} {\bibinfo {author} {\bibfnamefont {X.}~\bibnamefont
			{Gu}}, \bibinfo {author} {\bibfnamefont {I.-T.}\ \bibnamefont {Lin}}, \ and\
		\bibinfo {author} {\bibfnamefont {J.-M.}\ \bibnamefont {Liu}},\ }\bibfield
	{title} {\enquote {\bibinfo {title} {Extremely confined terahertz surface
				plasmon-polaritons in graphene-metal structures},}\ }\href {\doibase
		10.1063/1.4818660} {\bibfield  {journal} {\bibinfo  {journal} {Applied
				Physics Letters}\ }\textbf {\bibinfo {volume} {103}},\ \bibinfo {pages}
		{071103} (\bibinfo {year} {2013})}\BibitemShut {NoStop}%
	\bibitem [{\citenamefont {Zhao}\ \emph {et~al.}(2015)\citenamefont {Zhao},
		\citenamefont {Guo}, \citenamefont {Xia},\ and\ \citenamefont
		{Wang}}]{Zhao2015}%
	\BibitemOpen
	\bibfield  {author} {\bibinfo {author} {\bibfnamefont {H.}~\bibnamefont
			{Zhao}}, \bibinfo {author} {\bibfnamefont {Q.}~\bibnamefont {Guo}}, \bibinfo
		{author} {\bibfnamefont {F.}~\bibnamefont {Xia}}, \ and\ \bibinfo {author}
		{\bibfnamefont {H.}~\bibnamefont {Wang}},\ }\bibfield  {title} {\enquote
		{\bibinfo {title} {Two-dimensional materials for nanophotonics
				application},}\ }\href {\doibase 10.1515/nanoph-2014-0022} {\bibfield
		{journal} {\bibinfo  {journal} {Nanophotonics}\ }\textbf {\bibinfo {volume}
			{4}} (\bibinfo {year} {2015}),\ 10.1515/nanoph-2014-0022}\BibitemShut
	{NoStop}%
	\bibitem [{\citenamefont {Conde}\ \emph {et~al.}(2014)\citenamefont {Conde},
		\citenamefont {Bao}, \citenamefont {Cui}, \citenamefont {Baptista},\ and\
		\citenamefont {Tian}}]{Conde2014}%
	\BibitemOpen
	\bibfield  {author} {\bibinfo {author} {\bibfnamefont {J.}~\bibnamefont
			{Conde}}, \bibinfo {author} {\bibfnamefont {C.}~\bibnamefont {Bao}}, \bibinfo
		{author} {\bibfnamefont {D.}~\bibnamefont {Cui}}, \bibinfo {author}
		{\bibfnamefont {P.~V.}\ \bibnamefont {Baptista}}, \ and\ \bibinfo {author}
		{\bibfnamefont {F.}~\bibnamefont {Tian}},\ }\bibfield  {title} {\enquote
		{\bibinfo {title} {Antibody{\textendash}drug gold nanoantennas with raman
				spectroscopic fingerprints for in vivo tumour theranostics},}\ }\href
	{\doibase 10.1016/j.jconrel.2014.03.045} {\bibfield  {journal} {\bibinfo
			{journal} {Journal of Controlled Release}\ }\textbf {\bibinfo {volume}
			{183}},\ \bibinfo {pages} {87} (\bibinfo {year} {2014})}\BibitemShut
	{NoStop}%
	\bibitem [{\citenamefont {Huang}\ and\ \citenamefont
		{El-Sayed}(2010)}]{Huang2010}%
	\BibitemOpen
	\bibfield  {author} {\bibinfo {author} {\bibfnamefont {X.}~\bibnamefont
			{Huang}}\ and\ \bibinfo {author} {\bibfnamefont {M.~A.}\ \bibnamefont
			{El-Sayed}},\ }\bibfield  {title} {\enquote {\bibinfo {title} {Gold
				nanoparticles: Optical properties and implementations in cancer diagnosis and
				photothermal therapy},}\ }\href {\doibase 10.1016/j.jare.2010.02.002}
	{\bibfield  {journal} {\bibinfo  {journal} {Journal of Advanced Research}\
		}\textbf {\bibinfo {volume} {1}},\ \bibinfo {pages} {13} (\bibinfo {year}
		{2010})}\BibitemShut {NoStop}%
	\bibitem [{\citenamefont {Abbas}\ \emph {et~al.}(2011)\citenamefont {Abbas},
		\citenamefont {Linman},\ and\ \citenamefont {Cheng}}]{Abbas2011}%
	\BibitemOpen
	\bibfield  {author} {\bibinfo {author} {\bibfnamefont {A.}~\bibnamefont
			{Abbas}}, \bibinfo {author} {\bibfnamefont {M.~J.}\ \bibnamefont {Linman}}, \
		and\ \bibinfo {author} {\bibfnamefont {Q.}~\bibnamefont {Cheng}},\ }\bibfield
	{title} {\enquote {\bibinfo {title} {New trends in instrumental design for
				surface plasmon resonance-based biosensors},}\ }\href {\doibase
		10.1016/j.bios.2010.09.030} {\bibfield  {journal} {\bibinfo  {journal}
			{Biosensors and Bioelectronics}\ }\textbf {\bibinfo {volume} {26}},\ \bibinfo
		{pages} {1815} (\bibinfo {year} {2011})}\BibitemShut {NoStop}%
	\bibitem [{\citenamefont {Yanase}\ \emph {et~al.}(2014)\citenamefont {Yanase},
		\citenamefont {Hiragun}, \citenamefont {Ishii}, \citenamefont {Kawaguchi},
		\citenamefont {Yanase}, \citenamefont {Kawai}, \citenamefont {Sakamoto},\
		and\ \citenamefont {Hide}}]{Yanase2014}%
	\BibitemOpen
	\bibfield  {author} {\bibinfo {author} {\bibfnamefont {Y.}~\bibnamefont
			{Yanase}}, \bibinfo {author} {\bibfnamefont {T.}~\bibnamefont {Hiragun}},
		\bibinfo {author} {\bibfnamefont {K.}~\bibnamefont {Ishii}}, \bibinfo
		{author} {\bibfnamefont {T.}~\bibnamefont {Kawaguchi}}, \bibinfo {author}
		{\bibfnamefont {T.}~\bibnamefont {Yanase}}, \bibinfo {author} {\bibfnamefont
			{M.}~\bibnamefont {Kawai}}, \bibinfo {author} {\bibfnamefont
			{K.}~\bibnamefont {Sakamoto}}, \ and\ \bibinfo {author} {\bibfnamefont
			{M.}~\bibnamefont {Hide}},\ }\bibfield  {title} {\enquote {\bibinfo {title}
			{Surface plasmon resonance for cell-based clinical diagnosis},}\ }\href
	{\doibase 10.3390/s140304948} {\bibfield  {journal} {\bibinfo  {journal}
			{Sensors}\ }\textbf {\bibinfo {volume} {14}},\ \bibinfo {pages} {4948}
		(\bibinfo {year} {2014})}\BibitemShut {NoStop}%
	\bibitem [{\citenamefont {Helmerhorst}\ \emph {et~al.}(2012)\citenamefont
		{Helmerhorst}, \citenamefont {Chandler}, \citenamefont {Nussio},\ and\
		\citenamefont {Mamotte}}]{Helmerhorst2012}%
	\BibitemOpen
	\bibfield  {author} {\bibinfo {author} {\bibfnamefont {E.}~\bibnamefont
			{Helmerhorst}}, \bibinfo {author} {\bibfnamefont {D.~J.}\ \bibnamefont
			{Chandler}}, \bibinfo {author} {\bibfnamefont {M.}~\bibnamefont {Nussio}}, \
		and\ \bibinfo {author} {\bibfnamefont {C.~D.}\ \bibnamefont {Mamotte}},\
	}\bibfield  {title} {\enquote {\bibinfo {title} {Real-time and label-free
				bio-sensing of molecular interactions by surface plasmon resonance: A
				laboratory medicine perspective},}\ }\href
	{https://www.ncbi.nlm.nih.gov/pmc/PMC3529553/} {\bibfield  {journal}
		{\bibinfo  {journal} {The Clinical biochemist. Reviews}\ }\textbf {\bibinfo
			{volume} {33}},\ \bibinfo {pages} {161} (\bibinfo {year} {2012})}\BibitemShut
	{NoStop}%
	\bibitem [{\citenamefont {Dodson}\ \emph {et~al.}(2015)\citenamefont {Dodson},
		\citenamefont {Cao}, \citenamefont {Zaribafzadeh}, \citenamefont {Li},\ and\
		\citenamefont {Xiong}}]{Dodson2015}%
	\BibitemOpen
	\bibfield  {author} {\bibinfo {author} {\bibfnamefont {S.~L.}\ \bibnamefont
			{Dodson}}, \bibinfo {author} {\bibfnamefont {C.}~\bibnamefont {Cao}},
		\bibinfo {author} {\bibfnamefont {H.}~\bibnamefont {Zaribafzadeh}}, \bibinfo
		{author} {\bibfnamefont {S.}~\bibnamefont {Li}}, \ and\ \bibinfo {author}
		{\bibfnamefont {Q.}~\bibnamefont {Xiong}},\ }\bibfield  {title} {\enquote
		{\bibinfo {title} {Engineering plasmonic nanorod arrays for colon cancer
				marker detection},}\ }\href {\doibase 10.1016/j.bios.2014.07.083} {\bibfield
		{journal} {\bibinfo  {journal} {Biosensors and Bioelectronics}\ }\textbf
		{\bibinfo {volume} {63}},\ \bibinfo {pages} {472} (\bibinfo {year}
		{2015})}\BibitemShut {NoStop}%
\end{thebibliography}
%

\end{document}